%% file: thermalwkb-arXiv-v1.tex
\documentclass[final,11pt]{article}
\pdfoutput=1
\usepackage{MRnotesBB}

\input{spectral-macros}

\title{
  Exact holographic thermal spectral functions: OPE, non-perturbative corrections, and black hole singularity
}
\author{Hewei Frederic Jia${}^{a}$,  Mukund Rangamani$^{b}$}

\affiliation[a]{Institute for Advanced Study, Tsinghua University, Beijing, 100084, China}
\affiliation[b]{Center for Quantum Mathematics and Physics (QMAP)\\
  Department of Physics \& Astronomy, University of California, Davis, CA 95616 USA}

\emailAdd{heweifred@gmail.com}
\emailAdd{mukund@physics.ucdavis.edu}

\abstract{  
 We study analytic properties of thermal spectral functions of holographic CFTs, examining both their (a) exact properties at finite momentum and (b) asymptotics at large momentum. For even-dimensional holographic CFTs on Minkowski spacetime and for scalar primaries with integer dimensions, we demonstrate that the exact spectral function at finite momentum factorizes into a perturbative/OPE piece and a non-perturbative piece. The former is controlled by stress tensor exchange and fixed by a near-boundary analysis. The latter encodes information about the bulk interior, including the black hole horizon and singularity. Utilizing the exact factorization, we obtain the full transseries expansion of the non-perturbative piece at large timelike momentum. This is achieved by employing exact WKB techniques to compute the monodromy of the bulk wave equation. Finally, we use these results to work out the singular loci of a spatially averaged thermofield double correlator in the complex time plane. These singular loci have been argued to provide imprints of the black hole curvature singularity in the dual CFT observables. Our result, which includes the case of non-vanishing momentum, gives a clear link between the non-perturbative spectral function and the black hole singularity.
}

\newpage

\begin{document}
\maketitle


\section{Introduction and summary of results}\label{sec:intro}

The dynamics of quantum systems at finite temperature captures a plethora of interesting physics ranging from equilibration, transport phenomena, and in the holographic context, the physics of black holes.
A natural set of observables of such quantum systems are thermal correlators. In quantum field theories, these are fascinating from both theoretical and phenomenological perspectives. They are somewhat more constrained than their vacuum counterparts since thermal correlators in equilibrium density states are supposed to respect the Kubo-Martin-Schwinger (KMS) condition. In the Euclidean signature, the KMS condition imposes a periodicity around the thermal circle, which can be made manifest from the thermofield double construction. When one analytically continues to physical Lorentz signature, the KMS condition can be rephrased as a statement of analyticity of the correlator within the thermal strip 
$\Im(t) \in (-\frac{\beta}{2}, \frac{\beta}{2})$.

Our interest here is to understand what happens beyond this strip for the simplest class of thermal observables, two-point functions of conformal primaries of a conformal field theory. Through the holographic AdS/CFT correspondence, one anticipates that the analytic structure has interesting information regarding the dual spacetime geometry. The equilibrium thermal density matrix of the CFT maps to the eternal black hole spacetime prepared in its Hartle-Hawking state. We therefore anticipate that the thermal observables can be leveraged to extract information regarding the semiclassical geometry of the black hole.

Motivated by these considerations, we undertake a study of analytic properties of thermal two-point functions in holographic \CFT{d} with a classical gravity dual. This restricts our attention to the class of CFTs which admit a large central charge limit, and additionally have a sparse low-lying spectrum of states in this limit. The former is required for the semiclassical geometry to be protected from quantum gravitational (Planckian) corrections, while the latter says that stringy effects are negligible. For simplicity of exposition, we will focus on the case of the two-point function of  identical scalar primaries $\expval{\phi\phi}_{\beta}$ with \CFT{d} on spatial $\R^{d-1}$.
The main observable of our focus will be the associated thermal spectral function $\rho(\omega,k)$ in momentum space.

\paragraph{\emph{Exact} holographic thermal correlators:} Since the inception of the AdS/CFT correspondence, holographic thermal correlators have been extensively studied by many authors for a wide variety of reasons. Originally,~\cite{Horowitz:1999jd} argued that the quasinormal spectrum of black holes captures information about thermalization rate. These insights were further bolstered by the understanding of how conserved current correlators can be used to extract transport data from the quasinormal spectrum~\cite{Policastro:2001yc}. Development of real-time techniques for holographic correlators made aspects of their analytic structure clear in momentum space~\cite{Son:2002sd}. Furthermore, analytic approaches to capturing the asymptotic behavior of quasinormal modes were developed in~\cite{Natario:2004jd,Festuccia:2008zx,Amado:2008hw}.

The holographic dictionary for thermal correlators involves solving a classical wave equation for the field dual to the CFT operator in the black hole geometry. While in Euclidean signature one can use the standard extrapolate dictionary, the Lorentzian variant is best viewed as solving a connection problem between the AdS boundary and black hole horizon where suitable  boundary conditions are imposed~\cite{Son:2002sd}.\footnote{From a modern perspective, this prescription can be carefully derived by improving the original discussion of~\cite{Herzog:2002pc} using the gravitational Schwinger-Keldysh contour~\cite{Glorioso:2018mmw} (see also~\cite{Jana:2020vyx}). }

For the most part, exact analytic treatment of the wave equation remained elusive; the wave equations of interest on AdS black hole backgrounds do not admit standard closed-form solution. However, the situation has dramatically improved in recent years ---  a plethora of new developments have led to a renewed analytic understanding of holographic thermal correlators. Let us highlight two salient aspects, which will play a role in our analysis:
\begin{itemize}[wide, left=0pt]
  \item \emph{The thermal OPE:} Similar to vacuum correlators, thermal correlators admit operator product expansion (OPE), with thermal OPE data constrained by KMS condition~\cite{El-Showk:2011yvt,Katz:2014rla,Iliesiu:2018fao}. In general, the OPE of holographic thermal two-point function is expected to split into two contributions: one comes from the stress-tensor sector and the other from the double-twist sector. It turns out that the stress-tensor OPE coefficients can be extracted by a near-boundary expansion without solving the connection problem~\cite{Fitzpatrick:2019zqz}. The double-twist OPE coefficients, on the other hand, require imposing a boundary condition at the horizon to be determined.  The near-boundary expansion was systematically analyzed in~\cite{Buric:2025anb,Buric:2025fye}.  In addition,  new exact expressions for position space two-point function were recently obtained in~\cite{Barrat:2025twb} by combining stress-tensor OPE data with dispersion relations.

  \item \emph{An exact solution of momentum space connection problem:} As noted above, the wave equations in black hole backgrounds are not easily amenable to analytic solutions. However, the correlator only requires the connection data, not the actual wavefunctions themselves. The connection problem in question involves a Fuchsian differential operator with four or more singular points.\footnote{A second order linear differential equation in one variable is Fuchsian if its singular points are all regular singular points.}
  Remarkably, a closed-form solution was recently obtained in terms of the semiclassical Virasoro block using fusion transformations of degenerate Virasoro block and taking the  large-$c$ limit~\cite{Aminov:2020yma,Bonelli:2022ten,Lisovyy:2022flm}.\footnote{Alternately, one can phrase this in terms of the Nekrasov partition function (in the Nekrasov-Shatashvili limit) of a  4d $\mathcal{N}=2$ theory on $\Omega$-background using the AGT relation cf.~\cite{Aminov:2020yma,Dodelson:2022yvn}.}  The result can be immediately applied to yield exact expressions for momentum space two-point function~\cite{Dodelson:2022yvn,Jia:2024zes}
\end{itemize}

A natural question in this context is, whether there is any relation between the two approaches. For the thermal spectral function, the stress-tensor OPE coefficients determine $\rho^{(\OPE)}_{\pert}(\omega,k)$, the vacuum spectral function and its perturbative corrections at large momentum. Double-twist exchanges do not contribute to the momentum space OPE. On the other hand, the exact expression $\rho_{\mathrm{exact}}(\omega,k)$ obtained using Virasoro block method is valid at finite momentum. In principle, it captures both perturbative and non-perturbative corrections in the large momentum limit. However, a technicality in the exact expression (related to the absence of a closed form expression for the Virasoro block) obstructs straightforward computation of its large momentum limit to recover $\rho^{(\OPE)}_{\pert}(\omega,k)$ and extract the remaining non-perturbative contribution $\rho_{\np}(\omega,k)$.

Our goal here is to improve the situation and answer the aforementioned question. Specifically, working in even spacetime dimensions, $d \in 2\,\mathbb{Z}_{>1}$ and with scalar primaries of integer dimension $\Delta_{\phi} = \frac{d}{2} + \mathbb{Z}_{\geq 0}$, we uncover a precise relation. In such cases, holographic spectral function enjoys special properties. The perturbative corrections from OPE truncate~\cite{Teaney:2006nc,Kovtun:2006pf,Caron-Huot:2009ypo}. Moreover, the exact expression takes a simpler factorized form. We claim that the exact spectral function obtained from the Virasoro block method admits a perturbative/non-perturbative factorization (\cref{conj:rho-factorized}):
\begin{equation}
\label{eq:factorization-intro}
\rho_{\mathrm{exact}}(\omega,k) = \rho^{(\OPE)}_{\pert}(\omega,k) \rho_{\np}(\omega,k),
\end{equation}
with the non-perturbative part given by
\begin{equation}
\rho_{\np}(\omega,k) = \frac{\sinh\prn{\frac{\beta\omega}{2}}}{\cosh\prn{\frac{\beta\omega}{2}} + (-1)^{\Delta_{\phi} - \Half{d}} \cos(2\pi\cmexp)} \,.
\end{equation}
The unknown $\sigma$ here controls the monodromy around AdS boundary and horizon of bulk wave equation, in fact,
$\Tr M_{\hor,\bdy} = -2\cos(2\pi\cmexp)$. We provide strong evidence for the claim by both i) an analytic argument using apparent singularity for general even $d$, and ii) direct numerical computations using semiclassical Virasoro blocks for $d=4$. We expect the exact perturbative/non-perturbative factorization to hold more generally beyond the case of planar black holes. This result was first advertised in our companion paper~\cite{Jia:2025jbi} and has also been discussed subsequently in~\cite{Giombi:2026kdz}.

\paragraph{Exact WKB analysis and non-perturbative corrections:} Since the pioneering works \cite{Voros1983ReturnQuarticOscillator,Delabaere:1997srq}, it is now understood that WKB analysis is not merely an approximation method, but can be promoted to an \emph{exact} method by considering Borel-sum of formal WKB solutions. In particular, it allows precise controls over non-perturbative effects. Exact WKB can also be used to analyze global properties of Schr\"odinger-type equations, including their monodromy data~\cite{KawaiTakei2005}. In general, it represents the monodromy data in terms of Borel-summed all-orders WKB periods, also known as Voros periods.
Exact WKB quantization has thus found applications in many physical contexts where Schr\"odinger-type equations appear, from quantum mechanics to supersymmetric quantum field theories~\cite{Gaiotto:2009hg, Grassi:2021wpw}.
It also has interesting connections with cluster algebra~\cite{Iwaki:2014vad} and wall-crossing~\cite{Gaiotto:2009hg}. Recently this method has also been applied to black hole problems in~\cite{Imaizumi:2022qbi,Miyachi:2025ptm}. For a nice overview of the developments, we refer the reader to the recent review~\cite{Iwaki:2025tmq} (see also~\cite{Dorigoni:2014hea} for a nice overview).

In light of the factorization~\eqref{eq:factorization-intro}, we perform an exact WKB analysis of the monodromy data in $\rho_{\np}(\omega,k)$ for $d=4$. We focus on the regime of large timelike momentum with non-vanishing spatial momentum. This turns out to be the generic case. We determine the monodromy in terms of Borel-summed all-orders WKB periods (cf.~\cref{tab:monodromy-phases}) for generic phases $\pha = \arg(\omega) = \arg(k)$. This generalizes previous discussions on large $\omega$ asymptotics with $k=0$ in~\cite{Natario:2004jd,Festuccia2007BlackHoleSingularities}, and large $\Delta_{\phi}$ or large $\omega = k$ asymptotics in~\cite{Festuccia:2008zx}. 

The case of large real momentum poses an additional subtlety: it corresponds to a critical phase where there exist Stokes lines connecting two turning points. In general, the all-orders WKB periods are not Borel-summable at critical phases. Across such phase boundaries, the Borel sums of the WKB periods exhibit a Stokes automorphism~\cite{Delabaere:1997srq,Iwaki:2014vad,Iwaki_2015}. For physical applications at  the critical phase, it is convenient to work with the median Borel-summed WKB periods defined using half Stokes automorphism~\cite{Delabaere:1997srq}.\footnote{A well-known example where critical phase appears is double-well potential in quantum mechanics, with $\arg(\hbar) = 0$ a critical phase. Using median Borel-summed WKB periods gives an exact quantization condition with a manifestly real spectrum~\cite{Delabaere:1997srq}. See also~\cite{Kamata:2021jrs,Misumi:2024gtf} and references therein for more recent discussions on the use of median Borel-summation in quantum mechanics applications.} In our context, this ensures a manifestly real result for $\rho_{\np}(\omega,k)$ (cf.~\eqref{eq:monodromy-real-phase}).

The exact WKB analysis of monodromy allows us to determine the full transseries expansion of $\rho_{\np}(\omega,k)$ (cf.~\eqref{eq:rhonp-transseries})
\begin{equation}
\label{eq:rhonp-transseries-intro}
\rho_{\np}(\omega,\mr\, \omega) \sim
1 + \sum^{\infty}_{r=1}\sum^{\infty}_{s\overset{2}{=}-r} e^{(-r\pi-s\vcl)\frac{\beta\omega}{2\pi}}\,  \sum^{r}_{q\overset{2}{=}r-2\floor{\Half{r}}} \,
\Re e^{i q(\pi-\vcl) \frac{\beta\omega}{2\pi}} P_{rsq}(\omega) \,,
\quad  \omega \to \infty \,,
\end{equation}
in the large timelike momentum regime with $0<\mr<1$. The parameter $\vcl$ is related to classical WKB period and is determined explicitly as a function of $\mr$ in~\eqref{eq:classical-A-period}. The parameters are summed in steps of two, cf.~\cref{fn:2step} for our convention.

The case of vanishing spatial momentum ($\mr=0$) is a degenerate limit where certain genericity assumptions of exact WKB analysis (e.g., as stated in~\cite{Iwaki_2015}) are not satisfied. In this case we perform an asymptotic analysis of monodromy similar to the one in the context of asymptotic quasinormal modes~\cite{Natario:2004jd,Musiri:2005ev}. The resulting $\rho_{\np}(\omega,k)$ asymptotics agree with the transseries expression recently obtained in~\cite{Afkhami-Jeddi:2025wra}.

\paragraph{Imprints of black hole singularity:} In addition to understanding the structure of the holographic thermal spectral function, part of our motivation for undertaking the exercise was to revisit the question of understanding the black hole geometry from the field theory perspective. One broad goal of the bulk reconstruction programme is to decipher the encoding of geometric structures in the dual field theory observables. In the semiclassical limit, one can, of course, trust the bulk geometry, and therefore may use the AdS/CFT dictionary to find how a specific feature of the geometry is captured by the field theory observables. Over the years, we have accumulated evidence of how field theory correlation functions, subregion entanglement entropy, etc. capture geometric data, say the local form of the bulk metric, presence of horizons, curvature singularities, etc. Among these, the most fascinating is, of course, the curvature singularity in the black hole interior. Not only is this data in the causally inaccessible domain of the spacetime, but understanding its encoding also has a major pay-off in the context of the holographic duality. Since the duality purports to allow a non-perturbative definition of the quantum gravitational theory in terms of a conventional quantum field theory, one might hope to be able to shed light on how the singularity gets resolved.

Therefore, the question of extracting imprints of black hole singularity from dual CFT thermal correlator is an intriguing one, and has been explored by many authors over the years. It was realized more than two decades ago in a seminal work~\cite{Fidkowski:2003nf}, that a convenient probe in this context is the two-sided thermofield double correlator
$\ts(t,\vb{x}) = \expval{\phi(0)\,\phi(t+i\beta/2,\vb{x})}_{\beta}$, whose momentum space representation is related to spectral function by $\ts(\omega,k) = \rho(\omega,k)/\sinh(\beta\omega/2)$.  These authors studied $C(t,0)$ at large $\Delta_{\phi}$ using the geodesic approximation. They found that the geodesics in question, which are spacelike geodesics connecting the two boundaries of an eternal black hole spacetime, are repelled by the black hole singularity. Moreover, such geodesics fail to exist beyond a critical time $t_c$. In the limit $t \to t_c = \Half{\beta} \cot(\pi/d)$, the spacelike geodesics are bounded by a pair of null geodesics, which crash into the singularity. However, the manner they do so is highly suggestive of this null geodesic bouncing off the singularity, leading to a simple mnemonic
\emph{bouncing null geodesics}.  This null limit of spacelike geodesics connecting two asymptotic regions and bouncing off the black hole singularity, leads to  a prediction that the two-sided correlator diverges on a secondary sheet at $t_c$. The singular behavior at $t=t_{c}$ is referred to as \emph{bouncing singularity}. The reason for the secondary sheet is that the bouncing geodesic in question is not on the steepest descent contour, which turns out to instead be governed by the Euclidean geodesic.  

Since then, the notion of bouncing singularity has been extended beyond the large $\Delta_{\phi}$ regime originally studied in~\cite{Fidkowski:2003nf}. In~\cite{Ceplak:2024bja}, it is found that at finite $\Delta_{\phi}$, the stress-tensor sector of $C(t,0)$ also exhibits singular behavior at $t=t_{c}$ from the large-order behavior of stress-tensor OPE coefficients. This unphysical singularity inside the fundamental strip $\Im(t) \in (-\frac{\beta}{2}, \frac{\beta}{2})$ is expected to be canceled by contribution from the double-twist sector.

Rather than examining the real space observable, one could also ask what the imprint  of black hole singularity is directly in the momentum space correlator. This, as noted above, was examined in~\cite{Festuccia:2005pi}, who noted that the information of the black hole singularity can be traced to the asymptotic behavior of the  quasinormal modes.
In fact, the connection between the quasinormal mode asymptotics, and its link to the spacetime geometry in the vicinity of the black hole singularity was first noticed in a WKB analysis of~\cite{Motl:2003cd}. They examined the real part of highly damped quasinormal modes for asymptotically flat black holes. Their discussion was generalized to AdS black holes  in~\cite{Natario:2004jd}, which too  manifests a relation with black hole singularity. The asymptotic quasinormal mode analysis can be adapted to yield non-perturbative corrections to $\rho(\omega,0)$ at large $\omega$~\cite{Festuccia2007BlackHoleSingularities}. The connection with complex time singularities is examined in the recent works~\cite{Afkhami-Jeddi:2025wra,Dodelson:2025jff}, which study the spatially non-local correlator $C(t,k=0)$ (it is smeared over all space). The non-local correlator exhibits complex time singularities outside the fundamental strip, with their real parts related to $t_{c}$. The location of these complex time singularities is controlled by non-perturbative corrections to $\rho(\omega,0)$ at large $\omega$. A careful WKB analysis of the correlator led~\cite{Afkhami-Jeddi:2025wra} to argue for a connection with the bouncing geodesic. Relatedly, using the thermal product formula~\cite{Dodelson:2023vrw}, the authors of~\cite{Dodelson:2025jff} argued that the non-perturbative corrections are controlled by the same data entering in asymptotic quasinormal mode spectrum. A priori  these complex time singularities are distinct from the bouncing singularities in the sense of~\cite{Fidkowski:2003nf}; however, their predictions for the location of the singularities coincide. In our companion paper~\cite{Jia:2025jbi} we extended these observations to correlators with non-vanishing momenta, as we elucidate below. Subsequently, several authors have extended these investigations for a wide class of examples. For instance,~\cite{Ceplak:2025dds,Dodelson:2025jff,AliAhmad:2026wem} examine the behavior of bouncing (timelike) geodesics off timelike singularities inside charged black holes. Furthermore,~\cite{Giombi:2026kdz} analyzed the encoding of spacelike singularities from the perspective of Wilson loop correlators. The recent paper~\cite{Arnaudo:2026der} explores the analytic structure of the correlator, while ~\cite{Grozdanov:2026cut} examines the connection to the bouncing geodesic.\footnote{In a related vein~\cite{Jia:2026pmv,Araya:2026shz} study the connection to bulk-cone singularities, which were originally motivated as diagnostics of horizon formation and gravitational time delay in~\cite{Hubeny:2006yu}.}

We use the non-perturbative corrections of $\rho(\omega,\mr \omega)$ at large $\omega$ to study the complex time singularities of  the non-local correlator\footnote{
  The observable we analyze, $\ts_\mr(t)$, is spatially averaged, which may be easily seen by splitting $\vb{x} $ into radial and angular components; see~\eqref{eq:Ctspatialav}.}
\begin{equation}
\ts_{\mr}(t) = \int^{\infty}_{0} d\omega \cos(\omega t) \,\ts(\omega,\mr\omega).
\end{equation}
We demonstrate that each non-perturbative sector in~\eqref{eq:rhonp-transseries-intro} leads to a complex time singularity at $t=\pm t_{rsq}$ for $\ts_\mr(t)$ with
\begin{equation}
\begin{aligned}
 & t_{rsq} = \frac{i \beta}{2} + \prn{r + s\frac{v}{\pi}} \frac{i\beta}{2} + q \frac{\beta}{2}\prn{1-\frac{v}{\pi}}, \\
 & r \in \Z_{\geq1}, \quad s\in -r + 2\Z_{\geq 0}, \quad q \in \{-r,-r+2,\dots, r\}.
\end{aligned}
\end{equation}
This expression, which we obtained in~\cite{Jia:2025jbi} generalizes the complex time singularities deduced in~\cite{Dodelson:2025jff} to non-zero $\mr$. The black hole singularity, being one of the branch points of WKB curve, is related to the $\vcl$ parameter~\eqref{eq:classical-A-period} appearing in the classical WKB period.\footnote{This period also appeared in the thermal product analysis~\cite{Dodelson:2023nnr}; we thank Matthew Dodelson for a discussion on this point.}

\paragraph{Organization of the paper:} Our goal in this paper is to elaborate on the salient points advertised in our companion paper~\cite{Jia:2025jbi}. In the process we will refine some statements, provide additional supporting evidence, and explain how we deduced the transseries expression for the thermal spectral function. We begin the discussion in~\cref{sec:thermal-OPE} with a review of the thermal OPE for the spectral function and discuss its properties in holographic \CFT{d}. Subsequently, in~\cref{sec:specholo}, we examine the exact expressions of holographic spectral functions obtained from the Virasoro block method. We provide strong evidence for the exact factorization in~\cref{conj:rho-factorized} by two independent arguments. First, we present  an analytic argument which exploits the presence of an apparent singularity at the spacetime boundary for general even $d$ and special operator dimensions. Second, we provide direct numerical evidence for our claim by using the semiclassical Virasoro blocks for $d=4$. Taken together these give us strong support of our claim. Our analysis also suggests some interesting features hidden in the semiclassical Virasoro block that would be worth investigating further.

These observations taken together, motivate our study of the non-perturbative contributions to the thermal spectral functions. As preparatory material,  we provide a self-contained review of the exact WKB method in~\cref{sec:EWKBreview}. We then use these techniques in~\cref{sec:EWKB-timelike-momentum} to  perform an exact WKB analysis for the monodromy data in $\rho_{\np}$ for large timelike momentum with non-vanishing spatial momentum. Our primary results are summarized in~\cref{tab:monodromy-phases} and~\eqref{eq:monodromy-real-phase}.

We can then exploit the results found in~\cref{sec:EWKB-timelike-momentum} to obtain full transseries expansion of $\rho_{\np}$, and study the complex time singularities associated with non-perturbative corrections. This is undertaken in~\cref{sec:complex-time-singularity}, and completes the derivation of the results released in~\cite{Jia:2025jbi}. One element that is missing from our general WKB analysis is the case of vanishing spatial momentum. This turns out to be a singular case from our perspective. For completeness, in~\cref{sec:WKB-zero-momentum}, we perform an independent asymptotic WKB analysis for the monodromy data in $\rho_{\np}$ for vanishing spatial momentum. Using this analysis we recover known results on non-perturbative corrections in spectral function and asymptotic quasinormal modes. What is worth highlighting here is that while the leading WKB asymptotics are smooth between non-vanishing and vanishing spatial momentum, the perturbative corrections to the saddle have very different transseries expansions: the general case of $\mr \neq 0$ involves perturbative corrections in integer powers of inverse frequency, while the degenerate case has fractional powers appearing.

We conclude with a discussion of open problems in~\cref{sec:discussion}. The four appendices comprise some technical results which we used in the course of our analysis. In~\cref{app:Lorentz-integral} we outline the evaluation of an integral appearing in the Lorentzian OPE calculation.  The details of how we map the scalar wave equation in planar \SAdS{d+1} geometries to a canonical Schr\"odinger form is provided in~\cref{app:Schrodinger-form}. The proof of our factorization claim involves numerical evaluation of Virasoro blocks using the Zamolodchikov recursion, which we briefly review in~\cref{app:block-recursion}. Finally, the details for computing the period integrals on the black hole WKB curve are provided in~\cref{app:WKB}.


\section{Thermal OPE and the spectral function}\label{sec:thermal-OPE}

We will begin with an overview of thermal spectral functions in CFTs.
First we highlight general features that are expected to hold from the thermal
OPE analysis, and then turn to simplifications for holographic CFTs.
Our goal here is to collect salient facts that will inform our analysis in the sequel by helping motivate a universal form for the holographic spectral functions. Thermal OPE in position space is discussed in~\cite{Iliesiu:2018fao}, and in momentum space in~\cite{Manenti:2019wxs,Dodelson:2023vrw}. The relation between the OPE and spectral function asymptotics has been discussed in~\cite{Caron-Huot:2009ypo}.  

\subsection{Generalities: The spectral function block}

To anchor our discussion we start with the Euclidean thermal OPE, which can then be
analytically continued to obtain the real-time spectral function.
Our analysis will be restricted to correlation functions of scalar primary operators for simplicity.

\paragraph{Euclidean thermal OPE:}
Consider the thermal two-point function,
$\gE(\tau,\vb{x}) \equiv \expval{\phi(\tau,\vb{x})\,\phi(0)}_{\beta}$,
of identical scalar primaries $\phi$ of a Euclidean \CFT{d},
on $\vb{S}^{1}_{\beta} \times \R^{d-1}$ (we will often set  $\beta=1$ for convenience below).
Let $x^{\mu} = (\tau,\vb{x})$. For $\abs{x} < \beta$, the two-point function can be
computed using the flat-space OPE
\begin{equation}
\phi(x)\times \phi(0)
=
\sum_{\Op} \frac{f_{\phi\phi\Op}}{c_{\Op}}\,  \abs{x}^{\Delta-2\Delta_{\phi}-J}\,
x_{\mu_{1}} \cdots x_{\mu_{J}} \, \Op^{\mu_{1}\cdots\mu_{J}} + (\text{descendants})\,.
\label{eq:flat-space-OPE}
\end{equation}
Here the sum is over primaries: $\{\Delta,J\}$ are the dimension and spin of the
exchanged operator $\Op$, while $c_{\Op}$ and $f_{\phi\phi\Op}$ are the vacuum two-point and three-point function coefficients, respectively. The thermal one-point function of a spin-$J$ primary takes the form
\begin{equation}
\label{eq:thermal-1pt}
\expval{\Op^{\mu_{1}\cdots\mu_{J}}}_{\beta}
=
\frac{b_{\Op}}{\beta^{\Delta}} \,
e^{\{\mu_1}\, \cdots e^{\mu_J\}} \,.
\end{equation}
Here $b_{\Op}$ is the thermal one-point function coefficient, $e^{\mu} = (1,\vb{0})$ is the unit vector in Euclidean time direction, and $\stt{\cdot}$ denotes the symmetric traceless part of a tensor.
To be explicit, for rank-$2$ tensors one would define
\begin{equation}
a^{\{\mu} \,b^{\nu\}} = \frac{a^\mu\, b^\nu + a^\nu\, b^\mu}{2}  - \frac{1}{d}\, \delta^{\mu\nu}\, a\cdot b \,, \qquad a\cdot b \equiv a^{\alpha}\, b_{\alpha}\,.
\end{equation}

Thermal one-point functions of descendants vanish. When computing the two-point function using OPE, the following identity is needed for index contraction
\begin{equation}
\prn{x^{\mu_{1}} \cdots x^{\mu_{J}}} \,
e_{\{\mu_{1}} \cdots e_{\mu_{J}\}}
=
\Pjnorm_{J} \, \prn{x^2}^{\Half{J}} \, P_{J}\prn{\frac{x \cdot e}{\sqrt{x^2}}} \,.
\label{eq:STT-contraction-identity}
\end{equation}
Here $P_{J}$ is a degree $J$ polynomial
\begin{equation}
P_{J}(z)
=
\begin{dcases}
C^{\prn{\Half{d-2}}}_{J}(z)
 &
\quad d>2\,,
\\
T_{J}(z)
 &
\quad d=2\,,
\end{dcases}
\end{equation}
where $C^{(\lambda)}_{J}(z)$ is the Gegenbauer polynomial, and $T_{J}(z)$ is the Chebyshev polynomial of the first kind. Finally,  the normalization factor is
\begin{equation}
\Pjnorm_{J} =
\begin{dcases}
\frac{J!}{2^{J}\prn{\Half{d-2}}_{J}}
 &
\quad d>2 \,,
\\
\frac{2^{1-J}}{1+\delta_{J,0}}
 &
\quad d=2\,.
\end{dcases}
\end{equation}
The normalization $\Pjnorm_{J}$ is defined such that the highest degree term on RHS of~\eqref{eq:STT-contraction-identity} has unit coefficient. We use the standard notation for the Pochhammer symbol $(\alpha)_m = \alpha\, (\alpha+1) \, \cdots\, (\alpha + m -1)$.

From~\cref{eq:flat-space-OPE,eq:thermal-1pt,eq:STT-contraction-identity}, the thermal two-point function thus admits the following OPE expansion inside the unit disk $\sqrt{\tau^2 + \vb{x}^2} \leq \beta$:
\begin{equation}
\begin{aligned}
\gE(\tau,\vb{x})
 & =
\sum_{\Op \in \phi \times \phi} \, a_{\Op} \, G^{\Delta_{\phi}}_{\Delta,J}(\tau,\vb{x}),
\\
G^{\Delta_{\phi}}_{\Delta,J}(\tau,\vb{x})
 & =
P_{J} \prn{\frac{\tau}{\sqrt{\tau^2 + \vb{x}^2}}} \;
\prn{\tau^2 + \vb{x}^2}^{\frac{\Delta}{2} - \Delta_{\phi}}\,,
\end{aligned}
\label{eq:OPE-position-space}
\end{equation}
with
\begin{equation}
a_{\Op} = \frac{f_{\phi\phi\Op}\, b_{\Op}}{c_{\Op}} \, \Pjnorm_{J} \,.
\end{equation}
The function $G^{\Delta_{\phi}}_{\Delta,J}(\tau,\vb{x})$ is the \emph{thermal block}.
Due to the symmetry under $\tau \to -\tau$\footnote{If the CFT is not parity-invariant, sign flip of both $\tau$ and one of the spatial directions is a symmetry~\cite{Iliesiu:2018fao}. We restrict attention to parity invariant theories for simplicity.}, only operators with even spin appear in the OPE. This is consistent with $P_{J}(z)$ only containing even-degree terms for even $J$.

\paragraph{Lorentzian correlators:}
We will actually be interested in the Lorentzian two-point function,
in particular, in the commutator
\begin{equation}
g(t,\vb{x})
= \expval{\brk{\phi(x),\phi(0)}}_{\beta}
= \Disc\brk{\gE}\prn{t,\vb{x}}\,,
\end{equation}
where the discontinuity is defined as $\Disc\brk{f}(t) \coloneqq f(it +\epsilon) - f(it-\epsilon)$ for $f(\tau)$. The discontinuity of the thermal block in the OPE~\eqref{eq:OPE-position-space} is given by
\begin{equation}
\Disc\brk{G^{\Delta_{\phi}}_{\Delta,J}}(t,\vb{x})
= 2i \,\sin\prn{\pi \widetilde{\Delta}} \, \sgn(t) \,
\Theta(t^2-\vb{x}^2) \,
P_{J}\prn{\frac{t}{\sqrt{t^2-\vb{x}^2}}} \; (t^2-\vb{x}^2)^{\widetilde{\Delta}}
\end{equation}
with
\begin{equation}
\widetilde{\Delta} = \Half{\Delta} - \Delta_{\phi}.
\end{equation}
The resulting expansion of the commutator
\begin{equation}
\label{eq:commutator-ope}
g(t,\vb{x})
=
\sum_{\Op \in \phi \times \phi}\,
a_{\Op} \, \Disc\brk{G^{\Delta_{\phi}}_{\Delta,J}}(t,\vb{x})\,,
\end{equation}
is expected to be convergent for sufficiently small $t^2-\vb{x}^2$. The Heaviside step function $\Theta$ ensures that the commutator only has support for timelike separated points.

\paragraph{Momentum space OPE for spectral function:}
The spectral function can be defined as the Fourier transform of the commutator into momentum
space (NB: $d\vb{x} = d^{d-1}x$ for brevity)
\begin{equation}
\rho(\omega,\vb{k})
= \int d^dx\,  e^{i\omega t -i \,\vb{k}\cdot\vb{x}} \, g(t,\vb{x}) \,.
\end{equation}
The integration is over the entire Minkowski spacetime, with the convention
\begin{equation}
\int d^dx = \int_{\R^{d-1,1}} \, d\vb{x}\, dt\,.
\label{eq:Minkint}
\end{equation}
When both $\omega,k$ are large the Fourier transform is dominated by small $t,\vb{x}$. It is then  justified to use~\eqref{eq:commutator-ope} to compute the Fourier transform. Moreover, as explained in~\cite{Afkhami-Jeddi:2025wra}, since the commutator only has support inside the lightcone, taking only $\omega$ large at finite $k$ is also expected to be sufficient. Therefore, in the large frequency (and spatial momentum) limit, the spectral function admits a momentum space OPE expansion
\begin{equation}
\rho(\omega,k)
=
\sum_{\Op \in \phi \times \phi} \, a_{\Op} \,
\Gspec^{\Delta_{\phi}}_{\Delta,J}(\omega,k) + \order{e^{-\omega}}\,,
\qquad \omega \to \infty\,.
\label{eq:spectral-function-ope}
\end{equation}
This defines the associated \emph{spectral function block}, which is given by
\begin{equation}
\Gspec^{\Delta_{\phi}}_{\Delta,J}(\omega,k)
=
\int\, d^dx\,
e^{i\omega t -i \vb{k}\cdot\vb{x}}\,  \Disc\brk{G^{\Delta_{\phi}}_{\Delta,J}}(t,\vb{x}).
\end{equation}

The identity exchange in the OPE gives the leading vacuum spectral function
\begin{equation}
\Gspec^{\Delta_{\phi}}_{\mathbbm{1}}(\omega,k)
\propto
\sgn(\omega) \, \Theta\prn{\omega^2-k^2} \,
\prn{\omega^2 - k^2}^{\Delta_{\phi} - \Half{d}}\,.
\end{equation}
The contributions from non-identity operators scale as

\begin{equation}
\Gspec^{\Delta_{\phi}}_{\Delta,J}(\omega,k)
\sim
(\omega^2-k^2)^{\Delta_{\phi} - \Half{d}-\Half{\Delta}}\,,
\qquad \omega \to \infty\,,
\end{equation}
and give perturbative corrections to the vacuum spectral function. We will determine the spin dependence and analytic properties of non-identity blocks below.

To evaluate the thermal block for spectral function, we find it convenient to utilize the identity~\eqref{eq:STT-contraction-identity}, which holds for any metric $g_{\mu\nu}$ used for index contractions with $e^{\mu}$ unit-normalized $e^2 = 1$. From now on we use $x^{\mu} = (t,\vb{x})$ to denote the Lorentzian spacetime coordinate, and index contraction is performed using the Lorentzian metric $\eta_{\mu\nu} = \mathrm{diag}\prn{1,-\vb{1}}$. We denote the degree $J$ homogeneous polynomial as
\begin{equation}
H_{J}\prn{x^{\mu}}
= \prn{x^{\mu_{1}} \cdots x^{\mu_{J}}} \,e_{\{\mu_{1}} \cdots e_{\mu_{J}\}}\,,
\end{equation}
with $e^{\mu} = \prn{1,\vb{0}}$ the unit vector in the time direction.
From~\eqref{eq:STT-contraction-identity}, we can rewrite the discontinuity as
\begin{equation}
\Disc\brk{G^{\Delta_{\phi}}_{\Delta,J}}(t,\vb{x})
=
2i\,\Pjnorm^{-1}_{J} \,\sin\prn{\pi \widetilde{\Delta}} \,\sgn(t) \,
\Theta(t^2-\vb{x}^2) \, H_{J} \prn{x^{\mu}} \, \prn{x^2}^{\widetilde{\Delta}-\Half{J}}.
\end{equation}

We can then compute the spectral function block by taking derivatives w.r.t.\ momentum of a
Lorentz-invariant integral
\begin{equation}
\Gspec^{\Delta_{\phi}}_{\Delta,J}(\omega,k)
=
2i\,(-1)^{\Half{J}}\,  \Pjnorm^{-1}_{J} \,
\sin\prn{\pi \widetilde{\Delta}} \, H_{J}\prn{\Tilde{\partial}^{\mu}} \, I_{\widetilde{\Delta}-J/2}(p)\,,
\end{equation}
where $p^{\mu} = (\omega,\vb{k})$, $\Tilde{\partial}_{\mu} = \prn{\partial_{\omega},\partial_{\vb{k}}}$.
The integral itself is given by
\begin{equation}
I_{\alpha}(p)
=
\int\,d^dx\,
e^{i p \cdot x} \, \sgn(t)\, \Theta(x^2) \,\prn{x^2}^{\alpha}.
\label{eq:Lorentz-integral}
\end{equation}
The integral evaluates to
\begin{equation}\label{eq:Lorentz-answer}
I_{\alpha}(p)
=
i \,\pi^{\Half{d-2}}\, 2^{2\alpha +d} \,
\Gamma(\alpha+1)\, \Gamma\prn{\alpha+\Half{d}} \,
\sin\brk{\pi\prn{\alpha+\Half{d}}}
\,\sgn(\omega) \, \Theta(p^2) \, \prn{p^{2}}^{-\alpha-\Half{d}}.
\end{equation}
Details on computing the integral can be found in~\cref{app:Lorentz-integral}.

The spectral function block can be simplified further. We start with the following important identity:
\begin{equation}
H_{J}\prn{\Tilde{\partial}^{\mu}} \prn{p^{2}}^{\gamma}
=
2^{J}\prn{-\gamma}_{J} \; H_{J}\prn{p^{\mu}} \prn{p^{2}}^{\gamma-J}.
\label{eq:HJ-derivative-identity}
\end{equation}
This holds because when $H_{J}\prn{\Tilde{\partial}^{\mu}} =  e_{\{\mu_{1}} \cdots e_{\mu_{J}\}}\,  \Tilde{\partial}^{\mu_{1}} \cdots \Tilde{\partial}^{\mu_{J}}$ acts on $\prn{p^{2}}^{\gamma}$, the only contributing derivative term is the one proportional to $p^{\mu_{1}} \cdots p^{\mu_{J}}$; all other derivative terms contain $\eta^{\mu\nu}$ and vanish when contracted with the symmetric traceless tensor. The momentum dependence of the spectral function block thus simplifies to
\begin{equation}
\Pjnorm^{-1}_{J} H_{J}\prn{p^{\mu}} \prn{p^{2}}^{-\widetilde{\Delta} - \Half{d}-\Half{J}} = \prn{p^{2}}^{-\widetilde{\Delta}-\Half{d}} P_{J}\prn{\frac{\omega}{\sqrt{p^{2}}}} \,,
\end{equation}
upon using the identity~\eqref{eq:STT-contraction-identity}.

After using reflection identity to simplify the Gamma function prefactors, we find the spectral function block to be
\begin{empheq}[box=\fbox]{equation}
  \begin{aligned}
  \Gspec^{\Delta_{\phi}}_{\Delta,J}(\omega,k)
   & =
  \sgn(\omega) \, \Theta\prn{\omega^2-k^2}  \, \prn{\omega^2-k^2}^{\Delta_{\phi}-\Half{d}}\,
  \mathcal{N}^{\Delta_{\phi}}_{\Delta,J} \, \GspecOp_{\Delta,J}(\omega,k)
  \\
  \mathcal{N}^{\Delta_{\phi}}_{\Delta,J}
   & =  \frac{\pi^{\Half{d+2}}\, 2^{\Delta-2\Delta_{\phi}+d+1}}{
    \Gamma\prn{-\Half{\Delta}+\Delta_{\phi}+\Half{J}}\,
    \Gamma\prn{-\Half{\Delta}+\Delta_{\phi} -\Half{d} -\Half{J}+1}}
  \\
  \GspecOp_{\Delta,J}(\omega,k)
   & =
  P_{J}\prn{\frac{\omega}{\sqrt{\omega^2-k^2}}} \, \prn{\omega^2-k^2}^{-\Half{\Delta}}\,.
  \end{aligned}
  \label{eq:spectral-function-block}
\end{empheq}
The spectral function block manifestly has support only for timelike momenta.
From~\eqref{eq:spectral-function-ope}, this implies that in the large momentum limit, the spectral function for any \CFT{d} only has support at timelike momentum up to exponentially small corrections; this is referred to as \emph{spectrum condition}~\cite{Manenti:2019wxs}. The spin dependence of our expression enjoys a similar structure with the Euclidean position space block. The expression given above agrees with the one presented in~\cite{Dodelson:2023vrw} in the context of two-sided correlator.\footnote{To see the agreement with~\cite[eq.(4.5)]{Dodelson:2023vrw}, we use the following identity for the Gegenbauer polynomial
\begin{equation}
C^{(\lambda)}_{J} \prn{\frac{1}{\sqrt{1-\mr^2}}}
= \frac{(2\lambda)_{J}}{J!}\,
(1-\mr^2)^{-\Half{J}} \, {}_{2}F_{1}\prn{-\Half{J},\Half{1-J};\lambda+\half;\mr^2}.
\end{equation}
The identity can be seen from the standard representation
\[
  C^{(\lambda)}_{J}(z) = \frac{(2\lambda)_{J}}{J!} \, {}_{2}F_{1}(-J,J+2\lambda;\lambda+1/2;(1-z)/2)
\]
together with a quadratic transformation of hypergeometric function
\[
  {}_{2}F_{1}(a,b;(a+b+1)/2;u) = (1-2u)^{-a} \, {}_{2}F_{1}(a/2,(a+1)/2;(a+b+1)/2;4u(u-1)/(1-2u)^2)\,.
\]
}

From the above arguments, we conclude that for a generic \CFT{d} the asymptotics of spectral function in the $\omega \to \infty$ limit takes the form
\begin{equation}
\begin{aligned}
\rho(\omega,k)
=
\sgn(\omega) \, \Theta\prn{\omega^2-k^2} \,  \prn{\omega^2 - k^2}^{\Delta_{\phi} - \Half{d}}
\bqty{1 + \sum_{\Op \in \phi \times \phi \setminus \{\id\}} \rhovacsub_{\Op}(\omega,k) }
+ \order{e^{-\omega}},
\label{eq:spectral-function-asymptotics-generic-CFT}
\end{aligned}
\end{equation}
with
\begin{equation}
\rhovacsub_{\Op}(\omega,k)
= a_{\Op}\, \frac{\GspecNorm^{\Delta_{\phi}}_{\Delta,J}}{\GspecNorm^{\Delta_{\phi}}_{\id}}\,
\GspecOp_{\Delta,J}(\omega,k).
\end{equation}
The leading contribution is the vacuum spectral function. The perturbative corrections are controlled by thermal OPE. The vacuum spectral function and the perturbative corrections only have support at timelike momentum. The non-perturbative corrections are not controlled by thermal OPE, and can have support at spacelike momentum.

\paragraph{Zeros of spectral function block:} From~\eqref{eq:spectral-function-block}, we see that the spectral function block vanishes if
\begin{equation}
\Delta-2\Delta_{\phi}-J \,\in\, 2\Z_{\geq0}\,,
\label{eq:first-type-zeros}
\end{equation}
or if
\begin{equation}
\Delta-2\Delta_{\phi}+J+d-2 \,\in\, 2\Z_{\geq0}\,.
\label{eq:second-type-zeros}
\end{equation}
The first type of zeros~\eqref{eq:first-type-zeros} precisely corresponds to double twist operators
\begin{equation}
\label{eq:double-twist-op}
\Op = \phi \,\Box^{n} \partial_{\mu_{1}} \cdots \partial_{\mu_{J}} \phi
\end{equation}
with $\Delta = 2\Delta_{\phi}+J+2n$. This type of zeros also exists in the Euclidean thermal block in momentum space~\cite{Manenti:2019wxs}. The second type of zeros~\eqref{eq:second-type-zeros} is absent in the Euclidean case, and specific to computing the commutator in momentum space. The fact that the polynomial $P_{J}(z)$ appearing in the thermal block~\eqref{eq:OPE-position-space} comes from contraction with a symmetric traceless tensor is important for seeing the second type of zeros, due to the $(-\gamma)_{J}$ term in the identity~\eqref{eq:HJ-derivative-identity}. If we were to replace $P_{J}(z)$ by a generic even-degree polynomial, then~\eqref{eq:second-type-zeros} would be replaced by\footnote{If $H_{J}$ in~\eqref{eq:HJ-derivative-identity} is replaced by a generic homogeneous polynomial, then the zeros from taking derivatives, the $(-\gamma)_{J}$ factor appearing in that equation, would get replaced by $(-\gamma)_{J/2}$. }
\begin{equation}
\label{eq:second-type-zeros-wrong}
\Delta-2\Delta_{\phi}+d-2 \in 2\Z_{\geq0},
\end{equation}
and some zeros would be missing. The distinction will be important for seeing that the spectral function for $\Delta=d$ in holographic \CFT{d} has no perturbative corrections, which would not be the case for~\eqref{eq:second-type-zeros-wrong}.

\subsection{Specialization to holographic CFTs}

It is expected that in a holographic \CFT{d} with classical gravity dual
(viz., the class of CFTs admitting a large central charge limit with a sparse low-lying spectrum),
the scalar thermal OPE only receives contribution from identity, the mean field theory (MFT) or double-twist sector, and the stress-tensor sector. To wit,
\begin{equation}
\phi \times \phi  = \{\id\} \cup \brk{\phi\phi} \cup \brk{T}.
\label{eq:scalar-ope-holo}
\end{equation}
The double-twist sector $\brk{\phi\phi}$ consists of double twist/trace operators of the form~\eqref{eq:double-twist-op} with dimension $\Delta = 2\Delta_{\phi}+J+2n$ and spin $J$; the stress-tensor sector $\brk{T}$ includes stress-tensor and multi-stress tensors. The multi-stress tensors $T^{n}$ have dimension $\Delta = nd$ and even spin $0 \leq J \leq 2n$.

As a concrete example to keep in mind we can let $\phi$ belong to the family of scalar chiral primary (1/2-BPS) operators $\Op_{p}$ in $\mathcal{N}=4$ SYM. These operators are labeled by integer $p \geq 2$ and live in $[0,p,0]$ representation of $SU(4)_{R}$, and have dimension $\Delta_{\phi}=p$.
As explained in~\cite{Iliesiu:2018fao}, in this case the statement~\eqref{eq:scalar-ope-holo} can be precisely justified by exploiting the R-symmetry considerations and an emergent $U(1)_{Y}$ bonus symmetry.

Let us now consider the implication of~\eqref{eq:scalar-ope-holo} for the spectral function in holographic \CFT{d}. The double-twist sector in general doesn't contribute to the thermal OPE of the spectral function~\eqref{eq:spectral-function-ope}, since the block $\Gspec^{\Delta_{\phi}}_{\Delta,J}$ in general vanishes for double twist operators. Therefore, the OPE for the holographic spectral function only receives contribution from the stress-tensor sector, thereby simplifying~\eqref{eq:scalar-ope-holo} to read in the
$\omega\to \infty$ limit
\begin{equation}
\rho(\omega,k)
=
\sgn(\omega) \, \Theta\prn{\omega^2-k^2} \prn{\omega^2 - k^2}^{\Delta_{\phi} - \Half{d}} \,
\bqty{1 + \sum^{\infty}_{n=1} \rhovacsub_{T^{n}}(\omega,k) } + \order{e^{-\omega}}\,.
\end{equation}
Here $\rhovacsub_{T^{n}}(\omega,k)$ is the contribution from multi-stress tensor, $T^{n}$, sector
\begin{equation}
\rhovacsub_{T^{n}}(\omega,k)
= \sum^{2n}_{J=0} \,  a_{T^{n}}(J) \,
\frac{\GspecNorm^{\Delta_{\phi}}_{nd,J}}{\GspecNorm^{\Delta_{\phi}}_{\id}} \,
\GspecOp_{nd,J}(\omega,k)\,.
\label{eq:PTn}
\end{equation}
We have denoted the OPE coefficient for $n$-composite stress-tensor exchange with spin $J$ as $a_{T^{n}}(J)$.

Furthermore, for even $d$ and $\Delta_{\phi} \in \Z$, due to the second type of zeros~\eqref{eq:second-type-zeros} for the block, the perturbative corrections from the stress-tensor sector truncate further in the
$\omega \to \infty$ limit
\begin{equation}
\rho(\omega,k)
= \sgn(\omega)\,  \Theta\prn{\omega^2-k^2} \, \prn{\omega^2 - k^2}^{\Delta_{\phi} - \Half{d}} \,
\prn{1 + \sum^{n_{\max}}_{n=1} \rhovacsub_{T^{n}}(\omega,k) } + \order{e^{-\omega}},
\label{eq:truncation-rho-holoCFT}
\end{equation}
with
\begin{equation}
n_{\max}=
\begin{dcases}
0,
 & \quad
\dfrac{d}{2}\leq \Delta_{\phi}\leq d,
\\
1,
 & \quad
d+1\le \Delta_{\phi} \leq \dfrac{3d}{2}-1,
\\
\left\lfloor \dfrac{2\Delta_{\phi}}{d}\right\rfloor - 1,
 & \quad
\Delta_{\phi}\ge\dfrac{3d}{2}.
\end{dcases}
\end{equation}
In particular, the first case above with $n_{\max}=0$ corresponds to spectral function having no perturbative corrections:
\begin{equation}
\rho(\omega,k) =
\sgn(\omega) \, \Theta\prn{\omega^2-k^2} \,  \prn{\omega^2 - k^2}^{\Delta_{\phi} - \Half{d}}
+ \order{e^{-\omega}}\,, \qquad \omega \to \infty\,.
\end{equation}
Consequently, this class of operators hold promise for isolating the non-perturbative contributions to the
spectral functions in the high frequency limit in a relatively straightforward manner. They will, therefore,
be the focus of our attention in the sequel.

\paragraph{Exact examples in \CFT{2}:}
The aforementioned properties can be seen explicitly in \CFT{2}, where thermal two-point function on $\vb{S}^{1}_{\beta} \times \R$ is fixed by conformal symmetry.\footnote{Unlike holographic CFT in higher dimensions, in \CFT{2} the only contribution to position space thermal two-point function is from the stress-tensor sector. This difference is inessential for the discussion here, as double-twist sector doesn't contribute in momentum space anyway.} Holographic computation on BTZ background recovers the universal answer. In terms of the dimensionless $d$-momentum $(\freq,\mom) = \frac{\beta}{2\pi} (\omega,k)$, the universal spectral function in \CFT{2} is given by~\cite{Son:2002sd}
\begin{equation}
\rho(\freq,\mom) =
\sinh(\pi\,\freq) \,\abs{\Gamma\prn{\Half{1+\nu} - \Half{i\freq} \pm \Half{i\mom}}}^{2}.
\end{equation}
For $\Delta_{\phi} = \nu +1 \in \Z$, this can be written as
\begin{equation}
\rho(\freq,\mom)
=
P_{2\nu}(i\freq/2,i\mom/2) \,
\frac{\sinh(\pi\,\freq)}{\cosh\prn{\pi\,\freq}+(-1)^{\nu}\cosh(\pi\mom)},
\label{eq:spectral-function-CFT2-integer-dimension}
\end{equation}
with $P_{2\nu}(i\freq/2,i\mom/2)$ a degree $2\nu$ polynomial in $\freq,\mom$
\begin{equation}
P_{2\nu}(i\freq/2,i\mom/2) =
\begin{dcases}
\prn{\half \pm \Half{i\freq} \pm \Half{i\mom}}_{m}
 & \quad
\nu = 2m \in 2\Z_{\geq0}\,,
\\[2ex]
\prn{\freq^2 - \mom^2}\prn{1 \pm \Half{i\freq} \pm \Half{i\mom}}_{m}
 & \quad
\nu = 2m+1 \in 2\Z_{\geq0} +1 \,.
\end{dcases}
\end{equation}
In the large momentum limit, the polynomial factor $P_{2\nu}(i\freq/2,i\mom/2)$ gives the leading vacuum spectral function and perturbative corrections, and the remaining factor gives non-perturbative corrections and leads to the spectrum condition. The truncation of perturbative corrections is seen explicitly. For $\Delta_{\phi} \in \{1,2\}$, $P_{2\nu}(i\freq/2,i\mom/2)$ reduces to vacuum spectral function without perturbative corrections.

\subsection{Perturbative corrections to the holographic spectral function}

We require the stress-tensor OPE coefficients to determine the perturbative corrections~\eqref{eq:PTn}.
Unfortunately, for a generic \CFT{d} there are no simple techniques to obtain them. However,
the stress-tensor sector OPE coefficients in holographic \CFT{d} can be extracted using the near-boundary expansion as explained in~\cite{Fitzpatrick:2019zqz}. In the following we give a brief review of this technique, which we refer to as the \emph{FH method}, adapting to our conventions.

\paragraph{The bulk wave equation:} Let $\Phi$ be a bulk scalar field dual to our holographic \CFT{d} scalar primary $\phi$.   For holographic computation of $\gE(\tau,\vb{x})$ one needs to solve the bulk-to-boundary propagator $G_{\BH}$ on planar \SAdS{d+1} background.\footnote{Since discussion will be limited to the canonical ensemble at fixed temperature, it suffices to focus on the dual neutral black hole spacetime. } We will work with the
standard form of the Lorentzian planar \SAdS{d+1} metric, which reads
\begin{equation}\label{eq:sdasmet}
\begin{split}
ds^{2}
 & =
-r^{2} f(r) dt^{2} + \frac{dr^{2}}{r^{2}f(r)} + r^{2} d\vb{x}^2_{d-1}
\\
f(r)
 & =
1 - \prn{\frac{\rp}{r}}^{d}.
\end{split}
\end{equation}
The solution is parameterized by $r_+$ which maps to the temperature
\begin{equation}
T = \frac{d\,\rp}{4\pi}.
\end{equation}
The Euclidean bulk-to-boundary propagator is obtained by solving the minimally coupled scalar wave equation for
$\Phi$ in the Wick rotated geometry $t \to -i \,\tau$
\begin{equation}
\prn{\nabla^2 - \Delta_{\phi}(\Delta_{\phi}-d)} \Phi(r,\tau,\vb{x}) = 0\,.
\end{equation}
The boundary conditions are canonical: $G_{\BH} \sim r^{\Delta_\phi - d} \,\delta(\tau) \, \delta^{(d-1)}(\vb{x})$
as $r \to \infty$ (i.e., at AdS boundary) and regularity of $G_{\BH}$ at horizon $r=r_+$.
The two-point function is then computed by
\begin{equation}
\gE(\tau,\vb{x}) = \lim_{r \to \infty} r^{\Delta_{\phi}} G_{\BH}(r,\tau,\vb{x}).
\label{eq:two-pt-func-from-propgator}
\end{equation}
The main problem is that the bulk-to-boundary propagator for $\Phi$ has no simple closed-form solution for $d>2$.
We will work in units where $T = \beta =1$ below.

\paragraph{Stress-tensor OPE coefficients:}
Motivated by thermal OPE for two-point function, the following ansatz is considered in~\cite{Fitzpatrick:2019zqz}
\begin{equation}
G_{\BH}(r,\tau,\vb{x}) = G_{\mathrm{AdS}}(r,\tau,\vb{x}) \, G_{\OPE}(r,w,\rho), \qquad
G_{\OPE} = \prn{1 + G_{\brk{T}} + G_{\brk{\phi\phi}}}.
\end{equation}
Here $G_{\mathrm{AdS}}(r,\tau,\vb{x})$ is the bulk-to-boundary propagator for $\Phi$ on the vacuum AdS
geometry
\begin{equation}
G_{\mathrm{AdS}}(r,\tau,\vb{x}) = \prn{\frac{r}{1+r^2\prn{\tau^2 + \vb{x}^2}}}^{\Delta_{\phi}}.
\end{equation}
The coordinates $w,\rho$ are defined as
\begin{equation}
w^2 = 1+r^2\prn{\tau^2+\vb{x}^2}, \quad \rho = r\abs{\vb{x}}.
\label{eq:coords-wrho}
\end{equation}
The ansatz is considered in the limit
\begin{equation}
r \to \infty, \quad \tau,\abs{\vb{x}} \to 0, \quad w,\rho \ \text{fixed},
\end{equation}
i.e., $\tau,\abs{\vb{x}} \sim 1/r$  as $r \to \infty$. In this limit, the stress-tensor sector contribution $G_{\brk{T}}$ has large $r$ expansion
\begin{equation}
G_{\brk{T}}(r,w,\rho) = \sum^{\infty}_{n=1} r^{-nd} G_{T^{n}}(w,\rho),
\end{equation}
and the double-twist sector contribution $G_{\brk{\phi\phi}}$ has large $r$ expansion
\begin{equation}
G_{\brk{\phi\phi}}(r,w,\rho) = \prn{\frac{w}{r}}^{2\Delta_{\phi}} \sum^{\infty}_{m=1} r^{-m} G_{\brk{\phi\phi},m}(w,\rho).
\end{equation}
One then proceeds to solve the bulk equation order by order in $r$ after substituting the ansatz.

For noninteger $\Delta_{\phi}$, the stress-tensor and double-twist sector only involve, respectively, integer and noninteger powers of $r$, and are therefore decoupled. Moreover, assuming noninteger $\Delta_{\phi}$, it turns out that $G_{T^{n}}(w,\rho)$ in the stress-tensor sector can be solved without using the regularity condition at the horizon. For even $d$, it is found that the solution takes the form
\begin{equation}
G_{T^{n}}(w,\rho) = \sum^{2n}_{i \overset{2}{=}0} \, \sum^{nd-i}_{j \overset{2}{=} -2n} \,
\alpha^{(n)}_{ij} \rho^{i} w^{j}.
\end{equation}
The convention being used in the sums above is to indicate that the indices $i,j$ advance in steps of two, as indicated by the $\overset{2}{=}$ symbol.\footnote{Since we repeatedly use this notation later we will highlight it here for convenience of the reader:
  $\overset{2}{=}$ in a summation index indicates that the parameter is stepped forward by 2 units.\label{fn:2step}} For odd $d$, the method can in principle still be applied, but $G_{T^{n}}$ no longer admits simple rational solutions. 

One can then compute the stress-tensor sector contribution to thermal two-point function using~\eqref{eq:two-pt-func-from-propgator}. When taking the $r\to\infty$ limit, $w,\rho$ are no longer treated as independent of $r$, but contain $r$ dependence via~\eqref{eq:coords-wrho}. The $T^{n}$ contribution is then given by
\begin{equation}
\begin{aligned}
\gE(\tau,\vb{x})\bigg|_{T^{n}}
 & =
\lim_{r \to \infty} \, r^{\Delta_{\phi}-nd}\, G_{\mathrm{AdS}}(r,\tau,\vb{x})\, G_{T^{n}}(w,\rho)
\\
 & =
(\tau^2+\vb{x}^2)^{-\Delta_{\phi}} \,
\sum^{2n}_{i \overset{2}{=} 0} \,
\alpha^{(n)}_{i,nd-i} \, \abs{\vb{x}}^{i} \, \prn{\tau^2 + \vb{x}^{2}}^{\Half{nd-i}}.
\end{aligned}
\end{equation}
Compared with thermal OPE~\eqref{eq:OPE-position-space}, this is identified as the contribution to two-point function from $T^{n}$ exchange. The OPE coefficients $a_{T^{n}}(J)$ are then solved from
\begin{equation}
\sum^{2n}_{i \overset{2}{=} 0} \alpha^{(n)}_{i,nd-i} \abs{\vb{x}}^{i} \prn{\tau^2 + \vb{x}^{2}}^{\Half{nd-i}}
= \sum^{2n}_{J \overset{2}{=} 0} a_{T^{n}}(J) P_{J} \prn{\frac{\tau}{\sqrt{\tau^2 + \vb{x}^2}}} \prn{\tau^2+\vb{x}^2}^{\Half{nd}}.
\end{equation}

For general $d$, the OPE coefficient for single stress tensor exchange can be obtained using the method of~\cite{Fitzpatrick:2019zqz}
\begin{equation}
a_{T} = \frac{2^{-2-d}\pi^{\half} \, \Gamma\prn{\Half{d}-1}}{\Gamma\prn{\Half{d+3}}}
\, f_0\, \, \Delta_{\phi} \,, \qquad
f_{0}
= \prn{\frac{\beta\, \rp}{\lads^2}}^d
= \prn{\frac{4\pi}{d}}^{d}.
\label{eq:aT}
\end{equation}

There are no known closed-form expressions in general $d$ for multi-stress-tensor OPE coefficients.  For $d=4$, the OPE coefficients for $T^{2}$ exchange are given by
\begin{equation}
\begin{aligned}
a_{T^{2}}(0)
 & =
\frac{f^2_{0}\,\Delta_{\phi}\left(7\Delta_{\phi}^{4}-45\Delta_{\phi}^{3}+100\Delta_{\phi}^{2}-80\Delta_{\phi}+48\right)}
{201600\,(\Delta_{\phi}-4)(\Delta_{\phi}-3)(\Delta_{\phi}-2)}
\\
a_{T^{2}}(2)
 & =
\frac{f^2_{0}\,\Delta_{\phi}\left(7\Delta_{\phi}^{3}-23\Delta_{\phi}^{2}+22\Delta_{\phi}+12\right)}
{201600\,(\Delta_{\phi}-3)(\Delta_{\phi}-2)}
\\
a_{T^{2}}(4)
 & =
\frac{f^2_{0}\,\Delta_{\phi}\left(7\Delta_{\phi}^{2}+6\Delta_{\phi}+4\right)}
{201600\,(\Delta_{\phi}-2)},
\end{aligned}
\label{eq:aT2-d-4}
\end{equation}
and for $T^{3}$ exchange they are given by
\begin{equation}
\begin{aligned}
a_{T^{3}}(0)
 & =
\frac{f_0^{3}\,(\Delta_{\phi}-8)\,\Delta_{\phi}\,
\Big(1001\,\Delta_{\phi}^{6}-6864\,\Delta_{\phi}^{5}+17115\,\Delta_{\phi}^{4}-10460\,\Delta_{\phi}^{3}+1584\,\Delta_{\phi}^{2}+21384\,\Delta_{\phi}+12960\Big)}
{10\,378\,368\,000\,(\Delta_{\phi}-6)\,(\Delta_{\phi}-5)\,(\Delta_{\phi}-4)\,(\Delta_{\phi}-3)\,(\Delta_{\phi}-2)}
\\[4pt]
a_{T^{3}}(2)
 & =
\frac{f_0^{3}\,\Delta_{\phi}\,
\Big(1001\,\Delta_{\phi}^{6}-6864\,\Delta_{\phi}^{5}+12615\,\Delta_{\phi}^{4}-3980\,\Delta_{\phi}^{3}-6156\,\Delta_{\phi}^{2}-11736\,\Delta_{\phi}-1440\Big)}
{3\,459\,456\,000\,(\Delta_{\phi}-5)\,(\Delta_{\phi}-4)\,(\Delta_{\phi}-3)\,(\Delta_{\phi}-2)}
\\[4pt]
a_{T^{3}}(4)
 & =
\frac{f_0^{3}\,\Delta_{\phi}\,
\Big(1001\,\Delta_{\phi}^{5}-2145\,\Delta_{\phi}^{4}-2760\,\Delta_{\phi}^{3}-2390\,\Delta_{\phi}^{2}+2244\,\Delta_{\phi}+2160\Big)}
{5\,189\,184\,000\,(\Delta_{\phi}-4)\,(\Delta_{\phi}-3)\,(\Delta_{\phi}-2)}
\\[4pt]
a_{T^{3}}(6)
 & =
\frac{f_0^{3}\,\Delta_{\phi}\,
\Big(1001\,\Delta_{\phi}^{4}+3575\,\Delta_{\phi}^{3}+7310\,\Delta_{\phi}^{2}+7500\,\Delta_{\phi}+3024\Big)}
{10\,378\,368\,000\,(\Delta_{\phi}-3)\,(\Delta_{\phi}-2)} \, .
\end{aligned}
\end{equation}
As seen from the examples above, for $n \geq 2$, $a_{T^{n}}(J)$ has poles at certain integer $\Delta_{\phi}$. In general, the location of poles for $a_{T^{n}}(J)$ is
\begin{equation}
\label{eq:aTn-poles}
\Delta_{\phi} \in \cbrk{\Half{d}, \Half{d}+1, \dots, \Half{nd-J}}.
\end{equation}

Given the stress-tensor OPE coefficients, the perturbative corrections of holographic spectral function are then determined from~\eqref{eq:PTn}.

\paragraph{Extrapolating to integer $\Delta_{\phi}$:} We would like to use the above data to
ascertain the perturbative corrections from the multi-stress tensor for $\Delta_\phi \in \Z$. A priori
one may question if the truncated perturbative corrections~\eqref{eq:truncation-rho-holoCFT} for integer $\Delta_{\phi}$ can be determined using the stress-tensor OPE coefficients obtained from FH method. The derivation in FH method assumes noninteger $\Delta_{\phi}$, to decouple the double-twist from the stress tensor.  Moreover, the OPE coefficients thus computed have poles at the integer $\Delta_{\phi}$ in~\eqref{eq:aTn-poles}. Despite these subtleties, we expect that it is still valid to use the OPE coefficients at integer $\Delta_{\phi}$ for the following reasons:
\begin{itemize}[wide,left=0pt]
  \item For $n\leq n_{\max}$, i.e., $nd-2\Delta_{\phi}+J+d\in2\Z_{\leq0}$ (cf.~\eqref{eq:second-type-zeros}) so that the spectral function block is finite. In particular, the condition~\eqref{eq:aTn-poles} is not satisfied. Therefore, using the OPE coefficients from FH method would not give divergent contributions to~\eqref{eq:truncation-rho-holoCFT}.
  \item For $n > n_{\max}$, the condition~\eqref{eq:aTn-poles} is satisfied for large enough $n$. However, if~\eqref{eq:aTn-poles} is satisfied, both~\eqref{eq:first-type-zeros} and~\eqref{eq:second-type-zeros} are satisfied. This implies that when the OPE coefficient $a_{T^{n}}(J)$ has a simple pole, the associated spectral function block has a double zero. Therefore, there is no contribution due to divergent OPE coefficients beyond $n_{\max}$.
\end{itemize}
We will therefore proceed to use the OPE coefficients from FH method to compute the truncated perturbative correction in~\eqref{eq:truncation-rho-holoCFT} for integer $\Delta_{\phi}$.

\section{Spectral function in holographic CFTs}\label{sec:specholo}

At this point we have developed some basic intuition about the form of the spectral function
from the thermal OPE. We would like to exploit this to understand the non-perturbative corrections to
the spectral function in holographic CFT$_d$ with $d>2$, where we cannot analytically solve the wave
equation in the bulk geometry. The semiclassical parameter that will be taken to be large will be the momenta.
Note that~\cite{Festuccia:2008zx,Amado:2008hw} analyzed the WKB limit of the holographic Green's functions,
but they largely did so for high dimension operators. The one exception was the analysis of~\cite{Festuccia:2008zx}
who examined the limit of large lightlike momenta for general (not necessarily large) $\Delta_\phi$.
We will on the other hand take $\Delta_\phi \sim \order{1}$ but take $\omega \gg 1$ (analyzing both $k \sim \omega$ and $k$ finite as we take the limit). To do so, we shall use the exact expression for the retarded Green's function derived
in~\cite{Jia:2024zes} using the semiclassical Virasoro block method developed in~\cite{Bonelli:2022ten,Lisovyy:2022flm}.
We will first motivate a specific exact form of the spectral function (in momentum space) using this method, which we will then argue is well-suited for an asymptotic expansion and thus can access the non-perturbative information we seek.
As noted above this is the piece of the spectral function that is not controlled by the thermal OPE.

\subsection{An exact expression for the spectral function}\label{sec:exactsf}

We will work in even dimensional CFTs, for which the bulk analysis is straightforward. Therefore, we take $d$ to be even henceforth. In this case the governing wave equation for minimally-coupled scalar on planar \SAdS{d+1} can be transformed to canonical Schr\"odinger form with $2+d/2$ regular singular points. These comprise the singularity, the AdS boundary, and
the $d/2$ horizons (one physical, with the others complex).\footnote{ This statement assumes that we have mapped the radial coordinate of \SAdS{d+1} by projecting down, cf.~\eqref{eq:rtoz} below, for otherwise, there are $d+2$ regular singular points. The projection pairwise merges the roots of the metric function $f(r)$ given in~\eqref{eq:sdasmet}. }

We will start by generalizing the exact (s-channel) expression obtained in~\cite{Jia:2024zes} from Virasoro block method for $d=4$ to arbitrary even $d$. The presence of additional singular points compared to the case $d=4$ is not an obstruction for applying the Virasoro block method, thanks to the locality of fusion transformation (this was already noted in~\cite{Lisovyy:2022flm}). The channel of the semiclassical block $\VBcl$ is chosen such that the two heavy external operators corresponding to the singular points at physical horizon and AdS boundary fuse into an intermediate state with momentum $\cmexp$;
the remaining details of the block don't explicitly appear in the connection formula.

We work with a canonical Schr\"odinger form of the wave equation (cf.~\cref{app:Schrodinger-form} for details on its derivation)
\begin{equation}
\label{eq:Schrodinger-form}
\psi^{\prime\prime}(z)+ t(z)\, \psi(z) = 0
\end{equation}
in coordinate
\begin{equation}\label{eq:rtoz}
z = \frac{\rp^2}{r^2}\,.
\end{equation}
The potential $t(z)$ is meromorphic with double poles (regular singular points) at
\begin{equation}
z_{\bdy} = 0, \quad z^{(n)}_{\hor} = e^{\frac{4\pi i}{d}n}, \quad n=0,\dots, \Half{d}-1, \quad z_{\sing} = \infty.
\end{equation}
Explicitly, it is given by
\begin{equation}
t(z)
= \sum_{z_{i} \in \{z_{\bdy}, z^{(n)}_{\hor}\}} \, \frac{\delta_{i}}{(z-z_{i})^2}
+ \frac{c_{i}}{z-z_{i}} \,.
\end{equation}

The classical dimensions associated with the singular points are
\begin{equation}
\delta_{\bdy} = \frac{1}{4} - \frac{\nu^2}{4}, \qquad
\delta^{(n)}_{\hor} = \frac{1}{4} + \frac{\freq^2}{4}e^{\frac{4\pi i}{d}n}, \qquad
\delta_{\sing} = \frac{1}{4}\,,
\end{equation}
with
\begin{equation}
\nu = \Delta - \Half{d}\,.
\end{equation}
The classical momentum $\mexp_{i}$ is related to classical dimension $\delta_{i}$ by 
\begin{equation}
\delta_{i} = 1/4 - \mexp^2_{i}\,. 
\end{equation}
In particular,
\begin{equation}
\mexp_{\bdy} = \Half{\nu}, \qquad \mexp_{\hor} = \Half{i\freq},
\end{equation}
where we will in general denote $(\cdot)_{\hor} \equiv (\cdot)^{(0)}_{\hor}$.
Finally, the $c_i$ are the accessory parameters, which are functions of the
dimensionless frequency and momentum
\begin{equation}
\freq = \frac{\beta\, \omega}{2\pi}, \qquad
\mom = \frac{\beta\, k}{2\pi}\,.
\end{equation}
They are obtained to be
\begin{equation}
\label{eq:accparam-general-d}
c_{\bdy} = \frac{d^2}{16}(\freq^2 - \mom^2), \quad c^{(n)}_{\hor} = \frac{d}{8}(\mom^2 - \freq^2) + \frac{\nu^2 - d/2}{2d}e^{-\frac{4\pi i}{d} n}.
\end{equation}

\paragraph{Relating the wave-equation to a 2d CFT BPZ equation:}
There is a well-known correspondence between Schr\"odinger equation of the form~\eqref{eq:Schrodinger-form} and 2d CFT. The Schr\"odinger equation arises from the large $c$ limit of the BPZ equation associated with a degenerate Virasoro block --- a light level two degenerate operator $V_{\expval{2,1}}$ with $\order{c^{0}}$ dimension is inserted at $z$, and heavy operators $V_{i}$ with dimensions $h_{i} \sim \frac{c}{6} \delta_{i}$ are inserted at $z_{i}$. The degenerate Virasoro block in the large $c$ limit factorizes into a wave function $\psi(z)$ associated with the light degenerate operator, and an exponentiated semiclassical block $e^{\frac{c}{6} \VBcl}$ associated with the heavy operators. The accessory parameters are related to the semiclassical block via the Zamolodchikov relation
\begin{equation}
c_{i} = \partial_{z_{i}} \VBcl.
\end{equation}
This relation holds for semiclassical block in any channel, and in principle determines the dimensions of the intermediate states in the block $\VBcl$ given a set of accessory parameters $\{c_{i}\}$.

The relation with 2d CFT can be used to obtain connection formula for the Schr\"odinger equation relevant for the computation of holographic thermal two point function. The connection formula is derived using the degenerate fusion transformation of the
corresponding degenerate Virasoro block. This derivation does depend on the choice of channel for the block $\VBcl$. As we are interested in the connection problem between boundary and horizon, we find it convenient to pick the channel where the operator $V_{\bdy}$ sequentially fuses with operators $V^{(0)}_{\hor}, V^{(1)}_{\hor}, \dots, V^{(d/2-1)}_{\hor}$ and then finally with $V_{\sing}$.
\begin{equation}
\mathcal{W}
=
V_{\mathrm{bdy}}\,
\begin{tikzpicture}[
    baseline={(0,0)},
    x=0.85cm,
    y=1cm,
    line cap=round,
    line width=0.8pt
  ]
  \draw (0,0) -- (4,0);
  \draw (5.5,0) -- (9.5,0);

  \foreach \x/\V in {
  1.3/{$V_{\mathrm{hor}}$},
  2.7/{$V_{\mathrm{hor}}^{(1)}$}
  }{
  \draw (\x,0) -- (\x,1.35);
  \node[above] at (\x,1.35) {\V};
  }

  \draw (7.5,0) -- (7.5,1.35);
  \node[above, xshift=3pt] at (7.5,1.35) {$V_{\mathrm{hor}}^{(d/2-1)}$};

  \node[anchor=north, yshift=-3pt] at (2.0,0) {$\sigma$};
  \node[anchor=north, yshift=-3pt] at (3.35,0) {$\sigma^{(1)}$};
  \node[anchor=north, yshift=-3pt] at (6.5,0) {$\sigma^{(d/2-2)}$};

  \node at (4.75,0) {$\cdots$};
\end{tikzpicture}
\,V_{\mathrm{sing}}\,.
\end{equation}

The intermediate momenta in the block are related to monodromy invariants of the Schr\"odinger equation~\eqref{eq:Schrodinger-form} as follows. Let $\mathcal{C}^{(n)}$ be monodromy contours surrounding $z_{\bdy},z^{(0)}_{\hor},\dots,z^{(n)}_{\hor}$ for $n=0, \dots, d/2-2$. Then\footnote{The monodromy matrix $M_{\mathcal{C}}$ is the transformation for a basis of solutions when continued along the contour $\mathcal{C}$, and its trace is basis independent.}
\begin{equation}
\Tr M_{\mathcal{C}^{(n)}} = -2\cos(2\pi\cmexp^{(n)}).
\end{equation}
The only intermediate momentum that will appear explicitly in connection formula is $\cmexp \equiv \cmexp^{(0)}$, which is related to the monodromy invariant around AdS boundary and physical horizon
\begin{equation}
\Tr M_{\bdy,\hor} = -2\cos(2\pi\cmexp).
\end{equation}

The accessory parameters in~\eqref{eq:accparam-general-d} are not independent.
At fixed external dimensions, or equivalently at fixed $\freq,\nu$, there is only one free parameter $\mom$.
We therefore anticipate that there is only one free parameter $\cmexp$ when inverting the Zamolodchikov relation, with $\cmexp^{(n)}$ with $n\neq 0$ fixed in terms of $\cmexp$. More explicitly, the semiclassical block needs to obey the following $d/2$ constraints
\begin{equation}\label{eq:block-constraints-higher-d}
\partial_{z_{\bdy}} \VBcl
=
\frac{d}{2} \partial_{z^{(n)}_{\hor}} \VBcl
- \frac{\nu^2-d/2}{4} e^{-\frac{4\pi i}{d} n}\,.
\end{equation}
Note that in general an $n$-point block only depends on $n-3$ cross-ratios.
Since we didn't consider the variation w.r.t.\ $z_{\sing}$, the condition above gives $d/2-2$ non-trivial constraints.
These constraints, in principle, fix the $d/2-2$ intermediate momenta $\cmexp^{(n)}$ with $n\neq 0$ in terms of $\cmexp$.
For $d=4$ where the relevant block is a four-point block, there are no such additional constraints.

\paragraph{The exact retarded Green's function \& spectral function:}
With this choice of channel for the block, the derivation in~\cite{Jia:2024zes} in the $d=4$ case can be immediately generalized to yield the following exact expression for retarded two-point function for arbitrary even $d$:
\begin{equation}
\label{eq:exact-Gret}
\begin{split}
\Gret(\freq,\mom) & = \mexp^{-1}_{\bdy} \frac{\Gamma\prn{-2\mexp_{\bdy}}}{\Gamma\prn{2\mexp_{\bdy}}}  \frac{\Gamma\prn{\half+\mexp_{\bdy}-\mexp_{\hor}\pm\cmexp}}{\Gamma\prn{\half-\mexp_{\bdy}-\mexp_{\hor}\pm\cmexp}} e^{\VBcl_{\bdy}}, \quad \VBcl_{\bdy} = -\partial_{\mexp_{\bdy}}\VBcl.
\end{split}
\end{equation}
As mentioned previously, the intermediate momentum $\cmexp$ is defined from the Zamolodchikov relation,
and in general depends both on frequency and momentum. The prefactor $\mexp^{-1} = (\Delta-d/2)^{-1}$ arises from
a careful holographic renormalization analysis~\cite{deHaro:2000vlm}, and is relevant for continuing the expression to $\Delta = d/2$.

The spectral function can be computed from
\begin{equation}
\rho(\freq,\mom) = -i \prnbig{\Gret(\freq,\mom) - \Gret(-\freq,\mom)}.
\end{equation}
Notice that both the block factor and $\cmexp$ are unchanged under the sign change of $\freq$.\footnote{If $\cmexp$ is a solution to Zamolodchikov relation with $\freq$, it is also a solution with $-\freq$.} Furthermore, we can simplify the
Gamma functions using
\begin{equation}
\begin{aligned}
 &
\frac{\Gamma\prn{\half+\mexp_{\bdy}-\mexp_{\hor}\pm\cmexp}}{\Gamma\prn{\half-\mexp_{\bdy}-\mexp_{\hor}\pm\cmexp}}
- \frac{\Gamma\prn{\half+\mexp_{\bdy}+\mexp_{\hor}\pm\cmexp}}{\Gamma\prn{\half-\mexp_{\bdy}+\mexp_{\hor}\pm\cmexp}}
\\
 & \qquad =
-\pi^{-2}\sin(2\pi\mexp_{\bdy})\sin(2\pi\mexp_{\hor}) \Gamma\prn{\half+\mexp_{\bdy} \pm \mexp_{\hor} \pm \cmexp}\,.
\end{aligned}
\end{equation}
Altogether, these two facts imply
\begin{empheq}[box=\fbox]{equation}\label{eq:spectral-function-exact-expression}
  \rho(\freq, \mom) = \mathsf{N}_\nu\, \sinh(\pi\,\freq) \Gamma\prn{\Half{1+\nu} \pm \Half{i \freq} \pm \cmexp} e^{\VBcl_{\bdy}}.
\end{empheq}
Here we factored out a piece $\mathsf{N}_\nu$  which only depends on the external operator dimension $\Delta = \frac{d}{2} + \nu$. It is given by
\begin{equation}
\mathsf{N}_\nu = \nu^{-1} \sin(\pi \nu) \frac{\Gamma(-\nu)}{\Gamma(\nu)} \,,
\end{equation}
and is regular at all integer $\nu$ including $\nu =0$ owing to the holographic renormalization factor. Since we can absorb it by rescaling the operators in question, we will take the opportunity to do so, and ensure that the vacuum result for the correlation function appears with a unit coefficient. Therefore, henceforth, we will drop $\mathsf{N}_\nu$ from the formulae with this choice of normalization understood.

At real frequency and momentum, we anticipate $\cos(2\pi\cmexp) \in \R$,\footnote{We have verified this numerically for $d=4,6$ with various choices of $\freq,\mom,\nu$.} and therefore $\cmexp \in \R \cup i \R \cup \Z_{\neq 0}/2 + i \R$. The expression is manifestly real for $\cmexp \in \R \cup i\R$ provided $\VBcl \in \R$. For $\cmexp \in \Z_{\neq 0}/2 + i \R$, the reality property is not manifest.

\paragraph{Specialization to integer conformal dimensions:} For integer $\nu$, the spectral function takes the following suggestive form
\begin{empheq}[box=\fbox]{equation}\label{eq:spectral-function-exact-expression-integer-dimension}
  \rho(\freq,\mom) = P_{2\nu}(i\freq/2,\cmexp) e^{\VBcl_{\bdy}} \frac{\sinh(\pi\,\freq)}{\cosh\prn{\pi\,\freq}+(-1)^{\nu}\cos(2\pi\cmexp)}
\end{empheq}
with $P_{2\nu}(\mexp,\cmexp)$ being a degree $2\nu$ polynomial in $\freq,\cmexp$ given by
\begin{equation}\label{eq:P2v-poly}
P_{2\nu}(\mexp,\cmexp)
=
\begin{dcases}
\prn{\half \pm \mexp \pm \cmexp}_{m}
 & \qquad
\nu = 2m \in 2\Z_{\geq0} \,,
\\
\prn{\cmexp^2 - \mexp^2}\prn{1 \pm \mexp \pm \cmexp}_{m}
 & \qquad
\nu = 2m+1 \in 2\Z_{\geq0} +1\,,
\end{dcases}
\end{equation}
where we have used $\Gamma(z)\Gamma(1-z) = \pi/\sin(\pi z)$ together with $\Gamma(z+n) = (z)_{n} \Gamma(z)$.

In the large momentum limit, the polynomial $P_{2\nu}(i\freq/2,\cmexp)$ has the correct scaling in $\freq$ associated with vacuum spectral function
$(\freq^2-\mom^2)^{\nu}$. For the special integer dimensions without perturbative corrections, it is then natural to anticipate that
$P_{2\nu}(i\freq/2,\cmexp) e^{\VBcl_{\bdy}}$ evaluates exactly to the vacuum spectral function $(\freq^2-\mom^2)^{\nu}$. On the other hand, the remaining factor ought to give the  non-perturbative corrections which are not captured by the vacuum answer. Moreover, for generic integer dimensions, it is also natural to expect $P_{2\nu}(i\freq/2,\cmexp) e^{\VBcl_{\bdy}}$ corresponds to the vacuum spectral function along with the finitely truncated perturbative correction. Therefore, we define
\begin{equation}
\rho^{(\Vir)}_{\pert}(\freq,\mom) \coloneqq
P_{2\nu}(i\freq/2,\cmexp)\, e^{\VBcl_{\bdy}}\,,
\end{equation}
and
\begin{equation}
\label{eq:rho-np-holo-CFT}
\rho_{\np}(\freq,\mom) \coloneq \frac{\sinh(\pi\,\freq)}{\cosh\prn{\pi\,\freq}+(-1)^{\nu}\cos(2\pi\cmexp)}.
\end{equation}
The superscript in $\rho^{(\Vir)}_{\pert}(\freq,\mom)$ emphasizes that this is defined from the Virasoro block method.

On the other hand, the perturbative correction determined from thermal OPE is (cf.~\eqref{eq:PTn})
\begin{equation}\label{eq:rho-OPE-holo-CFT}
\rho^{(\OPE)}_{\pert}(\freq,\mom) \coloneq (\freq^2 - \mom^2)^{\nu} \prn{1 + \sum^{n_{\max}}_{n=1} \rhovacsub_{T^{n}}(\freq,\mom)}.
\end{equation}
Based on the arguments given above, it is natural to expect that the equality
\begin{equation}
\label{eq:rho-pert-Vir-equals-OPE}
\rho^{(\Vir)}_{\pert}(\freq,\mom) \, \overset{!}{=} \,
\rho^{(\OPE)}_{\pert}(\freq,\mom)
\end{equation}
holds for any integer $\Delta$. Notice that we did not include the spectrum condition $\theta(\freq^2-\mom^2)$ in the definition of $\rho^{(\OPE)}_{\pert}(\freq,\mom)$. We expect this spectrum condition to arise from $\rho_{\np}(\freq,\mom)$ in holographic spectral function. This leads us to
\theorembox{
  \begin{claim}
    \label{conj:rho-factorized}
    For holographic \CFT{d} with even $d$, the exact spectral function of identical scalars with integer dimension factorizes into perturbative and non-perturbative contributions
    \begin{equation}
    \rho(\freq,\mom) = \rho^{(\OPE)}_{\pert}(\freq,\mom) \rho_{\np}(\freq,\mom),
    \end{equation}
    with $\rho^{(\OPE)}_{\pert}(\freq,\mom)$ given by~\eqref{eq:rho-OPE-holo-CFT}, and $\rho_{\np}(\freq,\mom)$ given by~\eqref{eq:rho-np-holo-CFT}.
  \end{claim}
}
\emph{Remark.} While the notion of perturbative and non-perturbative corrections is specific to the large momentum limit, the~\cref{conj:rho-factorized} concerns the exact spectral function at finite momentum. Thus, it may be viewed as the generalization of~\eqref{eq:spectral-function-CFT2-integer-dimension} for $d=2$ to arbitrary even $d$. The $\rho^{(\OPE)}_{\pert}(\freq,\mom)$ part is fixed by a near-boundary expansion, independent of the connection problem between boundary and horizon. Information about the connection problem is captured in $\rho_{\np}(\freq,\mom)$ via the monodromy $\cos(2\pi\cmexp)$.

\paragraph{The two-sided correlator:} Another natural object of interest in thermal systems is the two-sided thermofield double correlator,
\begin{equation}
C(t,\vb{x}) =
\expval{\mathcal{O}\pqty{t + \frac{i}{2}\,\beta,\vb{x}}\,
  \mathcal{O}\pqty{0,\vb{0}}}
\end{equation}
This observable is directly related to the spectral function; in fact in the  Fourier domain the relation is simply
\begin{equation}
C(\freq,\mom) = \frac{\rho(\freq,\mom)}{\sinh(\pi\,\freq)}.
\end{equation}
Thence the analog of~\cref{conj:rho-factorized} for two-sided correlator is
\begin{equation}
C(\freq,\mom)
= \rho^{(\OPE)}_{\pert}(\freq,\mom) C_{\np}(\freq,\mom)\,,
\end{equation}
with
\begin{equation}\label{eq:Cnpdef}
C_{\np}(\freq,\mom)
= \frac{1}{\cosh\prn{\pi\,\freq}+(-1)^{\nu}\cos(2\pi\cmexp)}.
\end{equation}
%

\subsubsection{Argument for~\cref{conj:rho-factorized} using apparent singularity}

In general, the spectral function has poles where the denominator of $\rho_{\np}(\freq,\mom)$ vanishes
\begin{equation}
\label{eq:pole-rhonp-condition}
\cosh\prn{\pi\,\freq}+(-1)^{\nu}\cos(2\pi\cmexp) =0.
\end{equation}
This is the exact quantization condition for quasinormal modes and anti-quasinormal modes, the poles of spectral function. However, we observe that~\eqref{eq:pole-rhonp-condition} is also satisfied for the special locus in $\{\freq,\mom\}$ where the boundary singular point is an apparent singularity. To see this, recall that an apparent singularity of order $\nu$ corresponds to the following heavy degenerate operator (cf.~\cite{Jia:2024zes} and references therein for more details)
\begin{equation}
V_{\bdy} = V_{\expval{1,\nu}}.
\end{equation}
The fusion rule of the degenerate operator implies
\begin{equation}
\cmexp \in \Half{i\freq} + \{ -\Half{\nu-1}, -\Half{\nu-1} +1, \dots, \Half{\nu-1}\},
\end{equation}
leading to~\eqref{eq:pole-rhonp-condition}. Since the spectral function does not have poles at the special locus of apparent singularity, it must be that $\rho^{(\Vir)}_{\pert}(\freq,\mom)$ vanishes there. Let $Q^{(\app)}_{\nu}(\alpha,\beta)$ be the degree $\nu$ polynomial with unit coefficient for $\alpha^{\nu}$ so that
\begin{equation}
\label{eq:Qpoly-apparent-singularity}
Q^{(\app)}_{\nu}(\p^2,\freq^2) = 0 \Longleftrightarrow \text{$z_{\bdy}$ is an apparent singularity of degree $\nu$},
\end{equation}
where $\p^2 = \freq^2-\mom^2$. We have $\rho^{(\Vir)}_{\pert} = F\prn{Q^{(\app)}_{\nu}}$ for some function $F(x)$ satisfying $F(0) = 0$. We determine $F(x)$ by comparing with the OPE contribution. Using the stress-tensor OPE coefficients and spectral function block discussed in~\cref{sec:thermal-OPE}, we obtain
\begin{equation}
\label{eq:rho-pert-Q-poly}
\rho^{(\OPE)}_{\pert}(\freq,\mom) \eqqcolon Q^{(\OPE)}_{\nu}(\p^2,\freq^2)
\end{equation}
with $Q^{(\OPE)}_{\nu}(\alpha,\beta)$ a degree $\nu$ polynomial. It turns out that the choice $F(x) = x$ precisely produces the OPE contribution:
\begin{claim}
  \label{claim:Qapp-equals-Qope}
  The following equality holds
  \begin{equation}
  Q^{(\app)}_{\nu}(\alpha,\beta) = Q^{(\OPE)}_{\nu}(\alpha,\beta).
  \end{equation}
\end{claim}
\noindent We have verified~\cref{claim:Qapp-equals-Qope} by direct computations up to $\nu=d/2+6$ for $d=4,6,8$. Explicit forms of $Q_{\nu}(\alpha,\beta)$ are given in~\cref{tab:Q-poly}. Note that as is the case with $Q^{(\OPE)}_{\nu}$, $Q^{(\app)}_{\nu}$ is also fixed by a near-boundary expansion.

\begin{claim}
  The following equality holds
  \begin{equation}
  \rho^{(\Vir)}_{\pert}(\freq,\mom) = Q^{(\app)}_{\nu}(\p^2,\freq^2).
  \end{equation}
\end{claim}

The previous discussion applies to $\nu\geq1$.\footnote{There is no apparent singularity for $\nu=0$.} For $\nu=0$, $\rho^{(\OPE)}_{\pert}$ is unity. That $\rho^{(\Vir)}_{\pert}$ also evaluates to unity follows from the fact that $\partial_{\mexp_{\bdy}}\VBcl |_{\mexp_{\bdy}=0}$ vanishes, as $\VBcl$ only depends on $\mexp_{\bdy}$ through $\mexp_{\bdy}^2$.
 
\begin{table}[h!]
  \centering
  \setlength{\tabcolsep}{4pt}
  \begin{tabular}{|>{\centering\arraybackslash}m{0.12\textwidth}|>{\centering\arraybackslash}m{0.36\textwidth}|>{\centering\arraybackslash}m{0.36\textwidth}|}
    \hline
    $\nu-\frac{d}{2}$                            & $d=4$                               & $d=6$ \\
    \hline\hline
    $\leq 0$                                     & \multicolumn{2}{c|}{$\alpha^{\nu}$}         \\
    \hline
    1                                            & $\alpha^3 - \alpha + 4\beta$        &
    $\alpha^4-\frac{128}{243}\alpha+\frac{256}{81}\beta$                                       \\
    \hline
    2                                            &
    $\alpha^4 - 6\alpha^2 + 24\alpha\beta + 9$   &
    $\alpha^5-\frac{1024}{243}\alpha^2+\frac{2048}{81}\alpha\beta$                             \\
    \hline
    3                                            &
    \begin{tabular}[c]{@{}c@{}}
      $\alpha^5 - 21\alpha^3 + 84\alpha^2\beta$ \\
      $+108\alpha + 144\beta$
    \end{tabular} &
    \begin{tabular}[c]{@{}c@{}}
      $\alpha^6-\frac{512}{27}\alpha^3$ \\
      $+\frac{1024}{9}\alpha^2\beta+\frac{409600}{6561}$
    \end{tabular}                             \\
    \hline
    4                                            &
    \begin{tabular}[c]{@{}c@{}}
      $\alpha^6-56\alpha^4+224\alpha^3\beta$ \\
      $+784\alpha^2+640\alpha\beta+1600\beta^2$
    \end{tabular}    &
    \begin{tabular}[c]{@{}c@{}}
      $\alpha^7-\frac{5120}{81}\alpha^4+\frac{10240}{27}\alpha^3\beta$ \\
      $+\frac{52428800}{59049}\alpha+\frac{26214400}{19683}\beta$
    \end{tabular}            \\
    \hline
  \end{tabular}
  \caption{Explicit forms of the polynomial $Q_{\nu}(\alpha,\beta)$. The two definitions~\eqref{eq:Qpoly-apparent-singularity} and~\eqref{eq:rho-pert-Q-poly} coincide.}
  \label{tab:Q-poly}
\end{table}

\subsubsection{Verification of~\cref{conj:rho-factorized} using Virasoro block ($d=4$)}

In the following we verify~\eqref{eq:rho-pert-Vir-equals-OPE} at various values of integer dimensions for $d=4$ by direct computation of semiclassical Virasoro blocks. In this case, we find it convenient to perform a further M\"obius transformation $z^{\prime} = (1-z)/2$ to map the singular points to $\{0,x,1,\infty\} = \{\hor,\bdy,\hor^{(1)},\sing\}$ with $x=1/2$ (cf.~\cref{app:Schrodinger-form}).\footnote{Under $z^{\prime} = f(z)$, the block transforms as $\VBcl(z^{\prime}_{i}) = \VBcl(z_{i}) - \sum_{i} \delta_{i} \log f^{\prime}(z_{i})$. This will only shift $\VBcl_{\bdy}$ by a $\freq,\mom$ independent constant (since $\sum_i\, \delta_i$ is $\freq$ independent), and therefore rescale $\Gret(\freq,\mom)$ by an overall constant.} The relevant four-point semiclassical block is
\begin{equation}
\mathcal{W}
=
\begin{tikzpicture}[
    baseline={(0,0)},
    x=0.80cm,
    y=1cm,
    line cap=round,
    line width=0.8pt
  ]
  \draw (0,0) -- (8,0);

  \fill (2.8,0) circle (0.7pt);
  \fill (5.6,0) circle (0.7pt);

  \draw (2.8,0) -- (2.8,1.35);
  \draw (5.6,0) -- (5.6,1.35);

  \node[above] at (2.8,1.35) {$V_{\mathrm{bdy}}(x)$};
  \node[above] at (5.6,1.35) {$V^{(1)}_{\mathrm{hor}}(1)$};

  \node[anchor=north, yshift=-3pt] at (4.2,0) {$\sigma$};

  \node[left] at (0,0) {$V_{\mathrm{hor}}(0)$};
  \node[right] at (8,0) {$V_{\mathrm{sing}}(\infty)$};
\end{tikzpicture}
\,.
\end{equation}
The accessory parameter is $\acc = -\p^2$ with $\p^2 = \freq^2 - \mom^2$. The intermediate momentum $\cmexp$ is determined from $\acc = x\partial_{x}\VBcl$.

In verifying~\eqref{eq:rho-pert-Vir-equals-OPE}, we take the following shortcut: instead of first solving $\cmexp$ from $\acc = x\partial_{x}\VBcl$ and then verifying the identity for any $\{\freq,\mom\}$, we may as well verify the identity holds for any $\{\freq,\cmexp\}$ to bypass the need for solving $\cmexp$. The key point is to utilize $\p^2 = -\acc = -x\partial_{x}\VBcl$, so that the identity to verify is
\begin{equation}
\label{eq:block-identity-to-check}
P_{2\nu}(i\freq/2,\cmexp) \exp\prn{-\partial_{\mexp_{\bdy}}\VBcl} \overset{!}{=} Q_{\nu}\prn{-x\partial_{x}\VBcl,\freq^2}.
\end{equation}

\begin{figure}[htbp!]
  \begin{subfigure}[b]{0.5\textwidth}
    \centering
    \includegraphics[width=\linewidth]{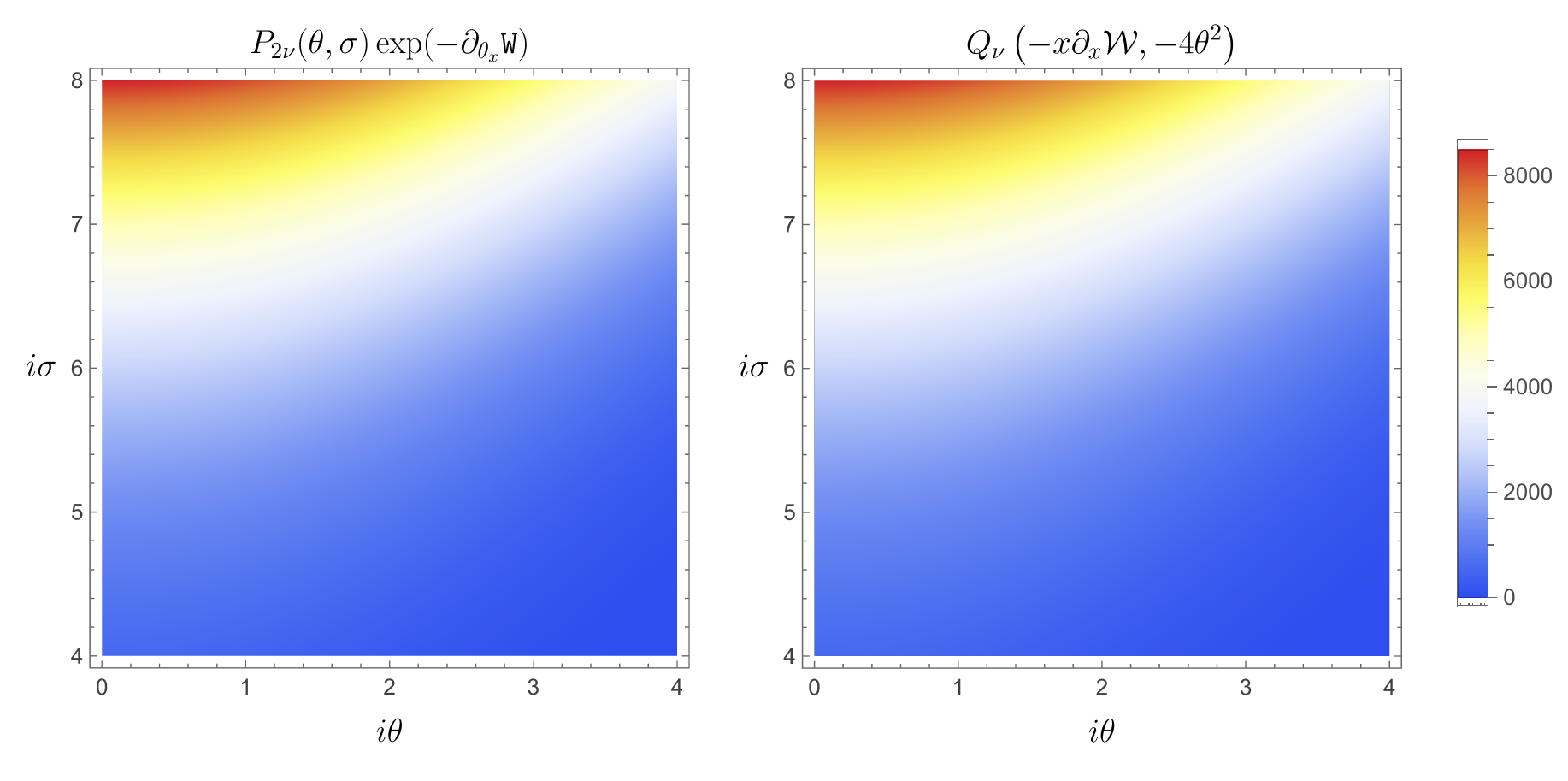}
    \caption{$\nu=2$}
  \end{subfigure}
  \hfill
  \begin{subfigure}[b]{0.5\textwidth}
    \centering
    \includegraphics[width=\linewidth]{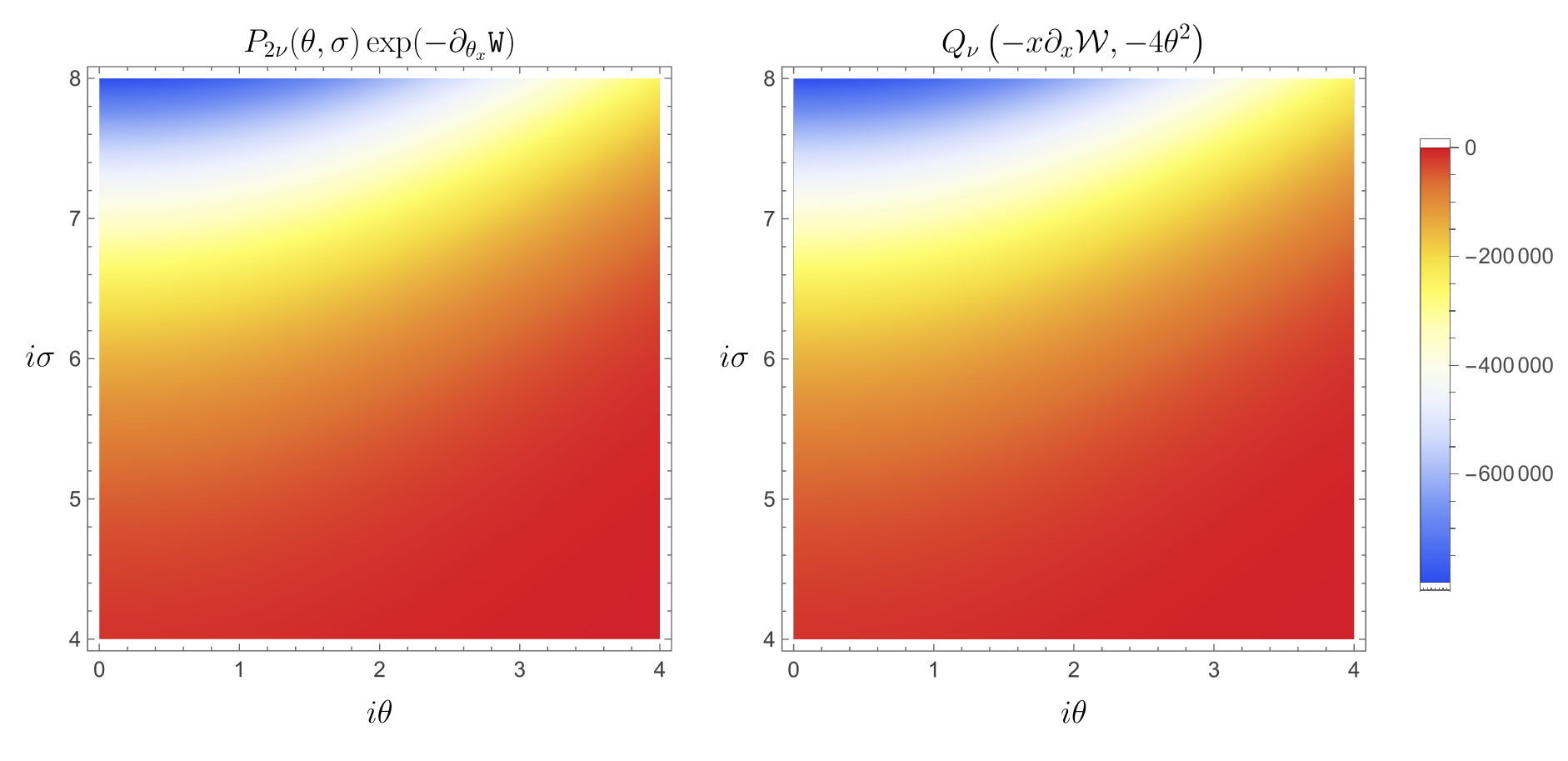}
    \caption{$\nu=3$}
  \end{subfigure}
  \hfill
  \begin{subfigure}[b]{0.5\textwidth}
    \centering
    \includegraphics[width=\linewidth]{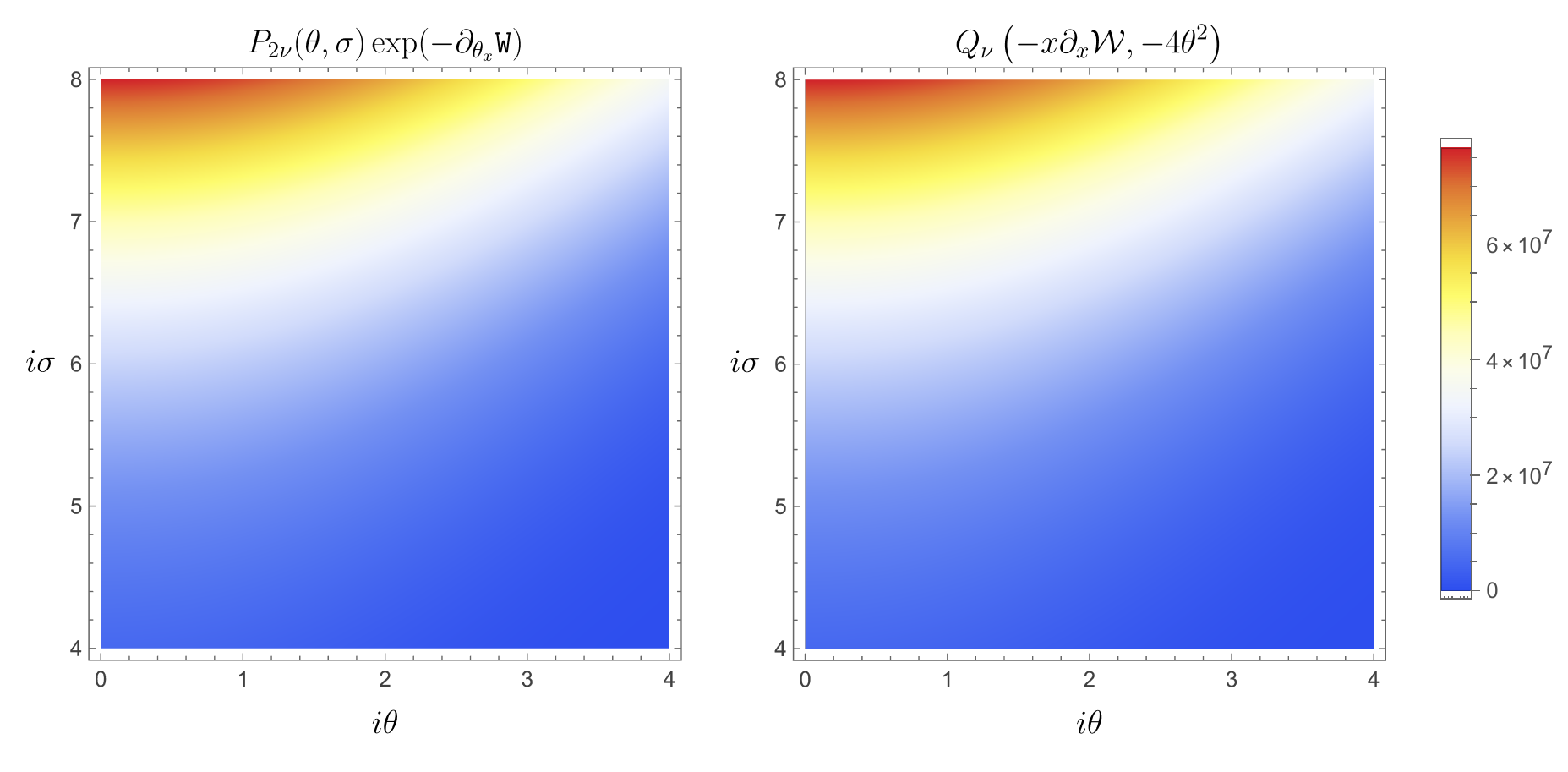}
    \caption{$\nu=4$}
  \end{subfigure}
  \hfill
  \begin{subfigure}[b]{0.5\textwidth}
    \centering
    \includegraphics[width=\linewidth]{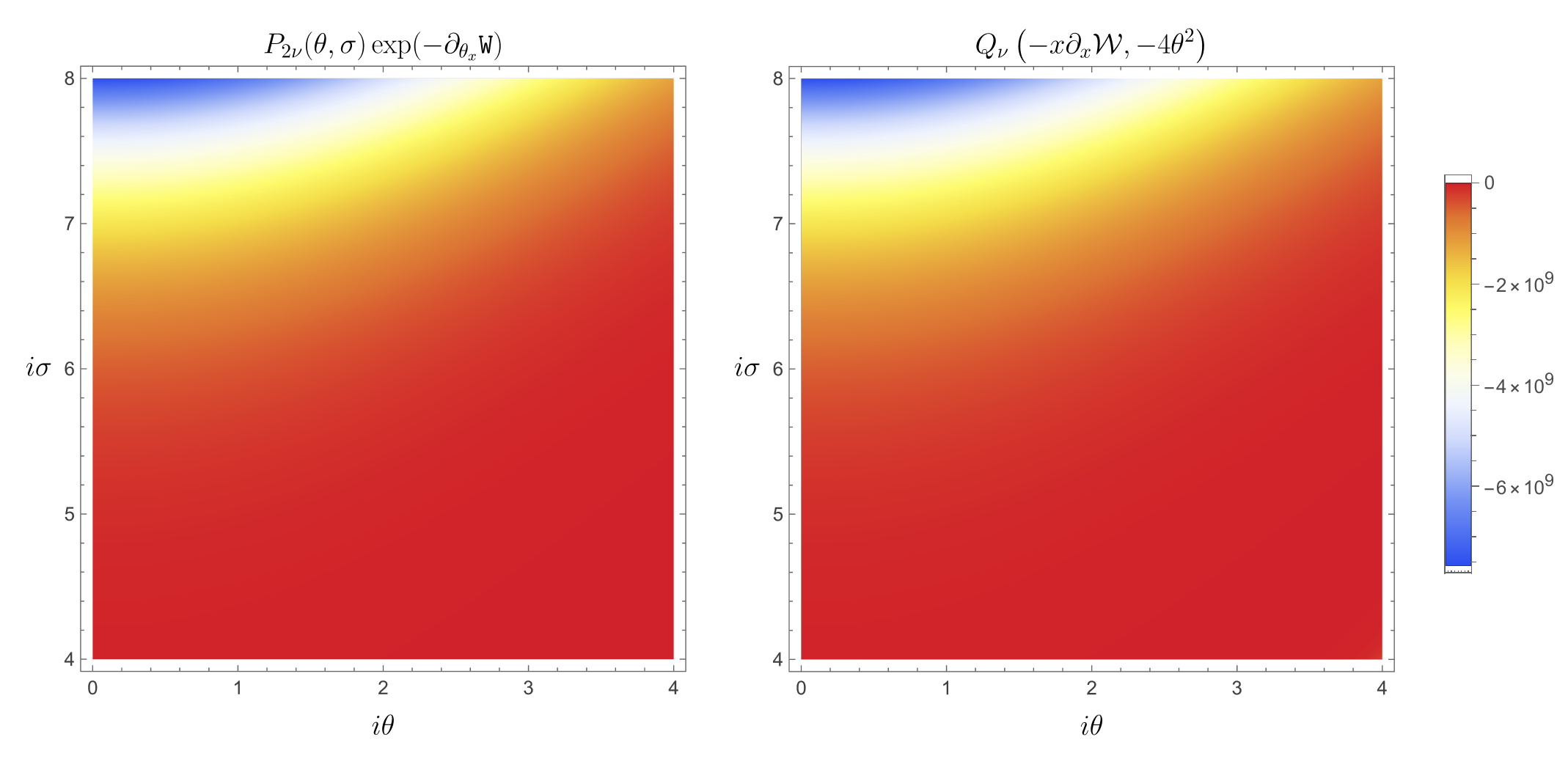}
    \caption{$\nu=5$}
  \end{subfigure}
  \caption{Numerical checks of~\cref{claim:block-identity}. The LHS and RHS of the identity in~\cref{claim:block-identity} are computed for purely imaginary $\{\cmexp,\mexp\}$, in the range $\Im(\cmexp) \in [4,8]$ and $\Im(\mexp) \in [0,4]$ with $100 \times 100$ data points. The semiclassical Virasoro blocks are computed using the Zamolodchikov recursion relation to 7th order in elliptic nome.}
  \label{fig:block-numerics}
\end{figure}

\begin{claim}
  \label{claim:block-identity}
  Let $\VBcl$ be spherical four-point $s$-channel semiclassical Virasoro block, with external momenta $\{\mexp_{0},\mexp_{x},\mexp_{1},\mexp_{\infty}\} = \{\mexp,\nu/2,i\mexp,0\}$, internal momentum $\cmexp$, and $\nu \in \Z_{\geq0}$. Let $\VBclpow$ be $\VBcl$ with the $log(x)$ piece removed, i.e., $\VBclpow = \VBcl - (\delta_{\cmexp} - \delta_{0} - \delta_{x})\log(x)$. The following identity holds at $x=\half$:
  \begin{equation}
  \label{eq:block-identity-general-nu}
  P_{2\nu}(\mexp,\cmexp) \exp\prn{-\partial_{\mexp_{x}}\VBclpow} = Q_{\nu}\prn{-x\partial_{x}\VBcl,-4\mexp^2},
  \end{equation}
  with the polynomial $P_{2\nu}(\mexp,\cmexp)$ defined in~\eqref{eq:P2v-poly}, and the polynomial $Q_{\nu}(\alpha,\beta)$ defined in~\eqref{eq:rho-pert-Q-poly}.
\end{claim}
Compared to~\eqref{eq:block-identity-to-check}, in the LHS of~\eqref{eq:block-identity-general-nu} we have removed the $\log(x)$ piece in $\VBcl$. This only changes the LHS of~\eqref{eq:block-identity-to-check} by an overall constant, and turns out to give the correct normalization for the identity to hold. To verify the identity, we compute the semiclassical block using the Zamolodchikov recursion relation (cf.~\cref{app:block-recursion} for details), and compare the LHS and RHS in the claimed identity at various values of $\{\mexp,\cmexp\}$ (cf.~\cref{fig:block-numerics}).

For $\nu=1$, it turns out that the identity in~\eqref{eq:block-identity-general-nu} holds for more generic external momenta and cross-ratio:
\begin{claim}
  Let $\VBcl$ be spherical four-point $s$-channel semiclassical Virasoro block, with external momenta $\{\mexp_{0},\mexp_{x},\mexp_{1},\mexp_{\infty}\}$, internal momentum $\cmexp$, and $\mexp_{x} = \half$. The following identity holds:
  \begin{equation}
  \label{eq:nu-1-identity}
  (\cmexp^2 - \mexp^2_{0})\exp\prn{-\partial_{\mexp_{x}}\VBclpow} = -x\partial_{x}\VBcl.
  \end{equation}
\end{claim}
Note that in this case $-x\partial_{x}\VBcl = (\cmexp^2 - \mexp^2_{0}) - x\partial_{x}\VBclpow$. We have verified this identity perturbatively in cross-ratio up to fifth order. That is, the LHS and RHS of~\eqref{eq:nu-1-identity} match as
\begin{equation}
\mathrm{LHS} = \mathrm{RHS} = (\cmexp^2 - \mexp^2_{0}) \prn{1 + \sum^{\infty}_{k=1} c_{k} x^{k}}.
\end{equation}
The first two coefficients read
\begin{equation}
c_{1} = \frac{\delta_{1}+\delta_{0}-\delta_{\infty}}{2\delta_{\cmexp}},
\end{equation}
and
\begin{equation}
\begin{split}
c_{2}
 & =
\frac{1}{8 \delta^{3}_{\cmexp}(3+4\delta_{\cmexp})} \left(-5 \delta _0 \delta _1^2 \delta
_{\sigma }+3 \delta _1^2 \delta
_{\sigma }+6 \delta _0 \delta _1
\delta _{\sigma }^2+12 \delta _1
\delta _{\sigma }^2-3 \delta _0
\delta _{\sigma }^2+3 \delta _0
\delta _1^2 \right. \\
 &
\quad \left.
{}+  6 \delta _0 \delta _{\sigma }^2
\delta _{\infty }-12 \delta
_{\sigma }^2 \delta _{\infty
}+10 \delta _0 \delta _1 \delta
_{\sigma } \delta _{\infty }-6
\delta _1 \delta _{\sigma }
\delta _{\infty }+13 \delta
_{\sigma }^4-\delta _0 \delta
_{\sigma }^3+18 \delta _1 \delta
_{\sigma }^3+9 \delta _{\sigma
}^3   \right.       \\
 &
\quad \left. {}
+ \delta _1^2 \delta _{\sigma}^2
-6 \delta _0 \delta _1 \delta
_{\infty }
-14 \delta _{\sigma }^3 \delta
_{\infty }+\delta _{\sigma }^2
\delta _{\infty }^2-2 \delta _1
\delta _{\sigma }^2 \delta
_{\infty }-5 \delta _0 \delta
_{\sigma } \delta _{\infty }^2+3
\delta _{\sigma } \delta
_{\infty }^2+3 \delta _0 \delta
_{\infty }^2 \right).
\end{split}
\end{equation}
%

\subsection{Exact position space correlator}

The Euclidean thermal two-point function can be obtained from the spectral
function via
\begin{equation}
\gE(\tau,\vb{x})
=  \int_{\R^{d-1}} d\vb{k} \, e^{i \vb{k} \cdot \vb{x}} \int^{\infty}_{0} d\omega \rho(\omega,k) \frac{\cosh\brk{\prn{\tau-\half}\omega}}{\sinh\prn{\Half{\omega}}}.
\end{equation}
From the exact holographic spectral function in~\cref{conj:rho-factorized}, we then have
\begin{equation}
\label{eq:exact-position-space-correlator}
\gE(\tau,\vb{x}) = F^{\Delta_{\phi}}_{\id}(\tau,\vb{x}) + \sum^{n_{\max}}_{n=1}\sum^{2n}_{J=0} a_{T^{n}}(J) F^{\Delta_{\phi}}_{nd,J}(\tau,\vb{x})
\end{equation}
with
\begin{equation}
F^{\Delta_{\phi}}_{\Delta,J}(\tau,\vb{x}) =  \int_{\R^{d-1}} d\vb{k} e^{i \vb{k} \cdot \vb{x}} \int^{\infty}_{0} d\omega \GspecNorm^{\Delta_{\phi}}_{\Delta,J}(\omega^2-k^2)^{\Delta_{\phi}-\Half{d}} \GspecOp_{\Delta,J}(\omega,k) \frac{\cosh\brk{\prn{\tau-\half}\omega}}{\cosh\prn{\Half{\omega}} + (-1)^{\Delta_{\phi}-\Half{d}} \cos(2\pi\cmexp) }.
\end{equation}
This is reminiscent of the results obtained in~\cite{Barrat:2025twb} for integer $\Delta_{\phi}$ at $\vb{x}=0$, where $\gE(\tau,\vb{x})$ is represented in terms of stress-tensor OPE data and generalized free field correlator. In the aforementioned result, however, the order of truncation in their principal part is different from our $n_{\max}$, and the divergent OPE coefficients do contribute in a regularized part. It would be interesting to understand the precise relation.

\subsection{Asymptotics in large momentum limit}

While motivated by the large momentum limit, we have so far focused on determining exact expressions for the spectral function in terms of the intermediate momentum $\cmexp$ in the semiclassical Virasoro block. To study the large momentum asymptotics using the exact expressions, the key question to address is: \emph{Does $\cmexp$ have well-defined asymptotics in the large momentum limit? If so, what is an efficient way of computing it?}

We find it hard to address this question from the definition of $\cmexp$ as a solution to the Zamolodchikov relation. Even in the large momentum limit, we are unaware of a closed-form expression of the relevant semiclassical Virasoro block, thereby lacking a non-perturbative solution of $\cmexp$. One can attempt to proceed by assuming that $\cmexp$ admits a cross-ratio expansion, and solve the Zamolodchikov relation perturbatively in cross-ratio. But a cross-ratio expansion of $\cmexp$ is not terribly useful for understanding asymptotics of $\cmexp$, unless one manages to find a way to resum the expansion. Furthermore, for planar black hole geometry, we are not in the OPE regime of Virasoro block, and this procedure of solving $\cmexp$ perturbatively in cross-ratio might not be valid in the first place.

On the other hand, we know that $\cmexp$ is related to the monodromy around physical horizon and AdS boundary, $\Tr M_{\bdy,\hor} = -2\cos(2\pi\cmexp)$. So we may as well study the asymptotics of monodromy directly. The governing black hole wave equation in the large momentum limit is amenable to exact WKB analysis. A general result (cf.~Theorem 3.5 of~\cite{KawaiTakei2005}) in exact WKB analysis states that the monodromy invariants of a second order Fuchsian equation can in general be computed in terms of the Borel-summed all-orders WKB periods (or Voros periods) $\BSV_{I}$ and local exponents $\theta_{i}$:
\begin{equation}
\Tr M = \Tr M \prn{\BSV_{I}, \theta_{i}}.
\end{equation}
This is an exact statement valid at finite momentum. Furthermore, the asymptotics of the Borel-summed periods recover the original asymptotic series of the all-order WKB periods. For the special integer dimensions whose spectral function only depends on the monodromy invariant $\cos(2\pi\cmexp)$, we thus have a systematic way of computing the large momentum asymptotics, along with an alternative exact description.

\section{Review of exact WKB method}\label{sec:EWKBreview}

Our discussion thus far has led us to motivate a study of the exact WKB asymptotics of the spectral function. In order to do so, we first give a self-contained review of exact WKB, which is a necessary preparation for computation of monodromy. We largely follow the exposition of exact WKB given in~\cite{Iwaki:2014vad,Iwaki_2015} where the reader can find additional details. In addition, we will exploit the recipe for computing monodromy of a second order Fuchsian equation using exact WKB as given in~\cite{KawaiTakei2005}.

\subsection{WKB solutions}

Consider the Schr\"odinger equation with a large parameter $\eta \in \R$ defined on a Riemann surface $\Sigma$:
\begin{equation}
\label{eq:Schrodinger-eqn}
\begin{split}
\psi^{\prime\prime}(z) - \eta^2 \,Q(z) \,\psi(z) =0, \quad Q(z) = Q_{0}(z) + \eta^{-2}\, Q_{2}(z).
\end{split}
\end{equation}
The exact WKB analysis is applicable when $\Sigma$ is a general compact Riemann surface. For our purpose, we focus on the case $\Sigma = \mathbb{P}^{1}$. In QM context, one has $\eta = \hbar^{-1}$ and $Q = Q_{0} = V-E$. When we adapt this to our holographic thermal Green's function problem, we will have $\eta = \freq$.

The formal WKB solutions of~\eqref{eq:Schrodinger-eqn} are
\begin{equation}
\begin{split}
\psi_{\pm}(z,\eta) = \frac{1}{\sqrt{\Podd(z,\eta)}} \exp\prn{\pm \int^{z}_{z_{0}} dz^{\prime} \Podd(z,\eta)},
\end{split}
\end{equation}
with $\Podd(z)$ being the odd part of all-orders “momentum” $P(z,\eta)$\footnote{
  In the standard quantum mechanical context one identifies the leading term with the classical momentum; $i P_{-1} = \sqrt{E-V}$.}
\begin{equation}
\begin{split}
 & P(z,\eta) = \sum^{\infty}_{n \geq -1} \eta^{-n} P_{n}(z)
\\
 & \Podd(z ,\eta) = \sum_{\text{odd} \ n \geq -1} \eta^{-n} P_{n}(z).
\end{split}
\end{equation}
The all-orders momentum $P(z,\eta)$ is defined order by order in $\eta$ from the Riccati equation associated with~\eqref{eq:Schrodinger-eqn}, viz.,
\begin{equation}
\frac{dP}{dz} + P^2 = \eta^2\, Q,
\end{equation}
with $P_{-1}(z) = \sqrt{Q_{0}(z)}$. Higher order $P_{n}(z)$ with $n\geq0$ are determined recursively from
\begin{equation}
P_{n} = \prn{2P_{-1}}^{-1}\prn{-\sum^{n-1}_{m=0} P_{m}P_{n-1-m} - \frac{dP_{n-1}}{dz} + \delta_{n,1} Q_{2}(z)}.
\end{equation}
For example,
\begin{equation}
P_{0} = -\frac{Q^{\prime}_{0}}{4Q_{0}}, \quad P_{1} = \frac{-5\prn{Q^{\prime}_{0}}^2 + 4 Q_{0}Q^{\prime\prime}_{0}}{32Q^{5/2}_{0}} + \frac{Q_{2}}{2 Q^{1/2}_{0}}.
\end{equation}
The even part of $P(z,\eta)$ is given by the logarithmic derivative of the odd part. The resulting formal WKB solutions are asymptotic series in $\eta$. The goal of the exact WKB analysis is to complete this into a transseries expression with the selfsame asymptotics.

We will need the notion of the  WKB curve, which is defined by
\begin{equation}
\Cwkb = \{(z,\nu) \mid  \nu^2 = Q_{0}(z) \} \,.
\end{equation}
The curve is a double cover of $\Sigma = \mathbb{P}^{1}$ branched at zeros and odd-order poles of the quadratic differential
\begin{equation}\label{eq:Strebel}
\phi = Q_{0}(z) \, dz^{2}.
\end{equation}
The all-orders WKB differential
\begin{equation}
\lambda = \Podd(z,\eta)\, dz
\end{equation}
is a single-valued one-form on $\Cwkb$.

\subsection{Borel sum of formal WKB solutions}

As asymptotic series in $\eta$, the formal WKB solutions take the form
\begin{equation}
\psi_{\pm}(z,\eta) = e^{\pm \eta\, s(z)} \,\eta^{-\half} \, \varphi_{\pm}(z,\eta),
\end{equation}
with
\begin{equation}
\begin{split}
\varphi_{\pm}(z,\eta) = \sum^{\infty}_{m \geq 0} \eta^{-m} \varphi_{\pm,m}(z)
\end{split}
\end{equation}
and
\begin{equation}
s(z) = \int^{z}_{z_0} P_{-1}(z).
\end{equation}

The Borel sum of formal WKB solutions can be defined in terms of Borel sum of $\varphi_{\pm}(z,\eta)$
\begin{equation}
\BS\brk{\psi_{\pm}}(z,\eta) \, \coloneqq\,
e^{\pm \eta s(z)} \, \eta^{-\half}\,  \BS\brk{\varphi_{\pm}}(z,\eta) \,.
\end{equation}
We recall that the Borel sum is in general defined by a Borel transform followed by a Laplace transform.
The Borel transform of $\varphi(z,\eta)$ is defined by (suppressing the $\pm$ subscript in $\varphi$)
\begin{equation}
\BT\brk{\varphi}(z,\bv) \coloneqq
\varphi_{0}(z)
+ \sum^{\infty}_{m \geq 1}\,  \frac{\varphi_{m}(z)}{(m-1)!} \, \bv^{m-1}.
\end{equation}
The Borel sum of $\varphi$ is defined by
\begin{equation}
\BS\brk{\varphi}(z,\eta)
\,\coloneqq\,
\varphi_{0}(z) + \int_{\R_{+}}\, d\bv \, e^{-\eta \,\bv} \, \BT\brk{\varphi}(z,\bv).
\end{equation}

The Borel sum is not well-defined when there is singularity on the integration contour for the Laplace transform.
In this case, one can deform the integration contour by an angle $\pha$ to define
\begin{equation}
\BS_{\pha}\brk{\varphi}(z,\eta)
\,\coloneqq\,
\varphi_{0}(z) + \int_{e^{-i\pha}\R_{+}}\,  d\bv e^{-\eta \,\bv} \, \BT\brk{\varphi}(z,\bv) \,.
\end{equation}
This is referred to as Borel sum in direction $\pha$. However, the Borel sum of $\varphi(z,\eta)$ in direction $\pha$
is equivalent to the usual Borel sum of $\varphi(z,\eta)$ with $\eta \to \eta \,e^{i \pha}$. To see this, define
\begin{equation}
\varphi^{(\pha)}\prn{z,\eta}
\,\coloneqq\, \varphi(z,\eta \,e^{i\pha})
= \sum^{\infty}_{m \geq 0} \,\eta^{-m} \, e^{-im\pha}\,  \varphi_{m}(z).
\end{equation}
One may verify from the definition that
\begin{equation}
\BT\brk{\varphi^{(\pha)}}(z,\bv)
= e^{-i\pha}\, \BT\brk{\varphi}(z,e^{-i\pha}\bv), \quad \BS_{\pha}\brk{\varphi}(z,\eta)
= \BS\brk{\varphi^{(\pha)}}\prn{z,\eta}.
\end{equation}
Therefore, one can define the Borel sum of WKB solutions in direction $\pha$ by
\begin{equation}
\BS_{\pha}\brk{\psi_{\pm}}(z,\eta)
\,\coloneqq\,
e^{\pm \eta e^{i\pha} s(z)} \prn{\eta e^{i \pha}}^{-\half} \, \BS_{\pha}\brk{\varphi_{\pm}}(z,\eta).
\end{equation}
This is then equivalent to the usual Borel sum of $\psi_{\pm}(z,\eta)$ with $\eta \to \eta \, e^{i\pha}$.

\subsection{Stokes graphs and Borel summability of WKB solutions}

The next question we need to address is how one ascertains the directions in which the Borel sum is well-defined.
The useful ingredient here is the \emph{Stokes graph}, which encodes the information regarding the Borel summability of formal WKB solutions. We now turn to its definition.

The Stokes graph $G(\phi)$ on $\Sigma$ for the Schr\"odinger equation is defined using the quadratic differential
$\phi = Q_{0}(z) \,dz^{2}$ introduced in~\eqref{eq:Strebel}, which is the Strebel differential for the WKB curve.
The edges of the Stokes graph  $G(\phi)$ are horizontal trajectories of $\phi$. The Stokes graph also coincides with the rank two case of spectral network of~\cite{Gaiotto:2009hg,Gaiotto:2012rg}.

Let $P$ be the zeros and poles of $\phi$, i.e.,
\begin{equation}
P = \mathrm{Crit}(\phi).
\end{equation}
These are the vertices in the Stokes graph $G(\phi)$. In our examples, these will consist of simple zeros (i.e., simple turning points) $P_{0}$, simple poles $P_{s}$, and double poles $P_{\infty}$
\begin{equation}
P = P_{0} \cup P_{s} \cup P_{\infty}.
\end{equation}
Consider the trajectory emanating from a branched point $v \in P_{0} \cup P_{s}$ defined by\footnote{Some older literature refer to the horizontal trajectories defined in~\eqref{eq:hortraj} to be the anti-Stokes lines. We will, however, stick to the modern usage.}
\begin{equation}\label{eq:hortraj}
\Im \int^{p}_{v} \sqrt{\phi} = 0.
\end{equation}
These trajectories are referred to as Stokes lines, and form the edges of the Stokes graph $G(\phi)$. Typically, the trajectory will end at a double pole in $P_{\infty}$. In a critical case, the trajectory can end at another branched point in $P_{0} \cup P_{s}$. Such a trajectory is referred to as a $\emph{saddle trajectory}$. The faces of the Stokes graph are referred to as Stokes regions.

As a simple example, let $\phi(z) = z\, dz^2$. This is the Airy differential, which controls the local behaviour at any simple turning point and constitutes the basis for textbook discussions of the Bohr-Sommerfeld formula. The Stokes graph comprises one vertex at the origin $z=0$ and three symmetric rays directed along the trajectories $\arg(z) = \frac{2}{3}\,\pi \, n$, for $n \in \{0,1,2\}$. The three Stokes domains correspond to the three different asymptotics of the Airy function.  

The Borel summability of formal WKB solutions in direction $\pha$ is related to the Stokes graph of $e^{2i\pha}\phi$
\begin{equation}
G_{\pha} \coloneqq G\prn{e^{2i\pha}\phi}
\end{equation}
as follows.
\begin{theorem}
  If the Stokes graph $G_{\pha}$ is free of saddle trajectories, the formal WKB solutions $\psi_{\pm}(z,\eta)$ are Borel summable in direction $\pha$ in each Stokes region. Their Borel sum  $\BS_{\pha}\brk{\psi_{\pm}}(z,\eta)$ gives analytic solutions of the Schr\"odinger equation~\eqref{eq:Schrodinger-eqn} with $\eta \to \eta\, e^{i\pha}$ in each Stokes region.
\end{theorem}

In general, if at a critical phase $\pha = \pha_{c}$ the Stokes graph $G_{\pha_{c}}$ contains saddle trajectories, the Stokes graphs at deformed phases $\pha^{\pm}_{c}$ are saddle-free. This is known as \emph{saddle reduction}. The topology of Stokes graphs changes across the critical phase.

When the formal WKB solutions are Borel-summable, we denote their Borel sums as
\begin{equation}
\Psi_{\pm,\pha}(z,\eta) = \BS_{\pha}\brk{\psi_{\pm}}(z,\eta).
\end{equation}
%

\subsection{Connection formulae}

Suppose the Stokes graph $G_{\pha}$ is saddle-free. Here we describe the connection formulae for the Borel sums $\Psi_{\pm,\pha}(z,\eta)$ across Stokes lines, and other ingredients that will be the basic building blocks in the computation of monodromy using exact WKB. In the following discussion, we drop the subscript $\pha$ in $\Psi_{\pm,\pha}$. We denote the formal WKB solutions normalized at point $p$ and their Borel sums as $\psi_{\pm,p}$ and $\Psi_{\pm,p}$.

\paragraph{1. Crossing Stokes lines:} The connection formula
\begin{equation}
\mqty(\Psi^{I}_{+,p} \\ \Psi^{I}_{-,p})
= C_{p} \, \mqty(\Psi^{II}_{+,p} \\ \Psi^{II}_{-,p}),
\end{equation}
for crossing a Stokes line from Stokes region I to II is given as follows.

For counter-clockwise crossing at $p_{c}$ of a Stokes line  emanating from a simple turning point $v_{t} \in P_{0}$ as measured w.r.t.\  $v_{t}$, the connection matrix for $\Psi_{\pm,v_{t}}$ is
\begin{equation}
C_{t\epsilon} =
\begin{dcases}
\mqty(1 & i \\ 0 & 1) & \quad  \epsilon = + \,,
\\
\mqty(1 & 0 \\ i & 1) &\quad \epsilon = - \,,
\end{dcases}
\end{equation}
with $\epsilon$ being the sign of $\int^{p_{c}}_{v_{t}} e^{i\pha} \sqrt{\phi}$ along the Stokes line. That is, the dominant and subdominant solutions are continued as $\Psi^{I}_{\mathrm{dom}} = \Psi^{II}_{\mathrm{dom}} + i \Psi^{II}_{\mathrm{sub}}$ and $\Psi^{I}_{\mathrm{sub}} = \Psi^{II}_{\mathrm{sub}}$. For clockwise crossing, the connection matrices are the inverses of those in counter-clockwise crossing.

For counter-clockwise crossing at $p_{c}$ of a Stokes line emanating from a simple pole $v_{s} \in P_{s}$, the connection matrix for $\Psi_{\pm,v_{s}}$ is
\begin{equation}
C_{s\epsilon}(\theta) =
\begin{dcases}
\mqty(1 & 2i \cos(2\pi\theta)
\\ 0 & 1) & \quad \epsilon = + \,,
\\
\mqty(1 & 0
\\ 2i \cos(2\pi\theta) & 1) &\quad  \epsilon = - \,.
\end{dcases}
\end{equation}
Here $\theta$ is defined by the local behaviour
\begin{equation}
Q_2(z) \sim \frac{\theta^2 - \frac{1}{4}}{(z-z_{s})^2}\,,
\end{equation}
near $v_s$, and $\epsilon$ is the sign of $\int^{p_{c}}_{v_{s}} e^{i\pha} \sqrt{\phi}$ along the Stokes line. That is, the dominant and subdominant solutions are continued as $\Psi^{I}_{\mathrm{dom}} = \Psi^{II}_{\mathrm{dom}} + 2i\cos(2\pi\mexp) \Psi^{II}_{\mathrm{sub}}$ and $\Psi^{I}_{\mathrm{sub}} = \Psi^{II}_{\mathrm{sub}}$. For clockwise crossing, the connection matrices are again the inverses of those in counter-clockwise crossing.

\paragraph{2. Crossing  a branch cut:} For $\Psi_{\pm,v}$ normalized at a branch point $v \in P_{0} \cup P_{s}$, crossing branch cut leads to the connection matrix
\begin{equation}
B_{\pm} = \mqty(0 & \pm i
\\ \pm i & 0),
\end{equation}
with $+(-)$ for counter-clockwise (clockwise) crossing of branch cut w.r.t.\  the branch point $v$.

\paragraph{3. Change of reference point:} In the process of continuing WKB solutions, one will often cross Stokes lines emanating from different branched points. This requires keeping track of the change of reference points in WKB solutions. Change of reference points transforms the formal WKB solutions as
\begin{equation}
\psi_{\pm,p} = e^{\pm V_{pq}} \,\psi_{\pm,q}, \qquad V_{pq} = \int_{\beta_{pq}} \lambda.
\end{equation}
The connection matrix associated with WKB solutions with different reference points is related by 
\begin{equation}
\begin{split}
C_{p} = \mqty(\dmat{e^{V_{pq}},e^{-V_{pq}}}) C_{q} \mqty(\dmat{e^{-V_{pq}},e^{V_{pq}}})
\,\eqqcolon\,  \brk{C_{q}}(V_{pq}).
\end{split}
\end{equation}
The path of integration $\beta_{pq}$ from $p$ to $q$ will be specified in context. At this point $e^{V} $ is an asymptotic series in $\eta$ that needs to be Borel-summed. Properties of its Borel sum will be discussed in the next subsection.

\paragraph{4. Behavior near double pole and monodromy:} Let us examine the behaviour near a double pole with exponent $\mexp = \eta \,\hat{\mexp}$, i.e.,
\begin{equation}
Q_{0}(z) \sim \frac{\hat{\mexp}^2}{z^2}\,, \qquad Q_{2}(z) \sim -\frac{\frac{1}{4}}{z^2}\,.
\end{equation}
In this neighborhood the WKB differential behaves as
\begin{equation}
\lambda \sim \frac{\eta \, \hat{\theta}}{z} dz
\end{equation}
i.e., only $P_{-1}$ contains a simple pole while higher terms in $\Podd$ are regular. This can be verified from the recursion relation for $P_{n}$. Therefore, when going around the double pole, the WKB solutions need to be multiplied with diagonal monodromy
\begin{equation}
D(\mexp) = \mqty(\dmat{-e^{\oint \lambda},-e^{-\oint \lambda}}) = \mqty(\dmat{-e^{2 \pi i \mexp},-e^{-2 \pi i \mexp}}).
\end{equation}
%

\subsection{Voros symbol and Stokes automorphism}\label{subsubsec:Voros-symbol-SA}

As mentioned in previous subsection, repeated use of WKB connection formulae requires keeping track of the change of reference points, and results in factors $V_{pq} = \int_{\beta_{pq}} \lambda$. The reference points $p,q$ will be some branched points $v,w \in \zeros \cup \spoles$. Let $\beta^{*}_{vw}$ be the pre-image of $\beta_{vw}$ on the second sheet of $\Cwkb$, and define $\gamma_{vw} = \beta_{vw} - \beta^{*}_{vw}$. The contour $\gamma_{vw}$ is a cycle on $\Cwkb$, and satisfies
\begin{equation}
V_{\gamma_{vw}} =\oint_{\gamma_{vw}} \lambda \ = \ 2\, V_{vw}
\end{equation}
as $\int_{\beta} \lambda = \int_{-\beta^{*}} \lambda$ due to the sign change of $\lambda$ on second sheet.

In general, for a cycle $\gamma$ on $\Cwkb$, the period integral
\begin{equation}
V_{\gamma} = \oint_{\gamma} \lambda
\end{equation}
is referred to as \emph{Voros period} or \emph{all orders WKB period}, and
\begin{equation}
X_{\gamma} = e^{V_{\gamma}}
\end{equation}
is referred to as \emph{Voros symbol}. The Voros period $V_{\gamma}$ only depends on the equivalence class
\begin{equation}
\brk{\gamma} \in H_{1}\prn{\Cwkb \setminus \hat{P};\Z}
\end{equation}
where $\hat{P}$ is the pre-image of $P = \zeros \cup \spoles \cup \dpoles$ on $\Cwkb$.

\paragraph{Borel summability of Voros symbol:} The Voros symbol $X_{\gamma}(\eta)$ is an asymptotic series in $\eta$ of the form
\begin{equation}
X_{\gamma}(\eta) = e^{\eta \,v_{\gamma}} \sum^{\infty}_{m=0}\, X_{\gamma,m} \,\eta^{-m}\,,
\end{equation}
where $v_{\gamma} = \oint_{\gamma} \sqrt{\phi}$ is the classical WKB period.
Its Borel sum in direction $\pha$, $\BS_{\pha}\brk{X_{\gamma}}$, is defined similarly to the Borel sum of WKB solutions.
One can still define $X^{(\pha)}_{\gamma}(\eta) \coloneqq X_{\gamma}(\eta \,e^{i\pha})$,
and $\BS_{\pha}\brk{X_{\gamma}} = \BS\brk{X^{(\pha)}_{\gamma}}$.
The Borel summability can again be read off from the Stokes graph $G_{\pha}$.

Let us first introduce the notion of \emph{saddle cycle}.
Suppose the Stokes graph $G_{\pha_{c}}$ contains a saddle trajectory $l$ at a critical phase $\pha = \pha_{c}$.
Let $\gamma_{l}$ be a cycle on $\Cwkb$ whose projection to $\Sigma$ surrounds $l$.
The orientation of $\gamma_{l}$ is chosen such that $e^{i\pha_{c}} \,v_{\gamma_{l}} < 0$, i.e.,
$X^{(\pha_{c})}_{\gamma_{l}}$ is exponentially small.
The cycle $\gamma_{l}$ is referred to as a saddle cycle, and the equivalence class
$\brk{\gamma_{l}} \in H_{1}\prn{\Cwkb \setminus \hat{P};\Z}$ is referred to as a saddle class.

We denote by $\Sad\prn{G_{\pha_{c}}}$ the set of all saddle trajectories in $G_{\pha_{c}}$.

\begin{theorem}
  The Voros symbol $X_{\gamma}(\eta)$ is Borel summable in direction $\pha$ if $G_{\pha}$ is saddle-free. At critical phase $\pha = \pha_{c}$, $X_{\gamma}(\eta)$ is Borel summable in direction $\pha_{c}$ if $\expval{\gamma,\gamma_{l}} = 0$ for all $l \in \Sad\prn{G_{\pha_{c}}}$.
\end{theorem}

\paragraph{Stokes automorphism of Voros symbol:} When the Voros symbol $X_{\gamma}(\eta)$ is Borel summable in direction
$\pha$, we denote its Borel sum as
\begin{equation}
\BSX_{\gamma,\pha}(\eta) = \BS_{\pha}\brk{X_{\gamma}}(\eta).
\end{equation}
When $X_{\gamma}(\eta)$ is not Borel summable in direction $\pha = \pha_{c}$, we define
\begin{equation}
\BSX^{\pm}_{\gamma,\pha_{c}}(\eta) \coloneqq \lim_{\pha \to \pha^{\pm}_{c}} \BS_{\pha}\brk{X_{\gamma}}(\eta).
\end{equation}
The two limits differ because of Borel singularity in direction $\pha = \pha_{c}$, and are related by jump formula or Stokes automorphism. The explicit form of jump formula depends on the type of saddle trajectory $l$ that appears in $G_{\pha_{c}}$.

For a saddle trajectory $l$ connecting two distinct simple turning points $v,v^{\prime} \in \zeros$, the associated jump formula is given by
\begin{equation}
\label{eq:jump-DDP}
\BSX^{-}_{\gamma,\pha_{c}} = \BSX^{+}_{\gamma,\pha_{c}} \prn{1 + \BSX_{\gamma_{l}, \pha_{c}}}^{-\expval{\gamma_{l},\gamma}}.
\end{equation}
This is referred to as DDP formula~\cite{Delabaere:1997srq}. The multiplication by $\prn{1 + \BSX_{\gamma_{l}, \pha_{c}}}^{-\expval{\gamma_{l},\gamma}}$ is the action of Stokes automorphism $\SA_{\gamma_{l}}$ associated with saddle trajectory $l$
\begin{equation}
\SA_{\gamma_{l}}: \BSX_{\gamma,\pha_{c}} \mapsto \BSX_{\gamma,\pha_{c}} \prn{1 + \BSX_{\gamma_{l}, \pha_{c}}}^{-\expval{\gamma_{l},\gamma}},
\end{equation}
and the jump formula is equivalent to
\begin{equation}
\BS_{\pha^{-}_{c}} = \BS_{\pha^{+}_{c}} \circ \SA_{\gamma_{l}}\,,
\end{equation}
where $\BS_{\pha^{\pm}_{c}} = \lim_{\pha \to \pha^{\pm}_{c}} \BS_{\pha}$.

For a saddle trajectory $l$ connecting a simple pole and a simple turning point, the jump formula is given by~\cite{Iwaki_2015}
\begin{equation}
\label{eq:jump-simple-pole}
\BSX^{-}_{\gamma,\pha_{c}} = \BSX^{+}_{\gamma,\pha_{c}}
\prn{1 + 2\cos(2\pi\mexp) \, \BSX_{\gamma_{l},\pha_{c}} + \BSX^2_{\gamma_{l},\pha_{c}}}^{-\expval{\gamma_{l},\gamma}},
\end{equation}
with $\theta$ defined by $Q_2(z) \sim \frac{\theta^2 - \frac{1}{4}}{(z-z_{s})^2}$ near the simple pole $v_s$. That is, the Stokes automorphism in this case is given by
\begin{equation}
\SA_{\gamma_{l}}:
\BSX_{\gamma,\pha_{c}} \,\mapsto\,
\BSX_{\gamma,\pha_{c}} \, \prn{1 + 2\cos(2\pi\mexp)\, \BSX_{\gamma_{l},\pha_{c}}
+ \BSX^2_{\gamma_{l},\pha_{c}}}^{-\expval{\gamma_{l},\gamma}}.
\end{equation}

When there are multiple saddle trajectories at $\pha=\pha_{c}$, the action of Stokes automorphism $\SA$ is given by the composition of Stokes automorphisms  associated with all saddle trajectories in the Stokes graph $G_{\pha_{c}}$:
\begin{equation}
\SA = \prod_{l \in \mathcal{L}(G_{\pha_{c}})} \SA_{\gamma_{l}}.
\end{equation}

\paragraph{Median Borel sum of Voros symbol:} The $\eta \to \infty$ asymptotics of the lateral Borel sums $\BSX^{\pm}_{\gamma,\pha_{c}}(\eta)$ differ from the asymptotic series $X^{(\pha_{c})}_{\gamma}(\eta)$ by exponentially small terms, i.e., the asymptotics gives a transseries. For our purpose, it will be convenient to work with an object whose asymptotics recovers $X^{(\pha_{c})}_{\gamma}(\eta)$. The object with this property is the median Borel sum of $X^{(\pha_{c})}_{\gamma}(\eta)$, which we denote as $\BSXm_{\gamma,\pha_{c}}(\eta) = \BSm_{\pha_{c}}\brk{X_{\gamma}}(\eta)$.

The median Borel sum $\BSm_{\pha_{c}}$ in a critical direction $\pha = \pha_{c}$ is defined using half Stokes automorphism
\begin{equation}
\BSm_{\pha_{c}}
\,\coloneqq\,
\BS_{\pha^{-}_{c}} \circ \SA^{-\half}
= \BS_{\pha^{+}_{c}} \circ \SA^{\half}.
\end{equation}
For example, in the case of $\SA = \SA_{l}$ and $l$ being the saddle trajectory connecting two distinct simple turning points, taking half power of $\SA_{l}$ simply means setting $\expval{\gamma_{l},\gamma} \to \half \expval{\gamma_{l},\gamma}$ in the multiplication factor $\prn{1 + \BSX_{\gamma_{l}, \pha_{c}}}^{-\expval{\gamma_{l},\gamma}}$. The median Borel sum $\BSXm_{\gamma,\pha_{c}}(\eta)$ in this case is then given by
\begin{equation}
\BSXm_{\gamma,\pha_{c}}
= \BSX^{-}_{\gamma,\pha_{c}} \, \prn{1 + \BSX_{\gamma_{l}, \pha_{c}}}^{\half\expval{\gamma_{l},\gamma}}
= \BSX^{+}_{\gamma,\pha_{c}} \prn{1 + \BSX_{\gamma_{l}, \pha_{c}}}^{-\half\expval{\gamma_{l},\gamma}}\,.
\end{equation}
%

\subsection{Computing monodromy using exact WKB}\label{subsubsec:monodromy-EWKB}

Having reviewed the ingredients that go into the WKB analysis, we are now ready to describe the recipe given in~\cite{KawaiTakei2005} for computing monodromy using exact WKB. We recall that this was the entire point of the discussion above, for in the physical problem we want to be able to compute the monodromy between the horizon and the asymptopia.

Suppose we want to compute the monodromy around a singular point $z_{i}$ along a monodromy contour $\mc_{i}$ with base point $z_{0}$. We work with the WKB solutions normalized at $z_{0}$
\begin{equation}
\psi_{\pm}(z,\eta) = \frac{1}{\sqrt{\Podd(z,\eta)}} \exp\prn{\pm \int^{z}_{z_{0}} dz^{\prime} \Podd(z,\eta)}.
\end{equation}
Let $\{z^{(\alpha)}\}$ be the set of intersection points between the monodromy contour $\mc_{i}$ and Stokes lines. That is,
\begin{equation}
z^{(\alpha)} = \mc_{i} \cap \Sl_{\alpha}
\end{equation}
for some Stokes line $\Sl_{\alpha}$ emanating from some branched point $v_{\alpha} \in P_{0} \cup P_{s}$. The connection matrix associated with crossing $\alpha$ is given by
\begin{equation}
C_{\alpha} = \brk{\prn{C_{v_{\alpha},\epsilon}}^{\epsilon^{\prime}}}\prn{V_{\alpha}}.
\end{equation}
The connection matrix $\prn{C_{v_{\alpha},\epsilon}}^{\epsilon^{\prime}}$ is the connection matrix for crossing $\alpha$ associated with WKB solutions normalized at $v_{\alpha}$
\begin{equation}
\psi^{(\alpha)}_{\pm}(z,\eta) = \frac{1}{\sqrt{\Podd(z,\eta)}} \exp\prn{\pm \int^{z}_{v_{\alpha}} dz^{\prime} \Podd(z,\eta)}.
\end{equation}
We recall our conventions described hitherto:
\begin{itemize}[wide,left=0pt]
  \item  The subscript $v_{\alpha} \in \{t,s\}$ in connection matrix denotes whether $v_{\alpha}$ is a turning point or simple pole.
  \item  $\epsilon = \pm$ is related to which one of the solutions $\psi^{(\alpha)}_{\pm}(z,\eta)$ is the dominant solution.
  \item  $\epsilon^{\prime} = \pm$ is related to the orientation of crossing w.r.t.\  $v_{\alpha}$.
\end{itemize}
The factor
\begin{equation}
V_{\alpha} = \int_{\beta_{\alpha}} \lambda
\end{equation}
takes into account the change of reference points between $z_{0}$ and $v_{\alpha}$. The integration path is
\begin{align}
\beta_{\alpha} = \beta(z_{0},z^{(\alpha)}) \cup \beta(z^{(\alpha)},v_{\alpha})
\end{align}
with $\beta(z_{0},z^{(\alpha)}) \in \mc_{i}$ the path from $z_{0}$ to $z^{(\alpha)}$ along the monodromy contour $\mc_{i}$, and $\beta(z^{(\alpha)},v_{\alpha}) \in \Sl_{\alpha}$ the path from $z^{(\alpha)}$ to $v_{\alpha}$ along the Stokes line $\Sl_{\alpha}$. The monodromy along $\mc_{i}$ is then computed as
\begin{equation}
M_{\mc_{i}} = \prn{\prod_{\alpha} C_{\alpha}} D(\mexp_{i}),
\end{equation}
with the product over all crossings of Stokes lines. The monodromy $M_{\mc_{i}}$ is thereby represented in terms of integrals of WKB differential $\lambda$ between the reference point $z_{0}$ and branched points $v_{\alpha}$. The monodromy invariant $\Tr M_{\mc_{i}}$, which is independent of the reference point, is represented in terms of WKB integrals between branched points. These are in turn related to WKB periods. Therefore, computing the latter allows one to extract the non-perturbative information in the monodromy, which was what we sought to achieve.

\section{An exact WKB analysis of monodromy: non-zero spatial momentum}\label{sec:EWKB-timelike-momentum}

We have now set the stage for understanding the non-perturbative behavior of the spectral function in holographic CFTs.
It will turn out to be useful to split the discussion into two parts. In the first part, which we explore in this section, we will
assume that the spatial momentum is non-vanishing and being scaled commensurately with the frequency. Later in~\cref{sec:WKB-zero-momentum} we will tackle the case where the spatial momentum vanishes.

\subsection{The wave equation and its WKB curve}

\paragraph{Set-up:} The Schr\"odinger form of scalar wave equation on the planar \SAdS{5} background is obtained in~\cref{app:Schrodinger-form} for $d=4$. In this case, we will use the  coordinate
parameterization $z = (r^2 - \rp^2)/(2\,r^2)$.
We take the WKB parameter to be the absolute value of the frequency, $\eta = \abs{\freq}$,
and let $\arg(\freq) = \arg(\mom) = \pha \in [\pi/2,-\pi/2)$. Finally, we introduce
\begin{equation}
\abs{\mom} = \mr \,\abs{\freq} \,.
\end{equation}

It will be convenient to work with the redefined coordinate
\begin{equation}
u = z^{-1} = \frac{2r^2}{r^2-\rp^2}
\end{equation}
so that all branched points are finite. In this coordinate, $u=0,1,2,\infty$ corresponds to singularity, complex horizon, AdS boundary, and physical horizon, respectively.  The potential is given by
\begin{equation}\label{eq:4dQpot}
\begin{aligned}
 &
Q_{0}(u)
= \frac{u^2 - 4\,(u-1)\, \mr^2}{4\,(2-u)\,(u-1)^2 \,u},
\\
 &
Q_{2}(u)
= -\frac{(u^2-2\,u+2)^2}{4\,u^2\, (u-1)^2\, (u-2)^2} + \frac{\nu^2}{4\, (u-2)^2(u-1)} \,.
\end{aligned}
\end{equation}
The quadratic differential is, therefore,  $\phi = Q_{0}(u)\,du^{2}$.
It has simple zeros $P_{0}$, simple poles $P_{s}$ and double poles $P_{\infty}$
\begin{equation}
\begin{split}
P_{0}
 & = \cbrk{u_{\pm}}, \quad  P_{s} = \cbrk{0,2}, \quad P_{\infty} = \cbrk{1,\infty},
\\
u_{\pm}
 & = 2 \,\mr^2 \pm 2i\, \mr\,  \sqrt{1-\mr^2}.
\end{split}
\end{equation}
We will focus on \emph{timelike momentum}, so  $0<\mr<1$.

\paragraph{Branch cut and cycles:} We choose the first sheet to be the branch where
\begin{equation}
\Res_{u=1} e^{i\pha} \sqrt{\phi} = \half e^{i\pha}
\end{equation}
with branch cuts as in~\cref{fig:branch-cut-cycles}. With this choice, the residue at $u=\infty$ is
\begin{equation}
\Res_{u=\infty} e^{i\pha} \sqrt{\phi} = \Half{i} e^{i\pha}.
\end{equation}
Our choice of $A,B$ cycles is as in~\cref{fig:branch-cut-cycles}, with $\expval{\gamma_{A},\gamma_{B}} = 1$. Our convention is $\expval{\rightarrow,\uparrow} = 1$. The other cycles are related by
\begin{equation}
\begin{split}
\gamma_{A'} & = \gamma_{A} + C(1) + C(\infty)                                                       \\
\gamma_{B'} & = \gamma_{B} - C(1) + C(\infty)                                                       \\
\gamma_{C}  & = \gamma_{A^\prime} + \gamma_{B} - 2C(1) = \gamma_{A} + \gamma_{B} - C(1) + C(\infty)
\end{split}
\end{equation}
Here $C(p)$ denotes positive-oriented loop around $p$ on the first sheet.

\begin{figure}[htbp!]
  \centering
  \begin{subfigure}[b]{0.4\textwidth}
    \centering
    \scalebox{.5}{\input{figures/timelike-tikz-codes/timelike-cycles-tikz}}
    \caption{Cycles on WKB curve appearing in the main text, including the $A,B$ cycles.}
  \end{subfigure}
  \hfill
  \begin{subfigure}[b]{0.4\textwidth}
    \centering
    \scalebox{.4}{\input{figures/timelike-tikz-codes/timelike-cycles-saddle0-tikz}}
    \caption{Saddle cycles $\gamma_{l}, \gamma_{l^{\prime}}$ surrounding the saddle connections $l,l^\prime$ at critical phase $\pha_{c}=0$.}
  \end{subfigure}
  \caption{Choice of branch cut on WKB curve and cycles. Dashed lines indicate going to the second sheet.}
  \label{fig:branch-cut-cycles}
\end{figure}

\begin{figure}[htbp!]
  \begin{subfigure}[b]{0.3\textwidth}
    \centering
    \includegraphics[width=\linewidth]{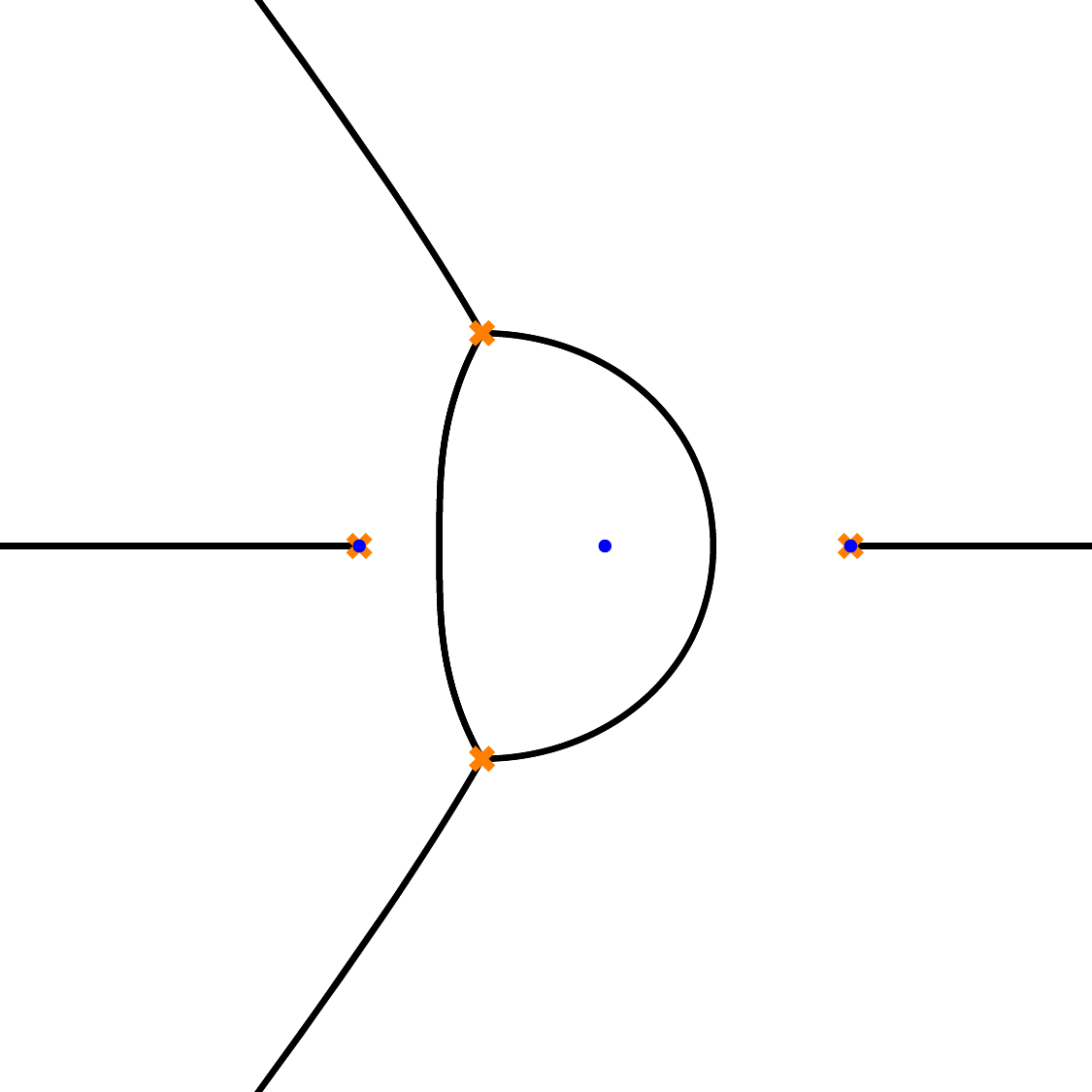}
    \caption{$\pha = \frac{\pi}{2}$}
  \end{subfigure}
  \hfill
  \begin{subfigure}[b]{0.3\textwidth}
    \centering
    \includegraphics[width=\linewidth]{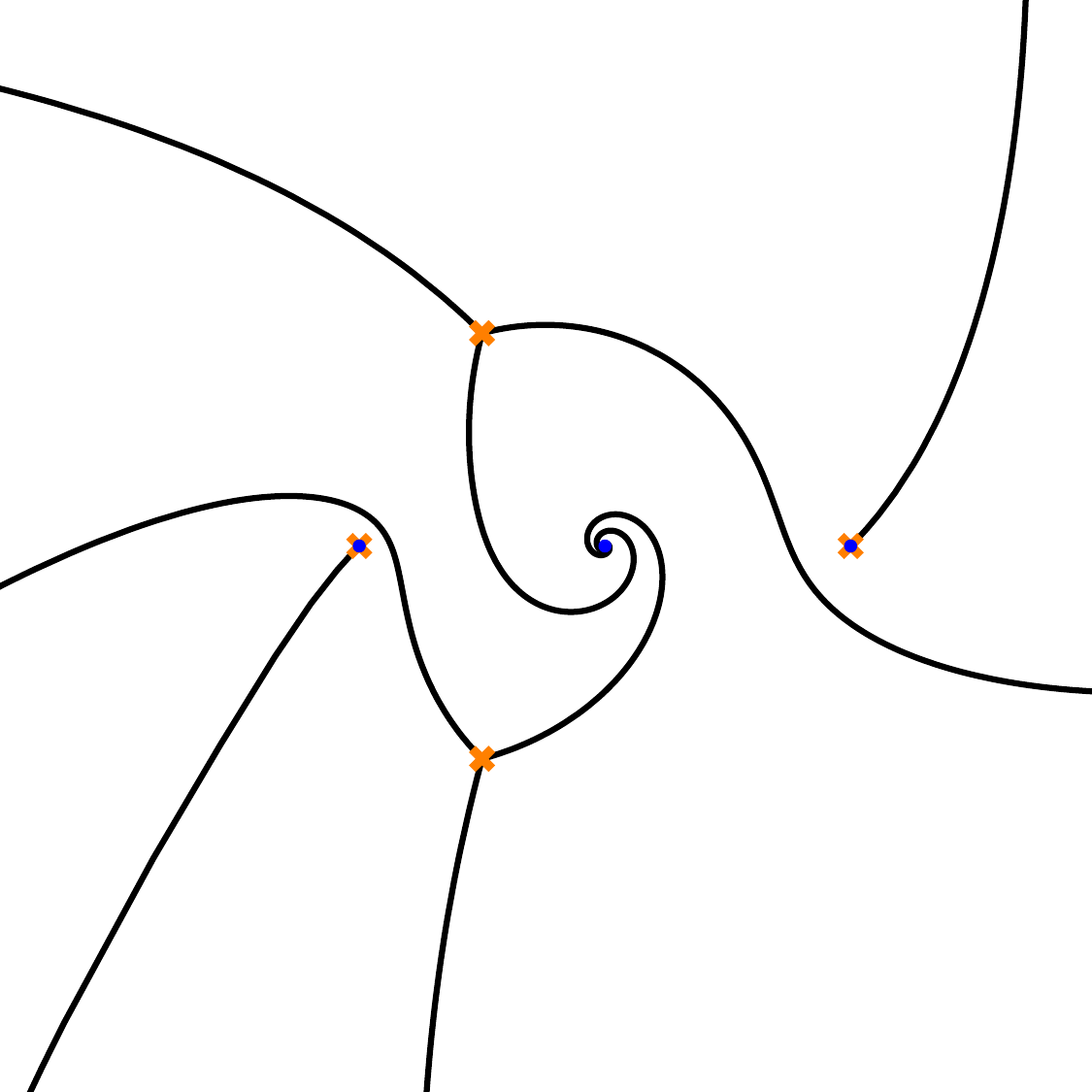}
    \caption{$\pha = \frac{3\pi}{8}$}
  \end{subfigure}
  \hfill
  \begin{subfigure}[b]{0.3\textwidth}
    \centering
    \includegraphics[width=\linewidth]{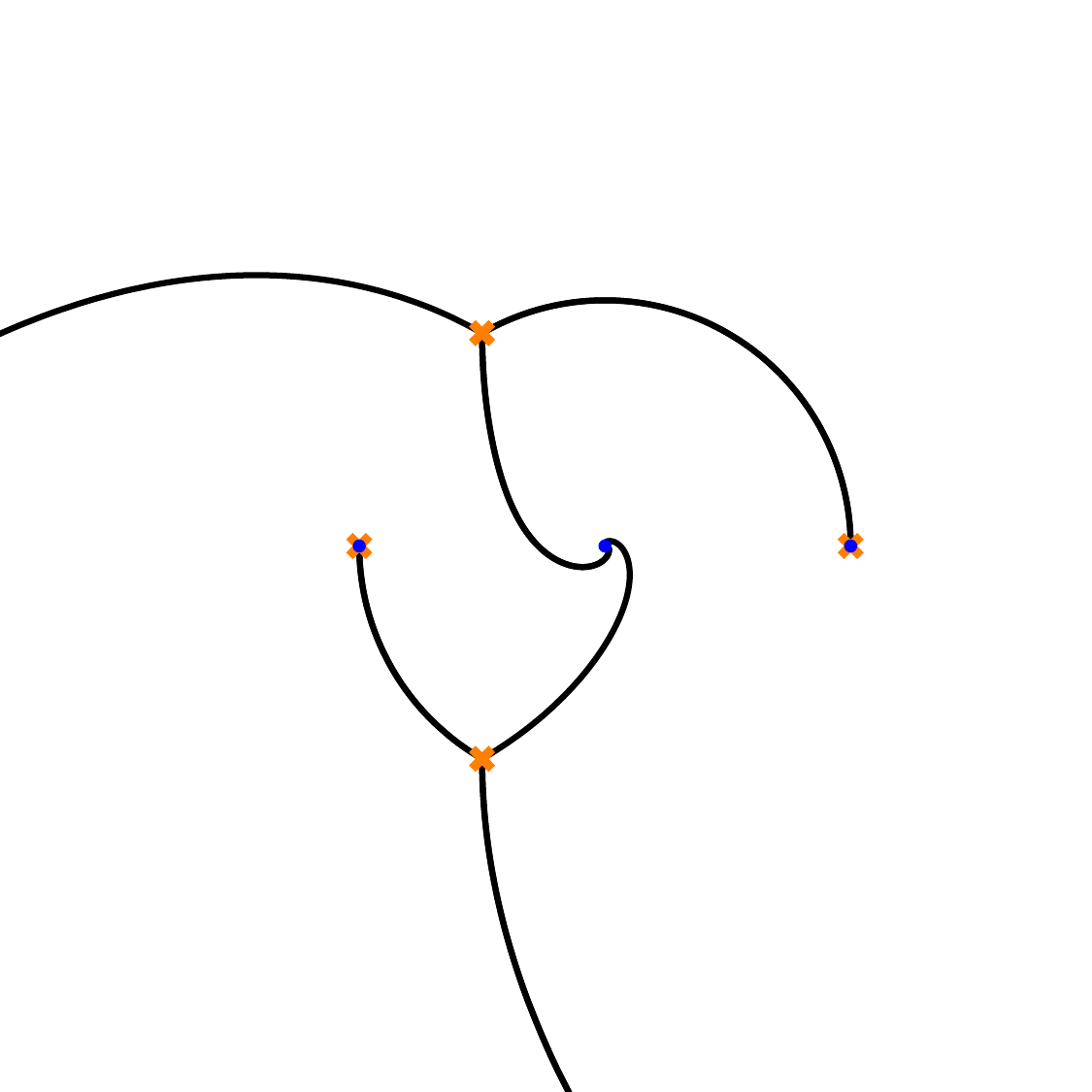}
    \caption{$\pha = \frac{\pi}{4}$}
  \end{subfigure}
  \hfill
  \begin{subfigure}[b]{0.3\textwidth}
    \centering
    \includegraphics[width=\linewidth]{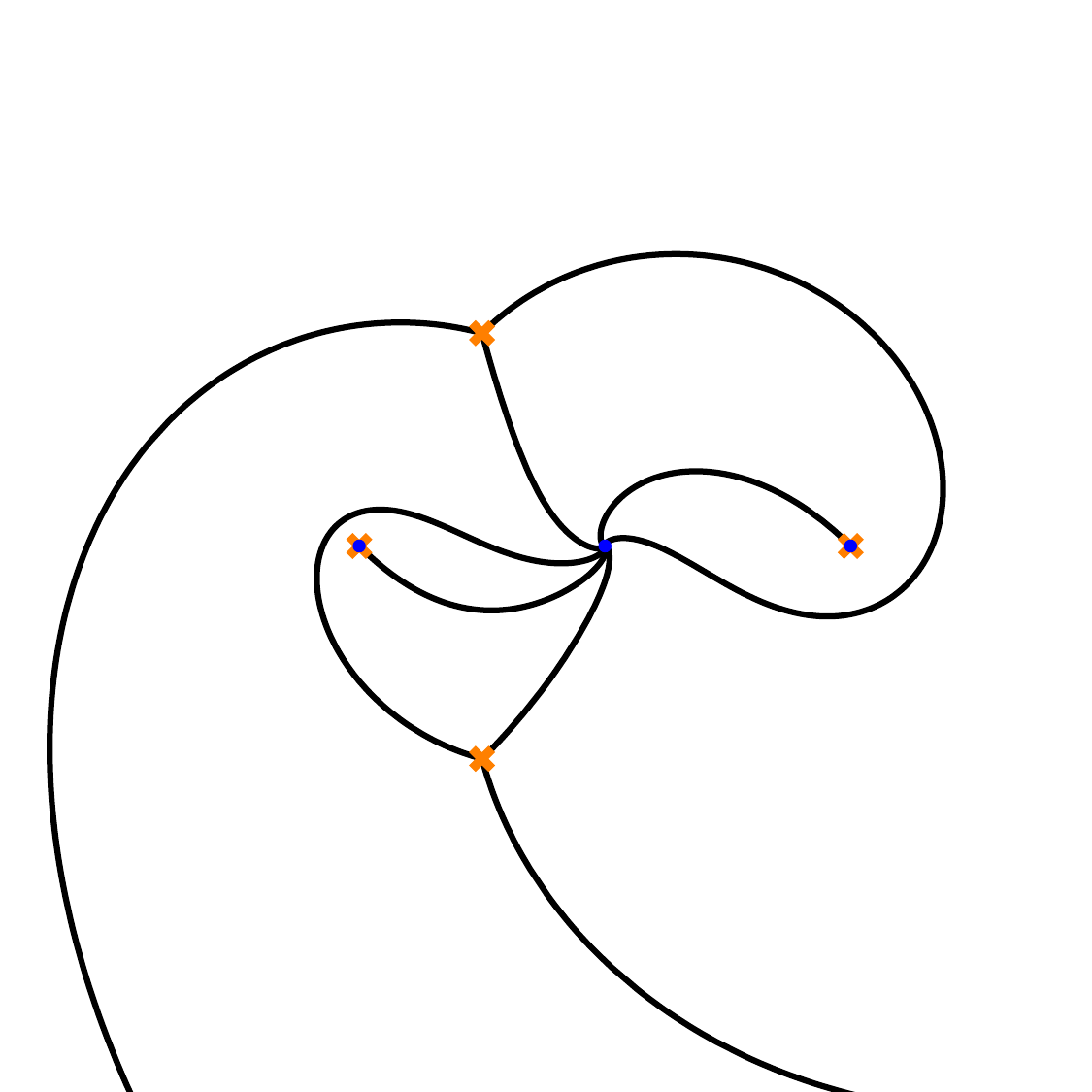}
    \caption{$\pha = \frac{\pi}{8}$}
  \end{subfigure}
  \hfill
  \begin{subfigure}[b]{0.3\textwidth}
    \centering
    \includegraphics[width=\linewidth]{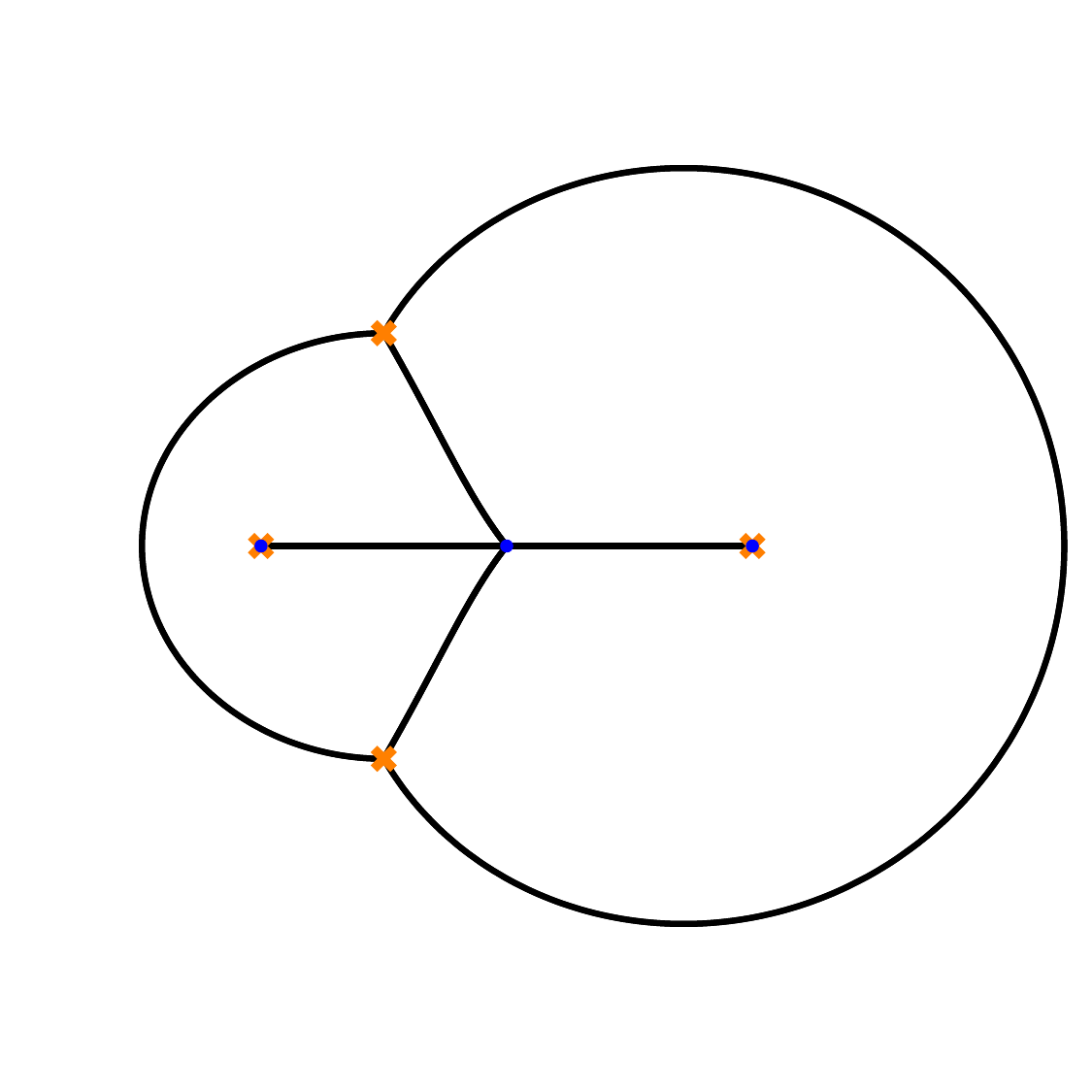}
    \caption{$\pha = 0$}
  \end{subfigure}
  \hfill
  \begin{subfigure}[b]{0.3\textwidth}
    \centering
    \includegraphics[width=\linewidth]{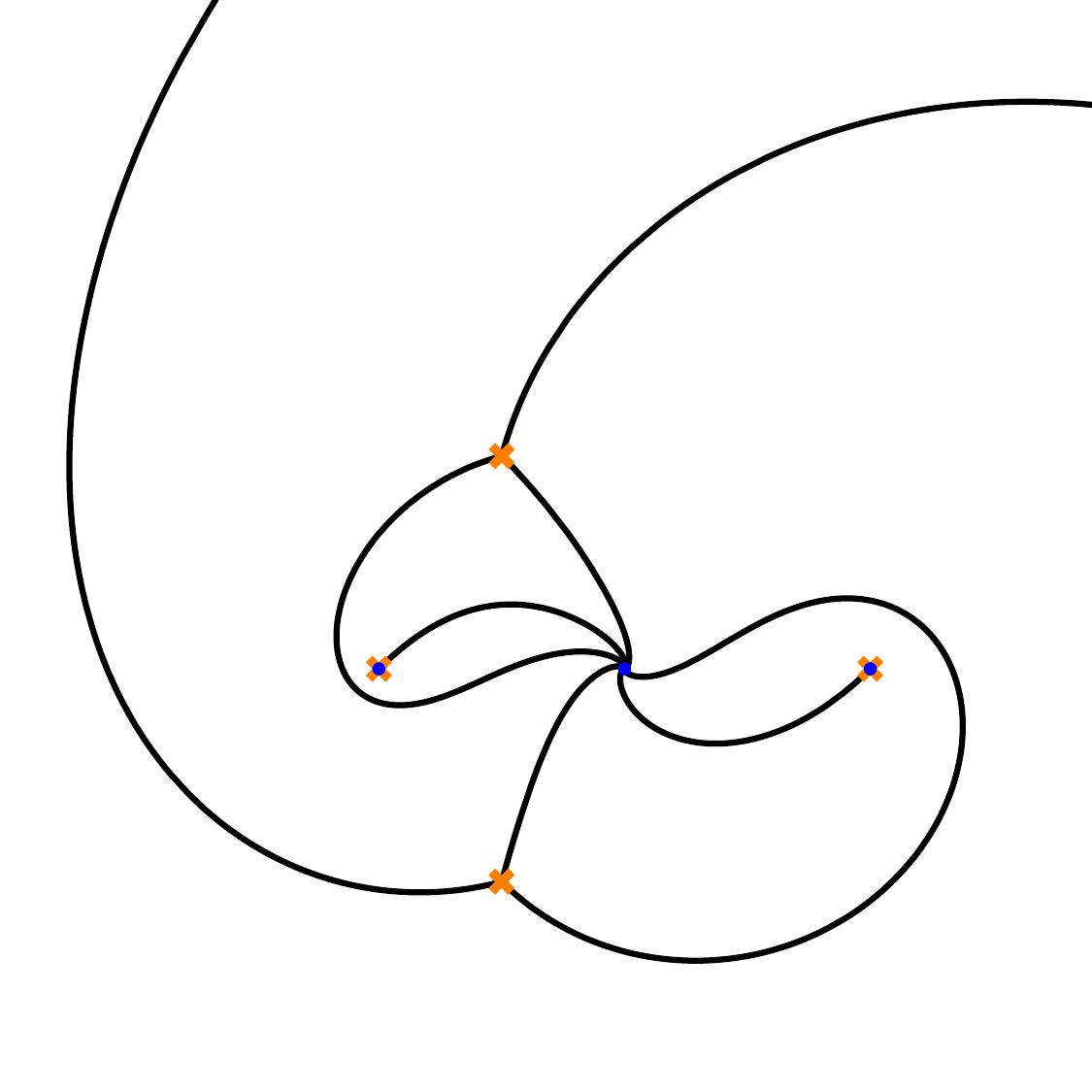}
    \caption{$\pha = -\frac{\pi}{8}$}
  \end{subfigure}
  \hfill
  \begin{subfigure}[b]{0.3\textwidth}
    \centering
    \includegraphics[width=\linewidth]{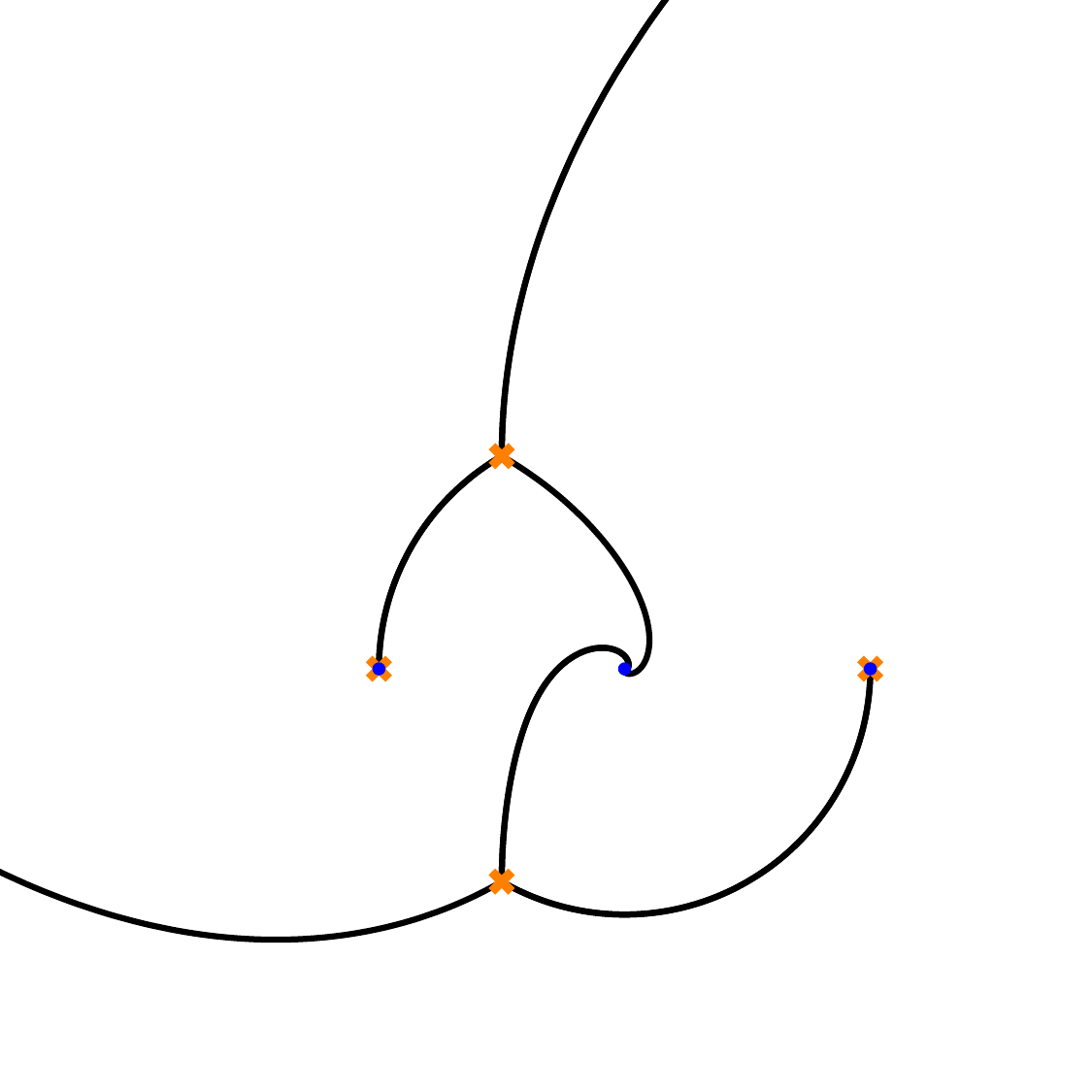}
    \caption{$\pha = -\frac{\pi}{4}$}
  \end{subfigure}
  \hfill
  \begin{subfigure}[b]{0.3\textwidth}
    \centering
    \includegraphics[width=\linewidth]{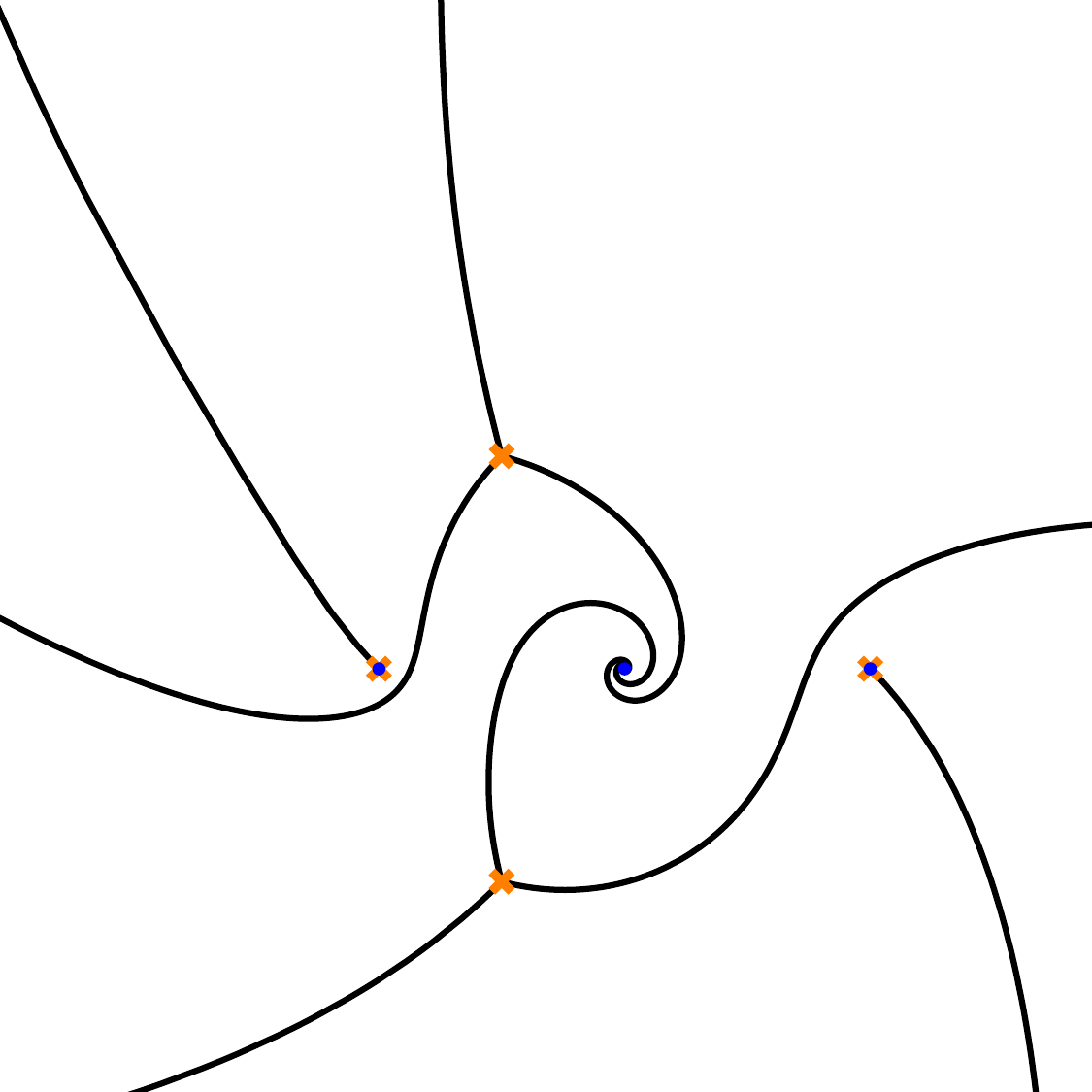}
    \caption{$\pha = -\frac{3\pi}{8}$}
  \end{subfigure}
  \hfill
  \begin{subfigure}[b]{0.3\textwidth}
    \centering
    \includegraphics[width=\linewidth]{figures/timelike-graphs/0.pdf}
    \caption{$\pha = -\frac{\pi}{2}$}
  \end{subfigure}
  \caption{Stokes graphs at different phases $\pha = \arg(\freq) = \arg(\mom)$. The critical phases are $\pha_{c} \in \{\pi/2, \pi/4, 0, -\pi/4\}$, where there exist Stokes lines connecting two branched points. The topology of Stokes graphs doesn't change between critical phases. }
  \label{fig:stokes-graph-timelike}
\end{figure}

\subsection{The Voros periods \& Stokes automorphisms}\label{sec:vorsto}

With the basic elements for the WKB analysis in place, it is useful to record here the ingredients which enter into the WKB computation of the monodromy. We outline the salient results, relegating some details of our analysis to~\cref{app:WKB}.

\paragraph{Voros periods:} The Voros periods for $A,B$ cycles are complex conjugate of each other.
The classical $A$ period is given by
\begin{equation}
\label{eq:classical-A-period}
\begin{split}
V^{(-1)}_{A} = (1-i)\vcl\,, \qquad
\vcl = \frac{\sqrt{2\pi}\,\Gamma\prn{\frac{5}{4}}}{\Gamma\prn{\frac{7}{4}}}
\, \mr^{\Half{3}}\, {}_2 F_1\prn{\frac{1}{4},\frac{3}{4};\frac{7}{4};\mr^2}.
\end{split}
\end{equation}
The quantity $v$ increases monotonically from $0$ to $\pi$ as $\mr$ increases from $0$ to 1
\begin{equation}
0 < \vcl < \pi, \quad 0 < \mr < 1.
\end{equation}
Higher-order periods can be systematically computed as a linear combination of the periods $\Pi_{i}$ associated with a basis of differentials $\omega_{i}$, with the linear coefficients that can be determined algebraically. To wit,
\begin{equation}
\label{eq:higher-periods}
V^{(n)}_{A} = \sum^{2}_{i=1} \,b^{(n)}_{i} \, \Pi_{i,A}, \qquad n \geq 1,
\end{equation}
with the leading contributions being explicitly given by
\begin{equation}
\begin{aligned}
\Pi_{1,A}
 & = e^{i\frac{\pi}{4}} \frac{4\sqrt{\pi}\Gamma\prn{\frac{5}{4}}}{\Gamma\prn{\frac{3}{4}}} \mr^{-\half} \prn{1-\mr^2}^{-\frac{1}{4}}\,,
\\
\Pi_{2,A}
 & =
-e^{i\frac{\pi}{4}} \frac{2\sqrt{\pi}\Gamma\prn{\frac{5}{4}}}{\Gamma\prn{\frac{3}{4}}} \mr^{-\half} \prn{1-\mr^2}^{-\frac{3}{4}}.
\end{aligned}
\end{equation}
The coefficients for $n=1,3$ read
\begin{equation}
\begin{aligned}
b^{(1)}_{1}
 & = \frac{1}{16}\prn{-2 + \mr^{-2} + 2\nu^2},
 & \quad
b^{(1)}_{2}
 & = 2\,b^{(1)}_{1}\,,
\\
b^{(3)}_{1}
 & = \frac{11-\mr^2\,(26+4\nu^2)-12\,\mr^4(-1-4\,\nu^2+\nu^4)-8\,\mr^6(-1+\nu^2)^2}{3072\,\mr^4\,(1-\mr^2)^2},
 & \quad
b^{(3)}_{2}
 & = 0\,.
\end{aligned}
\end{equation}

\paragraph{Stokes automorphism across critical phases:}  The Stokes graphs in $u$ coordinate at different phases are given in~\cref{fig:stokes-graph-timelike}. The critical phases are $\pha_{c} \in \{\pi/2, \pi/4, 0, -\pi/4\}$, where there exist Stokes lines connecting two branched points. At each critical phase, there exist two saddle connections $l,l^{\prime}$, i.e., Stokes lines connecting two branched points (simple zeros or poles). As explained in~\cref{subsubsec:Voros-symbol-SA}, the jump of Borel-summed Voros periods across critical phase is given by the Stokes automorphism
\begin{equation}
\SA = \SA_{\gamma_{l}} \SA_{\gamma_{l'}}.
\end{equation}
We now analyze the behaviour near each critical phase separately.

\bigskip
\noindent
\underline{$\pha_{c}=0$:} The two saddle cycles are (cf.~\cref{fig:branch-cut-cycles})
\begin{equation}
\begin{aligned}
\gamma_{l}
 & = -\gamma_{A} - \gamma_{B},
 & \quad
\expval{\gamma_{l},\gamma_{A}}
 & = 1,
 & \quad
\expval{\gamma_{l},\gamma_{B}}
 & = -1\,,
\\
\gamma_{l'}
 & = \gamma_{A'} + \gamma_{B'} = \gamma_{A} + \gamma_{B} + 2\,C(\infty),
 & \quad
\expval{\gamma_{l'},\gamma_{A}}
 & = -1,
 & \quad
\expval{\gamma_{l'},\gamma_{B}}
 & =1\,,
\end{aligned}
\end{equation}
and $\expval{\gamma_{l},\gamma_{l'}}=0$. The Voros periods for the saddle cycles are then related to $A,B$ periods by
\begin{equation}
\begin{aligned}
V_{l}
 & =
-V_{A} - V_{B}
\\
V_{l^\prime}
 & =
V_{A} + V_{B} + 2\,\freq\, \cdot 2\pi i \Res_{u=\infty} \sqrt{\phi}
= V_{A} + V_{B} - 2\pi\,\freq.
\end{aligned}
\end{equation}
Note that there is no ordering issue as $\expval{\gamma_{l},\gamma_{l'}}=0$.
The explicit form of the jump for $A,B$ periods is given by the DDP formula~\eqref{eq:jump-DDP}:
\begin{equation}
\label{eq:jump-periods}
\begin{aligned}
\BSX^{-}_{A,0}
 & =
\BSX^{+}_{A,0} \, \frac{1+\BSX_{l',0}}{1+\BSX_{l,0}}
= \,
\BSX^{+}_{A,0} \,\frac{1+\BSX_{A,0} \BSX_{B,0} \,e^{-2\pi\,\freq}}{1 + \prn{\BSX_{A,0} \BSX_{B,0}}^{-1}}
\\
\BSX^{-}_{B,0}
 & = \BSX^{+}_{B,0} \, \frac{1+\BSX_{l,0}}{1+\BSX_{l',0}}
= \,
\BSX^{+}_{B,0}\, \frac{1 + \prn{\BSX_{A,0} \BSX_{B,0}}^{-1}}{1+\BSX_{A,0} \BSX_{B,0} \,e^{-2\pi\,\freq}},
\end{aligned}
\end{equation}
i.e., the combination $\BSX_{A} \BSX_{B}$ doesn't jump. The median symbols are then given by
\begin{equation}
\label{eq:median-symbols-phase-0}
\BSXm_{A,0}
= \BSX^{+}_{A,0} \, \sqrt{\frac{1+\BSX_{A,0} \BSX_{B,0} \,e^{-2\pi\,\freq}}{1 + \prn{\BSX_{A,0} \BSX_{B,0}}^{-1}}},
\qquad
\BSXm_{B,0} = \BSX^{+}_{B,0} \, \sqrt{\frac{1 + \prn{\BSX_{A,0} \BSX_{B,0}}^{-1}}{1+\BSX_{A,0} \BSX_{B,0} \,e^{-2\pi\,\freq}}}.
\end{equation}

\bigskip
\noindent
\underline{$\pha_{c} = \pi/4$:} For this critical phase the saddle cycles are
\begin{equation}
\gamma_{l} = -\gamma_{A}, \qquad
\gamma_{l^\prime} = \gamma_{A^\prime} = \gamma_{A} + C(1) + C(\infty)\,,
\end{equation}
with associated Voros periods
\begin{equation}
V_{l} = -V_{A}, \qquad V_{l^\prime} = V_{A} + i\pi\,\freq - \pi\,\freq\,.
\end{equation}
Using~\eqref{eq:jump-simple-pole}, we find the jump of $B$ period
\begin{equation}
\label{eq:jump-period-pos-pi-4}
\begin{aligned}
\BSX^{-}_{B,\frac{\pi}{4}}
 & =
\BSX^{+}_{B,\frac{\pi}{4}} \,
\frac{1 + 2\cos(\pi\nu) \BSX_{l,\frac{\pi}{4}} + \BSX^2_{l,\frac{\pi}{4}}}{1 + 2\cos(\pi\nu) \, \BSX_{l^{\prime},\frac{\pi}{4}} + \BSX^2_{l^{\prime},\frac{\pi}{4}}}
\\
 & =
\BSX^{+}_{B,\frac{\pi}{4}} \, \frac{1 + 2\cos(\pi\nu)\, \BSX^{-1}_{A,\frac{\pi}{4}} + \BSX^{-2}_{A,\frac{\pi}{4}}}{1 + 2\cos(\pi\nu) \, \BSX_{A,\frac{\pi}{4}}e^{-\pi\,\freq+i\pi\,\freq} + \BSX^{2}_{A,\frac{\pi}{4}}\, e^{-2\pi\,\freq+2i\pi\,\freq}}\,.
\end{aligned}
\end{equation}
The $A$ period doesn't jump across $\pha_{c} = \pi/4$.

\bigskip
\noindent
\underline{$\pha_{c} = -\pi/4:$} This case is similar to $\pha_{c} = \pi/4$ with $A,B$ exchanged. The saddle cycles are
\begin{equation}
\gamma_{l} = -\gamma_{B}, \qquad \gamma_{l^\prime} = \gamma_{B^\prime} = \gamma_{B} - C(1) + C(\infty)\,,
\end{equation}
with associated Voros periods
\begin{equation}
V_{l} = -V_{B}, \qquad V_{l^\prime} = V_{B} - i\pi\,\freq - \pi\,\freq\,.
\end{equation}
Using~\eqref{eq:jump-simple-pole}, we find the jump of $A$ period
\begin{equation}
\label{eq:jump-period-neg-pi-4}
\begin{aligned}
\BSX^{-}_{A,-\frac{\pi}{4}}
 & =
\BSX^{+}_{A,-\frac{\pi}{4}}\,  \frac{1 + 2\cos(\pi\nu) \BSX_{l^{\prime},-\frac{\pi}{4}} + \BSX^2_{l^{\prime},-\frac{\pi}{4}}}{1 + 2\cos(\pi\nu) \,\BSX_{l,-\frac{\pi}{4}} + \BSX^2_{l,-\frac{\pi}{4}}}
\\
 & =
\BSX^{+}_{A,-\frac{\pi}{4}}\,  \frac{1 + 2\cos(\pi\nu) \, \BSX_{B,-\frac{\pi}{4}} \, e^{-\pi\,\freq-i\pi\,\freq} + \BSX^{2}_{B,-\frac{\pi}{4}}e^{-2\pi\,\freq-2i\pi\,\freq}}{1 + 2\cos(\pi\nu) \,\BSX^{-1}_{B,-\frac{\pi}{4}} + \BSX^{-2}_{B,-\frac{\pi}{4}}}.
\end{aligned}
\end{equation}
The $B$ period doesn't jump across $\pha_{c} = -\pi/4$.

We have now assembled all the elements required for computing the monodromy, so we turn to doing so next.

\subsection{Monodromy}

We will first compute the monodromy $\Tr(M_{0}M_{1}) = -2\cos(2\pi\cmexp)$ around $u=0,1$ for $\pha \in \prn{\pi/4,0}$ and $\pha \in \prn{0,-\pi/4}$, and verify that the resulting expressions are related by Stokes automorphism across $\pha_{c} = 0$. We recall that the method for computing monodromy is reviewed in~\cref{subsubsec:monodromy-EWKB}.
Following this, we shall then infer the monodromy at other phases by using Stokes automorphism across $\pha_{c} = \pm \pi/4$. The monodromy changes continuously in $\pha$; it is the ``coordinates'' $\BSX_{\gamma,\pha}$ and the representation of monodromy in terms of such coordinates that change discontinuously across critical phases $\pha_{c}$.

\begin{figure}[htbp!]
  \begin{subfigure}[b]{0.4\textwidth}
    \centering
    \scalebox{.3}{\input{figures/timelike-tikz-codes/timelike-pos-tikz-crossing-labeled}}
    \caption{$\pha \in \prn{\pi/4,0}$}
  \end{subfigure}
  \hfill
  \begin{subfigure}[b]{0.5\textwidth}
    \centering
    \scalebox{.3}{\input{figures/timelike-tikz-codes/timelike-neg-tikz-crossing-labeled}}
    \caption{$\pha \in \prn{0,-\pi/4}$}
  \end{subfigure}
  \caption{Monodromy contours $\mathcal{C}_{0,1}$ used in computing monodromies $M_{0,1}$. The labels for Stokes lines crossings are also shown.}
  \label{fig:monodromy-contours}
\end{figure}
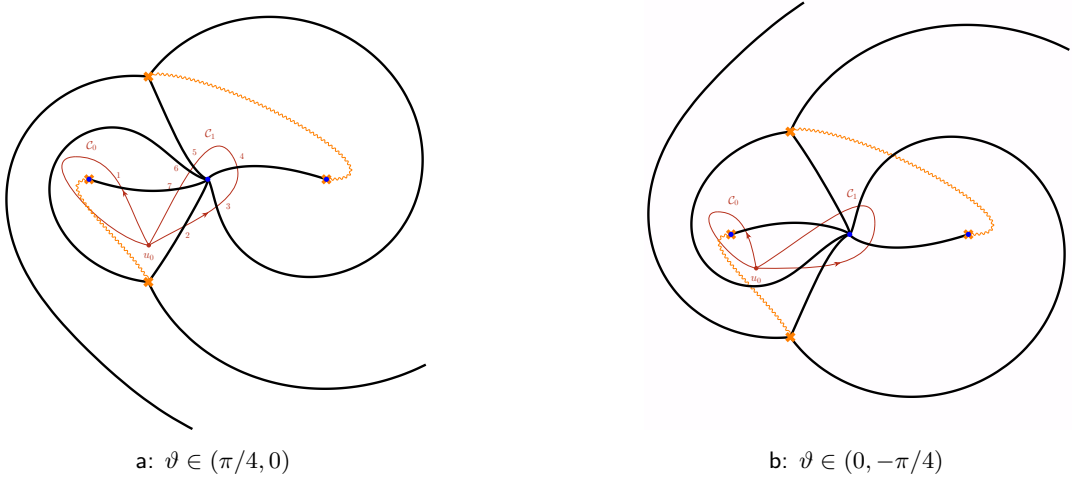

\paragraph{Computing the monodromy in different phases:}
We take the monodromy contours $\mathcal{C}_{0,1}$ to be as depicted in~\cref{fig:monodromy-contours}. For $\pha \in \prn{\pi/4,0}$, from repeated use of connection formulae, we compute the monodromies as
\begin{equation}
\begin{aligned}
M_{0}
& =
\brk{C_{s-}(0)B_{+}}\prn{V_{1}}
= \mqty(0 
& i e^{2 V_1}
\\
i e^{-2 V_1}
& -2)
\\
M_{1}
& =
\brk{C^{-1}_{t-}}\prn{V_{2}} \,\brk{C^{-1}_{t-}}\prn{V_{3}} \, \brk{C^{-1}_{s-}\prn{\frac{\nu}{2}}}\prn{V_{4}}\,
\brk{C^{-1}_{t-}}\prn{V_{5}}
\\
& \qquad \times \brk{C^{-1}_{t-}}\prn{V_{6}} \, \brk{C^{-1}_{s-}\prn{0}}\prn{V_{7}}\,
D\prn{\frac{\freq}{2}}                                                                            \\
& =
\mqty(-e^{i \pi\,\freq} 
& 0
\\
i e^{i \pi \freq} \prn{2 e^{-2 V_{4}} \cos\prn{\pi \nu} + 2e^{-2 V_{7}} +\sum_{i\in I_{t}} e^{-2 V_{i}}} 
& -e^{-i \pi \freq}),
\\
\Tr(M_{0}M_{1})
& = 
2 e^{-i\pi\,\freq} - e^{i\pi\,\freq} \prn{2\,e^{2\,V_{14}}\,\cos(\pi\nu) + 2\,e^{2\,V_{17}} +\sum_{i\in I_{t}} e^{2 \,V_{1i}}}
\end{aligned}
\end{equation}
with $I_{t} = \{2,3,5,6\}$, and $V_{ab} = V_{a} - V_{b}$. Here all connection matrices are of the form $C_{v-}$, i.e., the associated Stokes lines have negative orientations. This can be seen from $\Re \Res_{u=1}e^{i\pha} \sqrt{\phi} > 0$. The crossing of Stokes lines associated with $M_{0}$ is positive-oriented, and all the crossings associated with $M_{1}$ are negative-oriented. The subscript in $V_{a} = \int_{\beta_{a}} \lambda$ labels different crossings of Stokes lines. As explained in~\cref{subsubsec:monodromy-EWKB}, the integration path $\beta_{a}$ is from the reference point $u_{0}$ to the crossing point along the monodromy contour, and from the crossing point to the associated branch point along the Stokes line. The path $\beta_{ab} = \beta_{a} - \beta_{b}$ is a path between two branch points. We can define a cycle on $\Cwkb$ associated to $\beta_{ab}$ as
\begin{equation}
\gamma_{ab} = \beta_{ab} - \beta^{*}_{ab}
\end{equation}
where $\beta^{*}_{ab}$ is the pre-image of $\beta_{ab}$ on the other sheet. The period integral on the cycle satisfies
\begin{equation}
\oint_{\gamma_{ab}} \lambda = 2\,V_{ab}.
\end{equation}

The cycles that appear in the monodromy invariant can be decomposed into $A,B$ cycles and loops around double poles as
\begin{equation}
\begin{aligned}
\gamma_{12}
 & =
\gamma_{A}
\\
\gamma_{13}
 & =
2\prn{\gamma_{A^\prime} - 2\,C(1)} + \gamma_{B} = 2\gamma_{A} + \gamma_{B} + 2\,C(\infty)
\\
\gamma_{14}
 & =
\gamma_{C} = \gamma_{A} + \gamma_{B} - C(1) + C(\infty)
\\
\gamma_{15}
 & =
\gamma_{B} - 2\,C(1)
\\
\gamma_{16}
 & = -\gamma_{A} - 2C(1)
\\
\gamma_{17}
 & = -2\,C(1)\,.
\end{aligned}
\end{equation}
The Voros periods are then given by
\begin{equation}
\begin{split}
2\,V_{12}
 & =
2\,V_{A}
\\
2\,V_{13}
 & =
2\,V_{A} + V_{B} + 2\,\freq \cdot 2\pi i \Res_{u=\infty} \sqrt{\phi} = 2\,V_{A} + V_{B} - 2\pi\,\freq
\\
2\,V_{14}
 & =
V_{A} + V_{B} + \freq \cdot 2\pi i \prn{-\Res_{u=1} + \Res_{u=\infty}} \sqrt{\phi} = V_{A} + V_{B} -i\pi\,\freq - \pi\,\freq
\\
2\,V_{15}
 & =
V_{B} - 2\,\freq \cdot 2\pi i \Res_{u=1} \sqrt{\phi} = V_{B} - 2i\pi\,\freq
\\
2\,V_{16}
 & =
-V_{A} - 2\,\freq \cdot 2\pi i \Res_{u=1} \sqrt{\phi} = -V_{A} - 2i\pi\,\freq
\\
2\,V_{17}
 & =
-2\,\freq \cdot 2\pi i \Res_{u=1} \sqrt{\phi} = - 2i\pi\,\freq.
\end{split}
\end{equation}
The monodromy is then given by
\begin{equation}
\cos(2\pi\cmexp) = e^{i\pi\,\freq}\prn{e^{2\,V_{14}}\cos(\pi\nu) + \half \sum_{i=2,3,5,6} e^{2 V_{1i}}}.
\end{equation}
Plugging in the expressions for the other Voros periods, we obtain the result in~\cref{tab:monodromy-phases} for $\pha \in \prn{\pi/4,0}$.
It will be convenient to write its $\pha \to 0^{+}$ limit as
\begin{equation}
\label{eq:monodromy-phase-0-plus}
\begin{aligned}
\cos(2\pi\cmexp)\bigg|_{\pha = 0^{+}}
 & =
\BSX^{+}_{A,0} \, \BSX^{+}_{B,0} \, e^{-\pi\,\freq} \, \cos(\pi\nu) + \half\,  \BSX^{+}_{A,0} \,
\prn{1 + \BSX^{+}_{A,0} \, \BSX^{+}_{B,0} \, e^{-2\pi\,\freq}} \, e^{i \pi \freq}
\\
 & \qquad
+\; \half \, \BSX^{+}_{B,0}\, \prn{1 + \prn{\BSX^{+}_{A,0}\, \BSX^{+}_{B,0}}^{-1} } \, e^{-i \pi \freq} \,.
\end{aligned}
\end{equation}

We then perform a similar analysis for $\pha \in \prn{0,-\pi/4}$.
We note that the contributions associated with crossings $\{1,2,4,5,7\}$ are the same as in $\pha \in \prn{\pi/4,0}$.
The contributions associated with $\{3,6\}$ are modified as
\begin{equation}
\begin{aligned}
\gamma_{13'}
 & =
2\, \gamma_{B^\prime} +\gamma_{A} = \gamma_{A} + 2\gamma_{B} - 2\, C(1) + 2\, C(\infty)
\\
\gamma_{16'}
 & =
-\gamma_{B}\,.
\end{aligned}
\end{equation}
Therefore,
\begin{equation}
\begin{aligned}
V_{13'}
 & =
V_{A} + 2\, V_{B} + 2\,\freq\cdot 2\pi i \prn{-\Res_{u=1} + \Res_{u=\infty}} \sqrt{\phi}
= V_{A} + 2\, V_{B} - 2i\pi\,\freq - 2\pi\,\freq
\\
V_{16'}
 & =
-V_{B}.
\end{aligned}
\end{equation}
We thus obtain the result in~\cref{tab:monodromy-phases} for $\pha\in\prn{0,-\pi/4}$.
It is convenient to write its $\pha \to 0^{-}$ limit as
\begin{equation}
\label{eq:monodromy-phase-0-minus}
\begin{aligned}
\cos(2\pi\cmexp)\bigg|_{\pha = 0^{-}}
 & =
\BSX^{-}_{A,0} \, \BSX^{-}_{B,0} \, e^{-\pi\,\freq}\, \cos(\pi\nu) + \half \, \BSX^{-}_{A,0} \,
\prn{1 + \prn{\BSX^{-}_{A,0} \, \BSX^{-}_{B,0}}^{-1}} \, e^{i\pi\,\freq}
\\
 & \qquad
+\; \half\,  \BSX^{-}_{B,0}\,  \prn{1 + \BSX^{-}_{A,0} \, \BSX^{-}_{B,0} \, e^{-2\pi\,\freq}} \, e^{-i\pi\,\freq} \, .
\end{aligned}
\end{equation}
From the jump formula~\cref{eq:jump-periods} at $\pha_{c}=0$, we see that the continuity of the monodromy is indeed satisfied
\begin{equation}
\cos(2\pi\cmexp)\big|_{\pha = 0^{+}} = \cos(2\pi\cmexp)\big|_{\pha = 0^{-}}.
\end{equation}

\begin{table}[]
  \centering
  \begin{tabular}{|c|c|}
    \hline
    $\pha$
     & $\cos(2\pi\cmexp)$
    \\
    \hline
    $\in\prn{\frac{\pi}{2},\frac{\pi}{4}}$
     & $\half\prn{\BSX_{A} \,e^{i\pi\,\freq} + \BSX_{B}\, e^{-i\pi\,\freq} + \BSX^{-1}_{A}\,e^{-i\pi\,\freq}
      + \BSX^{-2}_{A}\,\BSX_{B}\,e^{-i\pi\,\freq} } + \cos(\pi\nu)\, \BSX^{-1}_{A}\,\BSX_{B}\,e^{-i\pi\,\freq} $
    \\
    \hline
    $\in\prn{\frac{\pi}{4},0}$
     & $\half\prn{\BSX_{A} e^{i\pi\,\freq} + \BSX_{B} e^{-i\pi\,\freq} + \BSX^{-1}_{A}e^{-i\pi\,\freq} + \BSX^{2}_{A}\BSX_{B}e^{-2\pi\,\freq+i\pi\,\freq} } + \cos(\pi\nu)\BSX_{A}\BSX_{B}e^{-\pi\,\freq} $
    \\
    \hline
    $\in\prn{0,-\frac{\pi}{4}}$
     & $\half\,\prn{\BSX_{A} \,e^{i\pi\,\freq} + \BSX_{B} \,e^{-i\pi\,\freq} + \BSX^{-1}_{B}\,e^{i\pi\,\freq}
      + \BSX_{A}\,\BSX^{2}_{B}\,e^{-2\pi\,\freq-i\pi\,\freq} } + \cos(\pi\nu)\,\BSX_{A}\,\BSX_{B}\,e^{-\pi\,\freq} $
    \\
    \hline
    $\in\prn{-\frac{\pi}{4},-\frac{\pi}{2}}$
     & $\half\prn{\BSX_{A}\, e^{i\pi\,\freq} + \BSX_{B} \,e^{-i\pi\,\freq} + \BSX^{-1}_{B}\,e^{i\pi\,\freq}
      + \BSX_{A}\BSX^{-2}_{B}\,e^{i\pi\,\freq} } + \cos(\pi\nu)\,\BSX_{A}\,\BSX^{-1}_{B}\,e^{i\pi\,\freq} $
    \\
    \hline
  \end{tabular}
  \caption{Monodromy at different phases $\pha = \arg(\freq) = \arg(\mom)$. Here we have the abbreviation $\BSX_{\gamma,\pha} = \BSX_{\gamma}$ for each range of $\pha$.}
  \label{tab:monodromy-phases}
\end{table}

We may now infer the monodromy at other phases using Stokes automorphism across $\pha_{c} = \pm \pi/4$ and exploit the continuity of monodromy. From the result for $\pha \in \prn{\pi/4,0}$, we write its $\pha \to (\pi/4)^{-}$ limit as
\begin{equation}
\begin{aligned}
\cos(2\pi\cmexp)\bigg|_{\pha = \prn{\frac{\pi}{4}}^{-}}
 & =
\half\BSX^{-}_{B,\frac{\pi}{4}} \, e^{-i\pi\,\freq}\,
\prn{1 + 2\BSX^{-}_{A,\frac{\pi}{4}} \, e^{-\pi\,\freq+i\pi\,\freq}\, \cos(\pi\nu)
+ \prn{\BSX^{-}_{A,\frac{\pi}{4}}}^2 \,e^{-2\pi\,\freq+2i\pi\,\freq}}
\\
 & \qquad +\;
\half\, \prn{\BSX^{-}_{A,\frac{\pi}{4}} \, e^{i\pi\,\freq} + \prn{\BSX^{-}_{A,\frac{\pi}{4}}}^{-1} \, e^{-i\pi\,\freq}} \,.
\end{aligned}
\end{equation}
Using the jump formula~\eqref{eq:jump-period-pos-pi-4} at $\pha_{c} = \pi/4$, we obtain the $\pha \to (\pi/4)^{+}$ limit of monodromy
\begin{equation}
\begin{aligned}
\cos(2\pi\cmexp)\bigg|_{\pha = \prn{\frac{\pi}{4}}^{+}}
 & =
\half\BSX^{+}_{B,\frac{\pi}{4}} \, e^{-i\pi\,\freq} \,
\prn{1 + 2\prn{\BSX^{+}_{A,\frac{\pi}{4}}}^{-1}\, \cos(\pi\nu)
+ \prn{\BSX^{+}_{A,\frac{\pi}{4}}}^{-2}}
\\
 & \qquad
+\;  \half\prn{\BSX^{+}_{A,\frac{\pi}{4}} \, e^{i\pi\,\freq} + \prn{\BSX^{+}_{A,\frac{\pi}{4}}}^{-1} \, e^{-i\pi\,\freq}}.
\end{aligned}
\end{equation}
We thus obtain the monodromy in~\cref{tab:monodromy-phases} for $\pha \in \prn{\pi/2,\pi/4}$. The monodromy for $\pha\in\prn{-\pi/4,-\pi/2}$ in~\cref{tab:monodromy-phases} can be similarly obtained using jump formula~\eqref{eq:jump-period-neg-pi-4} at $\pha_{c} = -\pi/4$.

\paragraph{Monodromy at $\pha=0$:} At $\pha=0$, we need $\cos(2\pi\cmexp)\in\R$ for the spectral function to be real-valued.
This property is not manifest from~\eqref{eq:monodromy-phase-0-plus} and~\eqref{eq:monodromy-phase-0-minus}, as the lateral Borel-summed Voros symbols are not related by conjugation, viz., 
$\BSX^{\pm}_{A,0} \neq \prn{\BSX^{\pm}_{B,0}}^{*}$.
On the other hand, the median symbols~\eqref{eq:median-symbols-phase-0} satisfy $\BSXm_{A,0} = \prn{\BSXm_{B,0}}^{*}$,
with their asymptotics recovering the asymptotic series $X_{A} = X^{*}_{B}$. In terms of the median symbols, the monodromy reads
\begin{equation}
\label{eq:mono-median-symbol}
\begin{aligned}
\cos(2\pi\cmexp)\bigg|_{\pha=0}
 & =
\BSXm_{A,0}\, \BSXm_{B,0} \, e^{-\pi\,\freq}\, \cos(\pi\nu)
\\
 &
+ \half \, \sqrt{\prn{1+\BSXm_{A,0}\,  \BSXm_{B,0}\,  e^{-2\pi\,\freq}}
  \, \prn{1 + \prn{\BSXm_{A,0} \BSXm_{B,0}}^{-1}}} \, \prn{\BSXm_{A,0}\, e^{i \pi \freq} + \BSXm_{B,0}\, e^{-i \pi \freq}}.
\end{aligned}
\end{equation}
This is indeed manifestly real after using $\BSXm_{A,0} = \BSXm^{*}_{B,0}$. Consequently, we can compute the monodromy at the critical phase using
\begin{empheq}[box=\fbox]{equation}
  \label{eq:monodromy-real-phase}
  \cos(2\pi\cmexp)\bigg|_{\pha=0}
  = \abs{\BSXm_{A,0}}^2 \, e^{-\pi\,\freq}\cos(\pi\nu)
  + \sqrt{\prn{1+\abs{\BSXm_{A,0}}^2 \, e^{-2\pi\,\freq}} \,  \prn{1 + \abs{\BSXm_{A,0}}^{-2}}} \,
  \Re\prn{\BSXm_{A,0} \, e^{i \pi \freq}}
\end{empheq}
This is the key result we sought, for it captures the complete transseries expression for the monodromy. Now it is a matter of finessing this to recover the transseries for the spectral function itself.

\section{The non-perturbative spectral function}\label{sec:non-pertsp}

We will first focus on the large real momentum ($\pha=0$) asymptotics of $\rho_{\np}$, in the timelike regime $0<\mr<1$
with non-zero spatial momentum. As noted earlier, the analysis for zero spatial momentum is deferred until~\cref{sec:WKB-zero-momentum}.

\paragraph{Transseries expansion of $\rho_{\np}$:} Our first task is to obtain a transseries expression for the non-perturbative part of the spectral function defined in~\eqref{eq:rho-np-holo-CFT}. We emphasize here that for $\nu \in \mathbb{Z}_{\geq 0}$ this is essentially the only correction to the vacuum spectral function, making it all the more interesting for one to be able to control its asymptotic behavior in these particular cases.

Let us begin by noting that the asymptotics of the median summed Voros symbol $\BSXm_{A,0}$ recovers the asymptotic series $X_{A}$.
Specifically,
\begin{equation}
\label{eq:median-symbol-asymptotics}
\BSXm_{A,0}(\freq) \sim  X_{A}(\freq) = e^{(1-i)\, \vcl\, \freq} \, P(\freq) \,, \qquad \freq \to \infty \,.
\end{equation}
Here, we already  used the explicit form of the classical $A$ period in~\eqref{eq:classical-A-period} and defined
\begin{equation}
P(\freq)
\,\coloneqq\,
\exp\prn{\sum_{\text{odd } n \geq 1} \, V^{(n)}_{A} \, \freq^{-n}} \,.
\end{equation}
The first few higher periods relevant for evaluating this are given by~\eqref{eq:higher-periods}.

From~\eqref{eq:monodromy-real-phase} and~\eqref{eq:median-symbol-asymptotics}, we find the monodromy asymptotics to be given by the transseries
\begin{equation}
\label{eq:cmexp-asymptotics}
\begin{aligned}
\cos(2\pi\cmexp)
 & \sim
e^{(2\,\vcl - \pi)\,\freq} \, \abs{P(\freq)}^2 \, \cos(\pi\nu)
+ \prn{1 + \sum^{\infty}_{n=1} \binom{\half}{n} e^{-2n\,(\pi - \vcl)\freq} \,\abs{P(\freq)}^{2n}}
\\
 &
\quad \times \prn{1 + \sum^{\infty}_{n=1} \binom{\half}{n} e^{-2n\,\vcl\,\freq} \, \abs{P(\freq)}^{-2n}} \,
e^{\vcl\, \freq} \, \Re e^{i(\pi-\vcl)\,\freq} \, P(\freq)\,, \qquad \freq \to \infty \,.
\end{aligned}
\end{equation}
Note that $\abs{\cos(2\pi\cmexp)} < e^{\pi\,\freq}$ as $\freq \to \infty$ so that $\rho_{\np} \sim 1$ to leading order, consistent with the spectrum condition in the timelike momentum regime. We can therefore expand $\rho_{\np}$ as
\begin{equation}
\label{eq:rhonp-asymptotics}
\begin{aligned}
\rho_{\np}(\freq,\mr\,\freq)
\sim
1 + 2\sum^{\infty}_{n=1} (-1)^{n\, (\nu+1)}\,  T_{n}\prn{\cos(2\pi\cmexp)} \, e^{-n\pi\,\freq} \,,
\qquad \freq \to \infty \,.
\end{aligned}
\end{equation}
Here $T_{n}(x)$ is the Chebyshev polynomial of the first kind, with $T_{n}\prn{\cos(2\pi\cmexp)} = \cos(2\pi n \cmexp)$. We obtain the full transseries expansion of $\rho_{\np}$ by substituting~\eqref{eq:cmexp-asymptotics} into~\eqref{eq:rhonp-asymptotics} (cf.~\cref{fn:2step}):
\begin{equation}
\label{eq:rhonp-transseries}
\rho_{\np}(\freq,\mr\,\freq)
\sim
1 + \sum^{\infty}_{r=1}\sum^{\infty}_{s\overset{2}{=}-r} \,
e^{(-r\pi-s\,\vcl)\,\freq} \sum^{r}_{q\overset{2}{=}r-2\floor{\Half{r}}} \, \Re e^{i q(\pi-\vcl) \,\freq} \,P_{rsq}(\freq)\,.
\end{equation}
We have defined here
\begin{equation}
\label{eq:Prsq-closed-form}
\begin{aligned}
P_{rsq}(\freq)
 & =
\frac{P(\freq)^{q}}{\abs{P(\freq)}^{q+s}} \,
\sum^{r}_{n=1} \sum^{\floor{\Half{n}}}_{j=0} \, \sum^{n-2j}_{\ell=0} \,
(2-\delta_{q,0})\,  (-1)^{n(\nu+1)+j} \,  \prn{2\, \cos(\pi\nu)}^{n-2j-\ell}  \mathfrak{P}(n,\ell,j)
\\
\mathfrak{P}(n,\ell,j)
 & =
\frac{n}{n-j} \binom{n-j}{j} \, \binom{n-2j}{\ell}\binom{\ell}{\Half{\ell+q}}  \binom{\Half{\ell}}{\Half{r-2n+2j+\ell}} \binom{\Half{\ell}}{\Half{r+s-2j}}.
\end{aligned}
\end{equation}
The convention for binomial symbol is $\binom{a}{n}=0$ for $n\notin\Z_{\geq 0}$. The leading non-perturbative sector is $(r,s,q)=(1,-1,1)$ with
\begin{equation}
P_{1,-1,1}(\freq) =  2(-1)^{\nu+1} P(\freq)\,.
\end{equation}
This is the analog of~\cite[eq.(23)]{Afkhami-Jeddi:2025wra} who examined $\freq \to \infty$ at fixed $\mom$. We, on the other hand, have focussed on taking  both $\freq,\mom \to \infty$ with their ratio fixed. There is a key difference in the transseries in the two cases: we have perturbative expansions involving integer powers of $\freq$, as opposed to fractional powers seen in each non-perturbative sector in~\cite{Afkhami-Jeddi:2025wra}. We will explain the origin of this difference in~\cref{sec:WKB-zero-momentum}.

Having understood the asymptotics of the spectral function, we can ask about the non-perturbative behaviour of the two-sided thermofield double correlator $\ts_{\np}$. Working out the asymptotics after stripping off the $\sinh \pi \freq$ factor, cf.~\eqref{eq:Cnpdef}, we obtain
\begin{equation}
\label{eq:twosidedexp-asymptotics}
\ts_{\np}(\freq,\mr\,\freq)
\sim e^{-\pi\,\freq} \prn{1 + \sum^{\infty}_{n=1} \, (-1)^{n(\nu+1)} \, U_{n}\prn{\cos(2\pi\cmexp)} \, e^{-n\pi\,\freq}} \,,
\qquad \freq \to \infty \,.
\end{equation}
Here $U_{n}(x)$ is the Chebyshev polynomial of the second kind. The full transseries expansion of $\ts_{\np}$ is obtained by substituting~\eqref{eq:cmexp-asymptotics} into~\eqref{eq:twosidedexp-asymptotics}, and takes the form (cf.~\cref{fn:2step} for notation)
\begin{equation}
\label{tsnp-transseries}
\ts_{\np}(\freq,\mr\,\freq)
\sim e^{-\pi\,\freq} \prn{1 + \sum^{\infty}_{r=1}\sum^{\infty}_{s\overset{2}{=}-r} e^{(-r\pi-s\vcl)\freq} \,
\sum^{r}_{q\overset{2}{=}r-2\floor{\Half{r}}} \Re e^{i q(\pi-\vcl) \freq} \, \widetilde{P}_{rsq}(\freq) },
\end{equation}
with
\begin{equation}
\label{eq:Ptilde-in-terms-of-P}
\widetilde{P}_{rsq}(\freq) = \sum^{\floor{\Half{r}}}_{m=0} P_{r-2m,s,q}(\freq),
\end{equation}
with the extension $P_{0,s,q}(\freq)=\delta_{s,0}\delta_{q,0}$.

\paragraph{\CFT{2}:} The analogue of~\eqref{tsnp-transseries} for \CFT{2} is (cf.~\eqref{eq:spectral-function-CFT2-integer-dimension})
\begin{equation}
\label{tsnp-transseries-CFT2}
    \ts_{\np}(\freq,\mr\,\freq)
\sim e^{-\pi\,\freq} \prn{1 + \sum^{\infty}_{r=1}\sum^{\infty}_{s\overset{2}{=}-r}  \Half{(-1)^{(\nu+1)r}} e^{(-r\pi-s\mr\pi)\freq} }.
\end{equation}
In this case, there are no oscillatory terms, and each non-perturbative sector contains no perturbative corrections. 

\section{Imprint of the black hole singularity at complex times}\label{sec:complex-time-singularity}

We have now obtained the asymptotic behaviour of the spectral function (and the two-sided correlator) in holographic CFTs.
Our result comprises the complete story for $\nu \in \mathbb{Z}_{\geq 0}$ given the discussion in~\cref{sec:thermal-OPE}. Note that from an exact WKB perspective we did not have to make any assumptions for determining the asymptotics of the monodromy. What is missing is the argument that this information by itself suffices to determine the asymptotics for \emph{any} $\nu \geq 0$.  We shall now proceed to analyze the implications of our result for signatures of the black hole singularity in the thermal observables following the analysis of earlier works described in~\cref{sec:intro}.

Let us examine the Fourier transform of the two-sided correlator with respect to frequency. This is a time domain function evaluated at fixed spatial momentum, and was considered in~\cite{Afkhami-Jeddi:2025wra,Dodelson:2025jff}:
\begin{equation}
C(t,k) = \int^{\infty}_{0} d\omega \,\cos(\omega t) \,C(\omega,k) \,.
\end{equation}
If we want to examine the case where the spatial momentum scales with frequency, we may generalize this to a spatially smeared observable by defining
\begin{equation}
\ts_{\mr}(t)
\,\coloneqq\,  \int^{\infty}_{0} d\omega \cos(\omega t) \, \ts(\omega,\mr\omega).
\end{equation}
To see the spatial smearing notice that we can write
%
%
\begin{equation}\label{eq:Ctspatialav}
    \ts_{\mr}(t) = \int_{\R^{d-1}} d\vb{x}\,C\big(t-\mr\hat{\vb{n}}\cdot\vb{x},\vb{x}\big)
\end{equation}
with $\hat{\vb{n}}$ a unit vector in an arbitrary spatial direction.

Assuming one can use the asymptotic form of $C(\omega)$ in performing the Fourier transform to find complex time singularities in $C(t)$, the singularities would arise from non-perturbative corrections in $C(\omega)$ asymptotics. This is seen from
\begin{equation}
\int^{\infty}_{0}d\omega \cos(\omega t) \,\omega^{2\nu} \,
e^{-\alpha \omega}= \Half{\Gamma(2\nu+1)} 
\, \brk{\prn{\alpha+i t}^{-2\nu-1} + 
\prn{\alpha-i t}^{-2\nu-1}}.
\end{equation}
Therefore, a non-perturbative sector with exponent $\alpha$ leads to complex time singularities at $t = \pm i \alpha$.

For the case with fixed spatial momentum, it is argued in~\cite{Dodelson:2025jff} that the exponents of all non-perturbative sectors can be determined from the directions of asymptotic quasinormal modes via the thermal product formula~\cite{Dodelson:2023vrw}. This leads to complex time singularities for $C(t,k)$ at $t = \pm t_{rq}$ with\footnote{The $r,q$ label here is related to the $n,m$ label in~\cite{Dodelson:2025jff} by $r=n+m$ and $q=n-m$.}
\begin{equation}
\label{eq:tc-mr-zero}
t_{rq} = \frac{i\beta}{2} + r \frac{i\beta}{2} + q \frac{\beta}{2}\cot\prn{\frac{\pi}{d}}\,,
\qquad r \in \Z_{\geq 1}, \quad q \in \{-r,-r+2,\dots, r\} \,.
\end{equation}

For the case with spatial momentum scaling with frequency, the exponents of all non-perturbative sectors are seen from~\eqref{tsnp-transseries}. This gives singularities at $t=\pm t_{rsq}$ with
\begin{equation}
\label{eq:tc-mr-non-zero}
\begin{split}
 & t_{rsq} = \frac{i \beta}{2} + \prn{r + s\frac{v}{\pi}} \frac{i\beta}{2} + q \frac{\beta}{2}\prn{1-\frac{v}{\pi}}, \\
 & r \in \Z_{\geq1}, \quad s\in -r + 2\Z_{\geq 0}, \quad q \in \{-r,-r+2,\dots, r\}.
\end{split}
\end{equation}
The analogue of~\eqref{eq:tc-mr-non-zero} for \CFT{2} is (cf.~\eqref{tsnp-transseries-CFT2})
\begin{equation}
\label{eq:tc-mr-non-zero-CFT2}
\begin{split}
 & t_{rs} = \frac{i \beta}{2} + \prn{r + s\mr} \frac{i\beta}{2}, \\
 & r \in \Z_{\geq1}, \quad s\in -r + 2\Z_{\geq 0}.
\end{split}
\end{equation}

A few comments are in order regarding this result:
\begin{itemize}[wide,left=0pt]
  \item
  Clearly,~\eqref{eq:tc-mr-non-zero} agrees with~\eqref{eq:tc-mr-zero} for it reduces to it
  at $\mr = \vcl = 0$ in $d=4$. However, it is generally incorrect to continue the $\mr \neq 0$ asymptotics to ascertain the $\mr = 0$ asymptotics for reasons that will become clear in~\cref{sec:WKB-zero-momentum}.  The reason for the match is that the location of the complex time singularities only care about the classical WKB periods, which have a smooth behaviour as
  a function of $\mr$. The difference lies in the asymptotic series --- we will show that for $\mr \to 0$ one has fractional powers of frequency $\freq^{-\frac{4}{3}}$, which agrees with the predictions of~\cite{Afkhami-Jeddi:2025wra}.
  \item Note that the singularity locus is not confined to $\Im(t) = n\,\frac{\beta}{2}$ for $\mr \neq 0$. It is unclear if~\eqref{eq:tc-mr-non-zero} in this case has a precise relation with the time shift of geodesics. The structure is actually more akin to that seen for the correlators of an uncharged  scalar primary around a charged black hole~\cite{Dodelson:2025jff,Ceplak:2025dds}. The effective potential for spacelike geodesics on a neutral black hole background with non-vanishing momentum does have similar qualitative features as that on a charged black hole background with vanishing momentum. In both cases, spacelike geodesics with energy above an energy barrier are attracted to the black hole singularity. 
  \item The real (resp.~imaginary) parts of~\eqref{eq:tc-mr-non-zero} are associated with oscillating (resp.~exponential) terms in the $\rho_{\np}$ asymptotics~\eqref{eq:rhonp-transseries}. There is no oscillation term in $\rho_{\np}$ asymptotics of \CFT{2} for any $\mr$. In our \CFT{4} example, the oscillation decreases as one approaches the null momentum limit. We anticipate the oscillation is absent in the large spacelike momentum limit. 
\end{itemize}

\section{Asymptotic analysis of monodromy: vanishing spatial momentum}\label{sec:WKB-zero-momentum}
Due to additional subtleties with exact WKB analysis at vanishing spatial momentum, we perform an asymptotic WKB analysis of monodromy in this case to obtain the first two perturbative terms in the leading non-perturbative sector of $\rho_{\np}$. The method is similar to the one used in asymptotic quasinormal mode  analysis~\cite{Natario:2004jd,Musiri:2005ev}. Our analysis provides an alternative derivation of the result in~\cite{Afkhami-Jeddi:2025wra} (in the case of integer dimension), and makes manifest the relation with asymptotic quasinormal modes: the $\rho_{\np}$ asymptotics and asymptotic quasinormal modes are controlled by the same data in monodromy asymptotics. As explained in~\cite{Dodelson:2025jff}, this is also anticipated from the thermal product formula.

Before diving into the details, let us understand conceptually the issues arising when $\mr = 0$.
Harken back to the starting point of our analysis, the Schr\"odinger equation, which is controlled in $d=4$ by the potential~\eqref{eq:4dQpot}, which we reproduce here for the special case of $\mr = 0$.
\begin{equation}\label{eq:4dQpotzeta0}
Q_{0}(u)
= \frac{u}{4\,(2-u)\,(u-1)^2} \,, \quad
Q_{2}(u)
= -\frac{(u^2-2\,u+2)^2}{4\,u^2\, (u-1)^2\, (u-2)^2} + \frac{\nu^2}{4\, (u-2)^2(u-1)} \,.
\end{equation}
The problem is clear now: the turning points at $u_\pm$ have merged with the singularity. In $Q_0$ the black hole singularity, which in these coordinates is at $u=0$, has transformed from a simple pole to a simple turning point. The issue is that $Q_2$ still retains a double pole at $u=0$. Consider the local model for the behavior near the origin, for which the Schr\"odinger equation can be written as
\begin{equation}
\psi''(u) - \bqty{\frac{\freq^2}{8}\, u - \frac{1}{4\,u^2} + \frac{\nu^2}{16}} \psi(u) = 0 \,.
\end{equation}
We see clearly in this form there is an issue with treating $u=0$ as a turning point at leading order in $Q_0$ (this falls under the category of merging turning points and singularities, which have been analyzed in the literature). While this is accurate for the region where $\freq^2\, u^3 \gtrsim 1$, it precisely fails for the region close to the origin. In that limit, both the leading term from $Q_0$ and the double pole from $Q_2$ contribute equally. In particular, for $\freq^2\, u^3 \ll 1$ we should treat both terms on equal footing. The local problem can be straightforwardly solved  by
changing coordinates and analyzing the solution in two regions: the very near-singularity region, which has a different scaling behavior in $\freq$, and an outer region, which is  more traditional. We will have to match the two solutions across some common domain of validity (standard matched asymptotic technique): doing so will lead to fractional powers in the perturbative series in contrast with the integer powers encountered for $\mr \neq 0$. Rather than tackle the general analysis, we will take a simple approach to recover the results derived in the literature.   

\subsection{The near-singularity expansion and monodromy asymptotics}

To set up the analysis for $\mr = 0$ first consider the change of coordinate
\begin{equation}
\hat{u}(u) = \freq \int^{u}_{0} \sqrt{Q_{0}(u^\prime)} du^\prime,
\end{equation}
i.e., $\hat{u}$ is the leading WKB integral normalized at $u=0$, with path of integration to be specified. The potential in this coordinate is
\begin{equation}
\label{eq:Qhat-of-u}
\hat{Q}(\hat{u}) = 1 + \frac{-9 + 16\, \theta^{2}}{2\, u^{3}\, \freq^{2}}
-\frac{3(-9 + 16\, \theta^{2})}{4\, u^{2}\, \freq^{2}}
+\frac{1-4\, \nu^{2}}{8\, (u-2)\, \freq^{2}}
+\frac{-17 + 32\,\theta^{2} - 4\, \nu^{2}}{8\, u\,\freq^{2}}.
\end{equation}
Compared to~\eqref{eq:4dQpotzeta0}, we have deformed the exponent for black hole singularity at $u=0$ to $\mexp_{\mathrm{sing}} = \mexp$, and will take $\mexp \to 0$ at the end of monodromy computation. We find this more convenient than directly treating the logarithmic case. Furthermore, solutions in the two coordinates are related by
\begin{equation}
\psi(u) = \freq^{-\half}\,Q^{-\frac{1}{4}}_{0}(u) \,\hat{\psi}(\hat{u}).
\end{equation}
\paragraph{Near-singularity expansion:} Consider the near-singularity region where $\hat{u}$ is finite or $u = \order{\freq^{-2/3}}$, as anticipated. We want the large $\freq$ expansion of $\hat{Q}(\hat{u})$ in this region. We expand $\hat{u}$ as
\begin{equation}
\hat{u} = \freq \,u^{3/2} \prn{\frac{1}{3\sqrt{2}} + \frac{1}{4\sqrt{2}} \,u
  + \frac{43}{224\sqrt{2}} \,u^2 + \cdots} \,.
\end{equation}
Inverting the expression we find
\begin{equation}
u = \left(\frac{3\sqrt{2}\,\hat{u}}{\freq}\right)^{2/3}
-\frac{1}{2}\left(\frac{3\sqrt{2}\,\hat{u}}{\freq}\right)^{4/3}
+\frac{5}{28}\left(\frac{3\sqrt{2}\,\hat{u}}{\freq}\right)^2 + \cdots\,.
\end{equation}
Substituting this into~\eqref{eq:Qhat-of-u}, we have the potential in the $\hat{u}$ coordinate to be given by
\begin{equation}
\hat{Q}(\hat{u}) = \hat{Q}^{(0)}(\hat{u}) +
\freq^{-\frac{4}{3}}\,
\hat{Q}^{(1)}(\hat{u}) + \order{\freq^{-2}}
\end{equation}
with
\begin{equation}\label{eq:QhatA}
\begin{aligned}
\hat{Q}^{(0)}(\hat{u})
                       & =
\frac{-\frac{1}{4} + \frac{4}{9}\, \mexp^2}{\hat{u}^2} +1
\\
\hat{Q}^{(1)}(\hat{u}) & = \mathcal{A}_{\mexp} \,\hat{u}^{-\frac{2}{3}}\,, \qquad
\mathcal{A}_{\mexp} = \frac{4-4\,\mexp^2-7\,\nu^2}{2^{1/3} \,3^{2/3}\,14}\,.
\end{aligned}
\end{equation}

We will now solve the Schr\"odinger equation in $\hat{u}$ perturbatively at large $\freq$.  Clearly, the expansion parameter is now $\freq^{-\frac{4}{3}}$. Denoting the near-singularity solution as $\hat{\chi}(\hu)$, we therefore  consider a perturbation ansatz of the form 
\begin{equation}
\hat{\chi}_{\epsilon}(\hat{u})
= \hat{\chi}^{(0)}_{\epsilon}(\hat{u})
+ \freq^{-\frac{4}{3}} \, \hat{\chi}^{(1)}_{\epsilon}(\hat{u}).
\end{equation}
The zeroth order solution in $\hat{u}$ coordinate is easily obtained to be a Bessel function owing to the double pole at the origin. One finds
\begin{equation}
\hat{\chi}^{(0)}_{\epsilon}(\hat{u})
= \sqrt{\hat{u}} \, J_{\frac{2\epsilon\mexp}{3}}(i\hat{u}), \qquad \epsilon = \pm\,.
\end{equation}
Translating to the $u$ coordinate this would give
\begin{equation}
\chi_{\epsilon}^{(0)}(u)
= \freq^{-\half}\,  \sqrt{\hat{u}}\, Q^{-\frac{1}{4}}_{0}(u) \, J_{\frac{2\epsilon\mexp}{3}}(i\hat{u})\,.
\end{equation}
This basis of solution has diagonal monodromy at $u=0$ with exponents $\frac{1}{2}+\epsilon\, \mexp$.

The first-order correction can be found using standard variation of parameters method. It is
\begin{equation}
\hchi^{(1)}_{\epsilon}(\hat{u})
= -\hchi^{(0)}_{+}(\hat{u})\,  I_{-\epsilon}(\hat{u})
+ \hchi^{(0)}_{-}(\hat{u}) \, I_{+\epsilon}(\hat{u}) \,.
\end{equation}
Here
\begin{equation}
I_{\epsilon\epsilon^{\prime}}(\hat{u})
=
\int^{\hat{u}}_{0} d\hat{v} \; \frac{\hchi^{(0)}_{\epsilon}(\hat{v})\,
Q^{(1)}(\hat{v}) \,\hchi^{(0)}_{\epsilon^{\prime}}(\hat{v})}{W}\,, \qquad
W = -\frac{2}{\pi}\, \sin(\frac{2}{3}\pi\,\mexp)\,.
\end{equation}
Once again our translation says that the first-order correction in $u$ coordinate solution is
\begin{equation}
\chi^{(1)}_{\epsilon}(u)
= -\chi^{(0)}_{+}(u) \, I_{-\epsilon}(\hat{u}) + \chi^{(0)}_{-}(u) \, I_{+\epsilon}(\hat{u}).
\end{equation}

\begin{figure}[htbp!]
  \begin{subfigure}[b]{0.3\textwidth}
    \centering
    \includegraphics[width=\linewidth]{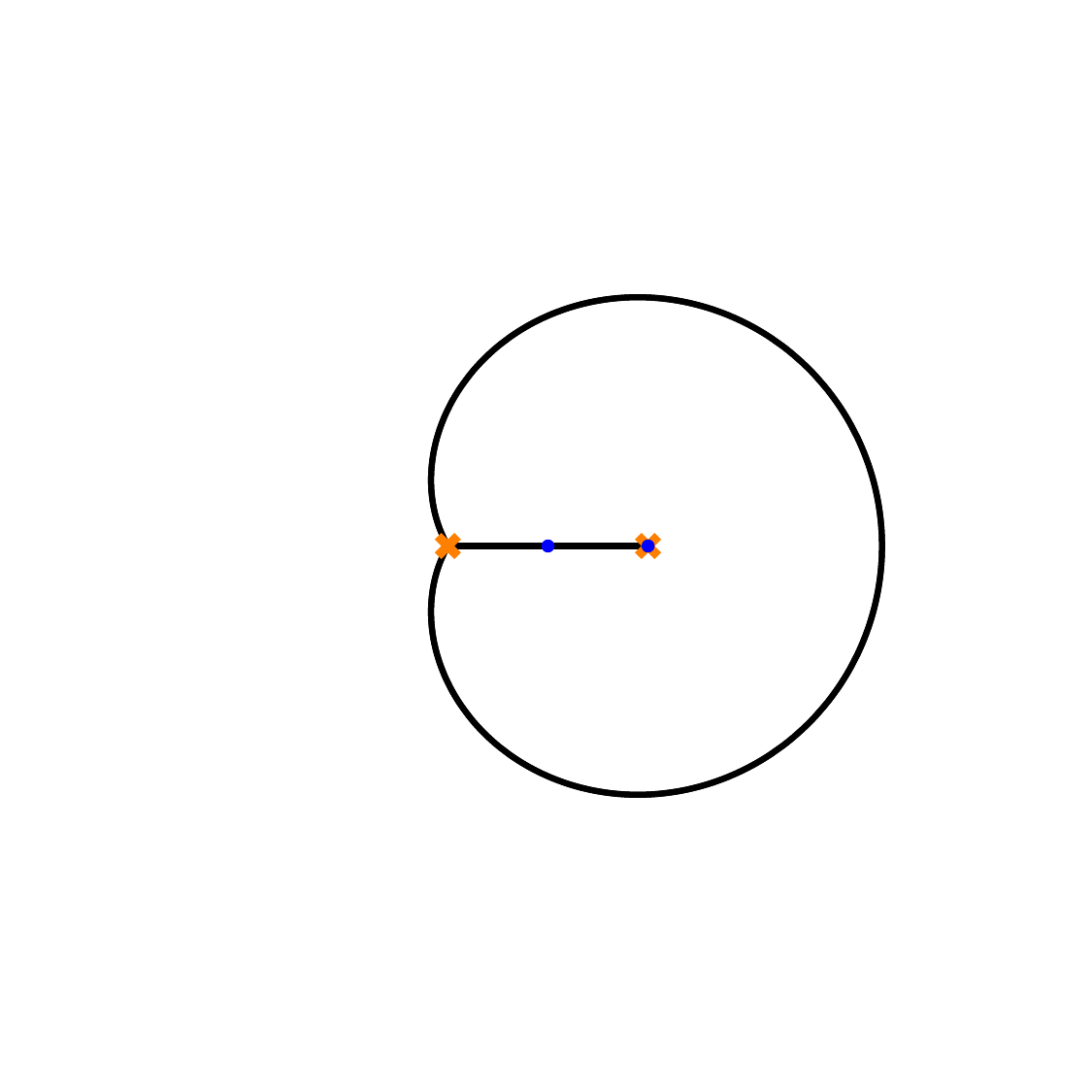}
    \caption{$\pha = \frac{\pi}{2}$}
  \end{subfigure}
  \hfill
  \begin{subfigure}[b]{0.3\textwidth}
    \centering
    \includegraphics[width=\linewidth]{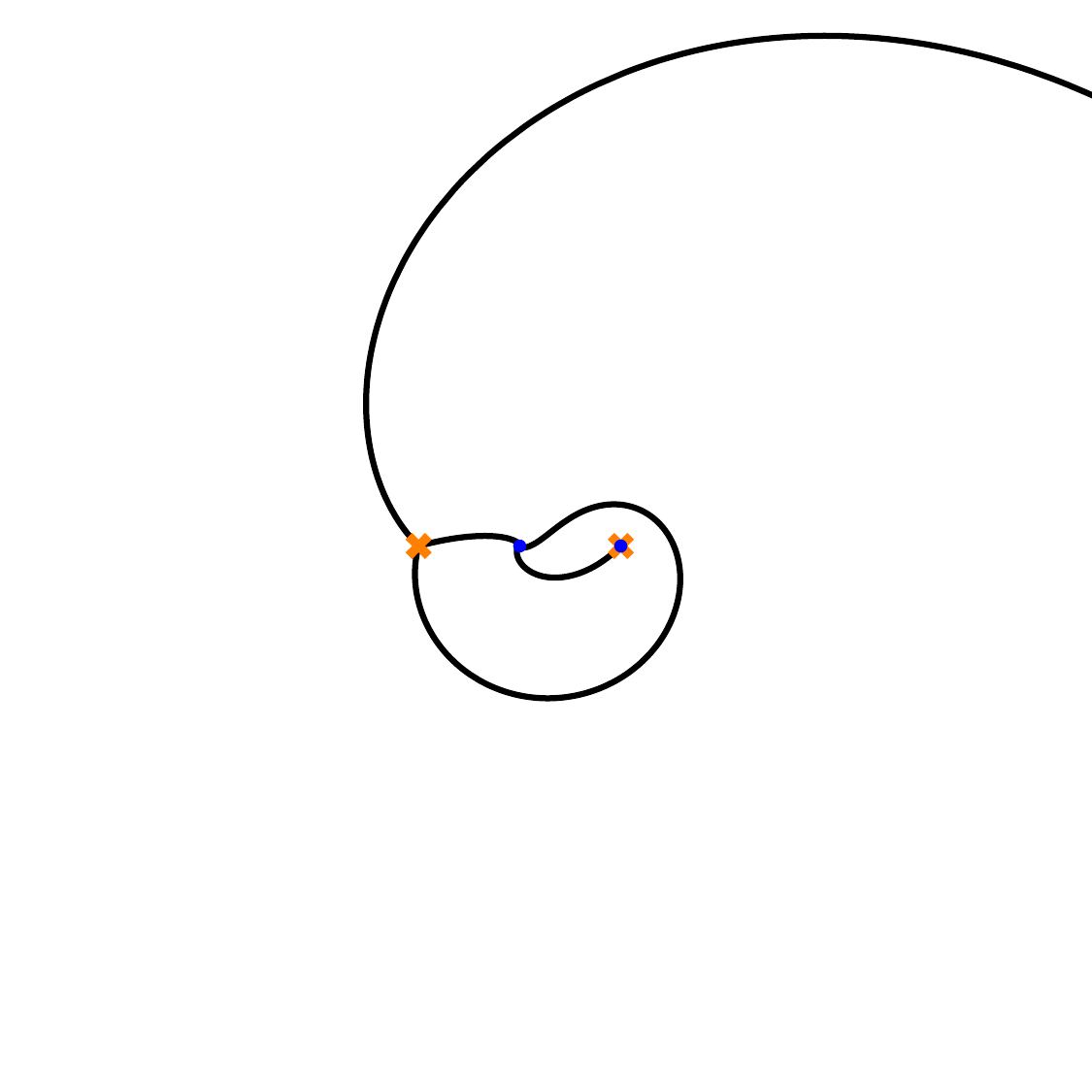}
    \caption{$\pha = \frac{3\pi}{8}$}
  \end{subfigure}
  \hfill
  \begin{subfigure}[b]{0.3\textwidth}
    \centering
    \includegraphics[width=\linewidth]{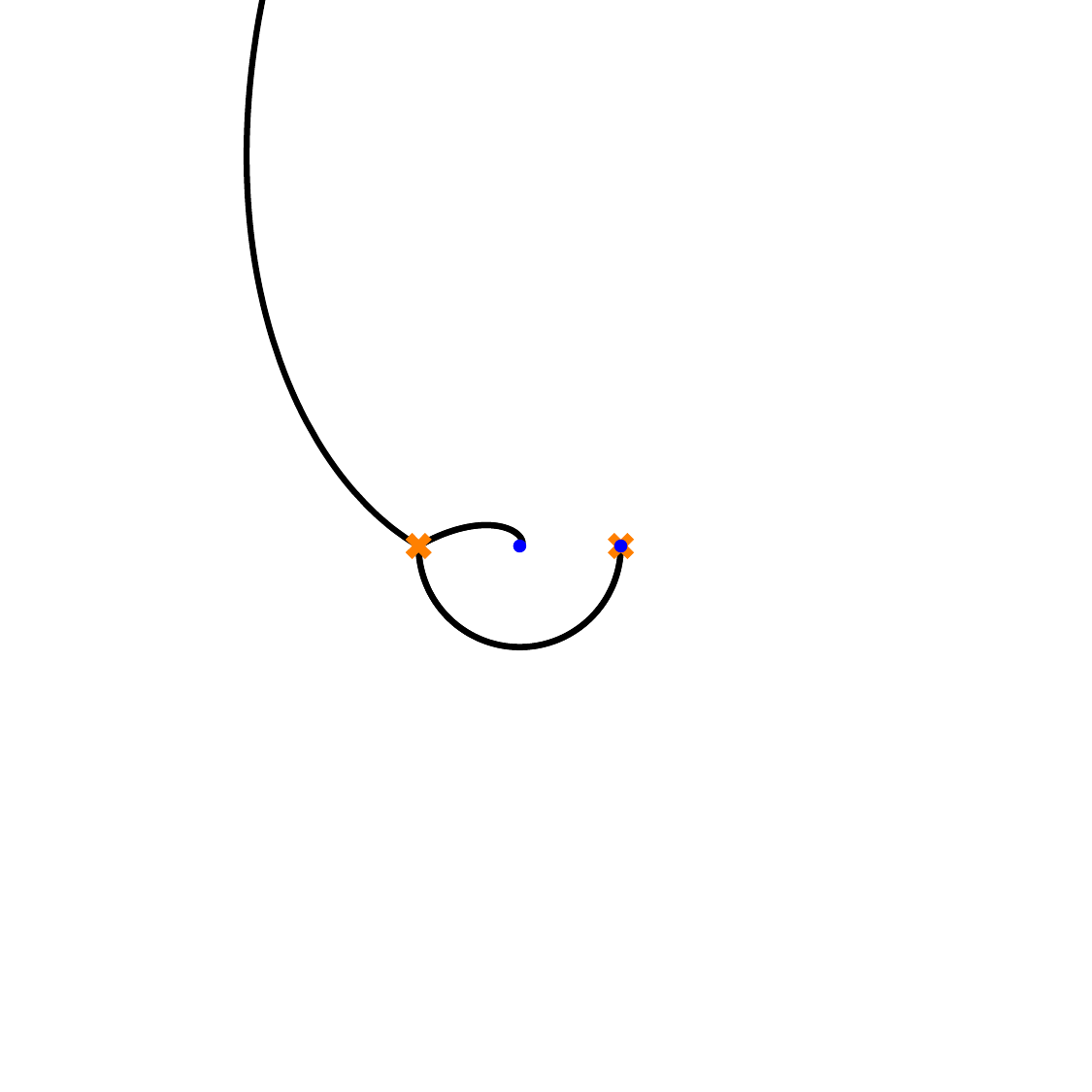}
    \caption{$\pha = \frac{\pi}{4}$}
  \end{subfigure}
  \hfill
  \begin{subfigure}[b]{0.3\textwidth}
    \centering
    \includegraphics[width=\linewidth]{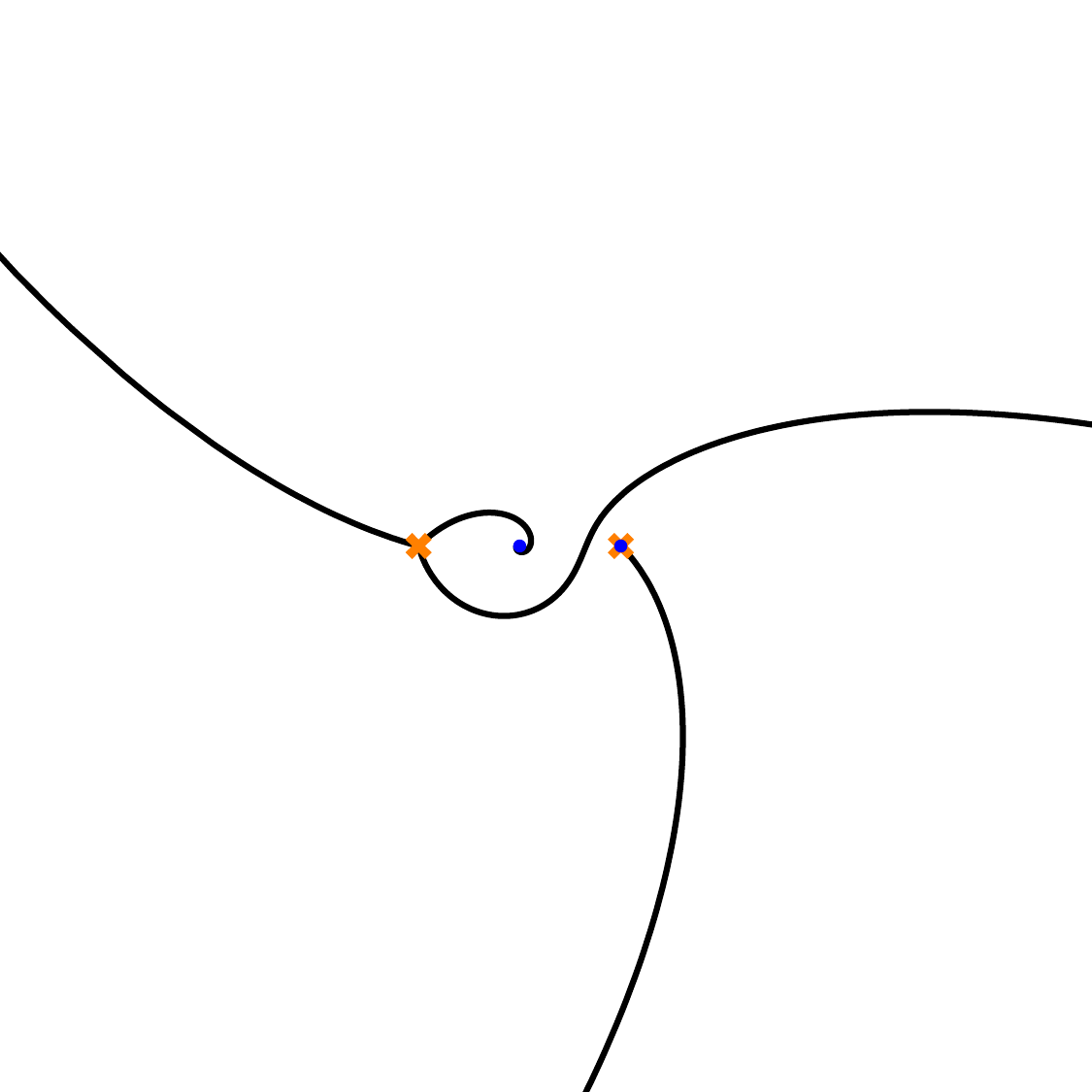}
    \caption{$\pha = \frac{\pi}{8}$}
  \end{subfigure}
  \hfill
  \begin{subfigure}[b]{0.3\textwidth}
    \centering
    \includegraphics[width=\linewidth]{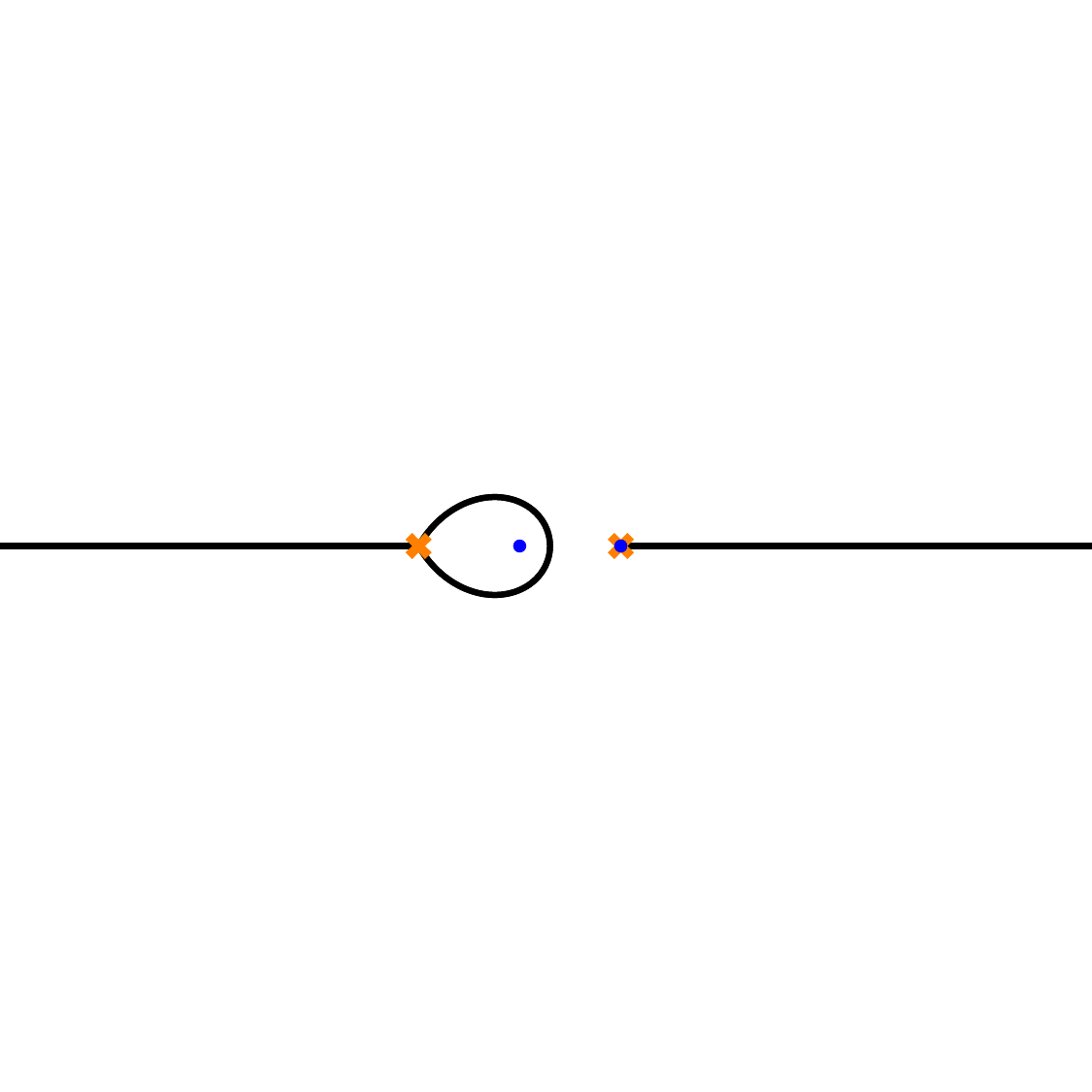}
    \caption{$\pha = 0$}
  \end{subfigure}
  \hfill
  \begin{subfigure}[b]{0.3\textwidth}
    \centering
    \includegraphics[width=\linewidth]{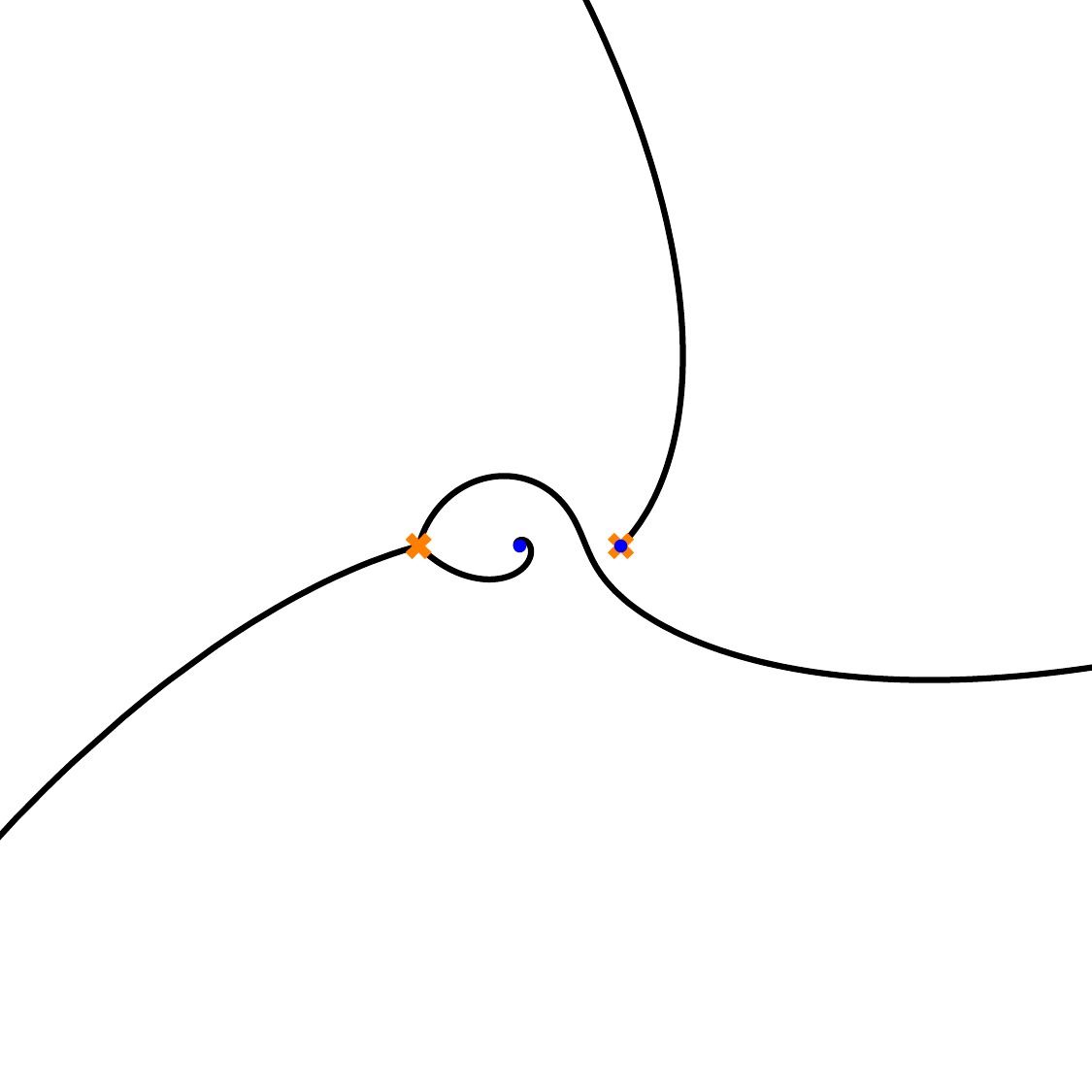}
    \caption{$\pha = -\frac{\pi}{8}$}
  \end{subfigure}
  \hfill
  \begin{subfigure}[b]{0.3\textwidth}
    \centering
    \includegraphics[width=\linewidth]{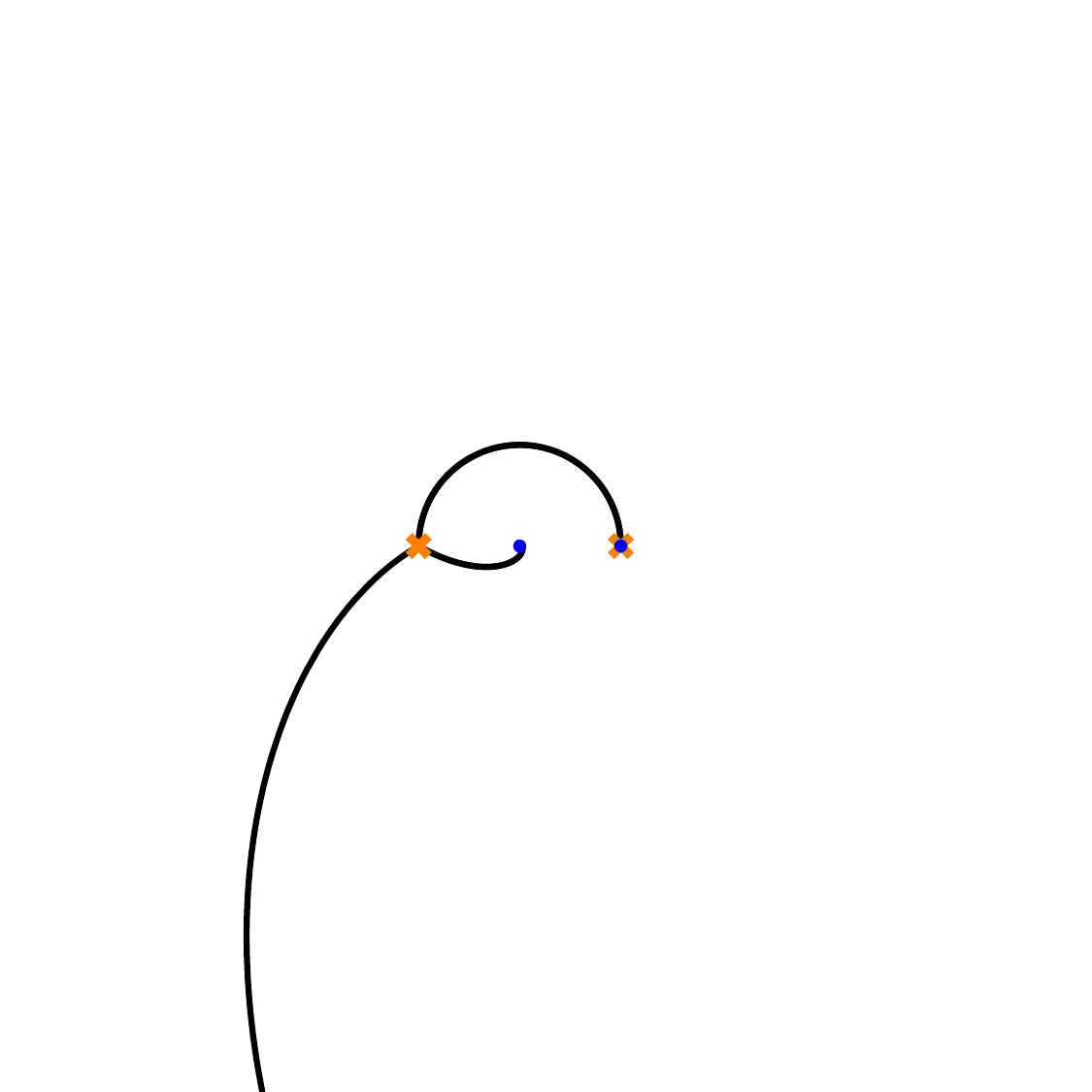}
    \caption{$\pha = -\frac{\pi}{4}$}
  \end{subfigure}
  \hfill
  \begin{subfigure}[b]{0.3\textwidth}
    \centering
    \includegraphics[width=\linewidth]{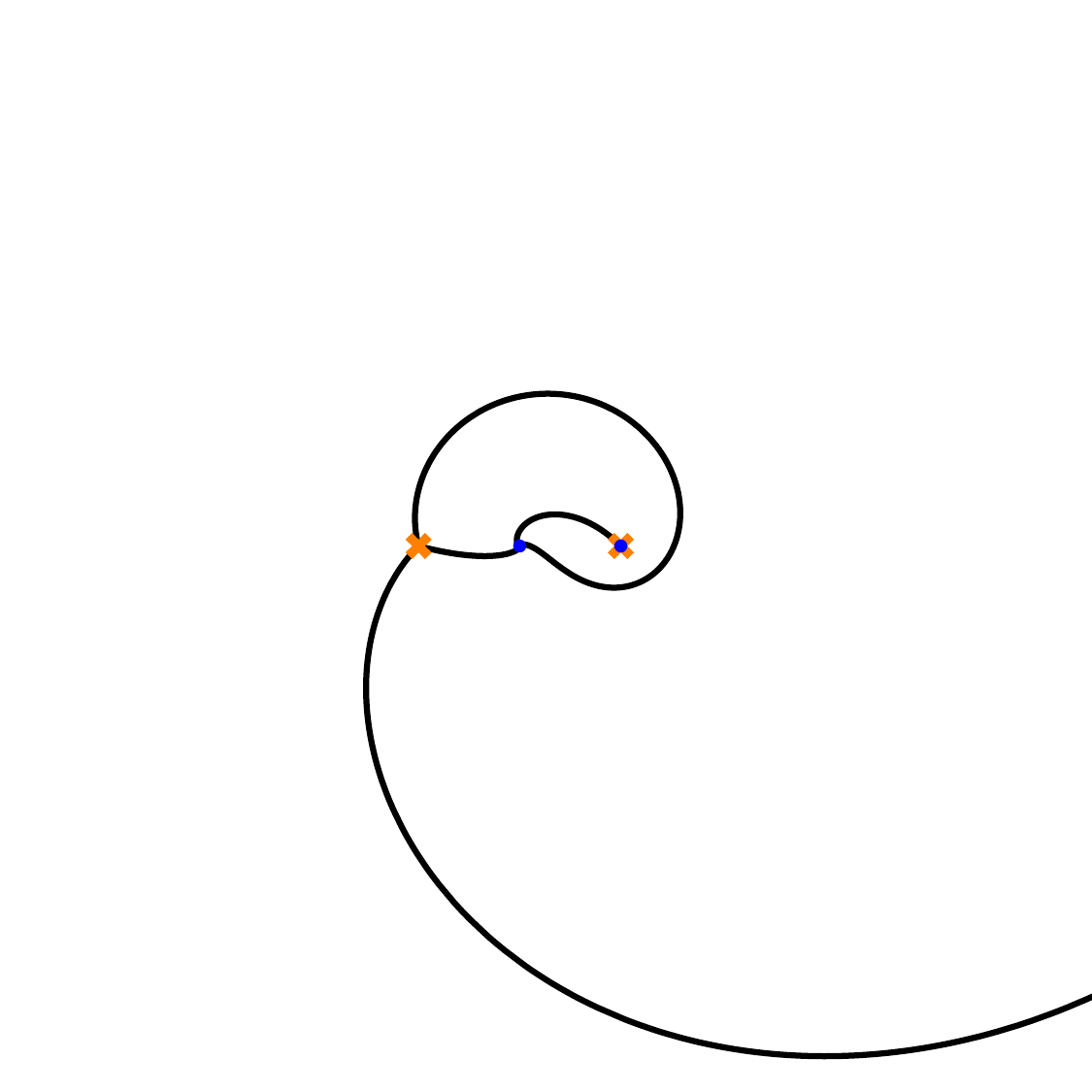}
    \caption{$\pha = -\frac{3\pi}{8}$}
  \end{subfigure}
  \hfill
  \begin{subfigure}[b]{0.3\textwidth}
    \centering
    \includegraphics[width=\linewidth]{figures/zero-momentum-anti-stokes/0.pdf}
    \caption{$\pha = -\frac{\pi}{2}$}
  \end{subfigure}
  \caption{Anti-Stokes graph at different phases $\pha = \arg(\freq)$. For $\pha \neq 0$, there exists an anti-Stokes line connecting $u=0$ (singularity) and $u=1$ (complex horizon). We match near-singularity solutions along this anti-Stokes line with the WKB solutions to compute monodromy around $u=0,1$. Note that the anti-Stokes lines are distinct from the Stokes lines that have appeared previously. The leading WKB solutions are purely oscillatory along anti-Stokes lines, while purely exponential along Stokes lines. }
  \label{fig:anti-stokes-zero-momentum}
\end{figure}

\paragraph{Matching with WKB solutions and monodromy ($\pha\neq0$):} We then continue this near-singularity solution from the very near-singularity region where $\hu$ is finite, to the outer region where $u$ is finite or $\hu = \order{\freq}$. The continuation is performed along anti-Stokes lines where
\begin{equation}
i\hu(u) \in \R.
\end{equation}
We wish to match near-singularity solutions with WKB solutions normalized at $u=1$ to compute monodromy around $u=0,1$. For $\pha \neq 0$, there exists an anti-Stokes line flowing from $u=0$ to $u=1$ (cf.~\cref{fig:anti-stokes-zero-momentum}). We therefore perform continuation along this anti-Stokes line. We will once again examine the different domains separated by critical phases.

\bigskip
\noindent\underline{$\pha \in \prn{0,-\pi/2}$:} In this case the anti-Stokes line of interest flows out of $u=0$ in direction $\arg(u) = -\pi/3 + 2\abs{\pha}/3$, and $i\hu \in \R_{+}$ along the anti-Stokes line.
From
\begin{equation}
J_{\alpha}(z) \sim \sqrt{\frac{2}{\pi z}} \cos(z-\frac{\pi\, \alpha}{2} - \frac{\pi}{4}) \,, \qquad
z \to \infty \,,
\end{equation}
we have in the outer region (along the anti-Stokes line)
\begin{equation}
\chi^{(0)}_{\epsilon}(u)
\sim
\freq^{-1/2}\, Q_{0}^{-1/4}(u) \, \cos\prn{i\hu -\epsilon \frac{\pi\, \mexp }{3} - \frac{\pi}{4}} \,.
\end{equation}
The WKB solutions normalized at $u=0$ are
\begin{equation}
\psi^{[0]}_{\epsilon}(u) \sim
\freq^{-1/2} \, Q_{0}^{-1/4}(u)\,  e^{\epsilon \,\freq \,\hu} \,.
\end{equation}
We have the matching condition
\begin{equation}
\chi^{(0)}_{\epsilon}(u)
\sim C^{(0)}_{\epsilon \epsilon^{\prime}} \, \psi^{[0]}_{\epsilon^{\prime}}(u) \,,
\qquad
C^{(0)}_{\epsilon \epsilon^{\prime}} = e^{i \epsilon^{\prime} \, \prn{\frac{\pi}{3} \, \epsilon\, \mexp - \frac{\pi}{4}}}.
\end{equation}
The first-order solution in outer region behaves as
\begin{equation}
\chi^{(1)}_{\epsilon}(u) \sim -\chi^{(0)}_{+}(u) I_{-\epsilon}(-i\infty) + \chi^{(0)}_{-}(u) I_{+\epsilon}(-i\infty).
\end{equation}
The matching including first-order correction, on the other hand, is
\begin{equation}
\label{eq:matching-first-order}
\chi_{\epsilon}(u)
\sim
C_{\epsilon \epsilon^{\prime}} \,\psi^{[0]}_{\epsilon^{\prime}}(u)\,,
\qquad
C_{\epsilon \epsilon^{\prime}} = C^{(0)}_{\epsilon \epsilon^{\prime}} + \freq^{-\frac{4}{3}} \, C^{(1)}_{\epsilon \epsilon^{\prime}}\,,
\end{equation}
with
\begin{equation}
C^{(1)}_{\epsilon\epsilon^{\prime}}
= -C^{(0)}_{+\epsilon^{\prime}} \, I_{-\epsilon}(-i\infty) + C^{(0)}_{-\epsilon^{\prime}} \, I_{+\epsilon}(-i\infty) \,.
\end{equation}

The WKB solutions normalized at $u=1$ have diagonal monodromy around $u=1$, and are related to the WKB solution normalized at $u=0$ by
\begin{equation}
\psi^{[0]}_{\pm}(u) = \exp(\pm \int^{1}_{0} \lambda) \; \psi^{[1]}_{\mp}(u).
\end{equation}
The monodromy around $u=0,1$ is therefore given by
\begin{equation}
M_{0,1} = D(\mexp) \,C \,D(-\freq/2) \,C^{-1}.
\end{equation}
Here $C=C_{\epsilon\epsilon^\prime}$ is as in~\eqref{eq:matching-first-order}, and $D(\mexp) = \mathrm{diag}\prn{-e^{2\pi i\mexp},-e^{-2\pi i\mexp}}$. 

We may now compute trace of the monodromy and take the $\mexp \to 0$ limit. Keeping only the exponentially large term and its leading perturbative correction, we have
\begin{equation}
\label{eq:monodromy-asymp-case-1}
2\cos(2\pi\cmexp)\bigg|_{\pha\in(0,-\pi/2)}
\sim e^{i\pi\,\freq}( c_{0} + c_{1}\, \freq^{-\frac{4}{3}} + \cdots) + \cdots \,,
\end{equation}
with
\begin{equation}\label{eq:c0c1zeta0}
\begin{aligned}
c_{0}
 & = 2 \,,
\\
c_{1}
 & =
3 \lim_{\mexp \to 0} \, \prnbig{I_{++}(-i\infty) - I_{--}(-i\infty)}
= e^{-2\pi i/3} \, \mathcal{A}_{0} \, \frac{6\, \pi^{3/2} \, \Gamma\prn{5/6}}{\Gamma\prn{1/3}^3}\,.
\end{aligned}
\end{equation}
We recall that $\mathcal{A}$ is defined in~\eqref{eq:QhatA}.

\bigskip
\noindent\underline{$\pha \in \prn{\pi/2,0}$:} In this case the anti-Stokes line of interest flows out of $u=0$ in direction
$\arg(u) = \pi/3 - 2\abs{\pha}/3$, and $i\hu \in \R_{-}$ along the anti-Stokes line. From
\begin{equation}
J_{\alpha}(z) \sim \sqrt{\frac{2}{\pi z}} \cos(z+\frac{\pi\, \alpha}{2} + \frac{\pi}{4}) \,, \qquad
z\to -\infty \,,
\end{equation}
we have in the outer region (along the anti-Stokes line)
\begin{equation}
\chi^{(0)}_{\epsilon}(u)
\sim \freq^{-1/2} \,  Q_{0}^{-1/4}(u) \, \cos\prn{i\hu +\epsilon \frac{\pi\,\mexp}{3} + \frac{\pi}{4}}.
\end{equation}
Consequently, the zeroth order matching is modified to
\begin{equation}
C^{(0)}_{\epsilon \epsilon^{\prime}} = e^{-i \epsilon^{\prime} \prn{\frac{\pi}{3}\, \epsilon \, \mexp - \frac{\pi}{4}}}\,,
\end{equation}
and first order correction to
\begin{equation}
C^{(1)}_{\epsilon\epsilon^{\prime}}
= -C^{(0)}_{+\epsilon^{\prime}} \,I_{-\epsilon}(i\infty) + C^{(0)}_{-\epsilon^{\prime}} \,I_{+\epsilon}(i\infty) \,.
\end{equation}
The monodromy asymptotics works out to be
\begin{equation}
\label{eq:monodromy-asymp-case-2}
2\cos(2\pi\cmexp)\big|_{\pha\in(\pi/2,0)} \sim e^{-i\pi\,\freq}(c_{0} + c^{*}_{1} \,\freq^{-\frac{4}{3}} + \cdots) + \cdots \,.
\end{equation}
Here $c_0$ is as in~\eqref{eq:c0c1zeta0}, and
\begin{equation}
c^{*}_{1}
= 3 \, \lim_{\mexp \to 0}\,  \prnbig{I_{++}(i\infty) - I_{--}(i\infty)}
= e^{2\pi i/3} \, \mathcal{A}_{0} \, \frac{6 \, \pi^{3/2} \Gamma\prn{5/6}}{\Gamma\prn{1/3}^3} \,.
\end{equation}

\paragraph{Monodromy at $\pha=0$:} One can now anticipate that the monodromy asymptotics at $\pha=0$, to the order of accuracy we keep track of, can be obtained by continuation in $\pha$ from the previous results at $\pha\neq0$. Continued to $\pha=0$, the asymptotics~\eqref{eq:monodromy-asymp-case-1} and~\eqref{eq:monodromy-asymp-case-2} are complex conjugate of each other. Hence, one may reasonably anticipate that the $\pha=0$ monodromy asymptotics is given by sum of these two terms
\begin{equation}
2\cos(2\pi\cmexp)\bigg|_{\pha=0}
\sim 2 \, \Re e^{i\pi\,\freq}\prn{c_{0} + c_{1} \, \freq^{-\frac{4}{3}} + \cdots} + \cdots \,.
\end{equation}
From~\eqref{eq:rho-np-holo-CFT}, this gives the first perturbative correction in the leading non-perturbative sector for $\rho_{\np}(\freq)$ at large real frequency
\begin{equation}
\label{eq:rhonp-asymptotics-zero-momentum}
\rho_{\np}(\freq) \sim 1 + 2(-1)^{\nu+1}\Re e^{i\pi\,\freq}\prn{c_{0} + c_{1} \freq^{-\frac{4}{3}} + \dots} e^{-\pi\,\freq} + \dots
\end{equation}
Happily, this agrees with the result obtained in~\cite{Afkhami-Jeddi:2025wra}. The $\rho_{\np}$ asymptotics~\eqref{eq:rhonp-asymptotics-zero-momentum} may be viewed as the $\mr=0$ analogue of the leading non-perturbative sector in~\eqref{eq:rhonp-transseries}, with the exponent related by continuation. The perturbative corrections in this leading sector, on the other hand, are not simply related by continuation.

\paragraph{Relation with asymptotic quasinormal modes:} An interesting cross-check of our analysis is to show that it reproduces known asymptotics of the quasinormal mode. At integer dimension, we have the following exact quantization condition for quasinormal modes and anti-quasinormal modes
\begin{equation}
\label{eq:quantization-condition}
\cosh(\pi\,\freq) + (-1)^{\nu}\cos(2\pi\cmexp) = 0.
\end{equation}
Consider $\abs{\freq} \to \infty$ with $\Re(\freq)>0$ and $\Im(\freq)<0$. The relevant monodromy asymptotics is~\eqref{eq:monodromy-asymp-case-1}. Solving~\eqref{eq:quantization-condition} with this asymptotics yields
\begin{equation}
\freq_{n} = (n+\epsilon_{\nu})(1-i) + \frac{1+i}{2\pi} \log(c_{0}) + \frac{c_{1}(1+i)}{2\pi c_{0}} \prn{n (1-i)}^{-\frac{4}{3}} + \cdots, \quad n\to\infty
\end{equation}
with $\epsilon_{\nu}=0$ for odd $\nu$ and $\epsilon_{\nu} = 1/2$ for even $\nu$. This agrees with known results on asymptotic quasinormal modes~\cite{Natario:2004jd,Dodelson:2023vrw}.


\section{Discussion}\label{sec:discussion}

We explored herein properties of the thermal spectral function for scalar primaries of holographic \CFT{d}. Our key result combines insights gleaned from the thermal OPE, the exact thermal correlators using the mapping to the connection problem in an auxiliary 2d CFT, to posit a factorization statement of the spectral function. Specifically, the spectral function factorizes  into an OPE part and a non-perturbative part~\cref{conj:rho-factorized}. While we were able to argue for this in the case of operators with integer conformal dimension (in even $d$) we believe the statement ought to hold in a more general fashion (see below). These details notwithstanding, we furthermore were able to exploit this factorization statement to provide an exact transseries expression for the asymptotics of the thermal spectral function. The latter heavily utilized the exact WKB formalism;  it relied on being able to control the asymptotics of the monodromy of the connection problem in the dual black hole spacetime.

Armed with the spectral function asymptotics, we can argue that the spatially smeared correlator $C_\mr(t)$ has a lattice of singularities in the complex time plane. These lie outside
the fundamental strip $\Im(t) \in (-\frac{\beta}{2}, \frac{\beta}{2})$ and are recorded in~\eqref{eq:tc-mr-non-zero}. Despite us not carrying out a geodesic analysis in the Lorentzian section,  these complex time singularities reproduce  the critical time $t_c$ which one would compute from the geodesic analysis (for $\mr =0$).   
What is however clear is that the monodromy between the horizon and the boundary is indeed controlled by the monodromy around the singularity and all other complex horizon locations (which for \SAdS{5} is just one other locus). This observation ties various lines of investigations together: the fact that the near-singularity behaviour controls the asymptotics of quasinormal modes~\cite{Motl:2003cd}, the bouncing singularity being tied to the direction of approach
of the quasinormal poles~\cite{Festuccia:2005pi} (which was also  recently explored in~\cite{Ceplak:2024bja,Dodelson:2025jff}).

We now outline some salient open questions that we believe are interesting to explore further.
These largely are from the perspective of bolstering the results derived herein to obtain a coherent statement about the asymptotics of the thermal spectral function in full generality.  

\paragraph{Proof of~\cref{conj:rho-factorized}:} We believe that it
should be possible to place the apparent singularity argument for~\cref{conj:rho-factorized} on a more rigorous footing by showing that $\rho^{(\Vir)}_{\pert}$ equals the coefficient of the logarithmic term in the near boundary expansion. This should involve taking the logarithmic limit of the connection formula in~\cite{Lisovyy:2022flm}.  
A related issue pertains to our numerical verification of the identity in~\cref{claim:block-identity}. Can one derive this directly from Virasoro block perspective analytically? 

\paragraph{Universality of~\cref{conj:rho-factorized}.} It would be interesting to investigate if the exact perturbative/non-perturbative factorization in~\cref{conj:rho-factorized} holds more generally beyond the case of planar neutral black holes, including spherical black holes or charged black holes. As explained in~\cite{Jia:2024zes}, the exact expression~\eqref{eq:exact-Gret} for retarded two-point function holds universally in such generalizations from applying $s$-channel connection formula. The exact expression~\eqref{eq:spectral-function-exact-expression-integer-dimension} for integer dimensions therefore also holds universally. One can then proceed as before to relate $\rho^{(\Vir)}_{\pert}$ to apparent singularity conditions at the boundary. One then needs to study if the apparent singularity conditions coincide with $\rho^{(\Vir)}_{\pert}$ and the OPE contributions (with OPE on spatial $\vb{S}^{d-1}$ and/or with chemical potentials). We hope to report on these generalizations in the near future.

\paragraph{Exact position space correlator:} The exact spectral function in~\cref{conj:rho-factorized} implies an exact expression~\eqref{eq:exact-position-space-correlator} for position space correlator in terms of stress-tensor OPE data and monodromy data of the bulk wave equation. This is reminiscent of the results obtained in~\cite{Barrat:2025twb} at vanishing spatial separation. It would be interesting to understand the 
precise relation and how the double-twist OPE  and monodromy data are linked. Both correspond to the behavior of the solution in the interior, and it would be useful to sharpen the observations already made in~\cite{Ceplak:2024bja} in this context.

\paragraph{Noninteger conformal dimension:} In this case, the spectral function no longer admits the exact perturbative/non-perturbative factorization. However, we expect its large momentum asymptotics still takes the factorized form discussed herein, viz., 
\begin{equation}
\rho(\freq,\mom) \sim \rho^{(\OPE)}_{\pert}(\freq,\mom) \rho_{\np}(\freq,\mom) \,,  \freq,\mom \to \infty \,,
\end{equation}
with $\rho_{\np}(\freq,\mom)$ to be determined. Can this property be seen from the exact expression~\eqref{eq:spectral-function-exact-expression}? Can $\rho_{\np}(\freq,\mom)$ be determined from $\rho^{(\OPE)}_{\pert}$ by resurgence? We anticipate the exponents in non-perturbative sectors coincide with the integer dimension case we have studied. This would give predictions for large order behavior of $\rho^{(\OPE)}_{\pert}$ with momentum dependence. The large order behavior in the case of vanishing momentum is studied in~\cite{Afkhami-Jeddi:2025wra}. One might also ask if a direct analysis using exact WKB techniques leads to insights into a factorized form for the case of the conformal dimensions being non-integral. 

\paragraph{Exact WKB analysis at vanishing spatial momentum:} We were only able to perform an exact WKB analysis of monodromy at non-zero spatial momentum. The vanishing momentum case is a degenerate limit where the two turning points merge with the simple pole at the black hole singularity. Despite this we were able to do a simpler analysis to obtain the leading asymptotic behavior and its perturbative corrections, exhibiting fractional scaling with the large parameter (frequency). We believe that one can make this work using a matched asymptotics --- the key element is to recognize that in the vicinity of turning points merging with singularities, the effective WKB parameter can change character.  Besides giving the full transseries expansion of $\rho_{\np}$, an exact WKB analysis of monodromy in this case would also keep track of $\order{e^{-n}}$ corrections in asymptotic quasinormal modes.

\paragraph{Geodesics interpretation of~\eqref{eq:tc-mr-non-zero}:} While the complex time singularities in~\eqref{eq:tc-mr-non-zero} associated with $\rho_{\np}$ asymptotics are natural from a momentum space perspective, their position space interpretation is somewhat obscure. The $\mr=0$ limit of~\eqref{eq:tc-mr-non-zero} does admit interpretation in terms of bouncing geodesics~\cite{Dodelson:2025jff}. A quantitative matching between~\eqref{eq:tc-mr-non-zero} and geodesic data would be desirable.

\section*{Acknowledgements}
It is a pleasure to thank Simon Caron-Huot,  Matthew Dodelson, Alba Grassi, Veronika Hubeny, Yue-Zhou Li, Andrei Parnachev, Steve Shenker and Zhenbin Yang for illuminating discussions.  M.R.~would like  to acknowledge support from the Simons Center for Geometry and Physics, Stony Brook University, during the initial stages of this project. M.R.~is also grateful to the Aspen Center for Physics, which is supported by National Science Foundation grant PHY-2210452, where some of this research was carried out. 
H.F.J.~acknowledges support from the Tsinghua Shuimu Scholar program.
M.R.~was supported by  U.S. Department of Energy grant DE-SC0009999 and by funds from the University of California.


\newpage

\appendix

\section{Evaluation of the spectral function block integral~\eqref{eq:Lorentz-integral}}\label[appendix]{app:Lorentz-integral}
In this appendix we elaborate on how the integral appearing in the spectral function block~\eqref{eq:Lorentz-integral} is evaluated. We begin by recalling the integral of interest
\begin{equation}
I_{\alpha}(p) = \int\, d^d x\,  e^{i p \cdot x} \, \sgn(t)\,  \Theta(x^2) \,  \prn{x^2}^{\alpha} \,.
\end{equation}
As the function $\sgn(t)\, \Theta(x^2) \,\prn{x^2}^{\alpha}$ is invariant under the restricted Lorentz group $SO^{+}\prn{1,d-1}$, its Fourier transform is also invariant under the restricted Lorentz group
\begin{equation}
I_{\alpha}(\Lambda p) = I_{\alpha}(p), \quad \Lambda \in SO^{+}\prn{1,d-1}.
\end{equation}
On the other hand, as the function is odd under $x \to -x$, its Fourier transform is also odd under $p \to -p$
\begin{equation}
I_{\alpha}(-p) = - I_{\alpha}(p).
\end{equation}
Since the sign of a spacelike vector can be flipped by an element of the restricted Lorentz group, this immediately implies that $I_{\alpha}(p)$ vanishes for spacelike momentum. It therefore takes the form
\begin{equation}
\label{eq:Lorentz-integral-general-form}
I_{\alpha}(p)  = \sgn(\omega)\, \Theta\prn{p^{2}} \, f(p^{2})\,
\end{equation}
We can therefore evaluate the integral in the rest frame
$p^{\mu} = (\omega,\vb{0})$ with $\omega>0$. We have
\begin{equation}
I_{\alpha}(\omega)
= 2i \, \Omega_{d-2} \,
\int^{\infty}_{0} \, dt\,  \sin\prn{\omega t} \,  \int^{t}_{0} \, dr \, r^{d-2} \, (t^2-r^2)^{\alpha},
\end{equation}
with
\begin{equation}
\Omega_{d-2} = \frac{2\pi^{\Half{d-1}}}{\Gamma\prn{\Half{d-1}}}.
\end{equation}
The radial integral evaluates to
\begin{equation}
\int^{t}_{0} \,dr\, r^{d-2} \,(t^2-r^2)^{\alpha}
= \frac{\Gamma\prn{\Half{d-1}} \,\Gamma\prn{\alpha+1}}{2\, \Gamma\prn{\alpha+\Half{d+1}}} \;
t^{2\alpha+d-1} \,.
\end{equation}
The remaining temporal integral then gives
\begin{equation}
\int^{\infty}_{0} dt \sin\prn{\omega t} \, t^{2\alpha+d-1}
=  \Gamma\prn{2\alpha+d} \, \sin\pqty{\pi\prn{\alpha + \Half{d}}} \omega^{-2\alpha-d} \,.
\end{equation}
After using duplication formula for Gamma function, we finally obtain
\begin{equation}
I_{\alpha}(\omega)
= i \pi^{\Half{d-2}}2^{2\alpha +d} \,  \Gamma(\alpha+1) \,
\Gamma\prn{\alpha+\Half{d}}  \, \sin\pqty{\pi\prn{\alpha+\Half{d}}} \, \omega^{-2\alpha-d}.
\end{equation}
The covariant formula is then inferred from~\eqref{eq:Lorentz-integral-general-form}, and is stated in the main text in~\eqref{eq:Lorentz-answer}.

\section{Schr\"odinger form of scalar wave equation on planar \SAdS{d+1}}\label[appendix]{app:Schrodinger-form}

The equation of motion for scalar field $\Phi = \phi(r) \, e^{-i\omega t + i \vb{k} \cdot \vb{x}}$
with mass $m^2 = \Delta(\Delta-d)$ is
\begin{equation}
\begin{aligned}
\phi^{\prime\prime}(r)
 & + p(r)\,  \phi^{\prime}(r) + q(r)\,  \phi(r)
= 0  \,,
\\
p(r) = \frac{d+1}{r} + \frac{f^{\prime}(r)}{f(r)}\,,
 & \qquad
q(r) = \frac{\omega^{2}}{r^{4}f(r)^{2}} - \frac{k^2}{r^4 f(r)} - \frac{m^2}{r^2 f(r)} \,,
\end{aligned}
\end{equation}
with $k = \abs{\vb{k}}$. The transformation $\phi(r) = \psi(r) \,\exp(-\half\int^{r}p(r^{\prime})dr^{\prime})$ recasts this into standard Schr\"odinger form
\begin{equation}
\begin{split}
 & \psi^{\prime\prime}(r)+ \,t(r) \,  \psi(r) = 0
\\
 & t(r) = q(r) - \frac{p(r)^2}{4} - \frac{p^{\prime}(r)}{2} \,.
\end{split}
\end{equation}
To perform further transformations in Schr\"odinger form, it is convenient to use the following  property:\footnote{This has a clear 2d CFT interpretation: $-\half$ is the dimension of the light level two degenerate operator in the large $c$ limit, and $\frac{c}{6}t$ transforms as a stress tensor.}
\begin{equation}
\begin{aligned}
\psi(r^\prime) \, (dr^{\prime})^{-\half}
 & =
\psi(r)\,  (dr)^{-\half}
\\
t(r^{\prime})
 & =
\prn{\frac{dr}{dr^{\prime}}}^2 t(r) + \half\{r,r^{\prime}\}
\end{aligned}
\end{equation}
where $\{\cdot,\cdot\}$ denotes the Schwarzian derivative.

For even $d$, we find it convenient to perform the change of variables
(which is a two-to-one map branched at $r=0,\infty$)
\begin{equation}
z = \frac{\rp^2}{r^2} \,.
\end{equation}
The coordinate change leads to a potential term which is given by
\begin{equation}
t(z) =
\frac{1}{4\, z^2} + \frac{d^2 + d^2\, \freq^2\, z}{16\, z^2\, \prn{z^{\Half{d}}-1}^2}
+ \frac{d^2 + 4\,\nu^2 + d^2 \,\mom^2 z}{16\,z^2\, \prn{z^{\Half{d}}-1}}.
\end{equation}
This potential has regular singular points at
\begin{equation}
z_{\bdy} = 0, \quad z_{\sing} = \infty, \quad z^{(n)}_{\hor} = e^{\frac{4\pi i}{d}n}, \quad n=0,\dots, \Half{d}-1.
\end{equation}
The classical dimensions associated with the singular points are
\begin{equation}
\delta_{\bdy} = \frac{1}{4} - \frac{\nu^2}{4}, \quad \delta^{(n)}_{\hor} = \frac{1}{4} + \frac{\freq^2}{4}e^{\frac{4\pi i}{d}n}, \quad \delta_{\sing} = \frac{1}{4}.
\end{equation}

It is convenient to write the result for $t(z)$ as
\begin{equation}
t(z) = \sum_{z_{i} \in \{z_{\bdy}, z^{(n)}_{\hor}\}} \,
\frac{\delta_{i}}{(z-z_{i})^2} + \frac{c_{i}}{z-z_{i}}
\end{equation}
with accessory parameters
\begin{equation}
c_{\bdy} = \frac{d^2}{16}\, (\freq^2 - \mom^2), \qquad
c^{(n)}_{\hor} = \frac{d}{8}\, (\mom^2 - \freq^2) + \frac{\nu^2 - d/2}{2d} \, e^{-\frac{4\pi i}{d} n}.
\end{equation}

In the case of $d=4$, it is efficacious to perform a different change of coordinates and use  instead
\begin{equation}
z = x \prn{1 - \frac{\rp^2}{r^2}}, \qquad x=\half \,.
\end{equation}
This is, of course, related to the general coordinate choice indicated above by a Möbius transformation.
In this $z$ coordinate, the potential is given by
\begin{equation}
t(z) = \frac{\delta_{0}}{z^{2}} + \frac{\delta_{1}}{(z-1)^2} + \frac{\delta_{x}}{(z-x)^2} + \frac{\delta_{\infty} - \delta_{0} - \delta_{1} - \delta_{x}}{z(z-1)} + \frac{(x-1)\, \acc}{z(z-1)(z-x)}\,,
\end{equation}
with
\begin{equation}
\delta_{i} = \frac{1}{4} -\mexp^2_{i}
\,, \qquad
\{\mexp_{0},\mexp_{x},\mexp_{1},\mexp_{\infty}\}
= \frac{1}{2}\, \{i\freq, \nu, \freq , 0\}\,,
\qquad
\acc = \mom^2-\freq^2 \,.
\end{equation}
%

\section{Computing semiclassical Virasoro block using recursion relation}\label[appendix]{app:block-recursion}

Here we explain how to compute semiclassical Virasoro block using Zamolodchikov's recursion relation. We will use the recursion relation in conformal dimension, which is an expansion in elliptic nome and converges faster than cross-ratio expansion. We use the conventions described in~\cite{Ribault:2014hia}, modulo one change: we use $h$ instead of $\Delta$ to denote the holomorphic weight.

The elliptic nome and cross-ratio are related by
\begin{equation}
\begin{aligned}
q  = \exp\prn{-\pi\frac{K(1-x)}{K(x)}},
 & \qquad
K(x) = \frac{\pi}{2}\,  {}_2 F_1 \prn{\half,\half;1;x} \\
x
 & =
\prn{\frac{\vartheta_{2}(q)}{\vartheta_{3}(q)}}^{4},
\end{aligned}
\end{equation}
with $q(x) = \frac{x}{16} + \order{x^2}$ at small $x$. We will need to take derivative w.r.t.\
cross-ratio in verifying the conjectured identities and therefore need a suitable translation.
The Jacobian factor between elliptic nome and cross-ratio is
\begin{equation}
x\partial_{x} = \frac{\pi^2}{4\,(1-x)\, K(x)^2} \, q\partial_{q} \,.
\end{equation}

The spherical four-point Virasoro block admits the following representation:
\begin{equation}
\mathcal{F}_{h_{\cmexp}}(h_{i}|x)
= (16\,q)^{h_{\cmexp}-\frac{Q^2}{4}} \, x^{\frac{Q^2}{4}-h_{0}-h_{x}}
\, (1-x)^{\frac{Q^2}{4}-h_{x}-h_{1}}\,
\vartheta_{3}(q)^{3Q^2-4(h_{0}+h_{x}+h_{1}+h_{\infty})}
\, H_{h_{\cmexp}}(h_{i}|q) \,.
\end{equation}
The function $H_{h_{\cmexp}}(h_{i}|q)$ can be computed using the recursion relation
\begin{equation}
H_{h_{\cmexp}}(h_{i}|q)
= 1 + \sum^{\infty}_{m,n=1} \, \frac{(16\, q)^{mn}}{h_{\cmexp} - h_{\expval{m,n}}} \,
R_{m,n} \, H_{h_{\expval{m,-n}}}(h_{i}|q) \,.
\end{equation}
The recursion uses the Virasoro representation theory, and the poles are at the degenerate representations. The residues are given by
\begin{equation}
R_{m,n} = \frac{2P_{\expval{0,0}} P_{\expval{m,n}}}{\prod_{r=1-m}^{m} \prod_{s=1-n}^{n} 2P_{\expval{r,s}}} \prod_{r\overset{2}{=}1-m}^{m-1}\prod_{s\overset{2}{=}{1-n}}^{n-1} \prod_{\pm,\pm} \prn{P_{0} \pm P_{x} + P_{\expval{r,s}}} \prn{P_{\infty} \pm P_{1} + P_{\expval{r,s}}}
\end{equation}
where $\overset{2}{=}$ denotes the index runs with step two (cf.~\cref{fn:2step}).

Using standard Liouville parameterization $c = 1 + 6\, (b +b^{-1})^2$, the semiclassical block is defined by the limit
\begin{equation}
\VBcl_{\delta_{\cmexp}}(\delta_{i}|x)
= \lim_{b \to 0} \, b^2 \, \log \mathcal{F}_{h_{\cmexp}}(h_{i}|x) \,,
\qquad P_{i} = b^{-1}\mexp_{i}, \quad P_{\cmexp} = b^{-1} \cmexp \,.
\end{equation}
From the recursive representation above, the non-trivial semiclassical limit to compute is
\begin{equation}
\Hcl_{\delta_{\cmexp}}(\delta_{i}|q) = \lim_{b \to 0} b^2 \log H_{h_{\cmexp}}(h_{i}|q).
\end{equation}
The semiclassical block is then given by
\begin{equation}
\begin{split}
\VBcl_{\delta_{\cmexp}}(\delta_{i}|x) & = -\cmexp^2 \log(16 q) + \prn{\frac{1}{4} - \delta_{x} - \delta_{0}} \log(x) + \prn{\frac{1}{4} - \delta_{x} - \delta_{1}}\log(1-x)   \\
+                                     & \brk{ 3 - 4\prn{\delta_{0} + \delta_{x} + \delta_{1} + \delta_{\infty}}} \log\vartheta_{3}(q) + \Hcl_{\delta_{\cmexp}}(\delta_{i}|q).
\end{split}
\end{equation}
In verifying the identity, it will be convenient to work with the power series part of the semiclassical block, $\VBclpow = \VBcl - (\delta_{\cmexp} - \delta_{0}-\delta_{x})\log(x)$, which is given by
\begin{equation}
\begin{split}
\VBclpow_{\delta_{\cmexp}}(\delta_{i}|x)
 &
= \cmexp^2 \prn{\log(x) - \log(16q)} + \prn{\frac{1}{4}
  - \delta_{x} - \delta_{1}}\log(1-x)
\\
 & \qquad
+  \brk{ 3 - 4\prn{\delta_{0} + \delta_{x} + \delta_{1} + \delta_{\infty}}} \log\vartheta_{3}(q) + \Hcl_{\delta_{\cmexp}}(\delta_{i}|q).
\end{split}
\end{equation}
%

\section{Details of WKB periods}\label[appendix]{app:WKB}
In this appendix we spell out the details of the evaluation of the Voros periods quoted in~\cref{sec:vorsto}.  We use these in our computation of the monodromy asymptotics.

\subsection{Classical periods}

The classical $A$ and $B$ periods for the cycles depicted in~\cref{fig:branch-cut-cycles} are given by
\begin{equation}
V^{(-1)}_{A/B} = 2 I_{\pm},
\end{equation}
with
\begin{equation}
I_{\pm} = -\int^{u{\pm}}_{0} \frac{\sqrt{(u-u_{-})\, (u-u_{+})}}{2\,(u-1)\,\sqrt{u\, (2-u)}} \,,
\qquad u_{\pm} = 2 \,\mr^2 \pm 2i\mr\, \sqrt{1-\mr^2} \,.
\end{equation}
We focus on the case with timelike momentum ($0<\mr<1$). After performing the change of coordinate
\begin{equation}
t(u) = \frac{\sqrt{1-\mr^2}}{\mr} \frac{u}{2-u}, \qquad t(u_{\pm})= \pm i,
\end{equation}
we have
\begin{equation}
\begin{split}
I_{\pm}
 & =
\mr^{\Half{3}}(1-\mr^2)^{-\frac{1}{4}} \int^{\pm i}_{0} dt \, t^{-\half} \,
(1+t^2)^{\half} \, (1-\lambda \, t^2)^{-1}, \qquad \lambda = \frac{\mr^2}{1-\mr^2}
\\
 & =
e^{\pm \frac{i\pi}{4}}\,  \mr^{\Half{3}}(1-\mr^2)^{-\frac{1}{4}} \,
\int^{1}_{0} d\abs{t} \, \abs{t}^{-\half} \, \prn{1-\abs{t}^2}^{\half} \,
\prn{1+\lambda \abs{t}^2}^{-1}
\\
 & =
\half \, e^{\pm \frac{i\pi}{4}} \, \mr^{\Half{3}}(1-\mr^2)^{-\frac{1}{4}} \,
B\prn{\frac{1}{4},\frac{3}{2}}\,  {}_2 F_1\prn{\frac{1}{4},1;\frac{7}{4};-\lambda}
\\
 & =
e^{\pm \frac{i\pi}{4}} \, \frac{\sqrt{\pi}\Gamma\prn{\frac{5}{4}}}{\Gamma\prn{\frac{7}{4}}} \,
\mr^{\Half{3}}\,  {}_2 F_1\prn{\frac{1}{4},\frac{3}{4};\frac{7}{4};\mr^2} \,.
\end{split}
\end{equation}
Here $B(x,y)$ is the Euler beta function.  In the third equality we have used the change of variable $\abs{t} = s^{1/2}$ to reduce the integral to a standard hypergeometric integral. In the last equality we have used
${}_2 F_1(a,b;c;z) = (1-z)^{-a} {}_2 F_1(a,c-b;c;z/(z-1))$.

\subsection{Higher-order periods}

Instead of directly evaluating the higher-order periods associated with higher-order WKB differentials $\lambda^{(n)} = P_{n}(w) \, dw$, we may decompose $\lambda^{(n)}$ into a basis of differentials up to total derivative pieces. The higher-order WKB differentials $\lambda^{(n)}$ with $n \geq 1$ have no simple poles,
i.e., they are meromorphic differentials of second kind.
In general, the space of such differentials modulo exact forms has dimension $2g$ on a genus $g$ surface.
For our genus one WKB curve, we can therefore decompose higher-order WKB differentials as
\begin{equation}
\label{eq:WKB-differential-decomposition}
\lambda^{(n)} = \sum^{2}_{i=1} b^{(n)}_{i} \,  \omega_{i} + df^{(n)}
\end{equation}
for some basis $\{\omega_{1},\omega_{2}\}$.
The higher-order WKB periods on cycle $\gamma$ are then given by
\begin{equation}
V^{(n)}_{\gamma}
= \sum^{2}_{i=1} b^{(n)}_{i}\,  \Pi_{i,\gamma}, \qquad
\Pi_{i,\gamma}
= \oint_{\gamma} \omega_{i} \,.
\end{equation}
We choose to work with the basis
\begin{equation}
\omega_{1} = \frac{du}{u\, (u-1)\, (u-2) \, \sqrt{Q_{0}(u)}},
\qquad
\omega_{2} = \frac{du}{u\, (u-1)\, (u-2)^2\, \sqrt{Q_{0}(u)}} \,.
\end{equation}
Here $\omega_{1}$ is a holomorphic differential, and $\omega_{2}$ a meromorphic differential with
double pole at $u=2$ on the WKB curve. The  $A$ and $B$ periods of the differentials
can be computed similarly to the classical periods. We find
\begin{equation}
\begin{aligned}
\Pi_{1,A/B}
 & =
e^{\pm i\frac{\pi}{4}} \, \frac{4\sqrt{\pi}\, \Gamma\prn{\frac{5}{4}}}{\Gamma\prn{\frac{3}{4}}} \mr^{-\half} \,  \prn{1-\mr^2}^{-\frac{1}{4}} \,,
\\
\Pi_{2,A/B}
 & =
-e^{\pm i\frac{\pi}{4}}  \, \frac{2\sqrt{\pi}\, \Gamma\prn{\frac{5}{4}}}{\Gamma\prn{\frac{3}{4}}} \,
\mr^{-\half}\, \prn{1-\mr^2}^{-\frac{3}{4}}.
\end{aligned}
\end{equation}

To find the decomposition in~\eqref{eq:WKB-differential-decomposition}, we make an ansatz for
$f^{(n)}$ of the form
\begin{equation}
f^{(n)}(u) = \frac{R^{(n)}(u)}{Q^{3n/2}_{0}(u)}\,,
\end{equation}
for some rational function $R^{(n)}(u)$ with undetermined coefficients. To wit,
\begin{equation}
R^{(n)}(u) = \sum^{2n}_{k = 1} \, \prn{ \frac{\alpha_{k}}{u^{k}} + \frac{\gamma_{k}}{(u-2)^{k}} }
+ \sum^{3n}_{k = 1} \frac{\beta_{k}}{(u-1)^k} \,.
\end{equation}
Multiplying both sides of~\eqref{eq:WKB-differential-decomposition} by
$Q^{(3n+2)/2}_{0} \, \prn{u\,(u-2)}^{2n+2}(u-1)^{3n+2}$, we find the undetermined coefficients in the rational function and $b^{(n)}_{i}$ by solving the resulting polynomial equation. Explicitly,
for $n=1,3$, we find
\begin{equation}
\begin{aligned}
b^{(1)}_{1}
 & =
\frac{1}{16}\prn{-2 + \mr^{-2} + 2\nu^2},
 & \qquad
b^{(1)}_{2}
 & = 2b^{(1)}_{1}
\\
b^{(3)}_{1}
 & =
\frac{11-\mr^2(26+4\nu^2)-12\mr^4(-1-4\nu^2+\nu^4)-8\mr^6(-1+\nu^2)^2}{3072\mr^4(1-\mr^2)^2},
 & \qquad
b^{(3)}_{2}
 & =
0\,.
\end{aligned}
\end{equation}
%

\printbibliography

\end{document}

%% file: spectral-macros.tex
\newcommand{\theorembox}[1]{%
  \begin{center}
    \fbox{\begin{minipage}{0.95\linewidth}
        #1
      \end{minipage}}%
  \end{center}
}

\newcommand{\prn}[1]{\left ( #1 \right )}
\newcommand{\prnbig}[1]{\Big( #1  \Big)}
\newcommand{\brk}[1]{\left [ #1 \right ]}

\newcommand{\cbrk}[1]{\left\{ #1 \right \} }

\newcommand{\gE}{g_{_\mathrm{E}}}
\newcommand{\Gret}{K}
\newcommand{\VBcl}{\mathcal{W}}
\newcommand{\VBclpow}{\mathtt{W}}
\newcommand{\mexp}{\theta}
\newcommand{\cmexp}{\sigma}
\newcommand{\bdy}{\mathrm{bdy}}
\newcommand{\hor}{\mathrm{hor}}
\newcommand{\sing}{\mathrm{sing}}
\newcommand{\acc}{\mathcal{E}}
\newcommand{\freq}{\mathfrak{w}}
\newcommand{\mom}{\mathfrak{q}}

\newcommand{\rp}{r_{+}}
\newcommand{\ts}{C} 
\newcommand{\pert}{\mathrm{pert}}
\newcommand{\np}{\mathrm{np}}
\newcommand{\Vir}{\mathrm{Vir}}
\newcommand{\OPE}{\mathrm{OPE}}
\newcommand{\app}{\mathrm{app}}
\renewcommand{\p}{\mathfrak{p}}

\newcommand{\Podd}{P_{\mathrm{odd}}}
\newcommand{\Cwkb}{\Sigma_{_\mathrm{WKB}}}
\newcommand{\BS}{\mathcal{S}}  
\newcommand{\BSm}{\hat{\mathcal{S}}}  
\newcommand{\BT}{\mathcal{B}}  
\newcommand{\bv}{\zeta}  
\newcommand{\pha}{\vartheta}
\newcommand{\zeros}{P_{0}}
\newcommand{\spoles}{P_{s}}
\newcommand{\dpoles}{P_{\infty}}
\newcommand{\Sad}{\mathcal{L}}
\newcommand{\BSX}{\mathcal{X}}
\newcommand{\BSV}{\mathcal{V}}
\newcommand{\BSXm}{\hat{\BSX}}

\newcommand{\SA}{\mathfrak{S}}  
\newcommand{\mc}{\mathcal{C}} 
\newcommand{\Sl}{\Gamma} 

\newcommand{\mr}{\zeta} 
\newcommand{\vcl}{v}

\tikzset{
  midarrow/.style={
      postaction={decorate},
      decoration={markings, mark=at position #1 with {\arrow{Stealth}}}
    },
  midarrow/.default=0.5
}

\newcommand{\hchi}{\hat{\chi}}

\newcommand{\hu}{\hat{u}}

\newcommand{\Op}{\mathcal{O}}
\newcommand{\stt}[1]{\cbrk{ #1 }}
\newcommand{\Pjnorm}{\mathtt{N}}
\newcommand{\Half}[1]{\frac{#1}{2}}

\newcommand{\Disc}{\mathrm{Disc}}
\newcommand{\Gspec}{\mathcal{G}}
\newcommand{\GspecOp}{\mathbb{G}}
\newcommand{\GspecNorm}{\mathcal{N}}
\newcommand{\sgn}{\mathrm{sgn}}

\newcommand{\id}{\mathbbm{1}}
\newcommand{\rhovacsub}{\mathcal{P}}
\newcommand{\BH}{\mathrm{BH}}

\newcommand{\Hcl}{\mathcal{H}}


%% file: figures/timelike-tikz-codes/timelike-cycles-tikz.tex
 
\definecolor{cfffeff}{RGB}{255,254,255}
\definecolor{cff7f00}{RGB}{255,127,0}

\def \globalscale {1.000000}
\begin{tikzpicture}[y=1cm, x=1cm, yscale=\globalscale,xscale=\globalscale, every node/.append style={scale=\globalscale}, inner sep=0pt, outer sep=0pt]

  \path[fill=cfffeff] (0.0, 19.05) rectangle (19.05, 0.0);

\coordinate (wp) at (6.3263, 12.5774);
\coordinate (wm) at (6.2163, 5.4046);
\coordinate (w0) at (4.2557, 8.8624);
\coordinate (w2) at (12.5382, 8.8624);
  
  \node[above] at ($(w0) + (-2,0)$) {\Large $u=0$}; 
  \node[below] at ($(w2) + (2,0)$)  {\Large $u=2$};
  \node[above] at ($(wp) - (0.7,-0.5)$)  {\huge $u_+$};
  \node[below] at ($(wm) - (0.5,0.5)$)  {\huge $u_-$};
  \node[below] at ($(w0)!0.5!(w2) - (0.4,0.3)$) {\Large $u=1$};

  \draw[line width = 1.2,
  draw=orange,
        decorate,
        decoration={snake, amplitude=.5mm, segment length=2mm}]
    (w0) to[out=-180, in=120] (wm);
    \draw[line width = 1.2,
  draw=orange,
        decorate,
        decoration={snake, amplitude=.5mm, segment length=2mm}]
    (w2) to[out=0, in=0] (wp);

\draw[line width = 1.4,midarrow,draw=DarkOrchid] ($ (wm)+(.5,-.5) $) to[out=0,in=0] ($ (w0) + (0,1)$);
\draw[line width = 1.4,draw=DarkOrchid] ($ (w0) + (0,1)$) to[out=180,in=180] ($ (wm)+(.5,-.5) $);
\node[text=DarkOrchid,font=\Large] at ($ (wm) + (-1.5,0)$) {$A$};

\draw[line width = 1.4,midarrow,draw=DarkOrchid, looseness=.7] ($ (w2)+(-.5,-.5) $) to[out=170,in=-120] ($ (wp) + (-.5,0)$);
\draw[line width = 1.4,draw=DarkOrchid, looseness=1.8] ($ (wp) + (-.5,0)$) to[out=60,in=-10] ($ (w2)+(-.5,-.5) $);
\node[text=DarkOrchid,font=\Large] at ($ (wp) + (6,0)$) {$A^\prime$};

\draw[line width = 1.4,midarrow,draw=ForestGreen] ($ (wp) + (1,-.1) $) to[out=-100,in=45] ($ (w0) + (0,-1.1)$);
\draw[line width = 1.4,draw=ForestGreen,dashed] ($ (w0) + (0,-1.1)$) to[out=-135,in=-130,looseness=2] ($ (w0) + (-.5,1) $) to[out=50,in=-120] ($ (wp) + (0,1) $) to[out=60, in=80,looseness=2] ($ (wp) + (1,-.1) $);
\node[text=ForestGreen,font=\Large] at ($ (wp) + (-1.5,0)$) {$B$};

\draw[line width = 1.4,midarrow,draw=ForestGreen] ($ (w2) + (.1,1.1) $) to[out=-150,in=45] ($ (wm) + (-.5,.5)$);
\draw[line width = 1.4,draw=ForestGreen,dashed,looseness=2] ($ (wm) + (-.5,.5)$) to[out=-135,in=-135,looseness=2] ($ (w2) + (0,-1) $) to[out=45, in=30,looseness=2] ($ (w2) + (.1,1.1) $); 
\node[text=ForestGreen,font=\Large] at ($ (wm) + (6,0)$) {$B^\prime$};

\draw[line width = 1.4,draw=Brown,dashed] ($ (w0) + (0,-1.1)$) to[out=150, in=180, looseness=2] ($ (w0) + (0,.5)$) to[out=0, in=180] ($(w0)!0.5!(w2) + (0,-.2)$) to[out=0,in=180] ($ (w2) + (.1,1.1) $);
\draw[line width = 1.4,draw=Brown,looseness=2] ($ (w2) + (.1,1.1) $) to[out=0, in=0] ($ (w2) + (0,-1) $); 
\draw[line width = 1.4,draw=Brown,midarrow] ($ (w2) + (0,-1) $) to[out=180,in=-30] ($ (w0) + (0,-1.1)$);
\node[text=Brown,font=\Large] at ($ (w2) + (0,-1.5)$) {$C$};

  \path[fill=cff7f00,even odd rule] (6.3263, 5.5046) -- (6.4405, 5.3902) -- 
  (6.3263, 5.276) -- (6.4405, 5.1616) -- (6.3263, 5.0473) -- (6.2119, 5.1616) --
   (6.0977, 5.0473) -- (5.9833, 5.1616) -- (6.0977, 5.276) -- (5.9833, 5.3902) 
  -- (6.0977, 5.5046) -- (6.2119, 5.3902) -- cycle(6.3263, 5.5046);

  \path[fill=cff7f00,even odd rule] (6.3263, 12.6774) -- (6.4405, 12.5632) -- 
  (6.3263, 12.4489) -- (6.4405, 12.3345) -- (6.3263, 12.2203) -- (6.2119, 
  12.3345) -- (6.0977, 12.2203) -- (5.9833, 12.3345) -- (6.0977, 12.4489) -- 
  (5.9833, 12.5632) -- (6.0977, 12.6774) -- (6.2119, 12.5632) -- cycle(6.3263, 
  12.6774);

  \path[fill=cff7f00,even odd rule] (4.2557, 9.0911) -- (4.3699, 8.9767) -- 
  (4.2557, 8.8624) -- (4.3699, 8.7481) -- (4.2557, 8.6338) -- (4.1413, 8.7481) 
  -- (4.027, 8.6338) -- (3.9127, 8.7481) -- (4.027, 8.8624) -- (3.9127, 8.9767) 
  -- (4.027, 9.0911) -- (4.1413, 8.9767) -- cycle(4.2557, 9.0911);

  \path[fill=cff7f00,even odd rule] (12.5382, 9.0911) -- (12.6525, 8.9767) -- 
  (12.5382, 8.8624) -- (12.6525, 8.7481) -- (12.5382, 8.6338) -- (12.424, 
  8.7481) -- (12.3096, 8.6338) -- (12.1954, 8.7481) -- (12.3096, 8.8624) -- 
  (12.1954, 8.9767) -- (12.3096, 9.0911) -- (12.424, 8.9767) -- cycle(12.5382, 
  9.0911);

  \path[fill=blue,even odd rule] (4.2557, 8.8624).. controls (4.2557, 8.8321) 
  and (4.2435, 8.803) .. (4.2222, 8.7815).. controls (4.2007, 8.7602) and 
  (4.1716, 8.7481) .. (4.1413, 8.7481).. controls (4.111, 8.7481) and (4.0819, 
  8.7602) .. (4.0605, 8.7815).. controls (4.039, 8.803) and (4.027, 8.8321) .. 
  (4.027, 8.8624).. controls (4.027, 8.8928) and (4.039, 8.9218) .. (4.0605, 
  8.9432).. controls (4.0819, 8.9647) and (4.111, 8.9767) .. (4.1413, 8.9767).. 
  controls (4.1716, 8.9767) and (4.2007, 8.9647) .. (4.2222, 8.9432).. controls 
  (4.2435, 8.9218) and (4.2557, 8.8928) .. (4.2557, 8.8624) -- cycle(4.2557, 
  8.8624);

  \path[fill=blue,even odd rule] (12.5382, 8.8624).. controls (12.5382, 8.8321) 
  and (12.5262, 8.803) .. (12.5047, 8.7815).. controls (12.4832, 8.7602) and 
  (12.4543, 8.7481) .. (12.424, 8.7481).. controls (12.3937, 8.7481) and 
  (12.3646, 8.7602) .. (12.3431, 8.7815).. controls (12.3216, 8.803) and 
  (12.3096, 8.8321) .. (12.3096, 8.8624).. controls (12.3096, 8.8928) and 
  (12.3216, 8.9218) .. (12.3431, 8.9432).. controls (12.3646, 8.9647) and 
  (12.3937, 8.9767) .. (12.424, 8.9767).. controls (12.4543, 8.9767) and 
  (12.4832, 8.9647) .. (12.5047, 8.9432).. controls (12.5262, 8.9218) and 
  (12.5382, 8.8928) .. (12.5382, 8.8624) -- cycle(12.5382, 8.8624);

  \path[fill=blue,even odd rule] (8.3969, 8.8624).. controls (8.3969, 8.8321) 
  and (8.3848, 8.803) .. (8.3635, 8.7815).. controls (8.342, 8.7602) and 
  (8.3129, 8.7481) .. (8.2826, 8.7481).. controls (8.2522, 8.7481) and (8.2232, 
  8.7602) .. (8.2018, 8.7815).. controls (8.1803, 8.803) and (8.1683, 8.8321) ..
   (8.1683, 8.8624).. controls (8.1683, 8.8928) and (8.1803, 8.9218) .. (8.2018,
   8.9432).. controls (8.2232, 8.9647) and (8.2522, 8.9767) .. (8.2826, 
  8.9767).. controls (8.3129, 8.9767) and (8.342, 8.9647) .. (8.3635, 8.9432).. 
  controls (8.3848, 8.9218) and (8.3969, 8.8928) .. (8.3969, 8.8624) -- 
  cycle(8.3969, 8.8624);

\end{tikzpicture}

%% file: figures/timelike-tikz-codes/timelike-cycles-saddle0-tikz.tex
 
\definecolor{cfffeff}{RGB}{255,254,255}
\definecolor{cff7f00}{RGB}{255,127,0}

\def \globalscale {1.000000}
\begin{tikzpicture}[y=1cm, x=1cm, yscale=\globalscale,xscale=\globalscale, every node/.append style={scale=\globalscale}, inner sep=0pt, outer sep=0pt]

  \path[fill=cfffeff] (0.0, 19.05) rectangle (19.05, 0.0);

\coordinate (wp) at (6.3263, 12.5774);
\coordinate (wm) at (6.2163, 5.4046);
\coordinate (w0) at (4.2557, 8.8624);
\coordinate (w2) at (12.5382, 8.8624);
  

  \draw[line width = 1.2,
  draw=orange,
        decorate,
        decoration={snake, amplitude=.5mm, segment length=2mm}]
    (w0) to[out=-180, in=120] (wm);
    \draw[line width = 1.2,
  draw=orange,
        decorate,
        decoration={snake, amplitude=.5mm, segment length=2mm}]
    (w2) to[out=0, in=0] (wp);

\draw[line width = 1.4,midarrow=.9,draw=DarkOrchid, looseness=3.5] ($ (wm)+(-.5,.5) $) to[out=0,in=-90] ($ (w0) + (-3,0)$);
\draw[line width = 1.4,draw=DarkOrchid, looseness=2] ($ (w0) + (-3,0)$) to[out=90,in=45] ($ (wp) + (.5,-.1)$);
\draw[line width = 1.4,draw=DarkOrchid, looseness=1,dashed] ($ (wp) + (.5,-.1)$) to[out=-135,in=90] ($ (w0) + (-1,0)$);
\draw[line width = 1.4,draw=DarkOrchid, looseness=1,dashed] ($ (w0) + (-1,0)$) to[out=-90,in=180] ($ (wm)+(-.5,.5) $);
\node[text=DarkOrchid,font=\Large] at ($ (w0) + (-3.5,0)$) {$\gamma_{l}$};
\node[font=\Large] at ($ (w0) + (-1.7,0)$) {$l$};

\draw[line width = 1.4,midarrow=.9,draw=ForestGreen, out looseness=2, in looseness=3] ($ (wp) + (.7,-.1)$) to[out=-135,in=180] ($ (w2) + (0,7)$);
\draw[line width = 1.4,draw=ForestGreen, out looseness=1, in looseness=1] ($ (w2) + (0,7)$) to[out=0,in=90] ($ (w2) + (6,0)$) to[out=-90, in=0] ($ (w2) + (0,-7)$) to[out=180, in=-135] ($ (wm)+(-.65,.7) $);
\draw[line width = 1.4,draw=ForestGreen, out looseness=1, in looseness=1, dashed] ($ (wm)+(-.65,.7) $) to[out=45,in=180] ($ (w2) + (0,-5)$) to[out=0,in=-90] ($ (w2) + (4,0)$) to[out=90,in=45] ($ (wp) + (.7,-.1)$);
\node[text=ForestGreen,font=\Large] at ($ (w2) + (6.5,0)$) {$\gamma_{l^\prime}$};
\node[font=\Large] at ($ (w2) + (4.7,0)$) {$l^\prime$};

  \path[draw=black,line cap=,line join=miter,line width=0.1058cm,miter 
  limit=3.25,shift={(-3.4925, 0.2646)}] (9.7044, 5.0114) -- (9.7711, 4.8991) -- 
  (9.8123, 4.8331) -- (9.8805, 4.7279) -- (9.9968, 4.5609) -- (10.0237, 4.5241) 
  -- (10.0502, 4.4887) -- (10.1018, 4.421) -- (10.2016, 4.2966) -- (10.226, 
  4.2674) -- (10.2503, 4.2386) -- (10.2985, 4.1829) -- (10.3936, 4.0771) -- 
  (10.5809, 3.8841) -- (10.6041, 3.8614) -- (10.6274, 3.8391) -- (10.674, 3.795)
   -- (10.7669, 3.7101) -- (10.9528, 3.5513) -- (10.9761, 3.5325) -- (10.9994, 
  3.5137) -- (11.0461, 3.4769) -- (11.1396, 3.4054) -- (11.3276, 3.2706) -- 
  (11.3531, 3.2531) -- (11.3787, 3.2359) -- (11.43, 3.2019) -- (11.5329, 3.136) 
  -- (11.7402, 3.0125) -- (11.7662, 2.9978) -- (11.7923, 2.9832) -- (11.8445, 
  2.9547) -- (11.9494, 2.8993) -- (12.1602, 2.7958) -- (12.1867, 2.7835) -- 
  (12.2133, 2.7714) -- (12.2663, 2.7477) -- (12.3729, 2.7018) -- (12.3996, 
  2.6906) -- (12.4263, 2.6797) -- (12.4798, 2.6582) -- (12.5871, 2.6168) -- 
  (12.6409, 2.5969) -- (12.6948, 2.5776) -- (12.8028, 2.5407) -- (13.0197, 
  2.4732) -- (13.0451, 2.4659) -- (13.0706, 2.4587) -- (13.1214, 2.4445) -- 
  (13.2232, 2.4178) -- (13.2489, 2.4114) -- (13.2744, 2.4051) -- (13.3254, 
  2.3928) -- (13.4277, 2.3697) -- (13.4532, 2.3642) -- (13.4789, 2.3588) -- 
  (13.5301, 2.3483) -- (13.6328, 2.3286) -- (13.6841, 2.3195) -- (13.7355, 
  2.3108) -- (13.8383, 2.2947) -- (13.864, 2.291) -- (13.8898, 2.2874) -- 
  (13.9412, 2.2804) -- (13.967, 2.2771) -- (13.9927, 2.2739) -- (14.0441, 
  2.2678) -- (14.0699, 2.2649) -- (14.0957, 2.2622) -- (14.1472, 2.257) -- 
  (14.173, 2.2545) -- (14.1988, 2.2521) -- (14.2503, 2.2477) -- (14.2759, 
  2.2457) -- (14.3017, 2.2439) -- (14.3275, 2.2419) -- (14.379, 2.2386) -- 
  (14.4048, 2.2372) -- (14.4562, 2.2345) -- (14.5077, 2.2323) -- (14.5335, 
  2.2313) -- (14.5592, 2.2305) -- (14.5849, 2.2297) -- (14.6106, 2.2291) -- 
  (14.6363, 2.2286) -- (14.6621, 2.228) -- (14.7125, 2.2275) -- (14.7377, 
  2.2273) -- (14.7628, 2.2273) -- (14.8132, 2.2276) -- (14.8383, 2.2279) -- 
  (14.8635, 2.2283) -- (14.8886, 2.2287) -- (14.9138, 2.2293) -- (14.9389, 
  2.2299) -- (14.9891, 2.2316) -- (15.0143, 2.2326) -- (15.0644, 2.2348) -- 
  (15.0894, 2.236) -- (15.1145, 2.2374) -- (15.1646, 2.2404) -- (15.1896, 
  2.2421) -- (15.2146, 2.2439) -- (15.2645, 2.2477) -- (15.2896, 2.2496) -- 
  (15.3146, 2.2519) -- (15.3644, 2.2564) -- (15.4639, 2.2667) -- (15.4887, 
  2.2696) -- (15.5137, 2.2725) -- (15.5633, 2.2786) -- (15.6622, 2.2921) -- 
  (15.687, 2.2957) -- (15.7117, 2.2993) -- (15.761, 2.307) -- (15.8594, 2.3235) 
  -- (15.884, 2.3278) -- (15.9085, 2.3322) -- (15.9574, 2.3414) -- (16.0551, 
  2.3609) -- (16.2494, 2.4044) -- (16.2756, 2.4107) -- (16.3018, 2.4172) -- 
  (16.354, 2.4304) -- (16.4582, 2.4581) -- (16.6648, 2.5188) -- (16.716, 2.535) 
  -- (16.7671, 2.5517) -- (16.869, 2.5863) -- (17.0706, 2.6603) -- (17.1205, 
  2.6799) -- (17.1702, 2.6999) -- (17.2693, 2.7409) -- (17.465, 2.828) -- 
  (17.8467, 3.0209) -- (17.8685, 3.0329) -- (17.8902, 3.0451) -- (17.9337, 
  3.0696) -- (18.0199, 3.1193) -- (18.1898, 3.223) -- (18.2109, 3.2363) -- 
  (18.2319, 3.2497) -- (18.2736, 3.2767) -- (18.3564, 3.3317) -- (18.5193, 
  3.4452) -- (18.5596, 3.4744) -- (18.5995, 3.5039) -- (18.6786, 3.5637) -- 
  (18.8339, 3.6869) -- (19.1325, 3.947) -- (19.1523, 3.9653) -- (19.1717, 
  3.9836) -- (19.2105, 4.0206) -- (19.2873, 4.0953) -- (19.437, 4.2485) -- 
  (19.4553, 4.2679) -- (19.4735, 4.2875) -- (19.5097, 4.3269) -- (19.5812, 
  4.4064) -- (19.72, 4.5689) -- (19.737, 4.5896) -- (19.7538, 4.6102) -- 
  (19.7873, 4.6518) -- (19.8531, 4.7358) -- (19.9806, 4.9069) -- (20.2176, 
  5.2609) -- (20.2314, 5.2833) -- (20.245, 5.3054) -- (20.2722, 5.3502) -- 
  (20.3251, 5.4402) -- (20.4265, 5.6227) -- (20.4388, 5.6458) -- (20.4511, 
  5.6688) -- (20.475, 5.7151) -- (20.5219, 5.8083) -- (20.611, 5.9969) -- 
  (20.6218, 6.0206) -- (20.6324, 6.0445) -- (20.6532, 6.0922) -- (20.6939, 
  6.1882) -- (20.7703, 6.382) -- (20.7794, 6.4064) -- (20.7884, 6.4308) -- 
  (20.8062, 6.4797) -- (20.8403, 6.5781) -- (20.9039, 6.7762) -- (20.9108, 
  6.7994) -- (20.9177, 6.8227) -- (20.9312, 6.8692) -- (20.9571, 6.9628) -- 
  (20.9634, 6.9862) -- (20.9695, 7.0097) -- (20.9816, 7.0567) -- (21.0046, 
  7.1509) -- (21.0101, 7.1745) -- (21.0155, 7.198) -- (21.0261, 7.2454) -- 
  (21.0462, 7.3403) -- (21.051, 7.364) -- (21.0557, 7.3878) -- (21.065, 7.4353) 
  -- (21.0821, 7.5307) -- (21.0862, 7.5547) -- (21.09, 7.5785) -- (21.0978, 
  7.6263) -- (21.112, 7.7223) -- (21.1186, 7.7702) -- (21.1248, 7.8183) -- 
  (21.1361, 7.9145) -- (21.141, 7.9627) -- (21.1459, 8.0109) -- (21.1481, 
  8.0351) -- (21.1501, 8.0592) -- (21.1541, 8.1074) -- (21.1577, 8.1556) -- 
  (21.1594, 8.1799) -- (21.1609, 8.204) -- (21.1624, 8.2283) -- (21.1638, 
  8.2524) -- (21.1663, 8.3008) -- (21.1674, 8.3269) -- (21.1685, 8.3533) -- 
  (21.1694, 8.3794) -- (21.1702, 8.4058) -- (21.1711, 8.4319) -- (21.1718, 
  8.4581) -- (21.1722, 8.4844) -- (21.1726, 8.5106) -- (21.173, 8.5369) -- 
  (21.1731, 8.5633) -- (21.1733, 8.5894) -- (21.1733, 8.6158) -- (21.1731, 
  8.642) -- (21.1729, 8.6683) -- (21.1725, 8.6945) -- (21.172, 8.7208) -- 
  (21.1715, 8.747) -- (21.1708, 8.7733) -- (21.17, 8.7995) -- (21.1691, 8.8258) 
  -- (21.1682, 8.852) -- (21.1669, 8.8781) -- (21.1658, 8.9045) -- (21.1645, 
  8.9307) -- (21.1614, 8.983) -- (21.158, 9.0355) -- (21.1541, 9.0879) -- 
  (21.1499, 9.1403) -- (21.1475, 9.1663) -- (21.145, 9.1925) -- (21.1399, 
  9.2447) -- (21.1372, 9.2709) -- (21.1343, 9.2969) -- (21.1282, 9.3492) -- 
  (21.125, 9.3752) -- (21.1217, 9.4013) -- (21.1149, 9.4533) -- (21.0997, 
  9.5574) -- (21.0914, 9.6092) -- (21.0827, 9.661) -- (21.0641, 9.7645) -- 
  (21.0592, 9.7903) -- (21.0541, 9.816) -- (21.0437, 9.8676) -- (21.0217, 
  9.9704) -- (21.0163, 9.9944) -- (21.0108, 10.0183) -- (20.9995, 10.066) -- 
  (20.9759, 10.1614) -- (20.9241, 10.3509) -- (20.9172, 10.3744) -- (20.9104, 
  10.3978) -- (20.896, 10.445) -- (20.8665, 10.5387) -- (20.803, 10.7247) -- 
  (20.7947, 10.7477) -- (20.7863, 10.7709) -- (20.7691, 10.8169) -- (20.7338, 
  10.9087) -- (20.6589, 11.0906) -- (20.492, 11.4474) -- (20.4699, 11.4904) -- 
  (20.4476, 11.5331) -- (20.402, 11.6184) -- (20.3069, 11.7866) -- (20.2824, 
  11.8283) -- (20.2574, 11.8697) -- (20.2067, 11.9521) -- (20.1014, 12.1146) -- 
  (20.0879, 12.1347) -- (20.0743, 12.1547) -- (20.0469, 12.1947) -- (19.9912, 
  12.2739) -- (19.8761, 12.4299) -- (19.6317, 12.7314) -- (19.6144, 12.7513) -- 
  (19.5971, 12.7711) -- (19.5621, 12.8107) -- (19.4914, 12.8888) -- (19.3459, 
  13.0416) -- (19.3273, 13.0605) -- (19.3087, 13.0791) -- (19.2712, 13.1163) -- 
  (19.1952, 13.1898) -- (19.0397, 13.3329) -- (19.02, 13.3506) -- (19.0, 
  13.3681) -- (18.9602, 13.4027) -- (18.8794, 13.4712) -- (18.7144, 13.6041) -- 
  (18.3715, 13.8538) -- (18.3509, 13.8676) -- (18.3304, 13.8815) -- (18.289, 
  13.9088) -- (18.2057, 13.9626) -- (18.0366, 14.0665) -- (18.0151, 14.079) -- 
  (17.9937, 14.0915) -- (17.9507, 14.1163) -- (17.8641, 14.165) -- (17.6885, 
  14.2584) -- (17.6441, 14.281) -- (17.5996, 14.3032) -- (17.5099, 14.3465) -- 
  (17.3286, 14.4292) -- (17.3057, 14.4391) -- (17.2828, 14.4489) -- (17.2369, 
  14.4683) -- (17.1446, 14.5062) -- (16.958, 14.5776) -- (16.9326, 14.5868) -- 
  (16.907, 14.5959) -- (16.8559, 14.614) -- (16.7532, 14.6485) -- (16.7275, 
  14.657) -- (16.7015, 14.6652) -- (16.6499, 14.6815) -- (16.5458, 14.7126) -- 
  (16.5198, 14.7202) -- (16.4936, 14.7276) -- (16.4412, 14.7421) -- (16.3361, 
  14.7698) -- (16.1243, 14.82) -- (16.0977, 14.8258) -- (16.0711, 14.8314) -- 
  (16.0176, 14.8423) -- (15.9106, 14.863) -- (15.8838, 14.8678) -- (15.8569, 
  14.8726) -- (15.8031, 14.8817) -- (15.6952, 14.8988) -- (15.6411, 14.9065) -- 
  (15.587, 14.914) -- (15.5327, 14.9208) -- (15.4784, 14.9272) -- (15.4511, 
  14.9302) -- (15.424, 14.9332) -- (15.3694, 14.9388) -- (15.3148, 14.9437) -- 
  (15.2603, 14.9483) -- (15.2065, 14.9524) -- (15.1528, 14.956) -- (15.099, 
  14.959) -- (15.072, 14.9604) -- (15.0451, 14.9618) -- (15.0181, 14.9629) -- 
  (14.9913, 14.964) -- (14.9643, 14.9649) -- (14.9372, 14.9658) -- (14.9102, 
  14.9665) -- (14.8832, 14.967) -- (14.8292, 14.9678) -- (14.8022, 14.9681) -- 
  (14.7752, 14.9683) -- (14.7482, 14.9683) -- (14.721, 14.9681) -- (14.694, 
  14.968) -- (14.6669, 14.9676) -- (14.6399, 14.9671) -- (14.6127, 14.9666) -- 
  (14.5857, 14.9659) -- (14.5586, 14.9651) -- (14.5316, 14.9643) -- (14.4773, 
  14.962) -- (14.4502, 14.9608) -- (14.4231, 14.9594) -- (14.396, 14.9579) -- 
  (14.3417, 14.9546) -- (14.2875, 14.9507) -- (14.2604, 14.9487) -- (14.2332, 
  14.9465) -- (14.1789, 14.9417) -- (14.1519, 14.9392) -- (14.1248, 14.9364) -- 
  (14.0705, 14.9308) -- (13.962, 14.918) -- (13.935, 14.9144) -- (13.9079, 
  14.9108) -- (13.8537, 14.9031) -- (13.7454, 14.8864) -- (13.7182, 14.882) -- 
  (13.6912, 14.8773) -- (13.6372, 14.8678) -- (13.5292, 14.8471) -- (13.5039, 
  14.842) -- (13.4787, 14.8368) -- (13.4284, 14.8262) -- (13.328, 14.8034) -- 
  (13.3029, 14.7975) -- (13.2778, 14.7914) -- (13.2277, 14.7789) -- (13.1276, 
  14.7527) -- (13.1027, 14.746) -- (13.0776, 14.7389) -- (13.0277, 14.7248) -- 
  (12.9281, 14.695) -- (12.7295, 14.6301) -- (12.7048, 14.6215) -- (12.68, 
  14.6127) -- (12.6307, 14.5949) -- (12.5322, 14.5579) -- (12.3359, 14.4782) -- 
  (12.3115, 14.4677) -- (12.2871, 14.4571) -- (12.2384, 14.4355) -- (12.1412, 
  14.3909) -- (11.9479, 14.2956) -- (11.9218, 14.2821) -- (11.8956, 14.2683) -- 
  (11.8437, 14.2405) -- (11.7399, 14.1829) -- (11.5338, 14.06) -- (11.5081, 
  14.044) -- (11.4825, 14.0276) -- (11.4314, 13.9945) -- (11.3294, 13.9263) -- 
  (11.1267, 13.7807) -- (11.1015, 13.7615) -- (11.0763, 13.7422) -- (11.026, 
  13.7029) -- (10.9255, 13.6218) -- (10.725, 13.4477) -- (10.6999, 13.4247) -- 
  (10.675, 13.4014) -- (10.6248, 13.354) -- (10.5244, 13.2552) -- (10.3211, 
  13.0384) -- (10.297, 13.011) -- (10.2729, 12.9833) -- (10.2241, 12.926) -- 
  (10.1244, 12.8035) -- (10.099, 12.7708) -- (10.073, 12.7372) -- (10.0203, 
  12.6667) -- (9.9066, 12.5062) -- (9.8753, 12.4598) -- (9.8421, 12.4091) -- 
  (9.8056, 12.3519) -- (9.7624, 12.2822) -- (9.7258, 12.1473) -- (9.7759, 12.06)
   -- (9.8107, 11.9991) -- (9.8397, 11.948) -- (9.8654, 11.9027) -- (9.8887, 
  11.8613) -- (9.9306, 11.7869) -- (10.0013, 11.6601) -- (10.0171, 11.6316) -- 
  (10.0324, 11.604) -- (10.0613, 11.5517) -- (10.1142, 11.4554) -- (10.2058, 
  11.2876) -- (10.3524, 11.0152) -- (10.3612, 10.9988) -- (10.3697, 10.9828) -- 
  (10.3865, 10.9513) -- (10.4189, 10.8906) -- (10.4789, 10.7776) -- (10.5842, 
  10.5789) -- (10.7546, 10.2576) -- (10.9987, 9.8065) -- (11.164, 9.5143) -- 
  (11.3009, 9.2848) -- (11.4001, 9.1265) -- (11.4771, 9.0089) -- (11.543, 
  8.9119) -- (11.5912, 8.843) -- (11.6324, 8.7857) -- (11.664, 8.7428) -- 
  (11.687, 8.712) -- (11.7067, 8.6859) -- (11.7212, 8.667) -- (11.7323, 8.6524) 
  -- (11.7419, 8.6402) -- (11.7489, 8.6312) -- (11.7548, 8.6236) -- (11.7591, 
  8.6181) -- (11.7624, 8.6138) -- (11.7653, 8.6103) -- (11.7673, 8.6076) -- 
  (11.7691, 8.6054) -- (11.7704, 8.6038) -- (11.7713, 8.6025) -- (11.7723, 
  8.6014) -- (11.7729, 8.6006) -- (11.7737, 8.5995) -- (11.774, 8.5992) -- 
  (11.7741, 8.5992) -- (11.7741, 8.599) -- (11.7742, 8.599) -- (11.7742, 8.5988);

  \path[draw=black,line cap=,line join=miter,line width=0.1058cm,miter 
  limit=3.25,shift={(-3.4925, 0.2646)}] (9.7044, 5.0114) -- (9.5743, 5.0132) -- 
  (9.4972, 5.0161) -- (9.4321, 5.0196) -- (9.3737, 5.0235) -- (9.3198, 5.0278) 
  -- (9.2694, 5.0325) -- (9.2214, 5.0374) -- (9.1755, 5.0425) -- (9.1314, 5.048)
   -- (9.0889, 5.0537) -- (9.0075, 5.0655) -- (8.9684, 5.0719) -- (8.9302, 
  5.0783) -- (8.8565, 5.0919) -- (8.8207, 5.0989) -- (8.7855, 5.106) -- (8.7169,
   5.1209) -- (8.5861, 5.1523) -- (8.5544, 5.1605) -- (8.5233, 5.1689) -- 
  (8.4621, 5.186) -- (8.344, 5.2217) -- (8.3152, 5.2309) -- (8.2867, 5.2401) -- 
  (8.2307, 5.2591) -- (8.1217, 5.2986) -- (8.0951, 5.3086) -- (8.0688, 5.3188) 
  -- (8.0167, 5.3396) -- (7.9153, 5.3823) -- (7.8905, 5.3932) -- (7.8658, 
  5.4043) -- (7.8171, 5.4264) -- (7.722, 5.4722) -- (7.5401, 5.5677) -- (7.5162,
   5.5811) -- (7.4927, 5.5946) -- (7.446, 5.6217) -- (7.3547, 5.6771) -- 
  (7.1804, 5.7923) -- (7.1595, 5.8071) -- (7.1385, 5.8219) -- (7.0973, 5.8518) 
  -- (7.0167, 5.9127) -- (6.863, 6.038) -- (6.8446, 6.054) -- (6.8261, 6.0701) 
  -- (6.7899, 6.1024) -- (6.719, 6.1678) -- (6.5844, 6.3019) -- (6.5681, 6.319) 
  -- (6.5522, 6.3361) -- (6.5205, 6.3705) -- (6.459, 6.44) -- (6.3426, 6.5815) 
  -- (6.3296, 6.5982) -- (6.3167, 6.615) -- (6.2913, 6.6488) -- (6.242, 6.7167) 
  -- (6.1491, 6.8546) -- (6.1379, 6.872) -- (6.127, 6.8895) -- (6.1055, 6.9245) 
  -- (6.0638, 6.9949) -- (5.9862, 7.1374) -- (5.9771, 7.1555) -- (5.968, 7.1734)
   -- (5.9504, 7.2095) -- (5.9163, 7.282) -- (5.9081, 7.3002) -- (5.9001, 
  7.3185) -- (5.8842, 7.355) -- (5.8539, 7.4283) -- (5.8396, 7.4652) -- (5.8257,
   7.5022) -- (5.7992, 7.5763) -- (5.7929, 7.5949) -- (5.7867, 7.6135) -- 
  (5.7747, 7.6509) -- (5.7519, 7.7257) -- (5.7467, 7.744) -- (5.7416, 7.7625) --
   (5.7317, 7.7993) -- (5.7131, 7.8731) -- (5.7087, 7.8917) -- (5.7045, 7.9102) 
  -- (5.6964, 7.9473) -- (5.6815, 8.0217) -- (5.6746, 8.0589) -- (5.6684, 
  8.0961) -- (5.6654, 8.1149) -- (5.6625, 8.1335) -- (5.6571, 8.1708) -- 
  (5.6545, 8.1895) -- (5.6522, 8.2083) -- (5.6476, 8.2456) -- (5.6455, 8.2644) 
  -- (5.6436, 8.2831) -- (5.64, 8.3206) -- (5.6384, 8.3393) -- (5.6369, 8.3581) 
  -- (5.6355, 8.3768) -- (5.6342, 8.3957) -- (5.633, 8.4144) -- (5.6311, 8.4519)
   -- (5.6303, 8.4708) -- (5.6289, 8.5083) -- (5.6285, 8.527) -- (5.628, 8.5459)
   -- (5.6278, 8.5646) -- (5.6276, 8.5834) -- (5.6276, 8.6023) -- (5.6279, 
  8.6397) -- (5.6282, 8.6586) -- (5.6286, 8.6774) -- (5.6291, 8.6961) -- 
  (5.6298, 8.715) -- (5.6315, 8.7525) -- (5.6325, 8.7712) -- (5.6336, 8.7901) --
   (5.6348, 8.8088) -- (5.6376, 8.8463) -- (5.6409, 8.8838) -- (5.6446, 8.9213) 
  -- (5.6466, 8.9415) -- (5.649, 8.9619) -- (5.6541, 9.0024) -- (5.6568, 9.0228)
   -- (5.6597, 9.0431) -- (5.6658, 9.0836) -- (5.6691, 9.1037) -- (5.6724, 
  9.124) -- (5.6797, 9.1644) -- (5.6957, 9.2451) -- (5.7, 9.2652) -- (5.7045, 
  9.2854) -- (5.7138, 9.3255) -- (5.7342, 9.4057) -- (5.7395, 9.4256) -- (5.745,
   9.4456) -- (5.7565, 9.4855) -- (5.7811, 9.565) -- (5.7941, 9.6047) -- 
  (5.8077, 9.6441) -- (5.8367, 9.7229) -- (5.8441, 9.7425) -- (5.8518, 9.762) --
   (5.8677, 9.8012) -- (5.9008, 9.879) -- (5.9094, 9.8983) -- (5.9181, 9.9177) 
  -- (5.936, 9.9563) -- (5.9735, 10.0331) -- (6.0548, 10.1851) -- (6.0649, 
  10.2026) -- (6.0751, 10.2201) -- (6.0958, 10.2551) -- (6.1386, 10.3245) -- 
  (6.2298, 10.4617) -- (6.2538, 10.4956) -- (6.2783, 10.5293) -- (6.3286, 
  10.5963) -- (6.4352, 10.728) -- (6.4491, 10.7442) -- (6.463, 10.7604) -- 
  (6.4914, 10.7927) -- (6.5495, 10.8566) -- (6.6716, 10.982) -- (6.7033, 
  11.0129) -- (6.7357, 11.0435) -- (6.8017, 11.1038) -- (6.9401, 11.2219) -- 
  (6.9594, 11.2376) -- (6.9789, 11.2532) -- (7.0185, 11.2842) -- (7.0995, 
  11.3454) -- (7.2693, 11.4636) -- (7.2912, 11.478) -- (7.3134, 11.4923) -- 
  (7.3582, 11.5207) -- (7.4501, 11.5763) -- (7.6426, 11.6831) -- (7.6675, 
  11.696) -- (7.6926, 11.7088) -- (7.7436, 11.7341) -- (7.8481, 11.7833) -- 
  (7.8748, 11.7955) -- (7.9017, 11.8074) -- (7.9562, 11.8309) -- (8.0684, 
  11.8766) -- (8.0972, 11.8878) -- (8.1262, 11.8987) -- (8.185, 11.9203) -- 
  (8.3065, 11.9619) -- (8.3378, 11.972) -- (8.3694, 11.9819) -- (8.4336, 
  12.0015) -- (8.5673, 12.0384) -- (8.6013, 12.0472) -- (8.6358, 12.0558) -- 
  (8.7063, 12.0723) -- (8.7424, 12.0804) -- (8.7792, 12.0883) -- (8.855, 
  12.1034) -- (8.894, 12.1108) -- (8.934, 12.1179) -- (8.9749, 12.1248) -- 
  (9.0168, 12.1314) -- (9.06, 12.1379) -- (9.1046, 12.1441) -- (9.1507, 12.15) 
  -- (9.1987, 12.1558) -- (9.2488, 12.1612) -- (9.3018, 12.1661) -- (9.3581, 
  12.171) -- (9.4192, 12.1752) -- (9.4871, 12.179) -- (9.5677, 12.1821) -- 
  (9.6996, 12.1843) -- (9.7704, 12.0694) -- (9.8083, 12.0031) -- (9.8393, 
  11.9486) -- (9.8913, 11.8568) -- (9.974, 11.7093) -- (9.9919, 11.6771) -- 
  (10.0091, 11.6462) -- (10.0414, 11.5878) -- (10.0998, 11.4818) -- (10.1989, 
  11.3002) -- (10.3553, 11.0098) -- (10.371, 10.9803) -- (10.3864, 10.9515) -- 
  (10.416, 10.896) -- (10.4713, 10.792) -- (10.5693, 10.607) -- (10.7298, 
  10.304) -- (10.9838, 9.8335) -- (11.1544, 9.5312) -- (11.2828, 9.3143) -- 
  (11.3916, 9.14) -- (11.4711, 9.018) -- (11.5387, 8.918) -- (11.5907, 8.8438) 
  -- (11.6289, 8.7908) -- (11.6614, 8.7463) -- (11.6852, 8.7143) -- (11.7055, 
  8.6874) -- (11.7209, 8.6672) -- (11.7323, 8.6526) -- (11.7419, 8.6402) -- 
  (11.7489, 8.6311) -- (11.7544, 8.6242) -- (11.7589, 8.6183) -- (11.7624, 
  8.6138) -- (11.7653, 8.6103) -- (11.7675, 8.6075) -- (11.769, 8.6054) -- 
  (11.7704, 8.6038) -- (11.7713, 8.6025) -- (11.7722, 8.6016) -- (11.7729, 
  8.6007) -- (11.7733, 8.6001) -- (11.7737, 8.5995) -- (11.7742, 8.599) -- 
  (11.7744, 8.5987) -- (11.7745, 8.5985) -- (11.7746, 8.5983) -- (11.7748, 
  8.5983) -- (11.7749, 8.5981) -- (11.7749, 8.598) -- (11.7751, 8.598) -- 
  (11.7751, 8.5979);

  \path[draw=black,line cap=,line join=miter,line width=0.1058cm,miter 
  limit=3.25,shift={(-3.4925, 0.2646)}] (9.7044, 5.0114) -- (9.7679, 5.1218) -- 
  (9.8042, 5.1851) -- (9.86, 5.2835) -- (9.945, 5.4346) -- (10.0711, 5.6618) -- 
  (10.2519, 5.9935) -- (10.4987, 6.4554) -- (10.8361, 7.0904) -- (11.0455, 
  7.4732) -- (11.1989, 7.7409) -- (11.3269, 7.9531) -- (11.4201, 8.1001) -- 
  (11.4993, 8.2199) -- (11.5599, 8.3082) -- (11.6045, 8.3712) -- (11.6425, 
  8.4237) -- (11.6703, 8.4614) -- (11.6921, 8.4904) -- (11.7106, 8.5148) -- 
  (11.7241, 8.5324) -- (11.7355, 8.5473) -- (11.7443, 8.5586) -- (11.7507, 
  8.5669) -- (11.7562, 8.5739) -- (11.7602, 8.5788) -- (11.7636, 8.5832) -- 
  (11.7661, 8.5864) -- (11.768, 8.5889) -- (11.7695, 8.5908) -- (11.7708, 
  8.5923) -- (11.7716, 8.5934) -- (11.7724, 8.5944) -- (11.7735, 8.5958) -- 
  (11.7738, 8.5962) -- (11.7741, 8.5965) -- (11.7744, 8.5969) -- (11.7748, 
  8.5973) -- (11.7748, 8.5974) -- (11.7751, 8.5977) -- (11.7751, 8.5979);

  \path[draw=black,line cap=,line join=miter,line width=0.1058cm,miter 
  limit=3.25,shift={(-3.4925, 0.2646)}] (9.7044, 12.1843) -- (9.7711, 12.2965) 
  -- (9.8123, 12.3627) -- (9.8805, 12.4677) -- (9.9968, 12.6347) -- (10.0237, 
  12.6715) -- (10.0502, 12.707) -- (10.1018, 12.7746) -- (10.2016, 12.899) -- 
  (10.226, 12.9283) -- (10.2503, 12.957) -- (10.2985, 13.0128) -- (10.3936, 
  13.1185) -- (10.5809, 13.3116) -- (10.6041, 13.3342) -- (10.6274, 13.3566) -- 
  (10.674, 13.4006) -- (10.7669, 13.4855) -- (10.9528, 13.6442) -- (10.9761, 
  13.6631) -- (10.9994, 13.6819) -- (11.0461, 13.7188) -- (11.1396, 13.7903) -- 
  (11.3276, 13.9249) -- (11.3531, 13.9424) -- (11.3787, 13.9597) -- (11.43, 
  13.9937) -- (11.5329, 14.0596) -- (11.7402, 14.1832) -- (11.7923, 14.2124) -- 
  (11.8445, 14.2411) -- (11.9494, 14.2963) -- (12.1602, 14.3998) -- (12.1867, 
  14.4121) -- (12.2133, 14.4242) -- (12.2663, 14.448) -- (12.3729, 14.4938) -- 
  (12.3996, 14.505) -- (12.4263, 14.516) -- (12.4798, 14.5375) -- (12.5871, 
  14.5788) -- (12.6409, 14.5987) -- (12.6948, 14.618) -- (12.8028, 14.6549) -- 
  (13.0197, 14.7224) -- (13.0451, 14.7297) -- (13.0706, 14.737) -- (13.1214, 
  14.7511) -- (13.2232, 14.7778) -- (13.2489, 14.7843) -- (13.2744, 14.7905) -- 
  (13.3254, 14.8027) -- (13.4277, 14.826) -- (13.4532, 14.8315) -- (13.4789, 
  14.8369) -- (13.5301, 14.8474) -- (13.6328, 14.867) -- (13.6584, 14.8717) -- 
  (13.6841, 14.8761) -- (13.7355, 14.8847) -- (13.8383, 14.9009) -- (13.8898, 
  14.9083) -- (13.9412, 14.9152) -- (13.9927, 14.9218) -- (14.0441, 14.9279) -- 
  (14.0699, 14.9308) -- (14.0957, 14.9335) -- (14.1472, 14.9388) -- (14.1988, 
  14.9434) -- (14.2503, 14.9479) -- (14.2759, 14.9499) -- (14.3017, 14.9519) -- 
  (14.3275, 14.9536) -- (14.379, 14.957) -- (14.4048, 14.9585) -- (14.4562, 
  14.9611) -- (14.4819, 14.9623) -- (14.5592, 14.9652) -- (14.5849, 14.9659) -- 
  (14.6106, 14.9666) -- (14.6363, 14.9671) -- (14.6621, 14.9676) -- (14.7125, 
  14.9681) -- (14.7377, 14.9683) -- (14.788, 14.9683) -- (14.8132, 14.968) -- 
  (14.8383, 14.9677) -- (14.8635, 14.9674) -- (14.8886, 14.9669) -- (14.9138, 
  14.9663) -- (14.964, 14.9649) -- (14.9891, 14.964) -- (15.0143, 14.963) -- 
  (15.0394, 14.962) -- (15.0644, 14.9608) -- (15.0894, 14.9596) -- (15.1145, 
  14.9582) -- (15.1646, 14.9552) -- (15.1896, 14.9535) -- (15.2146, 14.9517) -- 
  (15.2645, 14.948) -- (15.2896, 14.9459) -- (15.3146, 14.9437) -- (15.3644, 
  14.9392) -- (15.4639, 14.9288) -- (15.4887, 14.9261) -- (15.5137, 14.9231) -- 
  (15.5633, 14.917) -- (15.6622, 14.9036) -- (15.687, 14.9) -- (15.7117, 
  14.8963) -- (15.761, 14.8886) -- (15.8594, 14.8722) -- (15.9085, 14.8634) -- 
  (15.9574, 14.8541) -- (16.0551, 14.8347) -- (16.2494, 14.7913) -- (16.2756, 
  14.785) -- (16.3018, 14.7785) -- (16.354, 14.7653) -- (16.4582, 14.7374) -- 
  (16.6648, 14.6768) -- (16.6904, 14.6688) -- (16.716, 14.6605) -- (16.7671, 
  14.6439) -- (16.869, 14.6094) -- (17.0706, 14.5353) -- (17.0955, 14.5256) -- 
  (17.1205, 14.5157) -- (17.1702, 14.4959) -- (17.2693, 14.4547) -- (17.465, 
  14.3676) -- (17.8467, 14.1746) -- (17.8685, 14.1626) -- (17.8902, 14.1505) -- 
  (17.9337, 14.1261) -- (18.0199, 14.0762) -- (18.1898, 13.9726) -- (18.2109, 
  13.9594) -- (18.2319, 13.9459) -- (18.2736, 13.9189) -- (18.3564, 13.8639) -- 
  (18.5193, 13.7503) -- (18.5596, 13.7211) -- (18.5995, 13.6916) -- (18.6786, 
  13.6318) -- (18.8339, 13.5086) -- (19.1325, 13.2486) -- (19.1523, 13.2303) -- 
  (19.1717, 13.2119) -- (19.2105, 13.175) -- (19.2873, 13.1003) -- (19.437, 
  12.9471) -- (19.4553, 12.9277) -- (19.4735, 12.9081) -- (19.5097, 12.8688) -- 
  (19.5812, 12.7892) -- (19.72, 12.6267) -- (19.737, 12.606) -- (19.7538, 
  12.5853) -- (19.7873, 12.5437) -- (19.8531, 12.4598) -- (19.9806, 12.2887) -- 
  (20.2176, 11.9346) -- (20.2314, 11.9125) -- (20.245, 11.8901) -- (20.2722, 
  11.8455) -- (20.3251, 11.7554) -- (20.4265, 11.5729) -- (20.4388, 11.5499) -- 
  (20.4511, 11.5267) -- (20.475, 11.4804) -- (20.5219, 11.3873) -- (20.611, 
  11.1988) -- (20.6218, 11.1749) -- (20.6324, 11.1511) -- (20.6532, 11.1034) -- 
  (20.6939, 11.0075) -- (20.7703, 10.8137) -- (20.7794, 10.7894) -- (20.7884, 
  10.7648) -- (20.8062, 10.7159) -- (20.8403, 10.6176) -- (20.9039, 10.4193) -- 
  (20.9108, 10.3962) -- (20.9177, 10.3729) -- (20.9312, 10.3263) -- (20.9571, 
  10.2328) -- (20.9634, 10.2093) -- (20.9695, 10.1859) -- (20.9816, 10.1389) -- 
  (21.0046, 10.0448) -- (21.0101, 10.0211) -- (21.0155, 9.9975) -- (21.0261, 
  9.9503) -- (21.0462, 9.8553) -- (21.051, 9.8316) -- (21.0557, 9.8079) -- 
  (21.065, 9.7602) -- (21.0821, 9.6649) -- (21.0862, 9.641) -- (21.09, 9.6171) 
  -- (21.0978, 9.5692) -- (21.112, 9.4733) -- (21.1186, 9.4254) -- (21.1248, 
  9.3773) -- (21.1361, 9.2811) -- (21.1386, 9.257) -- (21.141, 9.233) -- 
  (21.1459, 9.1848) -- (21.1481, 9.1606) -- (21.1501, 9.1365) -- (21.1541, 
  9.0882) -- (21.1577, 9.0399) -- (21.1594, 9.0158) -- (21.1609, 8.9916) -- 
  (21.1624, 8.9674) -- (21.1638, 8.9432) -- (21.1663, 8.8948) -- (21.1685, 
  8.8425) -- (21.1694, 8.8161) -- (21.1702, 8.79) -- (21.1711, 8.7636) -- 
  (21.1718, 8.7375) -- (21.1722, 8.7111) -- (21.1726, 8.6849) -- (21.173, 
  8.6586) -- (21.1731, 8.6324) -- (21.1733, 8.6061) -- (21.1733, 8.5799) -- 
  (21.1731, 8.5536) -- (21.1729, 8.5274) -- (21.1725, 8.5011) -- (21.172, 
  8.4749) -- (21.1715, 8.4486) -- (21.1708, 8.4224) -- (21.17, 8.3961) -- 
  (21.1691, 8.3699) -- (21.1682, 8.3436) -- (21.1669, 8.3174) -- (21.1658, 
  8.2912) -- (21.1645, 8.2649) -- (21.1614, 8.2126) -- (21.158, 8.1602) -- 
  (21.1541, 8.1078) -- (21.1499, 8.0555) -- (21.1475, 8.0293) -- (21.145, 
  8.0031) -- (21.1399, 7.9509) -- (21.1372, 7.9247) -- (21.1343, 7.8986) -- 
  (21.1282, 7.8464) -- (21.125, 7.8204) -- (21.1217, 7.7943) -- (21.1149, 
  7.7422) -- (21.0997, 7.6383) -- (21.0956, 7.6123) -- (21.0914, 7.5864) -- 
  (21.0827, 7.5346) -- (21.0641, 7.4311) -- (21.0592, 7.4053) -- (21.0541, 
  7.3795) -- (21.0437, 7.328) -- (21.0217, 7.2252) -- (21.0163, 7.2012) -- 
  (21.0108, 7.1774) -- (20.9995, 7.1296) -- (20.9759, 7.0343) -- (20.9241, 
  6.8449) -- (20.9104, 6.7977) -- (20.896, 6.7507) -- (20.8665, 6.657) -- 
  (20.803, 6.471) -- (20.7947, 6.4478) -- (20.7863, 6.4247) -- (20.7691, 6.3787)
   -- (20.7338, 6.2869) -- (20.6589, 6.105) -- (20.492, 5.7483) -- (20.4699, 
  5.7054) -- (20.4476, 5.6625) -- (20.402, 5.5773) -- (20.3069, 5.4089) -- 
  (20.2824, 5.3673) -- (20.2574, 5.3258) -- (20.2067, 5.2434) -- (20.1014, 
  5.081) -- (20.0879, 5.061) -- (20.0743, 5.0409) -- (20.0469, 5.0009) -- 
  (19.9912, 4.9217) -- (19.8761, 4.7657) -- (19.6317, 4.4642) -- (19.6144, 
  4.4443) -- (19.5971, 4.4245) -- (19.5621, 4.3849) -- (19.4914, 4.3068) -- 
  (19.3459, 4.154) -- (19.3087, 4.1165) -- (19.2712, 4.0793) -- (19.1952, 
  4.0058) -- (19.0397, 3.8626) -- (19.02, 3.8451) -- (19.0, 3.8276) -- (18.9602,
   3.7929) -- (18.8794, 3.7244) -- (18.7144, 3.5914) -- (18.3715, 3.3419) -- 
  (18.3509, 3.328) -- (18.3304, 3.3142) -- (18.289, 3.2868) -- (18.2057, 3.233) 
  -- (18.0366, 3.1292) -- (18.0151, 3.1166) -- (17.9937, 3.104) -- (17.9507, 
  3.0792) -- (17.8641, 3.0306) -- (17.6885, 2.9372) -- (17.6663, 2.9259) -- 
  (17.6441, 2.9147) -- (17.5996, 2.8925) -- (17.5099, 2.8491) -- (17.3286, 
  2.7665) -- (17.2828, 2.7467) -- (17.2369, 2.7273) -- (17.1446, 2.6895) -- 
  (16.958, 2.618) -- (16.9326, 2.6088) -- (16.907, 2.5997) -- (16.8559, 2.5818) 
  -- (16.7532, 2.547) -- (16.7275, 2.5388) -- (16.7015, 2.5304) -- (16.6499, 
  2.5142) -- (16.5458, 2.4829) -- (16.5198, 2.4755) -- (16.4936, 2.4681) -- 
  (16.4412, 2.4535) -- (16.3361, 2.4258) -- (16.1243, 2.3756) -- (16.0977, 2.37)
   -- (16.0711, 2.3642) -- (16.0176, 2.3533) -- (15.9106, 2.3326) -- (15.8838, 
  2.3278) -- (15.8569, 2.323) -- (15.8031, 2.3139) -- (15.6952, 2.2969) -- 
  (15.6682, 2.2929) -- (15.6411, 2.2891) -- (15.587, 2.2818) -- (15.5598, 
  2.2782) -- (15.5327, 2.2749) -- (15.4784, 2.2684) -- (15.4511, 2.2654) -- 
  (15.424, 2.2625) -- (15.3694, 2.257) -- (15.3421, 2.2543) -- (15.3148, 2.2519)
   -- (15.2603, 2.2473) -- (15.2334, 2.2452) -- (15.1796, 2.2414) -- (15.1259, 
  2.2381) -- (15.099, 2.2366) -- (15.072, 2.2352) -- (15.0451, 2.2339) -- 
  (15.0181, 2.2327) -- (14.9913, 2.2317) -- (14.9643, 2.2308) -- (14.9372, 
  2.2299) -- (14.8832, 2.2286) -- (14.8292, 2.2277) -- (14.8022, 2.2275) -- 
  (14.7752, 2.2273) -- (14.7482, 2.2273) -- (14.721, 2.2275) -- (14.694, 2.2277)
   -- (14.6669, 2.228) -- (14.6399, 2.2284) -- (14.6127, 2.229) -- (14.5857, 
  2.2297) -- (14.5586, 2.2305) -- (14.5316, 2.2313) -- (14.4773, 2.2335) -- 
  (14.4502, 2.2348) -- (14.4231, 2.2361) -- (14.396, 2.2377) -- (14.3417, 2.241)
   -- (14.2875, 2.2448) -- (14.2604, 2.2469) -- (14.2332, 2.2491) -- (14.1789, 
  2.2539) -- (14.1519, 2.2564) -- (14.1248, 2.2592) -- (14.0705, 2.2648) -- 
  (13.962, 2.2778) -- (13.935, 2.2812) -- (13.9079, 2.2848) -- (13.8537, 2.2925)
   -- (13.7454, 2.3092) -- (13.7182, 2.3137) -- (13.6912, 2.3183) -- (13.6372, 
  2.3279) -- (13.5292, 2.3485) -- (13.5039, 2.3536) -- (13.4787, 2.3588) -- 
  (13.4284, 2.3695) -- (13.328, 2.3921) -- (13.2778, 2.4043) -- (13.2277, 
  2.4167) -- (13.1276, 2.4428) -- (13.1027, 2.4497) -- (13.0776, 2.4566) -- 
  (13.0277, 2.4708) -- (12.9281, 2.5006) -- (12.7295, 2.5655) -- (12.7048, 
  2.5742) -- (12.68, 2.5829) -- (12.6307, 2.6006) -- (12.5322, 2.6377) -- 
  (12.3359, 2.7174) -- (12.3115, 2.7278) -- (12.2871, 2.7384) -- (12.2384, 
  2.7601) -- (12.1412, 2.8047) -- (11.9479, 2.8999) -- (11.9218, 2.9136) -- 
  (11.8956, 2.9272) -- (11.8437, 2.9551) -- (11.7399, 3.0127) -- (11.5338, 
  3.1356) -- (11.5081, 3.1517) -- (11.4825, 3.168) -- (11.4314, 3.201) -- 
  (11.3294, 3.2694) -- (11.1267, 3.4151) -- (11.1015, 3.4341) -- (11.0763, 
  3.4534) -- (11.026, 3.4926) -- (10.9255, 3.5738) -- (10.725, 3.7479) -- 
  (10.6999, 3.7709) -- (10.675, 3.7942) -- (10.6248, 3.8416) -- (10.5244, 
  3.9405) -- (10.3211, 4.1573) -- (10.297, 4.1846) -- (10.2729, 4.2124) -- 
  (10.2241, 4.2697) -- (10.1244, 4.3922) -- (10.099, 4.4247) -- (10.073, 4.4584)
   -- (10.0203, 4.5289) -- (9.9066, 4.6893) -- (9.8753, 4.7358) -- (9.8421, 
  4.7865) -- (9.8056, 4.8438) -- (9.7624, 4.9134) -- (9.7258, 5.0483) -- 
  (9.7759, 5.1355) -- (9.8107, 5.1964) -- (9.8397, 5.2476) -- (9.8654, 5.2929) 
  -- (9.8887, 5.3344) -- (9.9306, 5.4087) -- (10.0013, 5.5354) -- (10.0171, 
  5.564) -- (10.0324, 5.5915) -- (10.0613, 5.644) -- (10.1142, 5.7404) -- 
  (10.2058, 5.9079) -- (10.3524, 6.1805) -- (10.3612, 6.1968) -- (10.3697, 
  6.2129) -- (10.3865, 6.2443) -- (10.4189, 6.3051) -- (10.4789, 6.4179) -- 
  (10.5842, 6.6167) -- (10.7546, 6.938) -- (10.9987, 7.3892) -- (11.164, 7.6813)
   -- (11.3009, 7.9108) -- (11.4001, 8.0691) -- (11.4771, 8.1868) -- (11.543, 
  8.2837) -- (11.5912, 8.3526) -- (11.6324, 8.4099) -- (11.664, 8.4527) -- 
  (11.687, 8.4838) -- (11.7067, 8.5098) -- (11.7212, 8.5287) -- (11.7323, 
  8.5431) -- (11.7419, 8.5554) -- (11.7489, 8.5644) -- (11.7548, 8.5719) -- 
  (11.7591, 8.5775) -- (11.7624, 8.5817) -- (11.7653, 8.5853) -- (11.7673, 
  8.5879) -- (11.7691, 8.5901) -- (11.7704, 8.5919) -- (11.7713, 8.5932) -- 
  (11.7723, 8.5941) -- (11.7729, 8.595) -- (11.7737, 8.5961) -- (11.774, 8.5965)
   -- (11.7741, 8.5965) -- (11.7741, 8.5966) -- (11.7742, 8.5966) -- (11.7742, 
  8.5968);

  \path[draw=black,line cap=,line join=miter,line width=0.1058cm,miter 
  limit=3.25,shift={(-3.4925, 0.2646)}] (9.7044, 12.1843) -- (9.7679, 12.0738) 
  -- (9.8042, 12.0104) -- (9.86, 11.912) -- (9.945, 11.761) -- (10.0711, 
  11.5338) -- (10.2519, 11.2022) -- (10.4987, 10.7402) -- (10.8361, 10.1053) -- 
  (11.0455, 9.7223) -- (11.1989, 9.4547) -- (11.3269, 9.2425) -- (11.4201, 
  9.0955) -- (11.4993, 8.9759) -- (11.5599, 8.8875) -- (11.6045, 8.8244) -- 
  (11.6425, 8.7719) -- (11.6703, 8.7341) -- (11.6921, 8.7052) -- (11.7106, 
  8.6808) -- (11.7241, 8.6632) -- (11.7355, 8.6483) -- (11.7443, 8.637) -- 
  (11.7507, 8.6287) -- (11.7562, 8.6218) -- (11.7602, 8.6167) -- (11.7636, 
  8.6125) -- (11.7661, 8.6092) -- (11.768, 8.6068) -- (11.7695, 8.6047) -- 
  (11.7708, 8.6032) -- (11.7716, 8.6021) -- (11.7724, 8.6012) -- (11.773, 
  8.6005) -- (11.7735, 8.5999) -- (11.7738, 8.5994) -- (11.7744, 8.5988) -- 
  (11.7745, 8.5985) -- (11.7748, 8.5983) -- (11.7748, 8.5981) -- (11.7749, 
  8.5981) -- (11.7749, 8.598) -- (11.7751, 8.598) -- (11.7751, 8.5979);

  \path[draw=black,line cap=,line join=miter,line width=0.1058cm,miter 
  limit=3.25,shift={(-3.4925, 0.2646)}] (9.7044, 12.1843) -- (9.5743, 12.1824) 
  -- (9.4972, 12.1795) -- (9.4321, 12.1761) -- (9.3737, 12.1721) -- (9.3198, 
  12.1678) -- (9.2694, 12.1631) -- (9.2214, 12.1583) -- (9.1755, 12.1531) -- 
  (9.1314, 12.1477) -- (9.0889, 12.142) -- (9.0075, 12.13) -- (8.9684, 12.1237) 
  -- (8.9302, 12.1172) -- (8.8565, 12.1037) -- (8.8207, 12.0967) -- (8.7855, 
  12.0895) -- (8.7169, 12.0748) -- (8.5861, 12.0434) -- (8.5544, 12.0351) -- 
  (8.5233, 12.0268) -- (8.4621, 12.0097) -- (8.344, 11.974) -- (8.3152, 11.9648)
   -- (8.2867, 11.9554) -- (8.2307, 11.9364) -- (8.1217, 11.897) -- (8.0951, 
  11.887) -- (8.0688, 11.8768) -- (8.0167, 11.856) -- (7.9153, 11.8134) -- 
  (7.8905, 11.8025) -- (7.8658, 11.7915) -- (7.8171, 11.7691) -- (7.722, 
  11.7234) -- (7.5401, 11.6279) -- (7.5162, 11.6145) -- (7.4927, 11.6012) -- 
  (7.446, 11.5739) -- (7.3547, 11.5185) -- (7.1804, 11.4033) -- (7.1595, 
  11.3885) -- (7.1385, 11.3736) -- (7.0973, 11.3437) -- (7.0167, 11.2828) -- 
  (6.863, 11.1576) -- (6.8261, 11.1256) -- (6.7899, 11.0932) -- (6.719, 11.0278)
   -- (6.5844, 10.8937) -- (6.5681, 10.8766) -- (6.5522, 10.8595) -- (6.5205, 
  10.8252) -- (6.459, 10.7557) -- (6.3426, 10.6141) -- (6.3296, 10.5974) -- 
  (6.3167, 10.5806) -- (6.2913, 10.5468) -- (6.242, 10.4789) -- (6.1491, 
  10.3411) -- (6.1379, 10.3236) -- (6.127, 10.3062) -- (6.1055, 10.2712) -- 
  (6.0638, 10.2008) -- (5.9862, 10.0582) -- (5.9771, 10.0402) -- (5.968, 
  10.0222) -- (5.9504, 9.9861) -- (5.9163, 9.9136) -- (5.9081, 9.8954) -- 
  (5.9001, 9.8772) -- (5.8842, 9.8407) -- (5.8539, 9.7673) -- (5.8467, 9.7488) 
  -- (5.8396, 9.7305) -- (5.8257, 9.6935) -- (5.7992, 9.6193) -- (5.7929, 
  9.6007) -- (5.7867, 9.5821) -- (5.7747, 9.5448) -- (5.7519, 9.47) -- (5.7467, 
  9.4516) -- (5.7416, 9.4332) -- (5.7317, 9.3963) -- (5.7131, 9.3224) -- 
  (5.7087, 9.304) -- (5.7045, 9.2854) -- (5.6964, 9.2483) -- (5.6815, 9.174) -- 
  (5.6781, 9.1554) -- (5.6746, 9.1367) -- (5.6684, 9.0995) -- (5.6654, 9.0807) 
  -- (5.6625, 9.0621) -- (5.6571, 9.0248) -- (5.6545, 9.006) -- (5.6522, 8.9874)
   -- (5.6476, 8.9499) -- (5.6455, 8.9312) -- (5.6436, 8.9125) -- (5.64, 8.875) 
  -- (5.6384, 8.8562) -- (5.6369, 8.8375) -- (5.6355, 8.8188) -- (5.6342, 8.8) 
  -- (5.633, 8.7811) -- (5.6311, 8.7437) -- (5.6303, 8.7249) -- (5.6296, 8.706) 
  -- (5.6289, 8.6873) -- (5.628, 8.6498) -- (5.6278, 8.6309) -- (5.6276, 8.6122)
   -- (5.6276, 8.5934) -- (5.6278, 8.5746) -- (5.6279, 8.5558) -- (5.6282, 
  8.5371) -- (5.6286, 8.5182) -- (5.6291, 8.4995) -- (5.6298, 8.4807) -- 
  (5.6307, 8.4618) -- (5.6315, 8.4431) -- (5.6325, 8.4244) -- (5.6336, 8.4056) 
  -- (5.6348, 8.3869) -- (5.6362, 8.368) -- (5.6376, 8.3493) -- (5.6409, 8.3118)
   -- (5.6446, 8.2743) -- (5.6466, 8.254) -- (5.649, 8.2336) -- (5.6541, 8.1931)
   -- (5.6568, 8.1729) -- (5.6597, 8.1526) -- (5.6658, 8.1121) -- (5.6724, 
  8.0716) -- (5.6797, 8.0312) -- (5.6957, 7.9506) -- (5.7, 7.9303) -- (5.7045, 
  7.9104) -- (5.7138, 7.8701) -- (5.7342, 7.7899) -- (5.7395, 7.7699) -- (5.745,
   7.7499) -- (5.7565, 7.7101) -- (5.7811, 7.6306) -- (5.7876, 7.6108) -- 
  (5.7941, 7.5911) -- (5.8077, 7.5515) -- (5.8367, 7.4728) -- (5.8441, 7.4531) 
  -- (5.8518, 7.4336) -- (5.8677, 7.3946) -- (5.9008, 7.3167) -- (5.9094, 
  7.2973) -- (5.9181, 7.278) -- (5.936, 7.2394) -- (5.9735, 7.1625) -- (6.0548, 
  7.0106) -- (6.0649, 6.993) -- (6.0751, 6.9755) -- (6.0958, 6.9405) -- (6.1386,
   6.871) -- (6.2298, 6.7339) -- (6.2538, 6.7) -- (6.2783, 6.6663) -- (6.3286, 
  6.5994) -- (6.4352, 6.4677) -- (6.463, 6.4352) -- (6.4914, 6.4029) -- (6.5495,
   6.339) -- (6.6716, 6.2136) -- (6.6875, 6.1981) -- (6.7033, 6.1828) -- 
  (6.7357, 6.1523) -- (6.8017, 6.0918) -- (6.9401, 5.9737) -- (6.9594, 5.9579) 
  -- (6.9789, 5.9424) -- (7.0185, 5.9114) -- (7.0995, 5.8503) -- (7.2693, 
  5.7321) -- (7.2912, 5.7176) -- (7.3134, 5.7033) -- (7.3582, 5.6749) -- 
  (7.4501, 5.6192) -- (7.6426, 5.5126) -- (7.6675, 5.4996) -- (7.6926, 5.4868) 
  -- (7.7436, 5.4616) -- (7.8481, 5.4122) -- (7.8748, 5.4001) -- (7.9017, 
  5.3883) -- (7.9562, 5.3647) -- (8.0684, 5.319) -- (8.0972, 5.3079) -- (8.1262,
   5.2969) -- (8.185, 5.2753) -- (8.3065, 5.2337) -- (8.3378, 5.2236) -- 
  (8.3694, 5.2137) -- (8.4336, 5.1942) -- (8.5673, 5.1572) -- (8.6013, 5.1485) 
  -- (8.6358, 5.1399) -- (8.7063, 5.1233) -- (8.7424, 5.1153) -- (8.7792, 
  5.1073) -- (8.855, 5.0921) -- (8.894, 5.0848) -- (8.934, 5.0778) -- (8.9749, 
  5.0708) -- (9.0168, 5.0642) -- (9.06, 5.0577) -- (9.1046, 5.0515) -- (9.1507, 
  5.0455) -- (9.1987, 5.0399) -- (9.2488, 5.0345) -- (9.3018, 5.0294) -- 
  (9.3581, 5.0247) -- (9.4192, 5.0205) -- (9.4871, 5.0166) -- (9.5677, 5.0134) 
  -- (9.6996, 5.0114) -- (9.7704, 5.1262) -- (9.8083, 5.1924) -- (9.8393, 
  5.2472) -- (9.8913, 5.3388) -- (9.974, 5.4862) -- (10.0091, 5.5495) -- 
  (10.0414, 5.6079) -- (10.0998, 5.7138) -- (10.1989, 5.8954) -- (10.3553, 
  6.1857) -- (10.3631, 6.2006) -- (10.371, 6.2152) -- (10.3864, 6.244) -- 
  (10.416, 6.2996) -- (10.4713, 6.4037) -- (10.5693, 6.5885) -- (10.7298, 
  6.8916) -- (10.9838, 7.3622) -- (11.1544, 7.6645) -- (11.2828, 7.8813) -- 
  (11.3916, 8.0556) -- (11.4711, 8.1776) -- (11.5387, 8.2776) -- (11.5907, 
  8.3517) -- (11.6289, 8.4049) -- (11.6614, 8.4493) -- (11.6852, 8.4813) -- 
  (11.7055, 8.5081) -- (11.7209, 8.5284) -- (11.7323, 8.5431) -- (11.7419, 
  8.5556) -- (11.7489, 8.5645) -- (11.7544, 8.5715) -- (11.7589, 8.5775) -- 
  (11.7624, 8.5817) -- (11.7653, 8.5853) -- (11.7675, 8.5881) -- (11.769, 
  8.5901) -- (11.7704, 8.5919) -- (11.7713, 8.5932) -- (11.7722, 8.5941) -- 
  (11.7729, 8.595) -- (11.7737, 8.5961) -- (11.774, 8.5965) -- (11.7744, 8.5969)
   -- (11.7745, 8.5972) -- (11.7748, 8.5974) -- (11.7749, 8.5974) -- (11.7749, 
  8.5977) -- (11.7751, 8.5977);

  \path[draw=black,line cap=,line join=miter,line width=0.1058cm,miter 
  limit=3.25,shift={(-3.4925, 0.2646)}] (7.6342, 8.5979) -- (11.7751, 8.5979);

  \path[draw=black,line cap=,line join=miter,line width=0.1058cm,miter 
  limit=3.25,shift={(-3.4925, 0.2646)}] (15.9161, 8.5979) -- (11.7752, 8.5979);

  \path[fill=cff7f00,even odd rule] (6.3263, 5.5046) -- (6.4405, 5.3902) -- 
  (6.3263, 5.276) -- (6.4405, 5.1616) -- (6.3263, 5.0473) -- (6.2119, 5.1616) --
   (6.0977, 5.0473) -- (5.9833, 5.1616) -- (6.0977, 5.276) -- (5.9833, 5.3902) 
  -- (6.0977, 5.5046) -- (6.2119, 5.3902) -- cycle(6.3263, 5.5046);

  \path[fill=cff7f00,even odd rule] (6.3263, 12.6774) -- (6.4405, 12.5632) -- 
  (6.3263, 12.4489) -- (6.4405, 12.3345) -- (6.3263, 12.2203) -- (6.2119, 
  12.3345) -- (6.0977, 12.2203) -- (5.9833, 12.3345) -- (6.0977, 12.4489) -- 
  (5.9833, 12.5632) -- (6.0977, 12.6774) -- (6.2119, 12.5632) -- cycle(6.3263, 
  12.6774);

  \path[fill=cff7f00,even odd rule] (4.2557, 9.0911) -- (4.3699, 8.9767) -- 
  (4.2557, 8.8624) -- (4.3699, 8.7481) -- (4.2557, 8.6338) -- (4.1413, 8.7481) 
  -- (4.027, 8.6338) -- (3.9127, 8.7481) -- (4.027, 8.8624) -- (3.9127, 8.9767) 
  -- (4.027, 9.0911) -- (4.1413, 8.9767) -- cycle(4.2557, 9.0911);

  \path[fill=cff7f00,even odd rule] (12.5382, 9.0911) -- (12.6525, 8.9767) -- 
  (12.5382, 8.8624) -- (12.6525, 8.7481) -- (12.5382, 8.6338) -- (12.424, 
  8.7481) -- (12.3096, 8.6338) -- (12.1954, 8.7481) -- (12.3096, 8.8624) -- 
  (12.1954, 8.9767) -- (12.3096, 9.0911) -- (12.424, 8.9767) -- cycle(12.5382, 
  9.0911);

  \path[fill=blue,even odd rule] (4.2557, 8.8624).. controls (4.2557, 8.8321) 
  and (4.2435, 8.803) .. (4.2222, 8.7815).. controls (4.2007, 8.7602) and 
  (4.1716, 8.7481) .. (4.1413, 8.7481).. controls (4.111, 8.7481) and (4.0819, 
  8.7602) .. (4.0605, 8.7815).. controls (4.039, 8.803) and (4.027, 8.8321) .. 
  (4.027, 8.8624).. controls (4.027, 8.8928) and (4.039, 8.9218) .. (4.0605, 
  8.9432).. controls (4.0819, 8.9647) and (4.111, 8.9767) .. (4.1413, 8.9767).. 
  controls (4.1716, 8.9767) and (4.2007, 8.9647) .. (4.2222, 8.9432).. controls 
  (4.2435, 8.9218) and (4.2557, 8.8928) .. (4.2557, 8.8624) -- cycle(4.2557, 
  8.8624);

  \path[fill=blue,even odd rule] (12.5382, 8.8624).. controls (12.5382, 8.8321) 
  and (12.5262, 8.803) .. (12.5047, 8.7815).. controls (12.4832, 8.7602) and 
  (12.4543, 8.7481) .. (12.424, 8.7481).. controls (12.3937, 8.7481) and 
  (12.3646, 8.7602) .. (12.3431, 8.7815).. controls (12.3216, 8.803) and 
  (12.3096, 8.8321) .. (12.3096, 8.8624).. controls (12.3096, 8.8928) and 
  (12.3216, 8.9218) .. (12.3431, 8.9432).. controls (12.3646, 8.9647) and 
  (12.3937, 8.9767) .. (12.424, 8.9767).. controls (12.4543, 8.9767) and 
  (12.4832, 8.9647) .. (12.5047, 8.9432).. controls (12.5262, 8.9218) and 
  (12.5382, 8.8928) .. (12.5382, 8.8624) -- cycle(12.5382, 8.8624);

  \path[fill=blue,even odd rule] (8.3969, 8.8624).. controls (8.3969, 8.8321) 
  and (8.3848, 8.803) .. (8.3635, 8.7815).. controls (8.342, 8.7602) and 
  (8.3129, 8.7481) .. (8.2826, 8.7481).. controls (8.2522, 8.7481) and (8.2232, 
  8.7602) .. (8.2018, 8.7815).. controls (8.1803, 8.803) and (8.1683, 8.8321) ..
   (8.1683, 8.8624).. controls (8.1683, 8.8928) and (8.1803, 8.9218) .. (8.2018,
   8.9432).. controls (8.2232, 8.9647) and (8.2522, 8.9767) .. (8.2826, 
  8.9767).. controls (8.3129, 8.9767) and (8.342, 8.9647) .. (8.3635, 8.9432).. 
  controls (8.3848, 8.9218) and (8.3969, 8.8928) .. (8.3969, 8.8624) -- 
  cycle(8.3969, 8.8624);

\end{tikzpicture}

%% file: figures/timelike-tikz-codes/timelike-pos-tikz-crossing-labeled.tex
\definecolor{cff7f00}{RGB}{255,127,0}
\definecolor{cff7d00}{RGB}{255,125,0}

\def \globalscale {1.000000}
\begin{tikzpicture}[y=1cm, x=1cm, yscale=\globalscale,xscale=\globalscale, every node/.append style={scale=\globalscale}, inner sep=0pt, outer sep=0pt]

\coordinate (wp) at (6.7994, 15.8213);
\coordinate (wm) at (6.6994, 6.7121);
\coordinate (w0) at (4.1697, 11.138);
\coordinate (w1) at (9.429, 11.138);
\coordinate (w2) at (14.6881, 11.138);
\coordinate (ref) at ($ (wm) + (0,1.5) $);

\fill[BrickRed] (ref) circle[radius=.1cm];
\draw[line width = 1.2,midarrow,draw=BrickRed] (ref) to[out=120,in=0] ($ (w0) + (-.7,1)$);
\draw[line width = 1.2,draw=BrickRed] ($ (w0) + (-.7,1)$) to[out=180,in=170] (ref);
\draw[line width = 1.2,midarrow,draw=BrickRed] (ref) to[out=30,in=-60] ($ (w1) + (1,1)$);
\draw[line width = 1.2,draw=BrickRed] ($ (w1) + (1,1)$) to[out=120,in=60] (ref);
\node[text=BrickRed,font=\Large] at ($ (ref) + (0,-.5)$) {$u_{0}$};
\node[text=BrickRed,font=\Large] at ($ (w0) + (0,1.5)$) {$\mathcal{C}_{0}$};
\node[text=BrickRed,font=\Large] at ($ (w1) + (0,2)$) {$\mathcal{C}_{1}$};
\node[text=BrickRed] at ($ (w0) + (1.2,.2)$) {$1$};
\node[text=BrickRed] at ($ (w1) + (-1,-2.5)$) {$2$};
\node[text=BrickRed] at ($ (w1) + (.8,-1.2)$) {$3$};
\node[text=BrickRed] at ($ (w1) + (1.4,1)$) {$4$};
\node[text=BrickRed] at ($ (w1) + (-.7,1.2)$) {$5$};
\node[text=BrickRed] at ($ (w1) + (-1.5,.5)$) {$6$};
\node[text=BrickRed] at ($ (w1) + (-1.8,-.3)$) {$7$};

  \draw[line width = 1.2,
  draw=orange,
        decorate,
        decoration={snake, amplitude=.5mm, segment length=2mm}]
    (w0) to[out=-180, in=120] (wm);
    \draw[line width = 1.2,
  draw=orange,
        decorate,
        decoration={snake, amplitude=.5mm, segment length=2mm}]
    (w2) to[out=0, in=0] (wp);

  \path[draw=black,line cap=,line join=miter,line width=0.1058cm,miter 
  limit=3.25,shift={(-3.4925, 0.2646)}] (10.1775, 6.3189) -- (10.2468, 6.168) --
   (10.2899, 6.0787) -- (10.362, 5.9358) -- (10.486, 5.7066) -- (10.5151, 
  5.6557) -- (10.5438, 5.6065) -- (10.5999, 5.5126) -- (10.709, 5.3381) -- 
  (10.7359, 5.2968) -- (10.7628, 5.2561) -- (10.8161, 5.177) -- (10.9221, 
  5.0258) -- (11.1337, 4.7461) -- (11.1602, 4.7129) -- (11.1868, 4.68) -- 
  (11.24, 4.6152) -- (11.3468, 4.4891) -- (11.5628, 4.25) -- (12.0067, 3.8148) 
  -- (12.0374, 3.7871) -- (12.0682, 3.7596) -- (12.1299, 3.7053) -- (12.2545, 
  3.5992) -- (12.5075, 3.3962) -- (12.5395, 3.3716) -- (12.5716, 3.3474) -- 
  (12.6361, 3.2993) -- (12.7659, 3.2052) -- (13.0294, 3.0256) -- (13.0627, 
  3.004) -- (13.096, 2.9825) -- (13.1632, 2.94) -- (13.2981, 2.8571) -- 
  (13.5717, 2.6993) -- (13.6063, 2.6803) -- (13.6409, 2.6615) -- (13.7105, 
  2.6243) -- (13.8504, 2.552) -- (14.1334, 2.4153) -- (14.1668, 2.4) -- 
  (14.2003, 2.3848) -- (14.2672, 2.3549) -- (14.4019, 2.2969) -- (14.674, 
  2.1876) -- (14.7084, 2.1745) -- (14.7427, 2.1616) -- (14.8114, 2.1362) -- 
  (14.9498, 2.0872) -- (15.2287, 1.996) -- (15.2638, 1.9851) -- (15.299, 1.9745)
   -- (15.3694, 1.9536) -- (15.5108, 1.9135) -- (15.5462, 1.9039) -- (15.5818, 
  1.8944) -- (15.6529, 1.8758) -- (15.7958, 1.8402) -- (15.8674, 1.8234) -- 
  (15.9392, 1.807) -- (16.0834, 1.776) -- (16.1196, 1.7687) -- (16.1557, 1.7614)
   -- (16.2282, 1.7474) -- (16.3736, 1.7209) -- (16.4093, 1.7148) -- (16.445, 
  1.7089) -- (16.5166, 1.6973) -- (16.6601, 1.6758) -- (16.696, 1.6707) -- 
  (16.7321, 1.6659) -- (16.8042, 1.6565) -- (16.9486, 1.6395) -- (16.9849, 
  1.6356) -- (17.021, 1.6317) -- (17.0936, 1.6246) -- (17.1298, 1.6213) -- 
  (17.1661, 1.6181) -- (17.2388, 1.612) -- (17.3116, 1.6065) -- (17.3481, 1.604)
   -- (17.3845, 1.6016) -- (17.421, 1.5994) -- (17.4574, 1.5973) -- (17.5306, 
  1.5934) -- (17.5671, 1.5918) -- (17.6036, 1.5903) -- (17.6403, 1.5887) -- 
  (17.6768, 1.5875) -- (17.7134, 1.5864) -- (17.7868, 1.5845) -- (17.8234, 
  1.5838) -- (17.8601, 1.5832) -- (17.8967, 1.5828) -- (17.9335, 1.5825) -- 
  (17.9702, 1.5824) -- (18.007, 1.5823) -- (18.0438, 1.5824) -- (18.0806, 
  1.5827) -- (18.1173, 1.5831) -- (18.1541, 1.5836) -- (18.1909, 1.5843) -- 
  (18.2277, 1.5852) -- (18.2645, 1.5861) -- (18.3013, 1.5872) -- (18.3382, 
  1.5885) -- (18.375, 1.5898) -- (18.412, 1.5914) -- (18.4488, 1.5931) -- 
  (18.4857, 1.5949) -- (18.5225, 1.5969) -- (18.5594, 1.5989) -- (18.5963, 
  1.6011) -- (18.6333, 1.6036) -- (18.707, 1.6087) -- (18.7471, 1.6118) -- 
  (18.7872, 1.6151) -- (18.8673, 1.6219) -- (18.9074, 1.6257) -- (18.9473, 
  1.6295) -- (19.0275, 1.6378) -- (19.0675, 1.6421) -- (19.1076, 1.6466) -- 
  (19.1877, 1.6563) -- (19.3479, 1.6775) -- (19.3879, 1.6831) -- (19.428, 
  1.6891) -- (19.5081, 1.7013) -- (19.6679, 1.7278) -- (19.7079, 1.7348) -- 
  (19.7478, 1.742) -- (19.8278, 1.757) -- (19.9873, 1.7888) -- (20.0272, 1.7971)
   -- (20.0671, 1.8056) -- (20.1468, 1.8233) -- (20.3058, 1.8605) -- (20.3455, 
  1.8701) -- (20.3852, 1.8801) -- (20.4646, 1.9003) -- (20.6229, 1.9429) -- 
  (20.6624, 1.9539) -- (20.7019, 1.9651) -- (20.7808, 1.988) -- (20.9383, 
  2.0359) -- (21.2517, 2.1395) -- (21.3245, 2.1652) -- (21.3972, 2.1915) -- 
  (21.542, 2.2458) -- (21.8302, 2.3613) -- (21.9018, 2.3916) -- (21.9732, 
  2.4224) -- (22.1157, 2.486) -- (22.3984, 2.6196) -- (22.4335, 2.637) -- 
  (22.4685, 2.6545);

  \path[draw=black,line cap=,line join=miter,line width=0.1058cm,miter 
  limit=3.25,shift={(-3.4925, 0.2646)}] (10.1775, 6.3189) -- (10.0137, 6.3386) 
  -- (9.9171, 6.3526) -- (9.8357, 6.3657) -- (9.7631, 6.3785) -- (9.6963, 
  6.3911) -- (9.6337, 6.4037) -- (9.5181, 6.4291) -- (9.464, 6.4419) -- (9.4117,
   6.4547) -- (9.3122, 6.4806) -- (9.1288, 6.534) -- (9.0854, 6.5475) -- 
  (9.0431, 6.5613) -- (8.9606, 6.5888) -- (8.804, 6.6456) -- (8.7664, 6.6599) --
   (8.7293, 6.6745) -- (8.6567, 6.7037) -- (8.5174, 6.7637) -- (8.4836, 6.7788) 
  -- (8.4501, 6.7941) -- (8.3847, 6.825) -- (8.2579, 6.8877) -- (8.02, 7.0171) 
  -- (7.9915, 7.0335) -- (7.9634, 7.0502) -- (7.908, 7.0835) -- (7.8001, 7.1512)
   -- (7.5962, 7.2897) -- (7.5697, 7.3088) -- (7.5434, 7.3279) -- (7.4917, 
  7.3664) -- (7.3911, 7.4442) -- (7.2014, 7.6028) -- (7.1786, 7.6228) -- 
  (7.1562, 7.6429) -- (7.1118, 7.6834) -- (7.0258, 7.7648) -- (6.8637, 7.9299) 
  -- (6.8444, 7.9507) -- (6.8253, 7.9717) -- (6.7875, 8.0134) -- (6.7145, 
  8.0976) -- (6.5775, 8.2677) -- (6.5613, 8.2892) -- (6.5451, 8.3105) -- 
  (6.5136, 8.3534) -- (6.4525, 8.4395) -- (6.3391, 8.6129) -- (6.3266, 8.6333) 
  -- (6.3142, 8.6535) -- (6.2901, 8.6942) -- (6.2433, 8.7758) -- (6.232, 8.7962)
   -- (6.2209, 8.8166) -- (6.1991, 8.8575) -- (6.1572, 8.9393) -- (6.1472, 
  8.9597) -- (6.1372, 8.9803) -- (6.1178, 9.0212) -- (6.0806, 9.1032) -- 
  (6.0133, 9.2672) -- (6.0055, 9.2877) -- (5.9979, 9.3082) -- (5.9832, 9.3492) 
  -- (5.9552, 9.431) -- (5.9486, 9.4516) -- (5.9421, 9.4719) -- (5.9294, 9.5129)
   -- (5.906, 9.5945) -- (5.9005, 9.6148) -- (5.8951, 9.6352) -- (5.8848, 9.676)
   -- (5.8798, 9.6963) -- (5.875, 9.7167) -- (5.8658, 9.7572) -- (5.8613, 
  9.7776) -- (5.8571, 9.7979) -- (5.8489, 9.8384) -- (5.8343, 9.9191) -- 
  (5.8312, 9.9388) -- (5.828, 9.9587) -- (5.8221, 9.9981) -- (5.8193, 10.0178) 
  -- (5.8168, 10.0374) -- (5.812, 10.0766) -- (5.8097, 10.0962) -- (5.8076, 
  10.1159) -- (5.8057, 10.1353) -- (5.8021, 10.1745) -- (5.8006, 10.1939) -- 
  (5.7991, 10.2135) -- (5.7977, 10.2329) -- (5.7966, 10.2523) -- (5.7955, 
  10.2716) -- (5.7945, 10.2911) -- (5.7937, 10.3103) -- (5.793, 10.3296) -- 
  (5.7925, 10.3489) -- (5.7916, 10.3875) -- (5.7913, 10.4258) -- (5.7916, 
  10.4641) -- (5.7919, 10.4832) -- (5.7923, 10.5023) -- (5.7929, 10.5213) -- 
  (5.7936, 10.5403) -- (5.7944, 10.5592) -- (5.7954, 10.5782) -- (5.7963, 
  10.5971) -- (5.7976, 10.616) -- (5.7988, 10.6347) -- (5.8002, 10.6536) -- 
  (5.8051, 10.7098) -- (5.8071, 10.7284) -- (5.8112, 10.7656) -- (5.8135, 
  10.7843) -- (5.8159, 10.8027) -- (5.821, 10.8396) -- (5.8237, 10.8581) -- 
  (5.8266, 10.8764) -- (5.8327, 10.9131) -- (5.8393, 10.9495) -- (5.8463, 
  10.9859) -- (5.8501, 11.0039) -- (5.8539, 11.022) -- (5.8619, 11.0579) -- 
  (5.8793, 11.1293) -- (5.8844, 11.1486) -- (5.8895, 11.1678) -- (5.9002, 
  11.2059) -- (5.9234, 11.2817) -- (5.9358, 11.3192) -- (5.9487, 11.3566) -- 
  (5.9762, 11.4304) -- (5.9908, 11.4671) -- (6.0058, 11.5034) -- (6.0375, 
  11.5752) -- (6.1071, 11.7159) -- (6.1164, 11.7333) -- (6.1258, 11.7504) -- 
  (6.145, 11.7847) -- (6.1849, 11.8522) -- (6.2056, 11.8856) -- (6.2268, 
  11.9185) -- (6.2708, 11.9837) -- (6.2819, 11.9998) -- (6.2934, 12.0158) -- 
  (6.3165, 12.0475) -- (6.3643, 12.1102) -- (6.4656, 12.2313) -- (6.4779, 
  12.2451) -- (6.4903, 12.2588) -- (6.5155, 12.286) -- (6.5669, 12.3394) -- 
  (6.6744, 12.4423) -- (6.6883, 12.4547) -- (6.7024, 12.4671) -- (6.7305, 
  12.4918) -- (6.7881, 12.5397) -- (6.9077, 12.6316) -- (6.9386, 12.6537) -- 
  (6.9697, 12.6753) -- (7.0331, 12.7175) -- (7.0491, 12.7278) -- (7.0652, 
  12.738) -- (7.0979, 12.7582) -- (7.1639, 12.7972) -- (7.1806, 12.8067) -- 
  (7.1974, 12.816) -- (7.2313, 12.8345) -- (7.2999, 12.8702) -- (7.4407, 
  12.9362) -- (7.4602, 12.9446) -- (7.4796, 12.9529) -- (7.5189, 12.969) -- 
  (7.5984, 12.9996) -- (7.6386, 13.0139) -- (7.679, 13.0278) -- (7.7608, 
  13.0537) -- (7.7815, 13.0599) -- (7.8022, 13.0659) -- (7.8437, 13.0773) -- 
  (7.8645, 13.0828) -- (7.8854, 13.0882) -- (7.9275, 13.0985) -- (7.9696, 
  13.1082) -- (8.0121, 13.1171) -- (8.0333, 13.1214) -- (8.0546, 13.1254) -- 
  (8.0974, 13.1331) -- (8.1187, 13.1367) -- (8.1402, 13.1401) -- (8.1832, 
  13.1465) -- (8.2049, 13.1495) -- (8.2263, 13.1523) -- (8.2696, 13.1572) -- 
  (8.2913, 13.1596) -- (8.313, 13.1616) -- (8.3347, 13.1636) -- (8.3564, 
  13.1652) -- (8.3781, 13.1669) -- (8.3998, 13.1683) -- (8.4216, 13.1695) -- 
  (8.4433, 13.1705) -- (8.465, 13.1714) -- (8.4868, 13.1721) -- (8.5086, 
  13.1727) -- (8.5302, 13.1729) -- (8.552, 13.1732) -- (8.5736, 13.1732) -- 
  (8.5954, 13.1729) -- (8.617, 13.1727) -- (8.6387, 13.1721) -- (8.6603, 
  13.1714) -- (8.6818, 13.1706) -- (8.7034, 13.1695) -- (8.7249, 13.1684) -- 
  (8.7464, 13.167) -- (8.7679, 13.1654) -- (8.7893, 13.1637) -- (8.8104, 
  13.1619) -- (8.8312, 13.1599) -- (8.8521, 13.1576) -- (8.8729, 13.1553) -- 
  (8.8937, 13.1528) -- (8.9144, 13.1502) -- (8.9557, 13.1444) -- (8.9763, 
  13.1413) -- (8.9967, 13.1379) -- (9.0375, 13.1311) -- (9.0577, 13.1273) -- 
  (9.0778, 13.1235) -- (9.1179, 13.1152) -- (9.1378, 13.1109) -- (9.1576, 
  13.1065) -- (9.197, 13.0973) -- (9.2746, 13.0772) -- (9.2938, 13.0719) -- 
  (9.3128, 13.0664) -- (9.3506, 13.0553) -- (9.4248, 13.0314) -- (9.4432, 
  13.0252) -- (9.4613, 13.0189) -- (9.4973, 13.0061) -- (9.5679, 12.9793) -- 
  (9.7032, 12.9221) -- (9.7357, 12.9073) -- (9.768, 12.8921) -- (9.8308, 
  12.8614) -- (9.9507, 12.7983) -- (9.9642, 12.7908) -- (9.9777, 12.7832) -- 
  (10.0042, 12.7682) -- (10.0561, 12.738) -- (10.1556, 12.6773) -- (10.3383, 
  12.5563) -- (10.65, 12.3256) -- (10.6596, 12.3182) -- (10.6689, 12.3107) -- 
  (10.6877, 12.296) -- (10.7245, 12.2668) -- (10.7952, 12.2098) -- (10.927, 
  12.1015) -- (11.1584, 11.9074) -- (11.5055, 11.6174) -- (11.7696, 11.4086) -- 
  (11.9943, 11.2455) -- (12.1597, 11.1368) -- (12.3013, 11.0528) -- (12.4103, 
  10.9951) -- (12.4908, 10.9568) -- (12.56, 10.9276) -- (12.611, 10.9086) -- 
  (12.6545, 10.8942) -- (12.688, 10.8849) -- (12.6888, 10.8846) -- (12.6897, 
  10.8844) -- (12.6915, 10.884) -- (12.701, 10.8815) -- (12.7127, 10.8789) -- 
  (12.713, 10.8788) -- (12.7134, 10.8788) -- (12.7141, 10.8786) -- (12.7156, 
  10.8782) -- (12.7185, 10.8777) -- (12.7239, 10.8766) -- (12.7338, 10.8748) -- 
  (12.7341, 10.8746) -- (12.7343, 10.8746) -- (12.7348, 10.8745) -- (12.7359, 
  10.8744) -- (12.7379, 10.8739) -- (12.7419, 10.8734) -- (12.7422, 10.8734) -- 
  (12.7425, 10.8733) -- (12.7429, 10.8733) -- (12.7439, 10.8731) -- (12.7492, 
  10.8723) -- (12.7496, 10.8723) -- (12.7501, 10.8721) -- (12.7509, 10.872) -- 
  (12.7525, 10.8719) -- (12.7556, 10.8715) -- (12.756, 10.8715) -- (12.7564, 
  10.8713) -- (12.7571, 10.8713) -- (12.7585, 10.8712) -- (12.7612, 10.8708) -- 
  (12.762, 10.8708) -- (12.7627, 10.8706) -- (12.7641, 10.8706) -- (12.7643, 
  10.8705) -- (12.7655, 10.8705) -- (12.7667, 10.8703) -- (12.7674, 10.8703) -- 
  (12.768, 10.8702) -- (12.7692, 10.8702) -- (12.7716, 10.8699) -- (12.7738, 
  10.8699) -- (12.7739, 10.8698) -- (12.7769, 10.8698) -- (12.7773, 10.8696) -- 
  (12.7885, 10.8696) -- (12.7888, 10.8698) -- (12.791, 10.8698) -- (12.7914, 
  10.8699) -- (12.7928, 10.8699) -- (12.7932, 10.8701) -- (12.7944, 10.8701) -- 
  (12.7949, 10.8702) -- (12.7958, 10.8702) -- (12.796, 10.8703) -- (12.7971, 
  10.8703) -- (12.7971, 10.8705) -- (12.7976, 10.8705) -- (12.7983, 10.8706) -- 
  (12.7993, 10.8708) -- (12.7995, 10.8708) -- (12.7997, 10.8709) -- (12.8002, 
  10.8709) -- (12.8009, 10.871) -- (12.8011, 10.871) -- (12.8012, 10.8712) -- 
  (12.8018, 10.8712) -- (12.8024, 10.8715) -- (12.8035, 10.8717) -- (12.8044, 
  10.872) -- (12.8051, 10.8722) -- (12.8055, 10.8725) -- (12.8059, 10.8726) -- 
  (12.8064, 10.8729) -- (12.8066, 10.873) -- (12.8067, 10.873) -- (12.8069, 
  10.8732) -- (12.807, 10.8732) -- (12.807, 10.8733) -- (12.8071, 10.8733) -- 
  (12.8071, 10.8734);

  \path[draw=black,line cap=,line join=miter,line width=0.1058cm,miter 
  limit=3.25,shift={(-3.4925, 0.2646)}] (10.1775, 6.3189) -- (10.2722, 6.45) -- 
  (10.3261, 6.5256) -- (10.409, 6.6431) -- (10.5348, 6.8242) -- (10.7203, 7.098)
   -- (10.9838, 7.5007) -- (11.337, 8.067) -- (11.8018, 8.8551) -- (12.0746, 
  9.3368) -- (12.264, 9.678) -- (12.4136, 9.9533) -- (12.5166, 10.148) -- 
  (12.5994, 10.3103) -- (12.6592, 10.433) -- (12.7006, 10.5226) -- (12.7336, 
  10.599) -- (12.7561, 10.6553) -- (12.7723, 10.6992) -- (12.785, 10.737) -- 
  (12.7933, 10.765) -- (12.7996, 10.7889) -- (12.7996, 10.7892) -- (12.7998, 
  10.7895) -- (12.7999, 10.7902) -- (12.8007, 10.7939) -- (12.8018, 10.7987) -- 
  (12.8038, 10.8073) -- (12.8038, 10.8078) -- (12.8039, 10.8082) -- (12.804, 
  10.8092) -- (12.8045, 10.811) -- (12.8051, 10.8146) -- (12.8062, 10.8209) -- 
  (12.8064, 10.8212) -- (12.8064, 10.8217) -- (12.8065, 10.8226) -- (12.8068, 
  10.8241) -- (12.8072, 10.8271) -- (12.8073, 10.8272) -- (12.8073, 10.8278) -- 
  (12.8075, 10.8286) -- (12.8076, 10.83) -- (12.808, 10.8326) -- (12.808, 
  10.8329) -- (12.8082, 10.8332) -- (12.8082, 10.8337) -- (12.8083, 10.835) -- 
  (12.8086, 10.8372) -- (12.8086, 10.8374) -- (12.8087, 10.8377) -- (12.8087, 
  10.8381) -- (12.8089, 10.8392) -- (12.809, 10.8412) -- (12.809, 10.8414) -- 
  (12.8091, 10.8417) -- (12.8091, 10.8421) -- (12.8093, 10.8432) -- (12.8093, 
  10.8442) -- (12.8094, 10.845) -- (12.8094, 10.846) -- (12.8096, 10.8468) -- 
  (12.8096, 10.8489) -- (12.8097, 10.8493) -- (12.8097, 10.8522) -- (12.8098, 
  10.8523) -- (12.8098, 10.8596) -- (12.8097, 10.86) -- (12.8097, 10.8624) -- 
  (12.8096, 10.8631) -- (12.8096, 10.8643) -- (12.8094, 10.8643) -- (12.8094, 
  10.8656) -- (12.8093, 10.8657) -- (12.8093, 10.8664) -- (12.8091, 10.8667) -- 
  (12.8091, 10.8671) -- (12.809, 10.8679) -- (12.809, 10.8682) -- (12.8089, 
  10.8686) -- (12.8087, 10.8691) -- (12.8087, 10.8697) -- (12.8083, 10.8709) -- 
  (12.8082, 10.8715) -- (12.8079, 10.8723) -- (12.8078, 10.8726) -- (12.8075, 
  10.8729) -- (12.8075, 10.873) -- (12.8073, 10.8731) -- (12.8073, 10.8733) -- 
  (12.8072, 10.8734);

  \path[draw=black,line cap=,line join=miter,line width=0.1058cm,miter 
  limit=3.25,shift={(-3.4925, 0.2646)}] (10.1775, 15.4281) -- (10.2432, 15.2802)
   -- (10.281, 15.1956) -- (10.3393, 15.0643) -- (10.4286, 14.8632) -- (10.5625,
   14.562) -- (10.7575, 14.1241) -- (11.0308, 13.5176) -- (11.4235, 12.6888) -- 
  (11.6845, 12.1919) -- (11.8874, 11.8474) -- (12.0662, 11.5781) -- (12.2029, 
  11.3954) -- (12.3242, 11.2504) -- (12.4207, 11.147) -- (12.4938, 11.0758) -- 
  (12.5583, 11.0191) -- (12.6067, 10.9801) -- (12.6453, 10.9517) -- (12.6791, 
  10.9289) -- (12.7044, 10.9134) -- (12.7262, 10.9011) -- (12.7433, 10.8924) -- 
  (12.7561, 10.8866) -- (12.767, 10.8821) -- (12.7752, 10.8791) -- (12.7823, 
  10.8769) -- (12.7876, 10.8753) -- (12.7878, 10.8752) -- (12.7882, 10.8752) -- 
  (12.7887, 10.8751) -- (12.7897, 10.8748) -- (12.7916, 10.8744) -- (12.7918, 
  10.8744) -- (12.7919, 10.8742) -- (12.7921, 10.8742) -- (12.7926, 10.8741) -- 
  (12.7951, 10.8737) -- (12.7952, 10.8737) -- (12.7954, 10.8735) -- (12.7958, 
  10.8735) -- (12.7965, 10.8734) -- (12.797, 10.8734) -- (12.7976, 10.8733) -- 
  (12.7978, 10.8733) -- (12.7981, 10.8731) -- (12.7991, 10.8731) -- (12.7996, 
  10.873) -- (12.8007, 10.873) -- (12.8009, 10.8729) -- (12.8051, 10.8729) -- 
  (12.8053, 10.873) -- (12.806, 10.873) -- (12.8061, 10.8731) -- (12.8065, 
  10.8731) -- (12.8067, 10.8733) -- (12.8069, 10.8733) -- (12.8071, 10.8734);

  \path[draw=black,line cap=,line join=miter,line width=0.1058cm,miter 
  limit=3.25,shift={(-3.4925, 0.2646)}] (10.1775, 15.4281) -- (10.0127, 15.4431)
   -- (9.9144, 15.4499) -- (9.8313, 15.4544) -- (9.7566, 15.4575) -- (9.6875, 
  15.4594) -- (9.6224, 15.4606) -- (9.5606, 15.461) -- (9.5013, 15.4609) -- 
  (9.4441, 15.4602) -- (9.3888, 15.459) -- (9.3351, 15.4573) -- (9.2829, 
  15.4552) -- (9.2318, 15.4529) -- (9.1819, 15.45) -- (9.1329, 15.4468) -- 
  (9.085, 15.4434) -- (9.0377, 15.4397) -- (8.9914, 15.4355) -- (8.9457, 
  15.4311) -- (8.9007, 15.4264) -- (8.8564, 15.4216) -- (8.8126, 15.4164) -- 
  (8.7267, 15.4052) -- (8.6845, 15.3993) -- (8.6428, 15.3932) -- (8.5606, 
  15.3803) -- (8.4013, 15.3518) -- (8.3623, 15.344) -- (8.3238, 15.3362) -- 
  (8.2476, 15.3199) -- (8.0987, 15.2852) -- (8.0622, 15.2761) -- (8.026, 
  15.2667) -- (7.9543, 15.2476) -- (7.8138, 15.2072) -- (7.7792, 15.1967) -- 
  (7.7449, 15.1861) -- (7.6768, 15.1643) -- (7.5432, 15.1189) -- (7.2853, 
  15.0208) -- (7.2512, 15.0067) -- (7.2173, 14.9925) -- (7.1504, 14.9637) -- 
  (7.0185, 14.904) -- (6.764, 14.7773) -- (6.733, 14.7607) -- (6.7021, 14.744) 
  -- (6.6409, 14.7103) -- (6.5209, 14.6408) -- (6.2888, 14.4952) -- (6.2606, 
  14.4763) -- (6.2325, 14.4573) -- (6.1768, 14.419) -- (6.0674, 14.3407) -- 
  (5.8565, 14.1777) -- (5.8309, 14.1567) -- (5.8054, 14.1356) -- (5.755, 
  14.0931) -- (5.6559, 14.0065) -- (5.4657, 13.8276) -- (5.4442, 13.8063) -- 
  (5.4227, 13.7848) -- (5.3804, 13.7415) -- (5.2973, 13.6537) -- (5.138, 
  13.4735) -- (5.1187, 13.4506) -- (5.0997, 13.4275) -- (5.0618, 13.3812) -- 
  (4.9878, 13.2872) -- (4.8467, 13.0949) -- (4.8296, 13.0706) -- (4.8128, 
  13.046) -- (4.7794, 12.9967) -- (4.7147, 12.8971) -- (4.592, 12.6938) -- 
  (4.5628, 12.6423) -- (4.5343, 12.5902) -- (4.4788, 12.4853) -- (4.375, 
  12.2718) -- (4.363, 12.2454) -- (4.351, 12.2188) -- (4.3277, 12.1655) -- 
  (4.2827, 12.058) -- (4.2718, 12.031) -- (4.261, 12.0038) -- (4.2399, 11.9494) 
  -- (4.1996, 11.8398) -- (4.1899, 11.8121) -- (4.1803, 11.7846) -- (4.1615, 
  11.729) -- (4.126, 11.6174) -- (4.1174, 11.5893) -- (4.109, 11.5612) -- 
  (4.0928, 11.5047) -- (4.0619, 11.3911) -- (4.0546, 11.3625) -- (4.0474, 
  11.3339) -- (4.0335, 11.2765) -- (4.0076, 11.161) -- (4.0015, 11.1321) -- 
  (3.9955, 11.103) -- (3.984, 11.0447) -- (3.963, 10.9276) -- (3.9581, 10.8981) 
  -- (3.9535, 10.8686) -- (3.9445, 10.8096) -- (3.9402, 10.78) -- (3.9361, 
  10.7502) -- (3.9284, 10.6908) -- (3.9247, 10.6609) -- (3.9212, 10.631) -- 
  (3.9147, 10.5712) -- (3.9117, 10.5412) -- (3.9089, 10.5111) -- (3.9037, 
  10.4509) -- (3.9011, 10.4181) -- (3.8987, 10.3853) -- (3.8965, 10.3525) -- 
  (3.8945, 10.3196) -- (3.8927, 10.2866) -- (3.891, 10.2537) -- (3.8896, 
  10.2206) -- (3.8884, 10.1876) -- (3.8873, 10.1543) -- (3.8865, 10.1211) -- 
  (3.8858, 10.0879) -- (3.8852, 10.0547) -- (3.885, 10.0214) -- (3.8848, 9.988) 
  -- (3.8848, 9.9545) -- (3.8851, 9.921) -- (3.8855, 9.8876) -- (3.8862, 9.8541)
   -- (3.887, 9.8204) -- (3.888, 9.7868) -- (3.8892, 9.7532) -- (3.8906, 9.7194)
   -- (3.8923, 9.6857) -- (3.8941, 9.6519) -- (3.896, 9.618) -- (3.8982, 9.5841)
   -- (3.9005, 9.5502) -- (3.9032, 9.5163) -- (3.9059, 9.4823) -- (3.9088, 
  9.4484) -- (3.9153, 9.3802) -- (3.9187, 9.346) -- (3.9226, 9.3118) -- (3.9306,
   9.2435) -- (3.9348, 9.2092) -- (3.9394, 9.1748) -- (3.9489, 9.1062) -- 
  (3.954, 9.0718) -- (3.9594, 9.0374) -- (3.9705, 8.9684) -- (3.9952, 8.8303) --
   (4.002, 8.7956) -- (4.0087, 8.761) -- (4.0232, 8.6917) -- (4.0542, 8.5527) --
   (4.0625, 8.5178) -- (4.071, 8.4831) -- (4.0885, 8.4133) -- (4.126, 8.2736) --
   (4.1352, 8.2409) -- (4.1446, 8.2083) -- (4.164, 8.143) -- (4.2047, 8.012) -- 
  (4.2948, 7.7495) -- (4.3068, 7.7167) -- (4.319, 7.6838) -- (4.3441, 7.6181) --
   (4.3964, 7.4865) -- (4.5094, 7.2227) -- (4.5394, 7.1568) -- (4.5703, 7.0908) 
  -- (4.6341, 6.9587) -- (4.7704, 6.6942) -- (5.0783, 6.1649) -- (5.1008, 6.129)
   -- (5.1235, 6.0931) -- (5.1697, 6.0213) -- (5.2647, 5.8777) -- (5.4649, 
  5.5904) -- (5.9056, 5.015) -- (5.9348, 4.979) -- (5.9643, 4.9429) -- (6.024, 
  4.8708) -- (6.1455, 4.7265) -- (6.3978, 4.4374) -- (6.9369, 3.8571) -- 
  (6.9714, 3.8214) -- (7.0059, 3.7857) -- (7.0755, 3.7144) -- (7.2164, 3.5715) 
  -- (7.5042, 3.2858) -- (8.1018, 2.7175) -- (8.1399, 2.6823) -- (8.1781, 
  2.6471) -- (8.255, 2.5768) -- (8.4096, 2.4369) -- (8.7228, 2.1599) -- (9.3647,
   1.6218) -- (9.4028, 1.5912) -- (9.4409, 1.5608) -- (9.5173, 1.5) -- (9.6709, 
  1.3797) -- (9.9815, 1.1442) -- (10.6164, 0.6952) -- (10.6568, 0.6681) -- 
  (10.6972, 0.6412) -- (10.778, 0.5879) -- (10.9408, 0.4824) -- (11.2699, 
  0.2782) -- (11.9419, -0.1038) -- (11.9881, -0.1283) -- (12.0344, -0.1527) -- 
  (12.1273, -0.2011);

  \path[draw=black,line cap=,line join=miter,line width=0.1058cm,miter 
  limit=3.25,shift={(-3.4925, 0.2646)}] (10.1775, 15.4281) -- (10.2767, 15.5607)
   -- (10.3374, 15.6383) -- (10.4374, 15.7609) -- (10.6059, 15.9535) -- 
  (10.6447, 15.9957) -- (10.6826, 16.0361) -- (10.7564, 16.1128) -- (10.8979, 
  16.2529) -- (10.9324, 16.2857) -- (10.9666, 16.3176) -- (11.0341, 16.3795) -- 
  (11.1667, 16.4959) -- (11.4249, 16.7049) -- (11.4567, 16.7291) -- (11.4884, 
  16.7529) -- (11.5517, 16.7997) -- (11.6776, 16.889) -- (11.9271, 17.0531) -- 
  (11.9581, 17.0724) -- (11.9891, 17.0914) -- (12.0509, 17.1287) -- (12.1746, 
  17.2005) -- (12.4208, 17.333) -- (12.454, 17.3499) -- (12.4874, 17.3666) -- 
  (12.5539, 17.3991) -- (12.6868, 17.4615) -- (12.72, 17.4766) -- (12.7532, 
  17.4914) -- (12.8195, 17.5204) -- (12.952, 17.5756) -- (12.9851, 17.589) -- 
  (13.0182, 17.6021) -- (13.0845, 17.6277) -- (13.2166, 17.6764) -- (13.4807, 
  17.7646) -- (13.5136, 17.7748) -- (13.5465, 17.7847) -- (13.6124, 17.8041) -- 
  (13.744, 17.8408) -- (13.7769, 17.8495) -- (13.8097, 17.858) -- (13.8755, 
  17.8745) -- (14.0067, 17.9055) -- (14.0393, 17.9128) -- (14.0721, 17.92) -- 
  (14.1376, 17.9338) -- (14.2683, 17.9593) -- (14.3009, 17.9653) -- (14.3335, 
  17.9711) -- (14.3987, 17.9823) -- (14.5288, 18.0026) -- (14.5592, 18.007) -- 
  (14.5896, 18.0111) -- (14.6501, 18.0192) -- (14.6804, 18.0231) -- (14.7106, 
  18.0267) -- (14.7711, 18.0337) -- (14.8012, 18.037) -- (14.8314, 18.04) -- 
  (14.8916, 18.046) -- (14.9218, 18.0487) -- (14.9519, 18.0513) -- (15.0119, 
  18.0562) -- (15.042, 18.0584) -- (15.072, 18.0604) -- (15.1019, 18.0624) -- 
  (15.1318, 18.0642) -- (15.1619, 18.0658) -- (15.1918, 18.0675) -- (15.2215, 
  18.0688) -- (15.2813, 18.0713) -- (15.3111, 18.0724) -- (15.3706, 18.0741) -- 
  (15.4004, 18.0748) -- (15.43, 18.0753) -- (15.4597, 18.0757) -- (15.4894, 
  18.076) -- (15.5189, 18.0761) -- (15.5485, 18.0761) -- (15.5782, 18.076) -- 
  (15.6077, 18.0759) -- (15.6666, 18.075) -- (15.6961, 18.0745) -- (15.7255, 
  18.0738) -- (15.7548, 18.073) -- (15.7842, 18.072) -- (15.8135, 18.0709) -- 
  (15.8429, 18.0697) -- (15.8721, 18.0684) -- (15.9013, 18.067) -- (15.9597, 
  18.0637) -- (15.9888, 18.0619) -- (16.0179, 18.06) -- (16.0761, 18.0559) -- 
  (16.105, 18.0537) -- (16.1341, 18.0512) -- (16.1918, 18.0461) -- (16.2207, 
  18.0435) -- (16.2496, 18.0406) -- (16.307, 18.0345) -- (16.4217, 18.0212) -- 
  (16.4496, 18.0176) -- (16.4776, 18.0139) -- (16.5334, 18.0061) -- (16.6446, 
  17.9895) -- (16.6723, 17.9851) -- (16.7, 17.9805) -- (16.7552, 17.9711) -- 
  (16.8651, 17.9512) -- (16.8925, 17.9458) -- (16.9198, 17.9404) -- (16.9744, 
  17.9294) -- (17.0829, 17.9061) -- (17.2978, 17.8546) -- (17.3246, 17.8477) -- 
  (17.3512, 17.8406) -- (17.4042, 17.8263) -- (17.5099, 17.7967) -- (17.7188, 
  17.7326) -- (17.7447, 17.7242) -- (17.7705, 17.7156) -- (17.822, 17.6983) -- 
  (17.9244, 17.6626) -- (18.1266, 17.5867) -- (18.1809, 17.5652) -- (18.2348, 
  17.5431) -- (18.342, 17.4979) -- (18.5529, 17.4027) -- (18.579, 17.3904) -- 
  (18.605, 17.378) -- (18.6567, 17.3528) -- (18.7594, 17.3013) -- (18.9613, 
  17.1936) -- (18.9861, 17.1799) -- (19.011, 17.1659) -- (19.0603, 17.1377) -- 
  (19.1582, 17.0802) -- (19.3503, 16.9612) -- (19.7188, 16.7069) -- (19.7396, 
  16.6915) -- (19.7604, 16.6759) -- (19.8016, 16.6446) -- (19.8834, 16.5814) -- 
  (20.0433, 16.4516) -- (20.0629, 16.435) -- (20.0824, 16.4184) -- (20.1213, 
  16.3851) -- (20.1982, 16.3176) -- (20.3482, 16.18) -- (20.3666, 16.1626) -- 
  (20.3849, 16.145) -- (20.4213, 16.1098) -- (20.4932, 16.0387) -- (20.633, 
  15.8939) -- (20.8968, 15.5944) -- (20.9139, 15.5738) -- (20.9309, 15.553) -- 
  (20.9646, 15.5112) -- (21.0309, 15.4273) -- (21.1585, 15.2567) -- (21.1741, 
  15.2352) -- (21.1895, 15.2135) -- (21.22, 15.1703) -- (21.2798, 15.0832) -- 
  (21.3946, 14.9069) -- (21.4085, 14.8847) -- (21.4223, 14.8624) -- (21.4496, 
  14.8178) -- (21.5029, 14.7281) -- (21.6045, 14.5468) -- (21.7879, 14.1785) -- 
  (21.7984, 14.1556) -- (21.8086, 14.1327) -- (21.8288, 14.087) -- (21.8682, 
  13.9951) -- (21.942, 13.8104) -- (21.9508, 13.7873) -- (21.9593, 13.764) -- 
  (21.9764, 13.7175) -- (22.0092, 13.6245) -- (22.0701, 13.4377) -- (22.0773, 
  13.4142) -- (22.0843, 13.3908) -- (22.0981, 13.344) -- (22.1244, 13.25) -- 
  (22.1307, 13.2265) -- (22.1371, 13.203) -- (22.1492, 13.156) -- (22.1724, 
  13.0619) -- (22.1834, 13.0147) -- (22.1939, 12.9676) -- (22.2138, 12.8734) -- 
  (22.2232, 12.8262) -- (22.2322, 12.7791) -- (22.2488, 12.6848) -- (22.2526, 
  12.6628) -- (22.2561, 12.6407) -- (22.263, 12.5968) -- (22.2757, 12.5089) -- 
  (22.2787, 12.487) -- (22.2816, 12.4649) -- (22.2871, 12.4211) -- (22.2971, 
  12.3333) -- (22.2994, 12.3113) -- (22.3016, 12.2893) -- (22.3058, 12.2455) -- 
  (22.3096, 12.2017) -- (22.3131, 12.158) -- (22.3161, 12.1142) -- (22.3176, 
  12.0924) -- (22.3188, 12.0705) -- (22.3201, 12.0487) -- (22.3213, 12.0268) -- 
  (22.3234, 11.9833) -- (22.3244, 11.9615) -- (22.3252, 11.9396) -- (22.3259, 
  11.9178) -- (22.3266, 11.8962) -- (22.3277, 11.8526) -- (22.3281, 11.8309) -- 
  (22.3284, 11.8092) -- (22.3286, 11.7875) -- (22.3289, 11.7442) -- (22.3289, 
  11.7224) -- (22.3286, 11.6791) -- (22.3284, 11.6577) -- (22.3275, 11.6144) -- 
  (22.327, 11.5929) -- (22.3264, 11.5712) -- (22.325, 11.5283) -- (22.3242, 
  11.5068) -- (22.3223, 11.4638) -- (22.3212, 11.4424) -- (22.3199, 11.4209) -- 
  (22.3173, 11.3782) -- (22.316, 11.3567) -- (22.3144, 11.3355) -- (22.3113, 
  11.2927) -- (22.3093, 11.2697) -- (22.3073, 11.2466) -- (22.303, 11.2006) -- 
  (22.3008, 11.1775) -- (22.2984, 11.1545) -- (22.2934, 11.1088) -- (22.2907, 
  11.0859) -- (22.288, 11.063) -- (22.2822, 11.0174) -- (22.2695, 10.9265) -- 
  (22.2661, 10.9037) -- (22.2626, 10.8811) -- (22.2553, 10.8359) -- (22.2396, 
  10.7459) -- (22.2355, 10.7236) -- (22.2312, 10.7012) -- (22.2224, 10.6565) -- 
  (22.2038, 10.5676) -- (22.199, 10.5454) -- (22.194, 10.5234) -- (22.1838, 
  10.4793) -- (22.1623, 10.3915) -- (22.1566, 10.3696) -- (22.151, 10.3478) -- 
  (22.1394, 10.3043) -- (22.115, 10.2177) -- (22.1088, 10.1961) -- (22.1024, 
  10.1746) -- (22.0894, 10.1318) -- (22.0623, 10.0464) -- (22.004, 9.8781) -- 
  (21.9968, 9.8585) -- (21.9896, 9.8391) -- (21.975, 9.8003) -- (21.9448, 
  9.7233) -- (21.881, 9.5712) -- (21.8726, 9.5524) -- (21.8644, 9.5335) -- 
  (21.8474, 9.4962) -- (21.8127, 9.4218) -- (21.7401, 9.2753) -- (21.7213, 
  9.2392) -- (21.7022, 9.2032) -- (21.6632, 9.1318) -- (21.5823, 8.9914) -- 
  (21.4082, 8.7201) -- (21.3971, 8.7038) -- (21.3858, 8.6877) -- (21.3632, 
  8.6556) -- (21.3173, 8.5919) -- (21.2229, 8.4672) -- (21.2108, 8.4518) -- 
  (21.1988, 8.4365) -- (21.1744, 8.406) -- (21.125, 8.3458) -- (21.0239, 8.228) 
  -- (21.0109, 8.2135) -- (20.9981, 8.1991) -- (20.9721, 8.1704) -- (20.9196, 
  8.1136) -- (20.8122, 8.003) -- (20.5887, 7.7928) -- (20.5733, 7.779) -- 
  (20.5576, 7.7654) -- (20.5263, 7.7384) -- (20.463, 7.6852) -- (20.3343, 
  7.5821) -- (20.3181, 7.5694) -- (20.2689, 7.5322) -- (20.2027, 7.4836) -- 
  (20.0686, 7.3899) -- (20.0517, 7.3786) -- (20.0346, 7.3671) -- (20.0006, 
  7.3448) -- (19.9319, 7.3008) -- (19.793, 7.2168) -- (19.5089, 7.063) -- 
  (19.4921, 7.0546) -- (19.4753, 7.0463) -- (19.4416, 7.0299) -- (19.3738, 
  6.998) -- (19.2374, 6.9373) -- (19.2203, 6.93) -- (19.2031, 6.9229) -- 
  (19.1688, 6.9085) -- (19.0997, 6.881) -- (18.961, 6.8289) -- (18.9436, 6.8227)
   -- (18.9261, 6.8166) -- (18.8913, 6.8045) -- (18.8212, 6.7812) -- (18.6808, 
  6.7378) -- (18.6632, 6.7327) -- (18.6103, 6.7178) -- (18.5396, 6.6989) -- 
  (18.3979, 6.6642) -- (18.3788, 6.6599) -- (18.3595, 6.6556) -- (18.321, 
  6.6472) -- (18.244, 6.6315) -- (18.2248, 6.6278) -- (18.2055, 6.6242) -- 
  (18.167, 6.6172) -- (18.0899, 6.604) -- (18.0513, 6.5979) -- (18.0128, 6.5921)
   -- (17.9935, 6.5894) -- (17.9742, 6.5867) -- (17.9357, 6.5815) -- (17.9164, 
  6.579) -- (17.8971, 6.5767) -- (17.8587, 6.5721) -- (17.7817, 6.564) -- 
  (17.7625, 6.5622) -- (17.7432, 6.5604) -- (17.7048, 6.5573) -- (17.6664, 
  6.5542) -- (17.628, 6.5516) -- (17.6088, 6.5504) -- (17.5705, 6.5482) -- 
  (17.5514, 6.5472) -- (17.5131, 6.5455) -- (17.4941, 6.5447) -- (17.4749, 
  6.544) -- (17.4559, 6.5435) -- (17.4367, 6.5429) -- (17.3987, 6.5421) -- 
  (17.3797, 6.5418) -- (17.3605, 6.5417) -- (17.3416, 6.5415) -- (17.3226, 
  6.5414) -- (17.3036, 6.5414) -- (17.2846, 6.5415) -- (17.2657, 6.5417) -- 
  (17.2467, 6.542) -- (17.2278, 6.5422) -- (17.1901, 6.5431) -- (17.1712, 
  6.5436) -- (17.1158, 6.5457) -- (17.0789, 6.5473) -- (17.0604, 6.5483) -- 
  (17.0236, 6.5505) -- (17.0053, 6.5516) -- (16.9869, 6.5528) -- (16.9504, 
  6.5556) -- (16.9321, 6.5571) -- (16.9139, 6.5586) -- (16.8774, 6.5618) -- 
  (16.841, 6.5654) -- (16.8048, 6.5692) -- (16.7867, 6.5712) -- (16.7687, 
  6.5732) -- (16.7326, 6.5777) -- (16.6967, 6.5823) -- (16.6608, 6.5872) -- 
  (16.5895, 6.5978) -- (16.554, 6.6036) -- (16.5185, 6.6095) -- (16.4481, 
  6.6223) -- (16.4306, 6.6256) -- (16.4131, 6.6291) -- (16.3783, 6.6361) -- 
  (16.3088, 6.651) -- (16.2744, 6.6587) -- (16.2399, 6.6668) -- (16.1717, 
  6.6836) -- (16.0369, 6.7203) -- (16.0058, 6.7294) -- (15.9749, 6.7387) -- 
  (15.9134, 6.7579) -- (15.7923, 6.7987) -- (15.7774, 6.8041) -- (15.7624, 
  6.8094) -- (15.7326, 6.8203) -- (15.6737, 6.8426) -- (15.5576, 6.8895) -- 
  (15.5433, 6.8956) -- (15.5291, 6.9016) -- (15.5006, 6.914) -- (15.4444, 
  6.9392) -- (15.3338, 6.9916) -- (15.1218, 7.1036) -- (15.1078, 7.1116) -- 
  (15.0941, 7.1195) -- (15.0665, 7.1356) -- (15.0121, 7.1681) -- (14.9061, 
  7.235) -- (14.893, 7.2435) -- (14.8802, 7.2521) -- (14.8546, 7.2693) -- 
  (14.804, 7.304) -- (14.7059, 7.3747) -- (14.6939, 7.3837) -- (14.6819, 7.3928)
   -- (14.6583, 7.4108) -- (14.6118, 7.4472) -- (14.5216, 7.5209) -- (14.3538, 
  7.6713) -- (14.3444, 7.6802) -- (14.3352, 7.689) -- (14.3169, 7.7068) -- 
  (14.2809, 7.7425) -- (14.2116, 7.8139) -- (14.083, 7.9568) -- (14.0754, 
  7.9657) -- (14.0678, 7.9746) -- (14.053, 7.9925) -- (14.0236, 8.0279) -- 
  (13.9674, 8.0987) -- (13.8639, 8.2383) -- (13.8574, 8.2477) -- (13.8508, 
  8.2571) -- (13.8378, 8.2757) -- (13.8126, 8.3127) -- (13.7645, 8.3859) -- 
  (13.677, 8.5287) -- (13.6668, 8.546) -- (13.6569, 8.5634) -- (13.6373, 8.5979)
   -- (13.6001, 8.6657) -- (13.5326, 8.7966) -- (13.5249, 8.8123) -- (13.5173, 
  8.8279) -- (13.5024, 8.8586) -- (13.4742, 8.9191) -- (13.4226, 9.0352) -- 
  (13.3364, 9.2493) -- (13.3342, 9.255) -- (13.332, 9.261) -- (13.3276, 9.2725) 
  -- (13.319, 9.2957) -- (13.3026, 9.3409) -- (13.2722, 9.4277) -- (13.2195, 
  9.5874) -- (13.1337, 9.8772) -- (13.0759, 10.0896) -- (13.0322, 10.2547) -- 
  (12.9946, 10.3954) -- (12.9657, 10.4984) -- (12.9395, 10.5858) -- (12.9177, 
  10.6521) -- (12.9005, 10.7006) -- (12.8847, 10.7415) -- (12.8721, 10.7712) -- 
  (12.8615, 10.7939) -- (12.8519, 10.8131) -- (12.8443, 10.8268) -- (12.8374, 
  10.8384) -- (12.832, 10.8467) -- (12.8276, 10.8529) -- (12.8237, 10.8581) -- 
  (12.8207, 10.8617) -- (12.8181, 10.8649) -- (12.8159, 10.8671) -- (12.8142, 
  10.8686) -- (12.8129, 10.87) -- (12.8118, 10.8708) -- (12.8108, 10.8715) -- 
  (12.8101, 10.872) -- (12.8096, 10.8724) -- (12.8094, 10.8724) -- (12.8094, 
  10.8726) -- (12.8091, 10.8726) -- (12.8091, 10.8727) -- (12.809, 10.8727);

  \path[draw=black,line cap=,line join=miter,line width=0.1058cm,miter 
  limit=3.25,shift={(-3.4925, 0.2646)}] (7.5485, 10.8733) -- (7.5488, 10.8731) 
  -- (7.549, 10.8731) -- (7.55, 10.8727) -- (7.5526, 10.8719) -- (7.5613, 
  10.8691) -- (7.5913, 10.8595) -- (7.7017, 10.8244) -- (7.7067, 10.8227) -- 
  (7.7116, 10.8212) -- (7.7217, 10.818) -- (7.7429, 10.8114) -- (7.7887, 
  10.7973) -- (7.8933, 10.7655) -- (8.1512, 10.6911) -- (8.1596, 10.6889) -- 
  (8.1853, 10.6819) -- (8.2199, 10.6724) -- (8.2914, 10.6532) -- (8.442, 
  10.6146) -- (8.4616, 10.6097) -- (8.4813, 10.6048) -- (8.521, 10.5952) -- 
  (8.6019, 10.576) -- (8.769, 10.5387) -- (8.79, 10.5343) -- (8.8532, 10.521) --
   (8.9385, 10.504) -- (8.96, 10.4998) -- (8.9816, 10.4958) -- (9.0249, 10.4877)
   -- (9.1119, 10.4721) -- (9.1228, 10.4702) -- (9.1338, 10.4684) -- (9.1555, 
  10.4647) -- (9.1994, 10.4574) -- (9.2873, 10.4436) -- (9.3093, 10.4403) -- 
  (9.3312, 10.4371) -- (9.3752, 10.4308) -- (9.4631, 10.4191) -- (9.475, 
  10.4176) -- (9.487, 10.416) -- (9.5107, 10.4131) -- (9.5581, 10.4075) -- 
  (9.5818, 10.4047) -- (9.6055, 10.4021) -- (9.6526, 10.3972) -- (9.6643, 
  10.3959) -- (9.676, 10.3948) -- (9.6995, 10.3925) -- (9.7462, 10.3882) -- 
  (9.8388, 10.3805) -- (9.8502, 10.3795) -- (9.8618, 10.3787) -- (9.8847, 
  10.377) -- (9.9301, 10.374) -- (9.9416, 10.3733) -- (9.9529, 10.3726) -- 
  (9.9753, 10.3713) -- (10.0203, 10.3688) -- (10.0313, 10.3682) -- (10.0425, 
  10.3677) -- (10.0646, 10.3667) -- (10.0757, 10.3662) -- (10.1087, 10.3649) -- 
  (10.1523, 10.3633) -- (10.1632, 10.363) -- (10.1739, 10.3626) -- (10.1956, 
  10.362) -- (10.2055, 10.3617) -- (10.2155, 10.3616) -- (10.2354, 10.3611) -- 
  (10.2552, 10.3608) -- (10.265, 10.3605) -- (10.2749, 10.3604) -- (10.2945, 
  10.3601) -- (10.3043, 10.3601) -- (10.3139, 10.36) -- (10.3237, 10.3598) -- 
  (10.3334, 10.3598) -- (10.343, 10.3597) -- (10.4098, 10.3597) -- (10.4192, 
  10.3598) -- (10.4287, 10.3598) -- (10.4381, 10.36) -- (10.4475, 10.36) -- 
  (10.4567, 10.3601) -- (10.4661, 10.3602) -- (10.503, 10.3608) -- (10.5212, 
  10.3611) -- (10.5394, 10.3615) -- (10.5485, 10.3617) -- (10.5574, 10.3619) -- 
  (10.5933, 10.363) -- (10.6109, 10.3635) -- (10.646, 10.3648) -- (10.6547, 
  10.3652) -- (10.6634, 10.3655) -- (10.6806, 10.3662) -- (10.7148, 10.3678) -- 
  (10.7484, 10.3695) -- (10.7816, 10.3713) -- (10.7906, 10.3718) -- (10.817, 
  10.3735) -- (10.8521, 10.3758) -- (10.8606, 10.3764) -- (10.8693, 10.3769) -- 
  (10.8864, 10.3781) -- (10.9201, 10.3808) -- (10.9285, 10.3813) -- (10.9369, 
  10.382) -- (10.9535, 10.3834) -- (10.9863, 10.3861) -- (11.0501, 10.3921) -- 
  (11.0579, 10.3928) -- (11.0658, 10.3936) -- (11.0812, 10.3951) -- (11.1118, 
  10.3983) -- (11.1716, 10.4047) -- (11.2849, 10.4187) -- (11.2916, 10.4195) -- 
  (11.2983, 10.4203) -- (11.3116, 10.4221) -- (11.3378, 10.4257) -- (11.3888, 
  10.4329) -- (11.4855, 10.4476) -- (11.4971, 10.4495) -- (11.5087, 10.4513) -- 
  (11.5314, 10.455) -- (11.5757, 10.4626) -- (11.6597, 10.4776) -- (11.6644, 
  10.4786) -- (11.6691, 10.4794) -- (11.6783, 10.4812) -- (11.6968, 10.4847) -- 
  (11.7328, 10.4917) -- (11.8011, 10.5055) -- (11.9247, 10.5326) -- (11.9286, 
  10.5334) -- (11.9326, 10.5344) -- (11.9403, 10.5362) -- (11.9554, 10.5396) -- 
  (11.9851, 10.5468) -- (12.0413, 10.5606) -- (12.1423, 10.5869) -- (12.296, 
  10.6316) -- (12.4117, 10.6695) -- (12.5076, 10.7046) -- (12.5761, 10.7324) -- 
  (12.6329, 10.7578) -- (12.6752, 10.7785) -- (12.7054, 10.7947) -- (12.7306, 
  10.8093) -- (12.7485, 10.8206) -- (12.7634, 10.8308) -- (12.7744, 10.839) -- 
  (12.7821, 10.8452) -- (12.7886, 10.8508) -- (12.7932, 10.8549) -- (12.7966, 
  10.8584) -- (12.7994, 10.8614) -- (12.8014, 10.8638) -- (12.8029, 10.8657) -- 
  (12.8042, 10.8673) -- (12.8056, 10.8696) -- (12.8064, 10.8709) -- (12.8065, 
  10.8715) -- (12.8068, 10.8719) -- (12.8069, 10.8722) -- (12.8069, 10.8724) -- 
  (12.8071, 10.8727) -- (12.8071, 10.8731) -- (12.8072, 10.8731) -- (12.8072, 
  10.8734);

  \path[draw=black,line cap=,line join=miter,line width=0.1058cm,miter 
  limit=3.25,shift={(-3.4925, 0.2646)}] (18.0658, 10.8737) -- (18.0655, 10.8737)
   -- (18.0606, 10.8753) -- (18.0495, 10.8789) -- (18.0103, 10.8916) -- (17.849,
   10.9425) -- (17.6113, 11.014) -- (17.3077, 11.0991) -- (16.9154, 11.1982) -- 
  (16.9094, 11.1997) -- (16.9033, 11.2011) -- (16.891, 11.204) -- (16.8664, 
  11.2097) -- (16.8166, 11.2211) -- (16.7156, 11.2436) -- (16.5079, 11.2867) -- 
  (16.5008, 11.2881) -- (16.4937, 11.2896) -- (16.4507, 11.2978) -- (16.3931, 
  11.3089) -- (16.277, 11.3298) -- (16.2624, 11.3323) -- (16.2478, 11.3349) -- 
  (16.2185, 11.3399) -- (16.16, 11.3495) -- (16.0426, 11.3679) -- (16.0353, 
  11.369) -- (16.0281, 11.3701) -- (16.0136, 11.3721) -- (15.9848, 11.3764) -- 
  (15.9271, 11.3845) -- (15.8117, 11.3995) -- (15.8046, 11.4004) -- (15.7974, 
  11.4013) -- (15.7831, 11.403) -- (15.7543, 11.4064) -- (15.697, 11.4129) -- 
  (15.583, 11.4246) -- (15.5698, 11.426) -- (15.5565, 11.4272) -- (15.5302, 
  11.4296) -- (15.4777, 11.4341) -- (15.4711, 11.4347) -- (15.4646, 11.4352) -- 
  (15.4515, 11.4362) -- (15.4253, 11.4383) -- (15.3735, 11.442) -- (15.3606, 
  11.4428) -- (15.3476, 11.4438) -- (15.3219, 11.4454) -- (15.3155, 11.4457) -- 
  (15.309, 11.4461) -- (15.2962, 11.4469) -- (15.2707, 11.4483) -- (15.2644, 
  11.4487) -- (15.2579, 11.449) -- (15.2452, 11.4497) -- (15.2199, 11.4509) -- 
  (15.2135, 11.4512) -- (15.2072, 11.4516) -- (15.1947, 11.4522) -- (15.1696, 
  11.4532) -- (15.1627, 11.4534) -- (15.1289, 11.4548) -- (15.1154, 11.4552) -- 
  (15.1087, 11.4555) -- (15.1019, 11.4556) -- (15.0886, 11.456) -- (15.0618, 
  11.4567) -- (15.0552, 11.4569) -- (15.0486, 11.4571) -- (15.0352, 11.4574) -- 
  (15.022, 11.4577) -- (15.0089, 11.4578) -- (14.9891, 11.4582) -- (14.9826, 
  11.4582) -- (14.976, 11.4584) -- (14.9695, 11.4584) -- (14.963, 11.4585) -- 
  (14.9564, 11.4585) -- (14.9499, 11.4587) -- (14.937, 11.4587) -- (14.9305, 
  11.4588) -- (14.8599, 11.4588) -- (14.8536, 11.4587) -- (14.8409, 11.4587) -- 
  (14.8346, 11.4585) -- (14.8282, 11.4585) -- (14.822, 11.4584) -- (14.8157, 
  11.4582) -- (14.8032, 11.4581) -- (14.797, 11.458) -- (14.7906, 11.458) -- 
  (14.7534, 11.4571) -- (14.7418, 11.4569) -- (14.7304, 11.4566) -- (14.7248, 
  11.4565) -- (14.719, 11.4563) -- (14.7077, 11.4559) -- (14.7019, 11.4558) -- 
  (14.6962, 11.4556) -- (14.6849, 11.4552) -- (14.6625, 11.4544) -- (14.6568, 
  11.4541) -- (14.6513, 11.454) -- (14.6401, 11.4534) -- (14.618, 11.4525) -- 
  (14.574, 11.4503) -- (14.5632, 11.4497) -- (14.5524, 11.449) -- (14.5309, 
  11.4478) -- (14.4883, 11.445) -- (14.4778, 11.4442) -- (14.4673, 11.4435) -- 
  (14.4465, 11.4419) -- (14.4053, 11.4385) -- (14.3954, 11.4377) -- (14.3855, 
  11.4368) -- (14.3658, 11.435) -- (14.3268, 11.4312) -- (14.322, 11.4307) -- 
  (14.3171, 11.4303) -- (14.3075, 11.4292) -- (14.2885, 11.4272) -- (14.2508, 
  11.423) -- (14.2462, 11.4224) -- (14.2416, 11.4219) -- (14.2322, 11.4208) -- 
  (14.2139, 11.4186) -- (14.1777, 11.4139) -- (14.1073, 11.4041) -- (14.0979, 
  11.4027) -- (14.0886, 11.4013) -- (14.0702, 11.3984) -- (14.0339, 11.3927) -- 
  (13.9639, 11.3805) -- (13.9597, 11.3798) -- (13.9554, 11.379) -- (13.9468, 
  11.3775) -- (13.93, 11.3743) -- (13.8971, 11.3679) -- (13.8333, 11.3548) -- 
  (13.8297, 11.3539) -- (13.826, 11.3532) -- (13.8188, 11.3516) -- (13.8046, 
  11.3484) -- (13.7765, 11.3421) -- (13.7225, 11.3293) -- (13.6218, 11.3031) -- 
  (13.6186, 11.3021) -- (13.6153, 11.3013) -- (13.5963, 11.2959) -- (13.5713, 
  11.2886) -- (13.5237, 11.2743) -- (13.4359, 11.2458) -- (13.4334, 11.2448) -- 
  (13.4309, 11.244) -- (13.4258, 11.2422) -- (13.4159, 11.2387) -- (13.3965, 
  11.2318) -- (13.3593, 11.2182) -- (13.2912, 11.1916) -- (13.1842, 11.1452) -- 
  (13.0934, 11.1) -- (13.0274, 11.0632) -- (12.9723, 11.0289) -- (12.9311, 
  11.0005) -- (12.9017, 10.9781) -- (12.8774, 10.958) -- (12.86, 10.9426) -- 
  (12.847, 10.9299) -- (12.8364, 10.9187) -- (12.829, 10.9102) -- (12.8229, 
  10.9026) -- (12.8185, 10.8967) -- (12.8155, 10.8923) -- (12.813, 10.8883) -- 
  (12.8113, 10.8854) -- (12.8101, 10.8831) -- (12.8091, 10.881) -- (12.8086, 
  10.8795) -- (12.808, 10.8782) -- (12.8078, 10.8773) -- (12.8075, 10.8764) -- 
  (12.8073, 10.8758) -- (12.8072, 10.8756) -- (12.8072, 10.8751) -- (12.8071, 
  10.8749) -- (12.8071, 10.8737);

  \path[fill=cff7f00,even odd rule] (6.7994, 6.8121) -- (6.9136, 6.6977) -- 
  (6.7994, 6.5834) -- (6.9136, 6.4692) -- (6.7994, 6.3548) -- (6.685, 6.4692) --
   (6.5708, 6.3548) -- (6.4564, 6.4692) -- (6.5708, 6.5834) -- (6.4564, 6.6977) 
  -- (6.5708, 6.8121) -- (6.685, 6.6977) -- cycle(6.7994, 6.8121);

  \path[fill=cff7f00,even odd rule] (6.7994, 15.9213) -- (6.9136, 15.8069) -- 
  (6.7994, 15.6927) -- (6.9136, 15.5783) -- (6.7994, 15.4641) -- (6.685, 
  15.5783) -- (6.5708, 15.4641) -- (6.4564, 15.5783) -- (6.5708, 15.6927) -- 
  (6.4564, 15.8069) -- (6.5708, 15.9213) -- (6.685, 15.8069) -- cycle(6.7994, 
  15.9213);

  \path[fill=cff7d00,even odd rule] (4.1697, 11.3666) -- (4.284, 11.2524) -- 
  (4.1697, 11.138) -- (4.284, 11.0238) -- (4.1697, 10.9094) -- (4.0554, 11.0238)
   -- (3.9412, 10.9094) -- (3.8268, 11.0238) -- (3.9412, 11.138) -- (3.8268, 
  11.2524) -- (3.9412, 11.3666) -- (4.0554, 11.2524) -- cycle(4.1697, 11.3666);

  \path[fill=cff7f00,even odd rule] (14.6881, 11.3666) -- (14.8025, 11.2524) -- 
  (14.6881, 11.138) -- (14.8025, 11.0238) -- (14.6881, 10.9094) -- (14.5739, 
  11.0238) -- (14.4595, 10.9094) -- (14.3452, 11.0238) -- (14.4595, 11.138) -- 
  (14.3452, 11.2524) -- (14.4595, 11.3666) -- (14.5739, 11.2524) -- 
  cycle(14.6881, 11.3666);

  \path[fill=blue,even odd rule] (4.1697, 11.138).. controls (4.1697, 11.1077) 
  and (4.1577, 11.0786) .. (4.1363, 11.0572).. controls (4.1148, 11.0357) and 
  (4.0857, 11.0238) .. (4.0554, 11.0238).. controls (4.0251, 11.0238) and 
  (3.996, 11.0357) .. (3.9747, 11.0572).. controls (3.9532, 11.0786) and 
  (3.9412, 11.1077) .. (3.9412, 11.138).. controls (3.9412, 11.1683) and 
  (3.9532, 11.1974) .. (3.9747, 11.2189).. controls (3.996, 11.2402) and 
  (4.0251, 11.2524) .. (4.0554, 11.2524).. controls (4.0857, 11.2524) and 
  (4.1148, 11.2402) .. (4.1363, 11.2189).. controls (4.1577, 11.1974) and 
  (4.1697, 11.1683) .. (4.1697, 11.138) -- cycle(4.1697, 11.138);

  \path[fill=blue,even odd rule] (14.6881, 11.138).. controls (14.6881, 11.1077)
   and (14.6761, 11.0786) .. (14.6546, 11.0572).. controls (14.6332, 11.0357) 
  and (14.6042, 11.0238) .. (14.5739, 11.0238).. controls (14.5435, 11.0238) and
   (14.5145, 11.0357) .. (14.493, 11.0572).. controls (14.4716, 11.0786) and 
  (14.4595, 11.1077) .. (14.4595, 11.138).. controls (14.4595, 11.1683) and 
  (14.4716, 11.1974) .. (14.493, 11.2189).. controls (14.5145, 11.2402) and 
  (14.5435, 11.2524) .. (14.5739, 11.2524).. controls (14.6042, 11.2524) and 
  (14.6332, 11.2402) .. (14.6546, 11.2189).. controls (14.6761, 11.1974) and 
  (14.6881, 11.1683) .. (14.6881, 11.138) -- cycle(14.6881, 11.138);

  \path[fill=blue,even odd rule] (9.429, 11.138).. controls (9.429, 11.1077) and
   (9.4168, 11.0786) .. (9.3955, 11.0572).. controls (9.374, 11.0357) and 
  (9.3449, 11.0238) .. (9.3146, 11.0238).. controls (9.2843, 11.0238) and 
  (9.2552, 11.0357) .. (9.2338, 11.0572).. controls (9.2123, 11.0786) and 
  (9.2003, 11.1077) .. (9.2003, 11.138).. controls (9.2003, 11.1683) and 
  (9.2123, 11.1974) .. (9.2338, 11.2189).. controls (9.2552, 11.2402) and 
  (9.2843, 11.2524) .. (9.3146, 11.2524).. controls (9.3449, 11.2524) and 
  (9.374, 11.2402) .. (9.3955, 11.2189).. controls (9.4168, 11.1974) and (9.429,
   11.1683) .. (9.429, 11.138) -- cycle(9.429, 11.138);

\end{tikzpicture}

%% file: figures/timelike-tikz-codes/timelike-neg-tikz-crossing-labeled.tex
\definecolor{cfffdff}{RGB}{255,253,255}
\definecolor{cff7f00}{RGB}{255,127,0}

\def \globalscale {1.000000}
\begin{tikzpicture}[y=1cm, x=1cm, yscale=\globalscale,xscale=\globalscale, every node/.append style={scale=\globalscale}, inner sep=0pt, outer sep=0pt]
  \path[fill=cfffdff,even odd rule] (0.0, 0.0353) -- (19.05, 0.0353) -- (19.05, 
  19.0853) -- (0.0, 19.0853) -- cycle;

\coordinate (wp) at (6.589, 13.402);
\coordinate (wm) at (6.489, 4.2928);
\coordinate (w0) at (3.9594, 8.7188);
\coordinate (w1) at (9.2185, 8.7188);
\coordinate (w2) at (14.4778, 8.7188);
\coordinate (ref) at ($ (w0) + (1,-1.5) $);

\fill[BrickRed] (ref) circle[radius=.1cm];
\draw[line width = 1.2,midarrow,draw=BrickRed] (ref) to[out=100,in=0] ($ (w0) + (-.7,1)$);
\draw[line width = 1.2,draw=BrickRed] ($ (w0) + (-.7,1)$) to[out=180,in=170] (ref);
\draw[line width = 1.2,midarrow=.6,draw=BrickRed] (ref) to[out=0,in=-90] ($ (w1) + (1,.5)$);
\draw[line width = 1.2,draw=BrickRed] ($ (w1) + (1,.5)$) to[out=90,in=30] (ref);
\node[text=BrickRed,font=\Large] at ($ (ref) + (0,-.5)$) {$u_{0}$};
\node[text=BrickRed,font=\Large] at ($ (w0) + (0,1.5)$) {$\mathcal{C}_{0}$};
\node[text=BrickRed,font=\Large] at ($ (w1) + (0,1.8)$) {$\mathcal{C}_{1}$};

  \draw[line width = 1.2,
  draw=orange,
        decorate,
        decoration={snake, amplitude=.5mm, segment length=2mm}]
    (w0) to[out=-180, in=120] (wm);
    \draw[line width = 1.2,
  draw=orange,
        decorate,
        decoration={snake, amplitude=.5mm, segment length=2mm}]
    (w2) to[out=0, in=0] (wp);

  \path[draw=black,line cap=,line join=miter,line width=0.1058cm,miter 
  limit=3.25,shift={(-3.4925, 0.2646)}] (9.9672, 3.8996) -- (10.0329, 4.0474) --
   (10.0706, 4.132) -- (10.129, 4.2634) -- (10.2183, 4.4644) -- (10.3521, 
  4.7657) -- (10.5471, 5.2036) -- (10.8205, 5.8102) -- (11.2132, 6.6388) -- 
  (11.4742, 7.1358) -- (11.6771, 7.4803) -- (11.856, 7.7495) -- (11.9927, 
  7.9323) -- (12.1138, 8.0772) -- (12.2102, 8.1807) -- (12.2836, 8.2518) -- 
  (12.3479, 8.3086) -- (12.3964, 8.3476) -- (12.435, 8.3761) -- (12.4688, 
  8.3987) -- (12.494, 8.4143) -- (12.5159, 8.4266) -- (12.533, 8.4352) -- 
  (12.5457, 8.441) -- (12.5567, 8.4456) -- (12.5648, 8.4486) -- (12.5718, 
  8.4508) -- (12.5772, 8.4523) -- (12.5774, 8.4523) -- (12.5774, 8.4525) -- 
  (12.5778, 8.4525) -- (12.5794, 8.4529) -- (12.5812, 8.4533) -- (12.5814, 
  8.4533) -- (12.5815, 8.4534) -- (12.5818, 8.4534) -- (12.5822, 8.4536) -- 
  (12.5831, 8.4537) -- (12.5847, 8.454) -- (12.5849, 8.454) -- (12.5851, 8.4541)
   -- (12.5853, 8.4541) -- (12.586, 8.4543) -- (12.5866, 8.4543) -- (12.5873, 
  8.4544) -- (12.5876, 8.4544) -- (12.5878, 8.4545) -- (12.5888, 8.4545) -- 
  (12.5892, 8.4547) -- (12.5903, 8.4547) -- (12.5904, 8.4548) -- (12.5949, 
  8.4548) -- (12.5949, 8.4547) -- (12.5957, 8.4547) -- (12.5957, 8.4545) -- 
  (12.5961, 8.4545) -- (12.5962, 8.4544) -- (12.5966, 8.4544) -- (12.5966, 
  8.4543) -- (12.5968, 8.4543);

  \path[draw=black,line cap=,line join=miter,line width=0.1058cm,miter 
  limit=3.25,shift={(-3.4925, 0.2646)}] (9.9672, 3.8996) -- (10.0663, 3.767) -- 
  (10.1269, 3.6894) -- (10.2271, 3.5668) -- (10.3955, 3.3741) -- (10.4344, 
  3.332) -- (10.4723, 3.2916) -- (10.546, 3.2148) -- (10.6875, 3.0748) -- 
  (10.722, 3.042) -- (10.7561, 3.01) -- (10.8237, 2.9482) -- (10.9562, 2.8317) 
  -- (11.2145, 2.6228) -- (11.2463, 2.5986) -- (11.2781, 2.5747) -- (11.3414, 
  2.5281) -- (11.4672, 2.4387) -- (11.7166, 2.2746) -- (11.7476, 2.2553) -- 
  (11.7786, 2.2363) -- (11.8407, 2.1989) -- (11.9641, 2.1271) -- (12.2104, 
  1.9947) -- (12.2437, 1.9778) -- (12.2769, 1.9611) -- (12.3435, 1.9286) -- 
  (12.4763, 1.8661) -- (12.5096, 1.8511) -- (12.5428, 1.8362) -- (12.609, 
  1.8073) -- (12.7416, 1.752) -- (12.7748, 1.7387) -- (12.8079, 1.7256) -- 
  (12.874, 1.6999) -- (13.0063, 1.6513) -- (13.2704, 1.5631) -- (13.3033, 
  1.5529) -- (13.3362, 1.543) -- (13.4021, 1.5236) -- (13.5337, 1.4869) -- 
  (13.5665, 1.4782) -- (13.5994, 1.4697) -- (13.665, 1.4531) -- (13.7962, 
  1.4221) -- (13.829, 1.4148) -- (13.8617, 1.4077) -- (13.9271, 1.3939) -- 
  (14.0579, 1.3684) -- (14.0906, 1.3623) -- (14.1231, 1.3565) -- (14.1883, 
  1.3454) -- (14.3185, 1.3251) -- (14.3488, 1.3207) -- (14.3791, 1.3166) -- 
  (14.4398, 1.3084) -- (14.47, 1.3046) -- (14.5003, 1.301) -- (14.5606, 1.294) 
  -- (14.621, 1.2876) -- (14.6813, 1.2817) -- (14.7114, 1.279) -- (14.7414, 
  1.2763) -- (14.8016, 1.2715) -- (14.8315, 1.2693) -- (14.8616, 1.2672) -- 
  (14.8916, 1.2653) -- (14.9215, 1.2635) -- (14.9813, 1.2602) -- (15.0112, 
  1.2588) -- (15.0411, 1.2576) -- (15.0709, 1.2564) -- (15.1304, 1.2544) -- 
  (15.1602, 1.2536) -- (15.19, 1.2529) -- (15.2196, 1.2524) -- (15.2494, 1.2519)
   -- (15.279, 1.2517) -- (15.3086, 1.2515) -- (15.3383, 1.2515) -- (15.3972, 
  1.2518) -- (15.4269, 1.2522) -- (15.4562, 1.2526) -- (15.4857, 1.2532) -- 
  (15.5152, 1.2539) -- (15.5445, 1.2547) -- (15.5739, 1.2557) -- (15.6033, 
  1.2568) -- (15.6325, 1.258) -- (15.6618, 1.2593) -- (15.691, 1.2606) -- 
  (15.7493, 1.2639) -- (15.7785, 1.2657) -- (15.8076, 1.2677) -- (15.8656, 
  1.2718) -- (15.8947, 1.274) -- (15.9236, 1.2765) -- (15.9815, 1.2816) -- 
  (16.0103, 1.2842) -- (16.0391, 1.2871) -- (16.0967, 1.2932) -- (16.2114, 
  1.3065) -- (16.2394, 1.3101) -- (16.2673, 1.3138) -- (16.3231, 1.3215) -- 
  (16.4343, 1.3382) -- (16.462, 1.3426) -- (16.4896, 1.3472) -- (16.5449, 
  1.3565) -- (16.6547, 1.3765) -- (16.6821, 1.3819) -- (16.7094, 1.3873) -- 
  (16.764, 1.3983) -- (16.8726, 1.4216) -- (17.0875, 1.4731) -- (17.1141, 1.48) 
  -- (17.1409, 1.487) -- (17.1939, 1.5014) -- (17.2995, 1.531) -- (17.5084, 
  1.5951) -- (17.5343, 1.6035) -- (17.5601, 1.612) -- (17.6116, 1.6294) -- 
  (17.714, 1.6651) -- (17.9163, 1.741) -- (17.9434, 1.7518) -- (17.9704, 1.7625)
   -- (18.0245, 1.7846) -- (18.1315, 1.8298) -- (18.3425, 1.925) -- (18.3687, 
  1.9372) -- (18.3946, 1.9496) -- (18.4464, 1.9749) -- (18.5491, 2.0264) -- 
  (18.7508, 2.134) -- (18.7758, 2.1478) -- (18.8006, 2.1617) -- (18.8499, 2.19) 
  -- (18.9479, 2.2474) -- (19.1398, 2.3665) -- (19.5083, 2.6208) -- (19.5291, 
  2.6362) -- (19.55, 2.6518) -- (19.5913, 2.683) -- (19.673, 2.7463) -- 
  (19.8329, 2.8761) -- (19.8526, 2.8926) -- (19.8721, 2.9093) -- (19.911, 
  2.9425) -- (19.9879, 3.01) -- (20.1378, 3.1477) -- (20.1563, 3.1651) -- 
  (20.1746, 3.1827) -- (20.211, 3.2179) -- (20.2828, 3.289) -- (20.4225, 3.4338)
   -- (20.6864, 3.7332) -- (20.7035, 3.7539) -- (20.7206, 3.7747) -- (20.7542, 
  3.8165) -- (20.8205, 3.9005) -- (20.9482, 4.071) -- (20.9637, 4.0925) -- 
  (20.9791, 4.1141) -- (21.0097, 4.1574) -- (21.0695, 4.2445) -- (21.1843, 
  4.4207) -- (21.1982, 4.4429) -- (21.212, 4.4653) -- (21.2392, 4.5099) -- 
  (21.2925, 4.5996) -- (21.3942, 4.7808) -- (21.5776, 5.1492) -- (21.5983, 
  5.1949) -- (21.6185, 5.2407) -- (21.6578, 5.3326) -- (21.7315, 5.5173) -- 
  (21.7403, 5.5404) -- (21.749, 5.5637) -- (21.766, 5.6101) -- (21.7989, 5.7031)
   -- (21.8597, 5.89) -- (21.8668, 5.9134) -- (21.8739, 5.9369) -- (21.8877, 
  5.9837) -- (21.9141, 6.0777) -- (21.9205, 6.1011) -- (21.9267, 6.1247) -- 
  (21.9388, 6.1717) -- (21.9619, 6.2658) -- (21.9676, 6.2894) -- (21.973, 
  6.3129) -- (21.9836, 6.3601) -- (22.0034, 6.4543) -- (22.0082, 6.4779) -- 
  (22.0128, 6.5014) -- (22.0217, 6.5486) -- (22.0384, 6.6428) -- (22.0421, 
  6.6649) -- (22.0457, 6.6869) -- (22.0526, 6.7309) -- (22.0654, 6.8188) -- 
  (22.0683, 6.8407) -- (22.0712, 6.8628) -- (22.0767, 6.9066) -- (22.0868, 
  6.9944) -- (22.089, 7.0164) -- (22.0912, 7.0383) -- (22.0974, 7.1041) -- 
  (22.0992, 7.126) -- (22.1026, 7.1697) -- (22.1043, 7.1916) -- (22.1058, 
  7.2135) -- (22.1072, 7.2353) -- (22.1086, 7.2572) -- (22.1098, 7.2789) -- 
  (22.1109, 7.3008) -- (22.1131, 7.3444) -- (22.1139, 7.3662) -- (22.1148, 
  7.3881) -- (22.1156, 7.4098) -- (22.1161, 7.4316) -- (22.1168, 7.4533) -- 
  (22.1177, 7.4968) -- (22.1181, 7.5184) -- (22.1182, 7.5402) -- (22.1185, 
  7.5618) -- (22.1185, 7.6053) -- (22.1182, 7.6485) -- (22.1179, 7.67) -- 
  (22.1177, 7.6917) -- (22.1172, 7.7133) -- (22.1167, 7.7348) -- (22.1161, 
  7.7564) -- (22.1154, 7.7779) -- (22.1138, 7.8209) -- (22.1119, 7.8639) -- 
  (22.1108, 7.8853) -- (22.1097, 7.9068) -- (22.107, 7.9495) -- (22.1055, 7.971)
   -- (22.1041, 7.9922) -- (22.1008, 8.0349) -- (22.0989, 8.0581) -- (22.097, 
  8.0811) -- (22.0927, 8.1271) -- (22.088, 8.1731) -- (22.0831, 8.2189) -- 
  (22.0775, 8.2646) -- (22.0718, 8.3103) -- (22.0591, 8.4012) -- (22.0556, 
  8.4239) -- (22.0522, 8.4465) -- (22.0449, 8.4917) -- (22.0292, 8.5817) -- 
  (22.025, 8.6041) -- (22.0208, 8.6265) -- (22.0121, 8.6712) -- (21.9935, 
  8.7601) -- (21.9885, 8.7822) -- (21.9836, 8.8043) -- (21.9734, 8.8484) -- 
  (21.9519, 8.9362) -- (21.9464, 8.9581) -- (21.9407, 8.9798) -- (21.929, 
  9.0234) -- (21.9047, 9.1099) -- (21.8984, 9.1316) -- (21.8921, 9.1531) -- 
  (21.879, 9.1959) -- (21.852, 9.2812) -- (21.7937, 9.4496) -- (21.7865, 9.4692)
   -- (21.7792, 9.4886) -- (21.7646, 9.5273) -- (21.7344, 9.6044) -- (21.6706, 
  9.7565) -- (21.6623, 9.7753) -- (21.6539, 9.7941) -- (21.637, 9.8315) -- 
  (21.6024, 9.9059) -- (21.5298, 10.0524) -- (21.5204, 10.0704) -- (21.5109, 
  10.0885) -- (21.4919, 10.1244) -- (21.4529, 10.1958) -- (21.3719, 10.3363) -- 
  (21.1978, 10.6076) -- (21.1866, 10.6238) -- (21.1755, 10.64) -- (21.1529, 
  10.6721) -- (21.107, 10.7357) -- (21.0125, 10.8605) -- (21.0005, 10.8759) -- 
  (20.9883, 10.8912) -- (20.964, 10.9216) -- (20.9146, 10.9819) -- (20.8135, 
  11.0997) -- (20.8007, 11.1142) -- (20.7877, 11.1286) -- (20.7618, 11.1573) -- 
  (20.7092, 11.2141) -- (20.6018, 11.3247) -- (20.3784, 11.5349) -- (20.3629, 
  11.5486) -- (20.3473, 11.5623) -- (20.3159, 11.5893) -- (20.2526, 11.6425) -- 
  (20.1239, 11.7456) -- (20.1076, 11.7582) -- (20.0584, 11.7955) -- (19.9924, 
  11.8441) -- (19.8582, 11.9378) -- (19.8413, 11.9492) -- (19.8243, 11.9605) -- 
  (19.7901, 11.9829) -- (19.7215, 12.0268) -- (19.5826, 12.1109) -- (19.2986, 
  12.2647) -- (19.2818, 12.2731) -- (19.265, 12.2814) -- (19.2312, 12.2978) -- 
  (19.1636, 12.3297) -- (19.0271, 12.3904) -- (19.0099, 12.3977) -- (18.9928, 
  12.405) -- (18.9584, 12.4192) -- (18.8895, 12.4467) -- (18.7507, 12.4988) -- 
  (18.7332, 12.505) -- (18.7158, 12.5111) -- (18.681, 12.5232) -- (18.611, 
  12.5465) -- (18.4704, 12.5899) -- (18.4528, 12.595) -- (18.4353, 12.6) -- 
  (18.4, 12.6099) -- (18.3293, 12.6288) -- (18.1876, 12.6635) -- (18.1683, 
  12.6678) -- (18.1492, 12.672) -- (18.1107, 12.6804) -- (18.0337, 12.6961) -- 
  (18.0144, 12.6999) -- (17.9951, 12.7034) -- (17.9567, 12.7105) -- (17.8795, 
  12.7237) -- (17.8409, 12.7298) -- (17.8025, 12.7356) -- (17.7832, 12.7383) -- 
  (17.7253, 12.7462) -- (17.706, 12.7486) -- (17.6868, 12.751) -- (17.6483, 
  12.7555) -- (17.5714, 12.7637) -- (17.5521, 12.7655) -- (17.5329, 12.7672) -- 
  (17.4945, 12.7704) -- (17.4752, 12.7719) -- (17.456, 12.7734) -- (17.4177, 
  12.7761) -- (17.3984, 12.7773) -- (17.3601, 12.7795) -- (17.3411, 12.7805) -- 
  (17.2836, 12.783) -- (17.2646, 12.7836) -- (17.2455, 12.7842) -- (17.2264, 
  12.7847) -- (17.2073, 12.7852) -- (17.1883, 12.7856) -- (17.1693, 12.7859) -- 
  (17.1122, 12.7863) -- (17.0932, 12.7863) -- (17.0743, 12.7861) -- (17.0553, 
  12.786) -- (17.0175, 12.7854) -- (16.9985, 12.785) -- (16.9796, 12.7846) -- 
  (16.9608, 12.7841) -- (16.9054, 12.782) -- (16.8684, 12.7803) -- (16.8501, 
  12.7794) -- (16.8133, 12.7772) -- (16.795, 12.7761) -- (16.7767, 12.7748) -- 
  (16.74, 12.7721) -- (16.7217, 12.7706) -- (16.7035, 12.769) -- (16.6671, 
  12.7659) -- (16.6307, 12.7623) -- (16.5945, 12.7584) -- (16.5764, 12.7565) -- 
  (16.5582, 12.7544) -- (16.5223, 12.75) -- (16.5042, 12.7477) -- (16.4863, 
  12.7453) -- (16.4505, 12.7405) -- (16.3791, 12.7299) -- (16.3435, 12.7241) -- 
  (16.3083, 12.7182) -- (16.2378, 12.7054) -- (16.2202, 12.7021) -- (16.2027, 
  12.6986) -- (16.1678, 12.6916) -- (16.0985, 12.6767) -- (16.0812, 12.6729) -- 
  (16.0639, 12.669) -- (16.0296, 12.6609) -- (15.9613, 12.6441) -- (15.8265, 
  12.6074) -- (15.8109, 12.6028) -- (15.7955, 12.5983) -- (15.7645, 12.5889) -- 
  (15.703, 12.5698) -- (15.582, 12.529) -- (15.567, 12.5236) -- (15.5521, 
  12.5182) -- (15.5224, 12.5074) -- (15.4634, 12.485) -- (15.3473, 12.4382) -- 
  (15.3187, 12.426) -- (15.2903, 12.4136) -- (15.2339, 12.3884) -- (15.1234, 
  12.3361) -- (14.9115, 12.224) -- (14.8976, 12.216) -- (14.8836, 12.2082) -- 
  (14.8561, 12.1921) -- (14.8016, 12.1595) -- (14.6957, 12.0927) -- (14.6698, 
  12.0756) -- (14.6441, 12.0584) -- (14.5936, 12.0237) -- (14.4954, 11.953) -- 
  (14.4835, 11.944) -- (14.4716, 11.9349) -- (14.4479, 11.9169) -- (14.4013, 
  11.8805) -- (14.3113, 11.8068) -- (14.1434, 11.6564) -- (14.1341, 11.6475) -- 
  (14.1249, 11.6386) -- (14.1066, 11.6209) -- (14.0706, 11.5852) -- (14.0011, 
  11.5138) -- (13.8726, 11.3709) -- (13.865, 11.3619) -- (13.8574, 11.3531) -- 
  (13.8425, 11.3352) -- (13.8132, 11.2998) -- (13.757, 11.2289) -- (13.6536, 
  11.0893) -- (13.647, 11.08) -- (13.6405, 11.0706) -- (13.6276, 11.052) -- 
  (13.6023, 11.0149) -- (13.5541, 10.9418) -- (13.4667, 10.799) -- (13.4615, 
  10.7903) -- (13.4565, 10.7816) -- (13.4465, 10.7643) -- (13.4269, 10.7298) -- 
  (13.3897, 10.662) -- (13.3222, 10.5311) -- (13.3145, 10.5154) -- (13.3069, 
  10.4998) -- (13.2921, 10.4691) -- (13.2638, 10.4086) -- (13.2122, 10.2924) -- 
  (13.1261, 10.0784) -- (13.1239, 10.0726) -- (13.1217, 10.0667) -- (13.1173, 
  10.0551) -- (13.1087, 10.032) -- (13.0922, 9.9868) -- (13.0617, 9.9) -- 
  (13.0092, 9.7402) -- (12.9232, 9.4504) -- (12.8655, 9.2381) -- (12.822, 9.073)
   -- (12.7842, 8.9323) -- (12.7553, 8.8292) -- (12.7291, 8.742) -- (12.7074, 
  8.6756) -- (12.6902, 8.6271) -- (12.6742, 8.5861) -- (12.6617, 8.5565) -- 
  (12.6512, 8.5338) -- (12.6414, 8.5146) -- (12.6339, 8.5008) -- (12.627, 
  8.4893) -- (12.6217, 8.481) -- (12.6173, 8.4748) -- (12.6133, 8.4696) -- 
  (12.6103, 8.466) -- (12.6077, 8.4628) -- (12.604, 8.4591) -- (12.6024, 8.4577)
   -- (12.6013, 8.4569) -- (12.6005, 8.4562) -- (12.5997, 8.4556) -- (12.5991, 
  8.4552) -- (12.599, 8.4551) -- (12.5989, 8.4551) -- (12.5987, 8.455) -- 
  (12.5986, 8.455);

  \path[draw=black,line cap=,line join=miter,line width=0.1058cm,miter 
  limit=3.25,shift={(-3.4925, 0.2646)}] (9.9672, 3.8996) -- (9.8023, 3.8846) -- 
  (9.7041, 3.8778) -- (9.621, 3.8733) -- (9.5462, 3.8702) -- (9.477, 3.8683) -- 
  (9.412, 3.8671) -- (9.3501, 3.8666) -- (9.2909, 3.8668) -- (9.2338, 3.8675) --
   (9.1786, 3.8687) -- (9.1248, 3.8704) -- (9.0725, 3.8724) -- (9.0215, 3.8748) 
  -- (8.9714, 3.8777) -- (8.9225, 3.8808) -- (8.8746, 3.8843) -- (8.8274, 3.888)
   -- (8.781, 3.8921) -- (8.7354, 3.8965) -- (8.6903, 3.9012) -- (8.6459, 
  3.9061) -- (8.6023, 3.9113) -- (8.5163, 3.9225) -- (8.4741, 3.9284) -- 
  (8.4325, 3.9344) -- (8.3504, 3.9474) -- (8.1909, 3.9759) -- (8.1521, 3.9836) 
  -- (8.1135, 3.9915) -- (8.0373, 4.0078) -- (7.8884, 4.0425) -- (7.8519, 
  4.0516) -- (7.8157, 4.0609) -- (7.7439, 4.0801) -- (7.6033, 4.1205) -- 
  (7.5687, 4.1309) -- (7.5344, 4.1416) -- (7.4663, 4.1633) -- (7.3328, 4.2088) 
  -- (7.075, 4.3069) -- (7.0409, 4.321) -- (7.007, 4.3352) -- (6.9399, 4.364) --
   (6.8081, 4.4236) -- (6.5535, 4.5504) -- (6.5225, 4.567) -- (6.4917, 4.5836) 
  -- (6.4306, 4.6175) -- (6.3105, 4.6868) -- (6.0784, 4.8325) -- (6.0501, 
  4.8514) -- (6.022, 4.8704) -- (5.9664, 4.9087) -- (5.8571, 4.987) -- (5.6461, 
  5.15) -- (5.6205, 5.171) -- (5.595, 5.192) -- (5.5445, 5.2346) -- (5.4456, 
  5.3212) -- (5.2553, 5.5) -- (5.2338, 5.5214) -- (5.2124, 5.5429) -- (5.17, 
  5.5862) -- (5.087, 5.6739) -- (4.9277, 5.8542) -- (4.9084, 5.8771) -- (4.8893,
   5.9002) -- (4.8514, 5.9467) -- (4.7774, 6.0405) -- (4.6363, 6.2327) -- 
  (4.6193, 6.2571) -- (4.6024, 6.2817) -- (4.5692, 6.331) -- (4.5044, 6.4306) --
   (4.3817, 6.6339) -- (4.367, 6.6596) -- (4.3525, 6.6854) -- (4.3239, 6.7375) 
  -- (4.2685, 6.8424) -- (4.1647, 7.0558) -- (4.1526, 7.0823) -- (4.1407, 
  7.1089) -- (4.1173, 7.1622) -- (4.0722, 7.2697) -- (4.0614, 7.2967) -- 
  (4.0506, 7.3239) -- (4.0295, 7.3783) -- (3.9893, 7.4878) -- (3.9795, 7.5155) 
  -- (3.97, 7.5431) -- (3.9512, 7.5986) -- (3.9156, 7.7103) -- (3.8987, 7.7665) 
  -- (3.8823, 7.823) -- (3.8516, 7.9367) -- (3.8443, 7.9652) -- (3.8371, 7.9937)
   -- (3.8232, 8.0512) -- (3.7972, 8.1667) -- (3.7911, 8.1956) -- (3.7852, 
  8.2247) -- (3.7738, 8.283) -- (3.7527, 8.4001) -- (3.7477, 8.4296) -- (3.743, 
  8.4591) -- (3.7341, 8.5181) -- (3.7298, 8.5477) -- (3.7257, 8.5775) -- 
  (3.7179, 8.637) -- (3.7144, 8.6668) -- (3.7109, 8.6967) -- (3.7044, 8.7565) --
   (3.7014, 8.7865) -- (3.6985, 8.8166) -- (3.6933, 8.8768) -- (3.6883, 8.9424) 
  -- (3.6861, 8.9752) -- (3.6842, 9.0081) -- (3.6824, 9.041) -- (3.6807, 9.074) 
  -- (3.6792, 9.107) -- (3.678, 9.1401) -- (3.6769, 9.1733) -- (3.6761, 9.2065) 
  -- (3.6754, 9.2397) -- (3.6748, 9.273) -- (3.6745, 9.3063) -- (3.6744, 9.3397)
   -- (3.6747, 9.4066) -- (3.6758, 9.4736) -- (3.6766, 9.5072) -- (3.6788, 
  9.5745) -- (3.6818, 9.642) -- (3.6836, 9.6758) -- (3.6878, 9.7436) -- (3.6901,
   9.7775) -- (3.6927, 9.8114) -- (3.6955, 9.8454) -- (3.6984, 9.8793) -- 
  (3.7048, 9.9475) -- (3.7084, 9.9817) -- (3.7122, 10.0159) -- (3.7201, 10.0842)
   -- (3.7244, 10.1185) -- (3.729, 10.1528) -- (3.7386, 10.2215) -- (3.7437, 
  10.2559) -- (3.7489, 10.2902) -- (3.7601, 10.3593) -- (3.7849, 10.4973) -- 
  (3.7915, 10.5321) -- (3.7984, 10.5667) -- (3.8127, 10.636) -- (3.8439, 10.775)
   -- (3.8522, 10.8099) -- (3.8606, 10.8446) -- (3.8781, 10.9143) -- (3.9157, 
  11.0541) -- (3.9249, 11.0867) -- (3.9343, 11.1194) -- (3.9536, 11.1847) -- 
  (3.9944, 11.3156) -- (4.0844, 11.5781) -- (4.0965, 11.6109) -- (4.1087, 
  11.6439) -- (4.1337, 11.7096) -- (4.1859, 11.8412) -- (4.2991, 12.105) -- 
  (4.3139, 12.1379) -- (4.3291, 12.1708) -- (4.3598, 12.2368) -- (4.4236, 
  12.369) -- (4.5601, 12.6334) -- (4.8679, 13.1627) -- (4.8905, 13.1987) -- 
  (4.9132, 13.2345) -- (4.9594, 13.3063) -- (5.0544, 13.4499) -- (5.2545, 
  13.7373) -- (5.6951, 14.3127) -- (5.7245, 14.3487) -- (5.754, 14.3848) -- 
  (5.8135, 14.4569) -- (5.9351, 14.6011) -- (6.1875, 14.8903) -- (6.7265, 
  15.4705) -- (6.7609, 15.5062) -- (6.7955, 15.5419) -- (6.8651, 15.6133) -- 
  (7.0059, 15.7562) -- (7.2939, 16.0419) -- (7.8913, 16.6102) -- (7.9295, 
  16.6453) -- (7.9678, 16.6806) -- (8.0446, 16.7509) -- (8.1992, 16.8908) -- 
  (8.5125, 17.1677) -- (9.1543, 17.7059) -- (9.1923, 17.7365) -- (9.2305, 
  17.767) -- (9.307, 17.8277) -- (9.4606, 17.948) -- (9.7712, 18.1835) -- 
  (10.4061, 18.6325) -- (10.4463, 18.6596) -- (10.4867, 18.6865) -- (10.5676, 
  18.7398);

  \path[draw=black,line cap=,line join=miter,line width=0.1058cm,miter 
  limit=3.25,shift={(-3.4925, 0.2646)}] (9.9672, 13.0088) -- (10.0365, 13.1597) 
  -- (10.0795, 13.249) -- (10.1516, 13.3919) -- (10.2758, 13.6211) -- (10.3048, 
  13.6719) -- (10.3334, 13.7211) -- (10.3894, 13.8151) -- (10.4987, 13.9896) -- 
  (10.5256, 14.0309) -- (10.5523, 14.0716) -- (10.6057, 14.1507) -- (10.7118, 
  14.3018) -- (10.9233, 14.5816) -- (10.9498, 14.6148) -- (10.9763, 14.6477) -- 
  (11.0295, 14.7126) -- (11.1363, 14.8386) -- (11.3526, 15.0777) -- (11.7963, 
  15.5129) -- (11.827, 15.5406) -- (11.8577, 15.5681) -- (11.9196, 15.6224) -- 
  (12.0441, 15.7285) -- (12.2972, 15.9315) -- (12.3292, 15.956) -- (12.3613, 
  15.9803) -- (12.4256, 16.0284) -- (12.5554, 16.1225) -- (12.8191, 16.3021) -- 
  (12.8524, 16.3237) -- (12.8858, 16.3452) -- (12.9527, 16.3876) -- (13.0878, 
  16.4706) -- (13.3615, 16.6284) -- (13.396, 16.6474) -- (13.4306, 16.6661) -- 
  (13.5001, 16.7033) -- (13.64, 16.7757) -- (13.9231, 16.9124) -- (13.9565, 
  16.9277) -- (13.9898, 16.9428) -- (14.0568, 16.9727) -- (14.1916, 17.0308) -- 
  (14.4638, 17.14) -- (14.4979, 17.1531) -- (14.5322, 17.1661) -- (14.6011, 
  17.1914) -- (14.7394, 17.2405) -- (15.0183, 17.3317) -- (15.0534, 17.3426) -- 
  (15.0886, 17.3532) -- (15.159, 17.374) -- (15.3004, 17.4141) -- (15.3359, 
  17.4238) -- (15.3713, 17.4333) -- (15.4426, 17.4519) -- (15.5853, 17.4874) -- 
  (15.6212, 17.496) -- (15.657, 17.5043) -- (15.7289, 17.5207) -- (15.8731, 
  17.5517) -- (15.9092, 17.559) -- (15.9454, 17.5663) -- (16.0178, 17.5803) -- 
  (16.1631, 17.6068) -- (16.1988, 17.6128) -- (16.2347, 17.6188) -- (16.3062, 
  17.6303) -- (16.4498, 17.6518) -- (16.4857, 17.6569) -- (16.5217, 17.6618) -- 
  (16.5938, 17.6711) -- (16.7382, 17.6882) -- (16.8107, 17.6959) -- (16.8832, 
  17.7031) -- (16.9194, 17.7064) -- (16.9558, 17.7096) -- (17.0284, 17.7156) -- 
  (17.0648, 17.7185) -- (17.1013, 17.7212) -- (17.1741, 17.7261) -- (17.2106, 
  17.7283) -- (17.2471, 17.7304) -- (17.3201, 17.7342) -- (17.3567, 17.7359) -- 
  (17.3933, 17.7374) -- (17.4298, 17.7389) -- (17.4665, 17.7402) -- (17.503, 
  17.7413) -- (17.5763, 17.7432) -- (17.613, 17.7439) -- (17.6498, 17.7444) -- 
  (17.6864, 17.7449) -- (17.7231, 17.7451) -- (17.7599, 17.7454) -- (17.7967, 
  17.7454) -- (17.8333, 17.7453) -- (17.8701, 17.745) -- (17.9069, 17.7446) -- 
  (17.9437, 17.744) -- (17.9805, 17.7433) -- (18.0173, 17.7425) -- (18.0541, 
  17.7416) -- (18.091, 17.7405) -- (18.1278, 17.7392) -- (18.1647, 17.7378) -- 
  (18.2015, 17.7363) -- (18.2385, 17.7345) -- (18.2753, 17.7327) -- (18.3122, 
  17.7308) -- (18.3491, 17.7287) -- (18.3859, 17.7265) -- (18.4229, 17.7241) -- 
  (18.4967, 17.719) -- (18.5367, 17.7159) -- (18.5768, 17.7126) -- (18.6568, 
  17.7057) -- (18.6969, 17.702) -- (18.737, 17.6981) -- (18.8171, 17.6899) -- 
  (18.8572, 17.6856) -- (18.8972, 17.6811) -- (18.9774, 17.6714) -- (19.1375, 
  17.6502) -- (19.1776, 17.6445) -- (19.2176, 17.6386) -- (19.2976, 17.6265) -- 
  (19.4576, 17.5999) -- (19.4976, 17.5929) -- (19.5375, 17.5857) -- (19.6173, 
  17.5707) -- (19.7771, 17.5388) -- (19.8169, 17.5306) -- (19.8567, 17.522) -- 
  (19.9364, 17.5044) -- (20.0954, 17.4672) -- (20.1351, 17.4575) -- (20.1748, 
  17.4476) -- (20.2541, 17.4274) -- (20.4125, 17.3848) -- (20.452, 17.3738) -- 
  (20.4916, 17.3626) -- (20.5704, 17.3397) -- (20.7279, 17.2918) -- (21.0413, 
  17.1881) -- (21.0778, 17.1753) -- (21.1142, 17.1625) -- (21.1868, 17.1362) -- 
  (21.3318, 17.0819) -- (21.6198, 16.9664) -- (21.6914, 16.9361) -- (21.7629, 
  16.9052) -- (21.9053, 16.8417) -- (22.1879, 16.708) -- (22.2231, 16.6907) -- 
  (22.2582, 16.6732) -- (22.3282, 16.6377);

  \path[draw=black,line cap=,line join=miter,line width=0.1058cm,miter 
  limit=3.25,shift={(-3.4925, 0.2646)}] (9.9672, 13.0088) -- (10.0619, 12.8776) 
  -- (10.1156, 12.8021) -- (10.1987, 12.6846) -- (10.3244, 12.5035) -- (10.51, 
  12.2297) -- (10.7734, 11.827) -- (11.1267, 11.2606) -- (11.5914, 10.4725) -- 
  (11.8644, 9.9911) -- (12.0537, 9.6497) -- (12.2032, 9.3744) -- (12.3062, 
  9.1797) -- (12.3891, 9.0173) -- (12.4488, 8.8947) -- (12.4903, 8.8051) -- 
  (12.5232, 8.7286) -- (12.5458, 8.6724) -- (12.5619, 8.6284) -- (12.5746, 
  8.5907) -- (12.5829, 8.5627) -- (12.5892, 8.5387) -- (12.5893, 8.5385) -- 
  (12.5893, 8.5382) -- (12.5895, 8.5375) -- (12.5899, 8.5363) -- (12.5904, 
  8.5338) -- (12.5915, 8.529) -- (12.5933, 8.5204) -- (12.5933, 8.5201) -- 
  (12.5935, 8.5199) -- (12.5935, 8.5194) -- (12.5938, 8.5185) -- (12.594, 
  8.5167) -- (12.5947, 8.5131) -- (12.596, 8.5068) -- (12.596, 8.5064) -- 
  (12.5961, 8.5059) -- (12.5962, 8.5051) -- (12.5965, 8.5036) -- (12.5969, 
  8.5006) -- (12.5969, 8.5002) -- (12.5971, 8.4999) -- (12.5971, 8.4992) -- 
  (12.5973, 8.4977) -- (12.5977, 8.4951) -- (12.5977, 8.4945) -- (12.5979, 
  8.494) -- (12.598, 8.4927) -- (12.5983, 8.4905) -- (12.5983, 8.4895) -- 
  (12.5984, 8.4884) -- (12.5987, 8.4865) -- (12.5987, 8.4855) -- (12.5989, 
  8.4844) -- (12.5989, 8.484) -- (12.599, 8.4835) -- (12.599, 8.4821) -- 
  (12.5991, 8.4817) -- (12.5991, 8.4809) -- (12.5993, 8.4792) -- (12.5993, 
  8.4773) -- (12.5994, 8.4769) -- (12.5994, 8.4665) -- (12.5993, 8.4664) -- 
  (12.5993, 8.4643) -- (12.5991, 8.464) -- (12.5991, 8.4629) -- (12.599, 8.4624)
   -- (12.599, 8.462) -- (12.5989, 8.4614) -- (12.5989, 8.461) -- (12.5987, 
  8.4606) -- (12.5987, 8.4598) -- (12.5986, 8.4598) -- (12.5986, 8.4591) -- 
  (12.5984, 8.4585) -- (12.5984, 8.4584) -- (12.5983, 8.4583) -- (12.5983, 
  8.458) -- (12.5982, 8.4576) -- (12.598, 8.4567) -- (12.5977, 8.4562) -- 
  (12.5975, 8.4554) -- (12.5973, 8.4551) -- (12.5973, 8.455) -- (12.5971, 
  8.4547) -- (12.5971, 8.4544) -- (12.5969, 8.4544) -- (12.5969, 8.4543) -- 
  (12.5968, 8.4543);

  \path[draw=black,line cap=,line join=miter,line width=0.1058cm,miter 
  limit=3.25,shift={(-3.4925, 0.2646)}] (9.9672, 13.0088) -- (9.8034, 12.9891) 
  -- (9.7066, 12.9751) -- (9.6255, 12.962) -- (9.5528, 12.9491) -- (9.486, 
  12.9366) -- (9.4234, 12.9239) -- (9.3077, 12.8986) -- (9.2535, 12.8858) -- 
  (9.2013, 12.8729) -- (9.1019, 12.847) -- (8.9184, 12.7937) -- (8.8751, 
  12.7802) -- (8.8327, 12.7664) -- (8.7503, 12.7389) -- (8.5936, 12.6821) -- 
  (8.556, 12.6678) -- (8.5189, 12.6531) -- (8.4464, 12.6239) -- (8.307, 12.564) 
  -- (8.2732, 12.5488) -- (8.2398, 12.5335) -- (8.1742, 12.5027) -- (8.0475, 
  12.44) -- (7.8096, 12.3107) -- (7.7812, 12.2942) -- (7.7531, 12.2775) -- 
  (7.6976, 12.2441) -- (7.5898, 12.1765) -- (7.3859, 12.038) -- (7.3593, 12.019)
   -- (7.3331, 11.9998) -- (7.2813, 11.9612) -- (7.1807, 11.8835) -- (6.9909, 
  11.7249) -- (6.9683, 11.7049) -- (6.9457, 11.6848) -- (6.9015, 11.6443) -- 
  (6.8155, 11.563) -- (6.6533, 11.3978) -- (6.634, 11.3769) -- (6.6149, 11.356) 
  -- (6.5772, 11.3142) -- (6.5041, 11.23) -- (6.3672, 11.06) -- (6.351, 11.0386)
   -- (6.3348, 11.0171) -- (6.3031, 10.9743) -- (6.2421, 10.8882) -- (6.1287, 
  10.7148) -- (6.1161, 10.6944) -- (6.1039, 10.6741) -- (6.0796, 10.6335) -- 
  (6.0329, 10.5519) -- (6.0106, 10.5111) -- (5.9887, 10.4702) -- (5.9469, 
  10.3883) -- (5.9369, 10.3679) -- (5.9268, 10.3474) -- (5.9074, 10.3065) -- 
  (5.8703, 10.2245) -- (5.8029, 10.0605) -- (5.7952, 10.04) -- (5.7876, 10.0194)
   -- (5.7727, 9.9785) -- (5.7448, 9.8967) -- (5.7382, 9.8761) -- (5.7317, 
  9.8557) -- (5.7191, 9.8148) -- (5.6957, 9.7332) -- (5.6902, 9.7128) -- 
  (5.6848, 9.6924) -- (5.6745, 9.6516) -- (5.6695, 9.6314) -- (5.6647, 9.611) --
   (5.6555, 9.5705) -- (5.6511, 9.5501) -- (5.6468, 9.5298) -- (5.6387, 9.4893) 
  -- (5.624, 9.4086) -- (5.6207, 9.3888) -- (5.6176, 9.369) -- (5.6118, 9.3296) 
  -- (5.609, 9.3099) -- (5.6064, 9.2903) -- (5.6016, 9.251) -- (5.5994, 9.2315) 
  -- (5.5973, 9.2118) -- (5.5952, 9.1923) -- (5.5935, 9.1728) -- (5.5918, 
  9.1532) -- (5.5901, 9.1338) -- (5.5888, 9.1142) -- (5.5874, 9.0948) -- 
  (5.5862, 9.0753) -- (5.5851, 9.0561) -- (5.5841, 9.0366) -- (5.5833, 9.0173) 
  -- (5.5826, 8.998) -- (5.582, 8.9787) -- (5.5816, 8.9595) -- (5.5813, 8.9402) 
  -- (5.5811, 8.921) -- (5.5811, 8.8827) -- (5.5812, 8.8635) -- (5.5816, 8.8445)
   -- (5.582, 8.8254) -- (5.5826, 8.8064) -- (5.5833, 8.7873) -- (5.5839, 
  8.7685) -- (5.5849, 8.7494) -- (5.5871, 8.7117) -- (5.5885, 8.6929) -- 
  (5.5899, 8.6741) -- (5.593, 8.6366) -- (5.5948, 8.6178) -- (5.5968, 8.5992) --
   (5.6009, 8.562) -- (5.6031, 8.5434) -- (5.6056, 8.525) -- (5.6107, 8.488) -- 
  (5.6134, 8.4696) -- (5.6163, 8.4512) -- (5.6224, 8.4146) -- (5.6256, 8.3964) 
  -- (5.629, 8.3782) -- (5.636, 8.3418) -- (5.6396, 8.3238) -- (5.6435, 8.3057) 
  -- (5.6515, 8.2697) -- (5.6688, 8.1984) -- (5.6739, 8.1791) -- (5.6792, 
  8.1599) -- (5.6899, 8.1217) -- (5.7131, 8.046) -- (5.7191, 8.0272) -- (5.7255,
   8.0085) -- (5.7383, 7.9711) -- (5.7658, 7.8973) -- (5.773, 7.8789) -- 
  (5.7803, 7.8606) -- (5.7953, 7.8242) -- (5.827, 7.7524) -- (5.8968, 7.6117) --
   (5.906, 7.5944) -- (5.9154, 7.5773) -- (5.9347, 7.543) -- (5.9745, 7.4754) --
   (5.9848, 7.4588) -- (5.9953, 7.4421) -- (6.0164, 7.4092) -- (6.0603, 7.344) 
  -- (6.0716, 7.3279) -- (6.0831, 7.3119) -- (6.1062, 7.2802) -- (6.154, 7.2175)
   -- (6.2553, 7.0963) -- (6.2676, 7.0826) -- (6.28, 7.0689) -- (6.3051, 7.0416)
   -- (6.3565, 6.9883) -- (6.4641, 6.8854) -- (6.4919, 6.8606) -- (6.5202, 
  6.8359) -- (6.5778, 6.7879) -- (6.6974, 6.696) -- (6.7127, 6.685) -- (6.7281, 
  6.674) -- (6.7594, 6.6523) -- (6.8227, 6.6102) -- (6.8388, 6.5998) -- (6.8549,
   6.5896) -- (6.8874, 6.5695) -- (6.9534, 6.5305) -- (6.9703, 6.521) -- 
  (6.9871, 6.5116) -- (7.0208, 6.4932) -- (7.0895, 6.4575) -- (7.2303, 6.3915) 
  -- (7.2497, 6.3831) -- (7.2693, 6.3748) -- (7.3086, 6.3587) -- (7.3879, 
  6.3281) -- (7.4282, 6.3138) -- (7.4687, 6.2998) -- (7.5504, 6.2739) -- 
  (7.5711, 6.2677) -- (7.5918, 6.2618) -- (7.6332, 6.2504) -- (7.6542, 6.2449) 
  -- (7.675, 6.2395) -- (7.717, 6.2291) -- (7.7592, 6.2195) -- (7.8016, 6.2105) 
  -- (7.8228, 6.2063) -- (7.8442, 6.2023) -- (7.8869, 6.1946) -- (7.9084, 6.191)
   -- (7.9298, 6.1875) -- (7.9729, 6.1812) -- (7.9944, 6.1782) -- (8.016, 
  6.1754) -- (8.0593, 6.1704) -- (8.081, 6.1681) -- (8.1026, 6.166) -- (8.1244, 
  6.1641) -- (8.146, 6.1624) -- (8.1678, 6.1608) -- (8.1894, 6.1594) -- (8.2112,
   6.1582) -- (8.2547, 6.1562) -- (8.2764, 6.1556) -- (8.2981, 6.155) -- 
  (8.3199, 6.1547) -- (8.3415, 6.1545) -- (8.3633, 6.1545) -- (8.4066, 6.155) --
   (8.4282, 6.1556) -- (8.4499, 6.1562) -- (8.4715, 6.1571) -- (8.493, 6.1582) 
  -- (8.5146, 6.1593) -- (8.5361, 6.1607) -- (8.5575, 6.1623) -- (8.579, 6.164) 
  -- (8.6209, 6.1678) -- (8.6417, 6.17) -- (8.6626, 6.1724) -- (8.6833, 6.1749) 
  -- (8.7041, 6.1775) -- (8.7453, 6.1833) -- (8.7658, 6.1864) -- (8.7864, 
  6.1897) -- (8.827, 6.1966) -- (8.8473, 6.2003) -- (8.8674, 6.2042) -- (8.9075,
   6.2125) -- (8.9275, 6.2167) -- (8.9473, 6.2212) -- (8.9866, 6.2304) -- 
  (9.0643, 6.2505) -- (9.0835, 6.2557) -- (9.1025, 6.2613) -- (9.1403, 6.2724) 
  -- (9.2145, 6.2963) -- (9.2327, 6.3025) -- (9.2509, 6.3088) -- (9.2869, 
  6.3216) -- (9.3574, 6.3483) -- (9.4928, 6.4055) -- (9.5092, 6.413) -- (9.5254,
   6.4204) -- (9.5575, 6.4356) -- (9.6204, 6.4663) -- (9.7404, 6.5294) -- 
  (9.7539, 6.537) -- (9.7673, 6.5444) -- (9.7939, 6.5595) -- (9.8458, 6.5896) --
   (9.9453, 6.6504) -- (10.128, 6.7714) -- (10.4396, 7.0021) -- (10.4586, 7.017)
   -- (10.4774, 7.0317) -- (10.514, 7.0609) -- (10.5847, 7.1178) -- (10.7166, 
  7.2262) -- (10.9481, 7.4203) -- (11.2952, 7.7103) -- (11.5593, 7.919) -- 
  (11.784, 8.0822) -- (11.9494, 8.1909) -- (12.0909, 8.2748) -- (12.1999, 
  8.3326) -- (12.2805, 8.3709) -- (12.3497, 8.4001) -- (12.4006, 8.4191) -- 
  (12.4442, 8.4335) -- (12.4776, 8.4428) -- (12.478, 8.443) -- (12.4785, 8.4431)
   -- (12.4794, 8.4432) -- (12.4843, 8.4445) -- (12.4907, 8.4461) -- (12.5023, 
  8.4488) -- (12.5027, 8.4489) -- (12.5031, 8.4489) -- (12.5038, 8.449) -- 
  (12.5051, 8.4494) -- (12.508, 8.45) -- (12.5134, 8.4511) -- (12.5233, 8.4529) 
  -- (12.5236, 8.453) -- (12.5239, 8.453) -- (12.5244, 8.4532) -- (12.5255, 
  8.4533) -- (12.5276, 8.4537) -- (12.5316, 8.4543) -- (12.5317, 8.4544) -- 
  (12.5324, 8.4544) -- (12.5353, 8.4548) -- (12.5388, 8.4554) -- (12.5392, 
  8.4554) -- (12.5396, 8.4555) -- (12.5404, 8.4556) -- (12.5421, 8.4558) -- 
  (12.5452, 8.4562) -- (12.5455, 8.4562) -- (12.5459, 8.4563) -- (12.5466, 
  8.4563) -- (12.5481, 8.4565) -- (12.5509, 8.4569) -- (12.5516, 8.4569) -- 
  (12.5523, 8.457) -- (12.5537, 8.4572) -- (12.555, 8.4572) -- (12.5564, 8.4573)
   -- (12.5567, 8.4573) -- (12.557, 8.4574) -- (12.5576, 8.4574) -- (12.5589, 
  8.4576) -- (12.5612, 8.4577) -- (12.5633, 8.4577) -- (12.5633, 8.4578) -- 
  (12.5663, 8.4578) -- (12.5666, 8.458) -- (12.5783, 8.458) -- (12.5789, 8.4578)
   -- (12.5805, 8.4578) -- (12.5811, 8.4577) -- (12.5827, 8.4577) -- (12.5836, 
  8.4576) -- (12.584, 8.4576) -- (12.5844, 8.4574) -- (12.5855, 8.4574) -- 
  (12.5856, 8.4573) -- (12.5866, 8.4573) -- (12.5867, 8.4572) -- (12.5873, 
  8.4572) -- (12.5878, 8.457) -- (12.5888, 8.4569) -- (12.5893, 8.4569) -- 
  (12.5898, 8.4567) -- (12.5906, 8.4566) -- (12.5907, 8.4566) -- (12.5907, 
  8.4565) -- (12.5913, 8.4565) -- (12.592, 8.4562) -- (12.5931, 8.4559) -- 
  (12.5939, 8.4556) -- (12.5946, 8.4555) -- (12.5951, 8.4552) -- (12.5955, 
  8.4551) -- (12.5958, 8.455) -- (12.596, 8.4548) -- (12.5962, 8.4547) -- 
  (12.5964, 8.4547) -- (12.5964, 8.4545) -- (12.5965, 8.4545) -- (12.5965, 
  8.4544) -- (12.5966, 8.4544);

  \path[draw=black,line cap=,line join=miter,line width=0.1058cm,miter 
  limit=3.25,shift={(-3.4925, 0.2646)}] (7.3381, 8.4544) -- (7.3383, 8.4545) -- 
  (7.3387, 8.4545) -- (7.3397, 8.455) -- (7.3423, 8.4558) -- (7.3509, 8.4585) --
   (7.3809, 8.4682) -- (7.4914, 8.5033) -- (7.4963, 8.505) -- (7.5012, 8.5065) 
  -- (7.5114, 8.5097) -- (7.5325, 8.5163) -- (7.5782, 8.5303) -- (7.683, 8.5622)
   -- (7.9409, 8.6366) -- (7.9494, 8.6388) -- (7.9578, 8.6411) -- (7.9748, 
  8.6458) -- (8.0096, 8.6553) -- (8.081, 8.6745) -- (8.2317, 8.7131) -- (8.2415,
   8.7155) -- (8.2511, 8.718) -- (8.2708, 8.7228) -- (8.3105, 8.7325) -- 
  (8.3914, 8.7516) -- (8.5587, 8.789) -- (8.5797, 8.7934) -- (8.6006, 8.7979) --
   (8.6429, 8.8066) -- (8.7282, 8.8237) -- (8.7497, 8.8279) -- (8.7712, 8.8318) 
  -- (8.8145, 8.84) -- (8.9014, 8.8555) -- (8.9125, 8.8575) -- (8.9233, 8.8593) 
  -- (8.9453, 8.863) -- (8.9891, 8.8703) -- (9.0769, 8.8841) -- (9.0989, 8.8874)
   -- (9.121, 8.8906) -- (9.1649, 8.8969) -- (9.2527, 8.9086) -- (9.2647, 
  8.9101) -- (9.2765, 8.9116) -- (9.3004, 8.9145) -- (9.3478, 8.9202) -- 
  (9.3715, 8.9229) -- (9.3951, 8.9256) -- (9.4422, 8.9305) -- (9.4539, 8.9318) 
  -- (9.4657, 8.9329) -- (9.4892, 8.9352) -- (9.5357, 8.9395) -- (9.6399, 
  8.9482) -- (9.6742, 8.9506) -- (9.7199, 8.9537) -- (9.7651, 8.9564) -- 
  (9.8098, 8.9589) -- (9.8322, 8.96) -- (9.8544, 8.961) -- (9.8654, 8.9615) -- 
  (9.8764, 8.9619) -- (9.8983, 8.9628) -- (9.9093, 8.9632) -- (9.942, 8.9644) --
   (9.9527, 8.9647) -- (9.9636, 8.9651) -- (9.9851, 8.9657) -- (9.9952, 8.9659) 
  -- (10.0051, 8.9662) -- (10.015, 8.9663) -- (10.0251, 8.9666) -- (10.0349, 
  8.9668) -- (10.0448, 8.9669) -- (10.0547, 8.9672) -- (10.0841, 8.9676) -- 
  (10.0939, 8.9676) -- (10.1036, 8.9677) -- (10.1133, 8.9679) -- (10.1229, 
  8.9679) -- (10.1326, 8.968) -- (10.1994, 8.968) -- (10.2089, 8.9679) -- 
  (10.2183, 8.9679) -- (10.2277, 8.9677) -- (10.237, 8.9677) -- (10.2464, 
  8.9676) -- (10.2926, 8.9669) -- (10.3017, 8.9668) -- (10.3109, 8.9666) -- 
  (10.3289, 8.9662) -- (10.338, 8.9659) -- (10.347, 8.9658) -- (10.3651, 8.9652)
   -- (10.3739, 8.965) -- (10.3828, 8.9647) -- (10.4006, 8.9641) -- (10.4356, 
  8.9629) -- (10.4443, 8.9625) -- (10.453, 8.9622) -- (10.4702, 8.9615) -- 
  (10.5044, 8.9599) -- (10.5128, 8.9595) -- (10.5213, 8.959) -- (10.538, 8.9582)
   -- (10.5712, 8.9564) -- (10.5891, 8.9553) -- (10.6068, 8.9542) -- (10.6416, 
  8.9519) -- (10.6503, 8.9513) -- (10.6588, 8.9508) -- (10.6759, 8.9495) -- 
  (10.7098, 8.9469) -- (10.7182, 8.9464) -- (10.743, 8.9443) -- (10.7758, 
  8.9415) -- (10.8398, 8.9356) -- (10.8475, 8.9349) -- (10.8554, 8.9341) -- 
  (10.8709, 8.9326) -- (10.9015, 8.9294) -- (10.9612, 8.9229) -- (11.0746, 
  8.909) -- (11.0812, 8.9082) -- (11.088, 8.9074) -- (11.1012, 8.9056) -- 
  (11.1274, 8.902) -- (11.1784, 8.8948) -- (11.2751, 8.8801) -- (11.281, 8.8791)
   -- (11.2868, 8.8781) -- (11.2983, 8.8764) -- (11.321, 8.8726) -- (11.3654, 
  8.8651) -- (11.4493, 8.85) -- (11.454, 8.8492) -- (11.4587, 8.8482) -- 
  (11.468, 8.8465) -- (11.4865, 8.843) -- (11.5223, 8.836) -- (11.5907, 8.8222) 
  -- (11.7143, 8.7952) -- (11.7183, 8.7942) -- (11.7221, 8.7933) -- (11.7299, 
  8.7916) -- (11.7452, 8.788) -- (11.7748, 8.7809) -- (11.831, 8.7671) -- 
  (11.9319, 8.7408) -- (12.0855, 8.6961) -- (12.2013, 8.6582) -- (12.2973, 
  8.6231) -- (12.3658, 8.5952) -- (12.4225, 8.5699) -- (12.4648, 8.5492) -- 
  (12.4951, 8.533) -- (12.5202, 8.5183) -- (12.5382, 8.507) -- (12.553, 8.4968) 
  -- (12.564, 8.4887) -- (12.5718, 8.4825) -- (12.5783, 8.4769) -- (12.5829, 
  8.4727) -- (12.5862, 8.4693) -- (12.5891, 8.4663) -- (12.591, 8.4639) -- 
  (12.5927, 8.462) -- (12.5938, 8.4603) -- (12.5946, 8.4592) -- (12.5951, 
  8.4581) -- (12.5957, 8.4574) -- (12.596, 8.4567) -- (12.5962, 8.4562) -- 
  (12.5964, 8.4558) -- (12.5966, 8.4552) -- (12.5966, 8.455) -- (12.5968, 
  8.4548) -- (12.5968, 8.4543);

  \path[draw=black,line cap=,line join=miter,line width=0.1058cm,miter 
  limit=3.25,shift={(-3.4925, 0.2646)}] (17.8555, 8.4541) -- (17.8552, 8.454) --
   (17.8551, 8.454) -- (17.8547, 8.4539) -- (17.8536, 8.4534) -- (17.8501, 
  8.4523) -- (17.8391, 8.4488) -- (17.7998, 8.4361) -- (17.6386, 8.3852) -- 
  (17.401, 8.3137) -- (17.0973, 8.2285) -- (16.7051, 8.1295) -- (16.6989, 
  8.1281) -- (16.6929, 8.1266) -- (16.6806, 8.1237) -- (16.6559, 8.118) -- 
  (16.6062, 8.1066) -- (16.5052, 8.0841) -- (16.2976, 8.041) -- (16.2905, 
  8.0396) -- (16.2833, 8.0381) -- (16.2403, 8.0298) -- (16.1827, 8.0188) -- 
  (16.0665, 7.9979) -- (16.0594, 7.9966) -- (16.0521, 7.9954) -- (16.0375, 
  7.9928) -- (16.0083, 7.9878) -- (15.9497, 7.9782) -- (15.8321, 7.9598) -- 
  (15.825, 7.9587) -- (15.8177, 7.9576) -- (15.8033, 7.9556) -- (15.7744, 
  7.9513) -- (15.7167, 7.9431) -- (15.6015, 7.9281) -- (15.5942, 7.9273) -- 
  (15.587, 7.9263) -- (15.5727, 7.9247) -- (15.544, 7.9212) -- (15.4867, 7.9148)
   -- (15.3726, 7.903) -- (15.3593, 7.9017) -- (15.3462, 7.9004) -- (15.3198, 
  7.8981) -- (15.2673, 7.8935) -- (15.2607, 7.893) -- (15.2542, 7.8924) -- 
  (15.2411, 7.8915) -- (15.2151, 7.8894) -- (15.1631, 7.8857) -- (15.1502, 
  7.8849) -- (15.1373, 7.8839) -- (15.1116, 7.8822) -- (15.1051, 7.882) -- 
  (15.0988, 7.8816) -- (15.0859, 7.8807) -- (15.0603, 7.8793) -- (15.054, 
  7.8789) -- (15.0476, 7.8787) -- (15.0349, 7.878) -- (15.0096, 7.8767) -- 
  (15.0033, 7.8765) -- (14.9969, 7.876) -- (14.9842, 7.8755) -- (14.9592, 
  7.8745) -- (14.9524, 7.8742) -- (14.9455, 7.874) -- (14.9185, 7.8729) -- 
  (14.905, 7.8725) -- (14.8982, 7.8722) -- (14.8916, 7.872) -- (14.8781, 7.8716)
   -- (14.8515, 7.8709) -- (14.8448, 7.8708) -- (14.8382, 7.8705) -- (14.8117, 
  7.87) -- (14.7985, 7.8698) -- (14.7919, 7.8697) -- (14.7854, 7.8696) -- 
  (14.7788, 7.8694) -- (14.7722, 7.8694) -- (14.7657, 7.8693) -- (14.7591, 
  7.8693) -- (14.7526, 7.8691) -- (14.7461, 7.8691) -- (14.7396, 7.869) -- 
  (14.7265, 7.869) -- (14.7201, 7.8689) -- (14.6495, 7.8689) -- (14.6432, 7.869)
   -- (14.6368, 7.869) -- (14.6305, 7.8691) -- (14.618, 7.8691) -- (14.6053, 
  7.8694) -- (14.5927, 7.8696) -- (14.5865, 7.8697) -- (14.5803, 7.8697) -- 
  (14.5678, 7.87) -- (14.543, 7.8705) -- (14.5373, 7.8707) -- (14.5316, 7.8708) 
  -- (14.5201, 7.8711) -- (14.5143, 7.8712) -- (14.5087, 7.8714) -- (14.4972, 
  7.8718) -- (14.4859, 7.872) -- (14.452, 7.8733) -- (14.4465, 7.8736) -- 
  (14.4409, 7.8737) -- (14.4297, 7.8742) -- (14.4075, 7.8752) -- (14.3637, 
  7.8774) -- (14.3582, 7.8777) -- (14.3528, 7.878) -- (14.3421, 7.8787) -- 
  (14.3204, 7.8799) -- (14.278, 7.8827) -- (14.2675, 7.8835) -- (14.2569, 
  7.8842) -- (14.2361, 7.8858) -- (14.195, 7.8891) -- (14.1901, 7.8895) -- 
  (14.185, 7.89) -- (14.1751, 7.8909) -- (14.1553, 7.8927) -- (14.1163, 7.8964) 
  -- (14.1115, 7.897) -- (14.1067, 7.8974) -- (14.0972, 7.8985) -- (14.078, 
  7.9004) -- (14.0404, 7.9047) -- (14.0359, 7.9053) -- (14.0312, 7.9058) -- 
  (14.0035, 7.9091) -- (13.9674, 7.9138) -- (13.897, 7.9236) -- (13.8782, 
  7.9263) -- (13.8599, 7.9292) -- (13.8237, 7.935) -- (13.7535, 7.9471) -- 
  (13.7492, 7.9478) -- (13.745, 7.9487) -- (13.7366, 7.9502) -- (13.7197, 
  7.9533) -- (13.6867, 7.9598) -- (13.6229, 7.9729) -- (13.6193, 7.9737) -- 
  (13.6157, 7.9744) -- (13.6085, 7.9761) -- (13.5942, 7.9793) -- (13.5662, 
  7.9856) -- (13.5121, 7.9984) -- (13.4115, 8.0246) -- (13.4082, 8.0256) -- 
  (13.405, 8.0264) -- (13.3985, 8.0282) -- (13.3858, 8.0319) -- (13.361, 8.0391)
   -- (13.3132, 8.0534) -- (13.2256, 8.0819) -- (13.223, 8.0829) -- (13.2205, 
  8.0837) -- (13.2155, 8.0855) -- (13.2056, 8.0889) -- (13.1862, 8.0958) -- 
  (13.149, 8.1095) -- (13.0808, 8.1361) -- (12.974, 8.1825) -- (12.883, 8.2277) 
  -- (12.8171, 8.2645) -- (12.762, 8.2988) -- (12.7208, 8.3272) -- (12.6913, 
  8.3495) -- (12.6669, 8.3697) -- (12.6497, 8.3851) -- (12.6367, 8.3978) -- 
  (12.6261, 8.4089) -- (12.6187, 8.4175) -- (12.6126, 8.425) -- (12.6082, 8.431)
   -- (12.6052, 8.4355) -- (12.6027, 8.4394) -- (12.6011, 8.4423) -- (12.5998, 
  8.4446) -- (12.5989, 8.4467) -- (12.5982, 8.4482) -- (12.5976, 8.4494) -- 
  (12.5973, 8.4504) -- (12.5971, 8.4512) -- (12.5969, 8.4519) -- (12.5969, 
  8.4522) -- (12.5968, 8.4523) -- (12.5968, 8.4532) -- (12.5966, 8.4532) -- 
  (12.5966, 8.454);

  \path[fill=cff7f00,even odd rule] (6.589, 4.3928) -- (6.7033, 4.2785) -- 
  (6.589, 4.1642) -- (6.7033, 4.0499) -- (6.589, 3.9355) -- (6.4747, 4.0499) -- 
  (6.3603, 3.9355) -- (6.2461, 4.0499) -- (6.3603, 4.1642) -- (6.2461, 4.2785) 
  -- (6.3603, 4.3928) -- (6.4747, 4.2785) -- cycle(6.589, 4.3928);

  \path[fill=cff7f00,even odd rule] (6.589, 13.502) -- (6.7033, 13.3876) -- 
  (6.589, 13.2734) -- (6.7033, 13.1592) -- (6.589, 13.0448) -- (6.4747, 13.1592)
   -- (6.3603, 13.0448) -- (6.2461, 13.1592) -- (6.3603, 13.2734) -- (6.2461, 
  13.3876) -- (6.3603, 13.502) -- (6.4747, 13.3876) -- cycle(6.589, 13.502);

  \path[fill=cff7f00,even odd rule] (3.9594, 8.9475) -- (4.0736, 8.8331) -- 
  (3.9594, 8.7188) -- (4.0736, 8.6045) -- (3.9594, 8.4902) -- (3.845, 8.6045) --
   (3.7308, 8.4902) -- (3.6165, 8.6045) -- (3.7308, 8.7188) -- (3.6165, 8.8331) 
  -- (3.7308, 8.9475) -- (3.845, 8.8331) -- cycle(3.9594, 8.9475);

  \path[fill=cff7f00,even odd rule] (14.4778, 8.9475) -- (14.592, 8.8331) -- 
  (14.4778, 8.7188) -- (14.592, 8.6045) -- (14.4778, 8.4902) -- (14.3634, 
  8.6045) -- (14.2492, 8.4902) -- (14.1348, 8.6045) -- (14.2492, 8.7188) -- 
  (14.1348, 8.8331) -- (14.2492, 8.9475) -- (14.3634, 8.8331) -- cycle(14.4778, 
  8.9475);

  \path[fill=blue,even odd rule] (3.9594, 8.7188).. controls (3.9594, 8.6885) 
  and (3.9473, 8.6595) .. (3.9259, 8.638).. controls (3.9044, 8.6166) and 
  (3.8753, 8.6045) .. (3.845, 8.6045).. controls (3.8147, 8.6045) and (3.7856, 
  8.6166) .. (3.7642, 8.638).. controls (3.7428, 8.6595) and (3.7308, 8.6885) ..
   (3.7308, 8.7188).. controls (3.7308, 8.7492) and (3.7428, 8.7782) .. (3.7642,
   8.7996).. controls (3.7856, 8.8211) and (3.8147, 8.8331) .. (3.845, 8.8331)..
   controls (3.8753, 8.8331) and (3.9044, 8.8211) .. (3.9259, 8.7996).. controls
   (3.9473, 8.7782) and (3.9594, 8.7492) .. (3.9594, 8.7188) -- cycle(3.9594, 
  8.7188);

  \path[fill=blue,even odd rule] (14.4778, 8.7188).. controls (14.4778, 8.6885) 
  and (14.4657, 8.6595) .. (14.4443, 8.638).. controls (14.4228, 8.6166) and 
  (14.3937, 8.6045) .. (14.3634, 8.6045).. controls (14.3331, 8.6045) and 
  (14.304, 8.6166) .. (14.2827, 8.638).. controls (14.2612, 8.6595) and 
  (14.2492, 8.6885) .. (14.2492, 8.7188).. controls (14.2492, 8.7492) and 
  (14.2612, 8.7782) .. (14.2827, 8.7996).. controls (14.304, 8.8211) and 
  (14.3331, 8.8331) .. (14.3634, 8.8331).. controls (14.3937, 8.8331) and 
  (14.4228, 8.8211) .. (14.4443, 8.7996).. controls (14.4657, 8.7782) and 
  (14.4778, 8.7492) .. (14.4778, 8.7188) -- cycle(14.4778, 8.7188);

  \path[fill=blue,even odd rule] (9.2185, 8.7188).. controls (9.2185, 8.6885) 
  and (9.2065, 8.6595) .. (9.185, 8.638).. controls (9.1637, 8.6166) and 
  (9.1346, 8.6045) .. (9.1043, 8.6045).. controls (9.074, 8.6045) and (9.0449, 
  8.6166) .. (9.0234, 8.638).. controls (9.002, 8.6595) and (8.9899, 8.6885) .. 
  (8.9899, 8.7188).. controls (8.9899, 8.7492) and (9.002, 8.7782) .. (9.0234, 
  8.7996).. controls (9.0449, 8.8211) and (9.074, 8.8331) .. (9.1043, 8.8331).. 
  controls (9.1346, 8.8331) and (9.1637, 8.8211) .. (9.185, 8.7996).. controls 
  (9.2065, 8.7782) and (9.2185, 8.7492) .. (9.2185, 8.7188) -- cycle(9.2185, 
  8.7188);

\end{tikzpicture}